\documentstyle[12pt,epsfig]{article}
\begin{document}
%\belowdisplayshortskip=9pt plus 3pt minus 5pt
\begin{titlepage}
\rightline{January 2014}
\vskip 2cm
\centerline{\Large \bf
Mirror dark matter:
}
\vskip 0.5cm
\centerline{\Large \bf
Cosmology, galaxy structure and direct detection
}

\vskip 1.5cm
\centerline{R. Foot\footnote{
E-mail address: rfoot@unimelb.edu.au}}

\vskip 0.7cm
\centerline{\it ARC Centre of Excellence for Particle Physics at the Terascale,}
\centerline{\it School of Physics, University of Melbourne,}
\centerline{\it Victoria 3010 Australia}
\vskip 2.6cm
\noindent
%Abstract here
%
%
A simple way to accommodate dark matter is to postulate the existence of a hidden sector. 
That is, a set of new particles and forces interacting with the known particles predominantly via gravity.
In general this leads to a large set of unknown parameters, however if the hidden sector 
is an exact copy of the standard model sector, then an enhanced symmetry arises. 
This symmetry, which can be interpreted as space-time parity, connects each ordinary particle 
($e, \ \nu, \ p, \ n, \ \gamma, ....)$ with a mirror partner ($e', \ \nu', \ p', \ n', \ \gamma', ...)$.  
If this symmetry is completely unbroken, then the mirror particles are degenerate with
their ordinary particle counterparts, and would interact amongst themselves with exactly 
the same dynamics that govern ordinary particle interactions.  
The only new interaction postulated is photon - mirror photon kinetic mixing, whose strength $\epsilon$, 
is the sole new fundamental (Lagrangian) parameter relevant for astrophysics and cosmology.
It turns out that such a theory, with suitably chosen initial conditions effective in the very early Universe,
can provide an adequate description of dark matter phenomena provided that $\epsilon \sim 10^{-9}$.
This review focusses on three main developments of this mirror dark matter theory during the last decade:
Early universe cosmology, galaxy structure and the application to direct detection experiments.

\end{titlepage}

%%%%%%%%%%%%%%%%%%%%%%%%%%%%%%%%%%%%%%%%%%%%%%%%%%%%%%%%%%%%%%%%%%%%%%%%%%%%%%%%%%%%%%%%%%%%%%%%%%%%%%%%%%
%%%%%%%%%%%%%b section one have gone through twice %%%%%%%%%%%%%%%%%%%55
%%%%%%%%%%%%%%%%%%%%%%%%%%%%%%%%%%%%%%%%%%%%%%%%%%%%%%%%%%%%%%%%%%%%%%%%

\section{Introduction and Overview}

\subsection{Introduction}

Astronomical observations provide a strong case for the existence of non-baryonic dark matter in the Universe.
The first evidence arose in the 1930's from observations of galaxies in clusters  
which showed unexpectedly high velocity dispersion \cite{zwicky,smith36}. Further evidence
followed from measurements of optical and radio emissions in spiral galaxies \cite{babcock,rubin,rag,roberts1,roberts,rubinb,bosma}
(for a review, and more detailed bibliography, see Ref.\cite{sr}). 
These observations allowed galactic rotation curves to be obtained, which
greatly strengthened the case for dark matter.
It was found that rotation curves 
in spiral galaxies were roughly flat near the observed edge of the galaxy, in sharp contrast to expectations 
from Newton's law of gravity applied to the inferred baryonic mass.

Dark matter is also needed to explain the observed Large-Scale Structure (LSS) of the Universe \cite{lssobs,lssobs2,weaklens} and also the  
anisotropies of the Cosmic Microwave Background radiation (CMB) \cite{planck}.  Such cosmological observations
have provided, perhaps, the strongest evidence yet for
dark matter in the Universe. These, and other measurements, 
can be explained within the Friedmann Robertson Walker (FRW) cosmological model. This model, with significant 
developments over the years, has a small number of parameters,
among which are $\Omega_b$, $\Omega_{dm}$, and $\Omega_{\Lambda}$ [$b$ = baryon, $dm$ = dark matter, $\Lambda$ = cosmological
constant]. Here
$\Omega_i \equiv \rho_i/\rho_c$, with critical density $\rho_c = 3H^2/8\pi G_N$ and with all densities evaluated
today (for a review, see e.g. \cite{dodelson}).
Comparison of the model with observations allows these parameters to be determined:
\begin{eqnarray}
\Omega_b \simeq 0.05, \ \Omega_{dm} \simeq 0.25, \ \Omega_\Lambda \simeq 0.70
\ .
\end{eqnarray}
That is, the energy density of dark matter in the Universe is currently around five times larger than
that of ordinary matter.

Most recently, evidence of dark matter direct detection in underground experiments has emerged\ \cite{dama,dama2,dama3,cogent,cresst,cdms/si}.
The strongest such dark matter signal is the measurement of an annually modulated event rate by the DAMA collaboration.
Due to the Earth's motion around the Sun,
the dark matter interaction rate in an Earth based detector should modulate with a period of one year
and have a maximum near $2^{nd}$ June\ \cite{spergel}. Such a modulation was observed by the 
DAMA/NaI \cite{dama} and DAMA/Libra \cite{dama2,dama3} experiments
and provides tantalizing evidence that dark matter particles may have been detected in the laboratory.

The non-gravitational interactions of the known elementary particles are well represented by the standard model.
This theory exhibits
$SU(3)_c \otimes SU(2) \otimes U(1)$ gauge symmetry, along with a host of space-time symmetries,
and can be described by a Lagrangian:
\begin{eqnarray}
{\cal L}_{SM} (e, u, d, \gamma, W, Z,...) 
\ .
\end{eqnarray}
This model together with Einstein's General Relativity theory provides an excellent description of the
elementary particles and their interactions.
Although the standard model is very successful, it contains no suitable dark matter candidate, 
so one is naturally led to consider new particle physics.

Perhaps the simplest way to accommodate dark matter is via a {\it hidden sector}.
This entails extending the standard model to include an additional set of matter particles $F_1,\ F_2, ...$ and gauge
fields $G_1,\ G_2$ (associated with a gauge group $G'$) so that:
\begin{eqnarray}
{\cal L} = {\cal L}_{SM} (e, u, d, \gamma, W, Z,...) + {\cal L}_{dark} (F_1, F_2, G_1, G_2,...) + {\cal L}_{mix}\ .
\end{eqnarray}
If the new particles do not interact with any of the standard 
$SU(3)_c \otimes SU(2) \otimes U(1)$ gauge fields then their properties are experimentally unconstrained.
They cannot be produced in colliders, unless some additional interactions, ${\cal L}_{mix}$, are assumed. It is for this
reason that such an additional set of particles is called a hidden sector.
The hidden sector can have accidental global or discrete symmetries stabilizing one or
more of the lightest particles: $F_1$, $F_2$, ... In this case these stable particles can potentially constitute 
the inferred dark matter in the Universe. 

From this perspective, the astrophysics of dark matter can be
simple or complex depending on the properties of the hidden sector. For example, if all the new gauge bosons 
are heavy, like the $W$ and $Z$ gauge bosons, then $F_1$, $F_2$, ... are essentially collisionless
particles, also called WIMPs in the literature \footnote{
Collisionless dark matter (WIMPs) can be motivated in other frameworks, such as in models with large extra dimensions
and supersymmetry. For a review, see for example \cite{hooper} or \cite{feng9}.}.
On the other hand, if one or more of the new gauge fields are  light or massless 
then the $F_1$, $F_2$,... can have significant self interactions which are also dissipative.
The case of collisionless dark matter has been very well studied in the literature, in part because the astrophysics is particularly
simple. However the alternative possibility, where dark matter has more complex astrophysical properties is equally simple from
a particle physics standpoint, and is also worth investigating. Naturally, the astrophysical implications of such
complex dark matter depends, to a significant extent, on the details of the particular model.

This review will focus on a very special hidden sector model - mirror dark matter, which
I will argue is exceptionally simple and well motivated from a particle physics perspective.
However, even if the reader does not share my enthusiasm for this particular model, this study 
at least serves to illustrate some of the rich dark matter phenomenology that is possible in generic hidden sector models.
It may thus (hopefully) provide useful insight if the astrophysical properties of dark matter are, in fact, non-trivial.

Mirror dark matter corresponds to the theoretically unique case where the
hidden sector is an exact {\it copy} of the standard model sector (up to an ambiguity concerning whether or not chirality
is flipped in the hidden sector, to be discussed in a moment). This means that each of the known particles has a 
{\it mirror} partner, denoted with a prime ($'$).  The mirror particles interact amongst themselves with exactly the same dynamics
that govern the ordinary-particle interactions. That is, ${\cal L}_{dark}$ is just the
standard model Lagrangian:
\begin{eqnarray}
{\cal L} = {\cal L}_{SM} (e, u, d, \gamma, W, Z, ...) + {\cal L}_{SM} (e', u',
d', \gamma', W', Z', ...) + {\cal L}_{mix}\ 
\label{lag1x}
\end{eqnarray}
where ${\cal L}_{mix}$ contains possible non-gravitational interactions coupling ordinary and mirror particles.
The mirror particles don't interact with any of the known particles except
via gravity and the terms in ${\cal L}_{mix}$.
This particular hidden sector theory is also tightly constrained: The only new parameters are 
those in ${\cal L}_{mix}$ (to be discussed shortly).
The vacuum structure of the mirror sector is presumed identical to the ordinary 
sector so that the mass and the lifetime
of each mirror particle is exactly identical to the corresponding ordinary matter
particle. That is $m_{e'} = m_{e}, \ m_{p'} = m_{p}, \ m_{\gamma'} = m_{\gamma} = 0, \ m_{W'} = m_{W}$ \ etc. 
\footnote{
In the presence of mass-mixing terms arising from ${\cal L}_{mix}$
massive neutral particles, such as the neutrinos and the Higgs boson, 
can have their particle - mirror particle degeneracy broken.}
and the mirror electrons and mirror protons (mirror neutrons in mirror nuclei) are stable and can constitute
the non-baryonic dark matter in the Universe.

Having the hidden sector isomorphic to the standard model sector is a sensible thing to do, not just because it
reduces the number of parameters, but also
because it increases the symmetry of the theory. There is an unbroken discrete $Z_2$ symmetry interchanging each
ordinary particle with its mirror partner. To appreciate the significance of this particular discrete symmetry
we need to remember some basic particle physics. Firstly, recall that
ordinary (Dirac) fermion fields are a combination of two chiralities: left and right-handed chiral fields.
These chiral states describe the two helicity states of the spin 1/2 fermion in a Lorentz covariant manner.
The two chiral states, together with the antiparticles are described by two complex fields, $\psi_L,\ \psi_R$.
That is, we have four degenerate physical states.
%%%%

With the presumption of a hidden sector isomorphic to the standard model sector, 
each fermion and its mirror partner forms now 
eight  degenerate chiral states. 
Considering for example the electron and its degenerate partners, we have four complex fields: $e_L, \ e_R,  \ e'_L, \ e'_R$.
Let us examine all possible discrete $Z_2$ transformations of these degenerate states.
A discrete symmetry which interchanges states of opposite chirality is possible if it 
also maps \ $(x,y,z,t) \to (-x,-y,-z,t)$.
Thus, for example, the conventional parity transformation maps \ $e_L \to e_R, \ e_R \to e_L$ and $(x,y,z,t) \to (-x,-y,-z,t)$ \footnote{
Technically, the transformation is $e_L \to \gamma_0 e_R$, where $\gamma_0$ is a Dirac gamma matrix.
As this detail is inessential for the purposes of this introductory discussion, the required $\gamma$ matrices
are taken as implicit.}.
However it is known that only
the left-handed ordinary fermion fields couple to the (charged current) weak interactions ($W$ bosons).
Thus a discrete symmetry which maps $e_L \to e_R$ cannot be an invariance of the full theory.
Since the fields are complex, the transformations: $e_L \to e_L^*, \ e_R \to e_R^*$,
or $e_L \to e_R^*, \ e_R \to e_L^*$, known as the CP and C transformation respectively, are both
possible, 
but again experiments have  shown that these cannot lead to an invariance
of the full theory. 
This leaves two possibilities.
Either $e_L \to e'_L, \ e_R \to e'_R$ or
$e_L \to e'_R, \ e_R \to e'_L$.  The latter case requires also  $(x,y,z,t) \to (-x,-y,-z,t)$ because
it interchanges chiralities. 
In both cases, the symmetry also interchanges the gauge bosons ($\gamma, W, ...$ etc.) with
their mirror partners ($\gamma', W',...$ etc.) and can be a full invariance of the theory \footnote{
The cases: $e_L \to {e'_R}^*, \ e_R \to {e'_L}^*$ and $e_L \to {e'_L}^*, \ e_R \to {e'_R}^*$
do not lead to any new theories, beyond the ones considered.}.

The conclusion is that, 
if the standard model is extended with an isomorphic hidden sector then
there are actually two (almost) phenomenologically equivalent theories,
depending on whether the chirality of the fermions are swapped in the mirror sector.
If the left and right-handed chiral fermion fields  
are in fact interchanged, 
%so that only the right-handed chiral mirror fields couple to mirror weak interactions ($W'$ bosons),  
then the $Z_2$ discrete symmetry can be interpreted as space-time parity symmetry as it also maps
$(x,y,z,t) \to (-x,-y,-z,t)$. The theory also exhibits an exact time reversal invariance, which means that the full Poincar$\acute{e}$ group
becomes an unbroken symmetry of the theory \cite{flv}. 

\vskip 0.4cm

Particle physics considerations have often been guided by symmetry principles, and space-time parity
appears to be as good a candidate as any for a fundamental symmetry of nature.
This was well recognized by our pre-1956 ancestors, who generally assumed fundamental interactions
were invariant under space inversion. Things changed in 1956:
Some experimental anomalies led  Lee and Yang to suggest that space-time parity might be broken
in nature \cite{lee}. Lee and Yang also pointed out that even if the interactions of the known particles 
were to violate
parity, the symmetry could be restored if a set of mirror particles existed. [Although at that time
it wasn't clear if every known particle had a mirror partner, or just some of them.] 
Shortly thereafter 
it was realized by Landau that the CP transformation could play the role of space-time parity \cite{landau},
and thereby argued that a mirror sector was not necessary. Space inversion accompanied by particles swapping with
antiparticles might be the mirror symmetry chosen by nature. However, following experiments
in 1964 which showed that CP was in fact violated \cite{cpgone}, Landau's former student Pomeranchuk 
and collaborators, influenced by Landau's strong belief in parity symmetry, reconsidered Lee and Yang's 
original idea \cite{pom}.
There they argued that a complete doubling of the known particles and forces (except gravity)
was necessary to realize Lee and Yang's vision.
Related ideas were also discussed around a decade later by Pavsic \cite{pavsic}.

The potential application to dark matter
was suggested in 1982 \cite{blin} and also independently in 1985 \cite{kolb} (the latter motivated not by space-time parity
but by $E_8 \otimes E_8$ anomaly free superstring theories \cite{green}). However, with the exception of two significant papers
in 1986, 1987 \cite{glas1,glas2}, the idea was not actively pursued.
So, perhaps surprisingly, the extension of the standard model
extended with such a mirror sector was not written down until 1991 \cite{flv}.
The 1991 work was independent of the earlier developments, and arose out of  
studies investigating a gauge model with non-standard parity symmetry which 
interchanged quarks with leptons
as well as $(x,y,z,t) \to (-x,-y,-z,t)$ \cite{ql}. \footnote{
The gauge model referred to here, called the quark-lepton symmetric model in \cite{ql}, 
extended the gauge symmetry of the standard model to include a gauged $SU(3)_{\ell}$ 
color symmetry for leptons.
This means that the leptons interact with an octet of leptonic gluons $G_{\ell}$,
in the same way in which the quarks interact with the familiar $SU(3)_c$ gluons, $G_q$.
%Consistency with experiments required the leptonic gluons to be heavy, so that $SU(3)_{\ell}$ was presumed to be spontaneously broken.
With an $SU(3)$ leptonic color, the Lagrangian can possess a discrete parity 
symmetry which not only maps $(x, y, z, t) \to (-x, -y, -z, t)$ but additionally interchanges 
left-handed (right-handed) leptons with right-handed (left-handed) quarks and also $G_{\ell}$ with $G_q$. 
Consistency with experiments requires the
vacuum such that $SU(3)_{\ell}$ is spontaneously broken, which means that the parity symmetry 
is also spontaneously broken. The non-degeneracy of the quarks and leptons can thereby be explained,
and a phenomenologically consistent model results.
With $SU(3)_{\ell}$ gauged, the leptons appear as a parity double of the quarks.
The jump from this model, to the model where the parity symmetry was completely unbroken followed
by assuming that all of the particles in the standard model have a parity double.
}
This review will mainly be concerned with the post-1991 evolution of the theory and its application to
dark matter. Readers wishing to know  more about pre-1991 work on the subject might 
consult Okun's articles \cite{okun,okun2}.

Returning to the Lagrangian of Eq.(\ref{lag1x}), we have yet to define the ${\cal L}_{mix}$ term.
This piece describes possible non-gravitational interactions coupling ordinary and mirror particles together.
It turns out that there are just two mixing terms consistent with the symmetries of the minimal theory and
which are also renormalizable \cite{flv,fh}:
\begin{eqnarray}
{\cal L}_{mix} = \frac{\epsilon}{2} F^{\mu \nu} F'_{\mu \nu} + 
\lambda \phi^{\dagger}\phi \phi'^{\dagger}\phi' \ ,
\label{mixy}
\end{eqnarray}
where $F_{\mu \nu}$ ($F'_{\mu \nu}$) is the ordinary (mirror) $U(1)$
gauge boson field strength tensor and
$\phi$ ($\phi'$) is the ordinary Higgs (mirror Higgs) field.
The two Lagrangian terms above involve two dimensionless parameters: $\epsilon$, $\lambda$ both of which are not
determined by the symmetries of the theory.
Of these two terms, only the first term, the $U(1)-U(1)'$ kinetic mixing
term will be important for the astrophysical and cosmological applications discussed in this review. 
The relevant particle physics
thus involves only one additional fundamental parameter, $\epsilon$. 

The physical effect of the kinetic mixing interaction is to induce a tiny ordinary electric charge for the
mirror proton and mirror electron of $\pm \epsilon e$ \cite{flv,holdom}.
Kinetic mixing can thereby lead to electromagnetic interactions of the form: $ \bar e e \to \bar e'  e'$.
Although the cross-section for such processes is suppressed by $\epsilon^2$, these kinetic mixing induced
interactions can still have extremely important astrophysical and 
cosmological implications. In particular,
such kinetic mixing can make supernovae - both ordinary and mirror varieties - play critical roles in
astrophysics and cosmology.
Recall that in standard theory, ordinary supernovae release almost all of their core collapse energy into neutrinos
since these particles can escape from the core due to their extremely weak interactions.
If photon - mirror photon kinetic mixing exists with strength $\epsilon \sim 10^{-9}$, then
around half of this energy can instead be released into light mirror particles: $\bar e', e', \gamma'$,
produced through processes such as $\bar e e \to \bar e' e'$ in the hot supernova core \cite{raffelt,sil}.
These light mirror particles, once produced, escape from the core and are injected into the region
around the supernova. Ultimately this energy is expected to be converted into mirror photons. These mirror photons,
as we will see, provide an excellent candidate for
the heat source responsible for stabilizing mirror-particle halos hosting spiral galaxies.
At an earlier epoch, mirror supernovae might also have played an important role. 
These supernovae can release a large fraction of 
their core collapse energy into ordinary photons.  A rapid period of mirror star formation 
at an early epoch:  $6 < z < 20$, might have been responsible
for the reionization of ordinary matter - inferred from CMB and other observations.

In the mirror dark matter scenario, it is supposed that all of the inferred non-baryonic dark matter in the Universe,
on both large and small scales, is comprised
of mirror particles, in one form or another.  At the particle level, dark matter consists 
of a spectrum of stable massive mirror particles which are not only self interacting but also dissipative. 
It turns out that this dark matter picture gives consistent early Universe cosmology, and predicts 
large-scale structure and CMB anisotropies which are compatible with observations. Furthermore, on smaller scales
the dissipative interactions lead to non-trivial halo dynamics.
The picture is that dark matter halos hosting spiral galaxies are composed predominately of a 
mirror-particle plasma containing: 
$e', H', He', O', Fe', ...$ \cite{sph}. The loss of energy
due to dissipative processes, such as thermal bremsstrahlung, is (currently) being replaced by a heat source,
with ordinary core-collapse supernovae, as briefly described above, posing the best available candidate. 
It turns out that this dynamics
leads to a satisfactory explanation of the inferred dark matter properties of
spiral galaxies, i.e. asymptotically flat rotation curves, cored density profile, empirical scaling relations and so on \cite{sph,r69,foot5,rf2013b}.

A key test of this dark matter theory comes from direct detection experiments.
Ordinary supernovae can only stabilize dark matter halos if the kinetic mixing interaction exists,
with $\epsilon \sim 10^{-9}$.  Such an interaction also implies that mirror particles can elastically
scatter off ordinary nuclei and thereby be observed in direct detection
experiments.  The impressive annual modulation signal recorded by the DAMA collaboration \cite{dama,dama2,dama3}, and the low energy excesses
observed by CoGeNT \cite{cogent}, CRESST-II \cite{cresst} and CDMS/Si \cite{cdms/si} can all be simultaneously 
explained in this framework \cite{foot05}. However the dust has not completely settled;
some tension with the null results of the XENON-100 \cite{xenon100} and LUX \cite{lux} experiments remain.

The purpose of this article is to review these developments, in a (hopefully) coherent and pedagogical manner.
This review is structured as follows: In the remainder of this section a qualitative overview of the mirror dark matter picture
is provided. Section 2 reviews the relevant particle physics of mirror matter.
Section 3 discusses early Universe cosmology: Big Bang Nucleosynthesis (BBN), mirror BBN, CMB and LSS.
Section 4 looks at small-scale structure, reviewing recent work on the nontrivial halo dynamics suggested by this dark matter candidate.
Section 5 examines the mirror dark matter interpretation of the direct detection experiments, especially DAMA, CoGeNT, CRESST-II and CDMS/Si.
Finally, some concluding remarks are given in section 6.

%%%%%%%%%%%%%%%%%%%%%%%%%%%%%%%%%%%%%%%%%%%%%%%%%

\subsection{Overview}

Cosmological observations indicate that the energy in the Universe consists of ordinary matter,
non-baryonic dark matter and the cosmological constant. This review is concerned with a particular
dark matter theory - mirror dark matter. 
The mirror dark matter hypothesis contains three main ingredients.
First, the particle physics Lagrangian is extended to include a hidden sector
exactly isomorphic to the ordinary matter sector. 
This provides stable massive particles which make up the presumed dark matter
in the Universe.
Second, we assume ordinary and mirror matter interact with each other via
gravity and also the photon  - mirror photon kinetic mixing interaction, with
$\epsilon \sim 10^{-9}$. 
This assumption is required to account for small-scale structure and also direct detection
experiments (as we will see).
Third, we need appropriate initial conditions arising in the very early Universe.
In addition to the usual assumptions of tiny adiabatic scalar perturbations which seed the structure
in the Universe,
we also have: $T' \ll T,\ \Omega_{b'} \approx 5\Omega_b$.\footnote{
The precise value for $\Omega_{b'}$ is set by fits to the CMB anisotropy spectrum, in the same
way in which cold dark matter density is determined in the $\Lambda$CDM model.}  
These initial conditions are required to explain large-scale structure and CMB anisotropies.

It is perhaps useful to first give a qualitative discussion of how these three ingredients might
combine to provide an adequate description of dark matter phenomena.
The subsequent sections will review  what is known quantitatively about the various parts of this picture.
Our starting point is the early Universe, around the time of the BBN epoch, $t {\sim} 1$ second.
By then, any antibaryons created in the early Universe have efficiently annihilated with
baryons. It follows that our existence today requires the generation of a baryon - antibaryon asymmetry in the Universe.
In a similar manner, any mirror antibaryons created in the early Universe would have efficiently annihilated with
mirror baryons, so it is safe to assume that dark matter is composed of mirror baryons, with a negligible mirror antibaryon component
(or vice versa).
The origin of the mirror-baryon asymmetry of the Universe is unknown,
although several mechanisms have been discussed, e.g. \cite{hodges,ber1,berasym,bellvolkas}. 
Clearly, the result that $\Omega_{b'} \sim \Omega_b$ does suggest that these asymmetries 
might be connected in some way \cite{fvasym}. This kind of asymmetric dark matter has also been examined
in the context of more generic hidden sector models. See the recent reviews \cite{admreview,admreview2}
and references therein for relevant discussions.

Of course radiation - not baryons - dominated the energy density during the BBN epoch. 
Since BBN arguments constrain the energy density of the Universe to be less than around one additional
neutrino at that time, the mirror particles and ordinary counterparts did not have the same temperature.
The mirror particles must have been cooler than the ordinary particles.
This is possible, if the interactions in ${\cal L}_{mix}$ which couple  
the two sectors together, are small enough.
In fact, we make the simple assumption that $T' \ll T$ holds at some early time before the BBN epoch
(our notation is that $T$ [$T'$] without subscript is the photon [mirror photon] temperature).
We take a similarly pragmatic approach to $\Omega_{b'}$. CMB observations (and others) constrain
$\Omega_{b'} \approx 5\Omega_b$.
We call these {\it effective} initial conditions since
it is certainly possible that they might have arisen from symmetric ones at an even earlier 
time. [This occurs, for instance, 
in chaotic inflation models where the
reheating of the ordinary and mirror sectors can be asymmetric \cite{kolb,hodges,others}.]

Even if the Universe started with $T' \ll T$, entropy in the mirror sector can be generated via kinetic
mixing induced interactions: $e \bar e \to e' \bar e'$ \cite{glas2}.
For $\epsilon \sim 10^{-9}$, the asymptotic value (i.e. $t \to \infty$) of the ratio $T'/T$ is $\sim 0.3$ \cite{paolo1,footplb2012}.  
Since $T' < T$, mirror nucleosynthesis 
would have occurred somewhat earlier than ordinary nucleosynthesis.
To understand what this means, let us first recall what happens in ordinary BBN. The nucleon 
number densities are determined by the two-body and three-body reactions:
\begin{eqnarray}
n + \bar e \leftrightarrow p + \bar \nu_e, \ \ n + \nu_e \leftrightarrow p + e, \ \ n \leftrightarrow p + e + \bar \nu_e
\ .
\end{eqnarray}
Initially these reactions  drive the neutron to proton ratio to unity but as
the temperature drops to around 1 MeV, the neutron - proton mass
difference leads to a larger proportion of protons. Eventually
the rate of these reactions became less frequent than the expansion rate of the Universe.
When this happens
the two-body reactions become infrequent enough to effectively {\it freeze}  the neutron/proton ratio.
The temperature where this occurs is $T = T_{freeze} \sim 0.8$ MeV. 
This ratio is then only further modified by neutron decays which occur until deuterium formation at
$T \sim 0.07$ MeV. The end product is that around 25\%  of the baryons end up in helium and 
75\% of the baryons in hydrogen, with
trace amounts of other light elements. Mirror nucleosynthesis is qualitatively similar,
but since it occurs earlier, the expansion rate is greater so that the mirror-neutron/mirror-proton ratio 
freezes out at a higher temperature: $T'_{freeze} > T_{freeze}$. For this reason, and also because there is less time
for mirror neutrons to decay, the mirror-neutron/mirror-proton ratio remains close to unity.
This means that there is a high proportion of mirror helium in the mirror sector \cite{ber1}.
For $\epsilon \sim 10^{-9}$, around 
90\% of mirror baryons are synthesized into mirror helium, with 10\% into mirror hydrogen \cite{paolo2}. 

At these early times the Universe is remarkably isotropic and homogeneous.
The Universe is not completely smooth though, tiny perturbations, possibly seeded by quantum
fluctuations and amplified by inflation, are present. Consider a perturbation to the matter
density: $\delta ({\bf x}) \equiv (\rho ({\bf x}) - \langle \rho \rangle)/\langle \rho \rangle$. 
In Fourier space such a perturbation is described by a wavevector
${\bf k}$:
\begin{eqnarray}
\delta ({\bf k}) \equiv \int {d^3 x \over (2\pi)^3}\ \delta ({\bf x}) \ e^{i{\bf k} \cdot {\bf x}} \ .
\end{eqnarray}
While these perturbations are small: $\delta ({\bf k}) \ll 1$, modes with different wavevector
evolve in time independently and linearly. This is the so-called {\it linear regime}.
The linear evolution of such modes is described by linearlized Boltzmann-Einstein equations. Qualitatively, the evolution
of these modes depends on their scale relative to the comoving horizon size at the time under consideration.
Large-scale modes with $k^{-1}$ much larger than the horizon are not influenced by causal physics; they remain
unchanged. Small-scale modes with $k^{-1}$ less than
the horizon can be influenced by the physical processes of gravity and potentially also pressure.
As the Universe expands, the comoving horizon 
increases; large-scale modes {\it enter} the horizon and are processed by causal physics
(the comoving wavelength $\sim k^{-1}$ remains constant).

Matter density perturbations can be divided into baryonic perturbations and mirror-baryonic ones.
For baryonic perturbations prior to hydrogen recombination, the photons are tighly coupled to electrons via Compton
scattering and electrons to protons via Coulomb scattering. At this time, the particles: $e, H, He, \gamma$ 
can be treated as a tightly coupled
fluid. The effects of gravity and pressure are well understood for this system: acoustic oscillations occur
and are responsible for the peaks in the CMB anisotropy spectrum. The physics of mirror baryonic perturbations is very similar.
Prior to mirror-hydrogen recombination, i.e. when $T' \stackrel{>}{\sim} 0.3$ eV,
the mirror particles: $e', H', He', \gamma'$ also form a tightly coupled fluid.
Fourier modes which are small enough to have entered the horizon at this epoch  
undergo acoustic oscillations due to the ($\gamma'$) radiation pressure; this suppresses
perturbations on scales smaller than the horizon at this time.
Only after mirror-hydrogen recombination can matter density perturbations grow.

Perhaps it is useful to pause here and compare this picture with that of collisionless dark matter.
Collisionless dark matter by definition has no pressure and therefore no acoustic oscillations. 
Mirror dark matter 
might therefore appear to be very different, however this
need not be the case.
In the limit $T'/T \to 0$, equivalently $\epsilon \to 0$ given the assumed initial condition $T' \ll T$,
mirror nuclei were always in neutral atoms. Mirror-baryonic acoustic oscillations 
would not then occur. In this limit therefore, mirror dark matter  
would be indistinguishable from collisionless cold dark matter
during the linear regime \cite{ber1,ig1}.
Clearly, for nonzero $T'/T$ departures from collisionless cold dark matter would be expected on small scales, 
smaller than a characteristic scale $L(\epsilon)$. 
Observations can then be used to yield an 
upper limit on $T'/T$, and hence also on $\epsilon$.

Within the mirror dark matter context, the formation  
and evolution of structure on scales larger than $L(\epsilon)$ should be similar to 
collisionless cold dark matter, at least in the linear regime. 
If $L(\epsilon)$ is small enough, then linear evolution of structures on 
galactic scales and larger can therefore be very similar to collisionless cold dark matter.
What about the early evolution of small-scale structure in the nonlinear regime?
Consider first
collisionless cold dark matter. In that model, 
halos hosting galaxies such as the Milky Way are believed to have formed hierarchically from
the merging of smaller structures \cite{tom} (see also \cite{mo} for an up-to-date review and more detailed bibliography). 
This picture would presumably need some revision
if  acoustic oscillations were effective at 
suppressing small-scale inhomogenities in the linear regime.
It could happen for instance that structure evolves hierarchically above a certain scale $\sim L(\epsilon)$
and top down below this scale [and some mixture of both mechanisms on scales near $L(\epsilon)$]. 

It is tempting to speculate that the suppression of small-scale structure below $L(\epsilon)$ 
might be connected with  the surprisingly small number of satellite galaxies that have been observed in the local group.
This ``missing satellites problem''  is considered to be a serious issue for the 
collisionless cold dark matter model (for a review and references to the original literature
see for example \cite{missing1}).
Mirror dark matter appears to have the potential to address this and other small-scale
shortcomings of collisionless cold dark matter, however much more work is needed \footnote{
Another small-scale puzzle of collisionless cold dark matter is the observed large proportion of bulgeless disk galaxies.
That is, pure disk galaxies with no evidence for merger-built bulges.
This is surprising given the level of hierarchical clustering anticipated if dark matter were collisionless. For relevant
discussions see  \cite{kenintro,kenintro2} and references therein. 
In fact \cite{kenintro2} describes this as the biggest problem in the theory of galaxy formation.
The suppression of small scale structure below $L(\epsilon)$ and also the early heating of
ordinary matter from mirror supernovae (to be discussed) may help address this issue.
Qualitatively, mergers should be less frequent, and importantly, the formation of the baryonic disk might be delayed due to the
early heating.}.
Suffice to say that the formation and early evolution of structure on galaxy scales
is a complex issue and is, at present, poorly understood in the mirror dark matter framework. 
Ideally, hydrodynamical simulations taking into account mirror dark matter interactions, 
both dissipative and non-dissipative, along with heating from supernovae 
in the presence of kinetic mixing (see below) could be attempted.
Alternatively, analytic or semi-analytic techniques could conceivably be developed using
the Press-Schechter formalism as a starting point \cite{press}.
At the present time though, such work has not yet been done.
In the absence of such computations or analytic studies, any discussion is
certainly speculative. Nevertheless, a self-consistent if not quantitative picture appears to be emerging.

Initially, mirror density perturbations evolve linearly and grow in both density and size as the Universe expands.
Consider now a particular galaxy-scale perturbation.
When the matter overdensity reaches $\delta \sim 1$ the evolution starts to become  nonlinear.
Around this time the perturbation breaks away from the expansion and can begin to collapse.
Mirror dark matter is collisional, however it is also dissipative, and if the
cooling time scale is faster than the free-fall time scale then the collapse of mirror-particle
perturbations are not impeded \cite{zurab2013}. The perturbation will collapse into a disk-like system on the free-fall time scale
(the size of the disk depending on details such as the amount of angular momentum)
\footnote{
The disk is not expected to be completely uniform and smaller scale perturbations on the edge of the disk might break away
from the main perturbation and collapse. Such perturbations might seed satellite galaxies and could potentially
explain why the bulk of the dwarf satellite galaxies of the Milky Way and M31 in the local group are aligned in a plane \cite{ibata,paw}.
Alternatively \cite{zurab2013} the dwarf satellite galaxies might have originated much later as tidal dwarf galaxies formed during
a merger event \cite{hammer}.}.
Mirror star formation
can occur during the free-fall phase and/or later in the collapsed disk.
Mirror supernovae are also expected to be occurring during this early time.
This is especially important assuming photon - mirror photon kinetic mixing interaction exists with
$\epsilon \sim 10^{-9}$. As briefly mentioned in subsection 1.1, mirror
supernovae would then influence ordinary matter by providing a huge heat source.
Basic  particle processes such as $e' \bar e' \to e \bar e$ in the mirror supernova's core 
would convert about half of the mirror supernova's core collapse energy into creation
of light ordinary particles $e, \bar e, \gamma$ \cite{sph,sil}. In the region around
each supernova ($\sim {\rm pc}^3$) this energy is converted (via complex and poorly understood
processes, e.g. generation of shocks etc.) into ordinary photons which are anticipated 
to have an energy spectrum in the X-ray region. 
These
photons would not only heat ordinary matter but might have been responsible for its reionization - inferred from observations to 
have occurred at early times at redshift: $6 < z < 20$.

Once the ordinary matter is ionized it can no longer efficiently absorb radiation.
This is because ordinary matter has very little metal content at this early time, and the Thomson scattering
cross-section is so small. [We adopt the astrophysics convention of describing every element heavier than helium as a metal.] 
Ordinary matter can now start to cool and accumulate in these mirror dark matter structures.
One expects, therefore, that the ordinary baryons will ultimately collapse potentially forming a separate disk \footnote{ 
Gravitational interactions between the baryonic disk and mirror baryonic one, should both form, could lead to their alignment cf. \cite{randal}.}.
Ordinary star formation can now begin and is expected to proceed extremely rapidly. 
In fact, the density of the baryonic gas ($n_{gas}$) in these collapsed structures would be very high, which
is known to be directly correlated with the star formation rate:
\begin{eqnarray}
\dot{\Sigma}_* \propto n_{gas}^N
\end{eqnarray}
where $N \sim 1-2$ \cite{s1959,ken}.
Thus leads inevitably to the production of ordinary supernovae. Now, the physics of ordinary supernovae, 
like mirror supernovae as we briefly described above, is extremely interesting if the kinetic mixing interaction exists.
% of strength $\epsilon \sim 10^{-9}$. 
Ordinary supernovae will produce a huge flux of mirror photons in the presence of the
kinetic mixing interaction of strength $\epsilon \sim 10^{-9}$.
These mirror photons can heat the mirror disk, which by now has a substantial mirror metal fraction.
[This energy is absorbed very efficiently because of 
the large photoionization cross-section of the mirror metal atoms.]
This huge energy input can potentially expand the gas in the mirror disk out into an approximately spherically 
distributed plasma.
This, it is presumed, is the origin of the roughly spherical
halos inferred to exist around spiral galaxies today.
Naturally, much work
needs to be done in order to check this qualitative picture of the early period of galaxy evolution.

The (current) structure of galactic halos 
appears to be a more tractable problem \cite{sph,r69,foot5,rf2013b}. 
As described above, the dark matter distribution in galaxies was once very compact 
until heating by ordinary supernovae occurred.
If the rate of supernovae became large enough, then the heating rate of the mirror-particle plasma could exceed its cooling rate 
(due to processes such as thermal bremsstrahlung)
in which case this plasma component will expand. 
The mirror star formation rate falls drastically at this time as the gas component heats up and its
mass density falls.
As the mirror dark matter expands, the ordinary star formation rate (and hence supernova rate) also falls as the 
ordinary matter densities drop in the weakening gravitational potential. 
The halo will continue to expand 
until the heating is balanced by cooling.
The end result is that at the current epoch
the halo should have evolved to a quasi-static equilibrium configuration 
where the energy being absorbed
in each halo volume element is balanced by the energy being emitted 
in the same volume element:
\begin{eqnarray}
{d^2 E_{in} \over dt dV}= 
{d^2 E_{out} \over dt dV} 
\ .
\label{meal8x}
\end{eqnarray}
Under the simplifying assumption of spherical symmetry,
the above dynamical condition, along with the hydrostatic equilibrium
equation,
\begin{eqnarray}
{dP \over dr} &=& -\rho (r) g(r) 
\label{butler}
\end{eqnarray}
can be used to determine the dark matter density and temperature profiles: $\rho (r), \ T(r)$.  
That is, the current bulk properties of the dark matter  
halo around spiral galaxies can be derived from this assumed model of halo dynamics. 

Numerically it has been shown that this dynamics requires dark matter 
to have an approximate quasi-isothermal
distribution \cite{foot5,rf2013b}:
\begin{eqnarray}
\rho (r) \simeq {\rho_0 r_0^2 \over r^2 + r_0^2 }
\end{eqnarray}
where $\rho_0,\ r_0$ 
is the dark matter central density and core radius.
Numerically it is also found that the 
core radius, $r_0$, scales with disk scale length, $r_D$, via $r_0 \simeq 1.4r_D $ 
and that the product $\rho_0 r_0$ is roughly $constant$, i.e. independent of galaxy size (the
$constant$ is set by the parameters of the model). 

Dark matter with this constrained distribution 
is known to provide an excellent description of galactic rotation curves in spiral galaxies \cite{salucci}.
Indeed, a result of all these scaling relations, together with baryonic relations
connecting the disk mass with the disk scale length, is that the ordinary and mirror dark
matter content of spiral galaxies are, roughly, specified by a single parameter. This
parameter can be taken to be $m_D, \ r_D$ or the galaxy's luminosity in some band, $L$.
This has important implications for the galaxy's rotation curve.
It should be roughly universal, i.e. completely fixed once $L$ is specified.
This is consistent with observations, which show just this behaviour 
\cite{rubin2,PS2,PSS}.
As should be clear from the above scaling relations, the agreement with observations
is not just qualitative, the dynamics allows  quantitative predictions
to be made, all of which appear to be consistent with the observations.

Another result of numerical solution to Eqs.(\ref{meal8x}), (\ref{butler}) is that the halo is approximately isothermal.
Numerical work and also some analytic arguments \cite{sph} indicate that the average halo temperature is approximately:
\begin{eqnarray}
T \approx {1 \over 2} \bar m v_{rot}^2
\label{10}
\end{eqnarray}
where $v_{rot}$ is the galactic rotational velocity (for the Milky Way $v_{rot} \approx 220$ km/s)
and $\bar m$ is the mean mass of the particles, $e', H', He',...$, constituting the plasma.
Arguments from early Universe cosmology (mirror BBN) indicate that $\bar m \approx 1.1$ GeV \cite{paolo2}.
This means that for the Milky Way the halo temperature is roughly: $T \sim 200$ eV, i.e. a few million
degrees kelvin \footnote{Unless otherwise indicated,
we use natural units with $\hbar = c = k_B = 1$ throughout.}.

The end result of all this, is that at the present time, spiral galaxies such as the Milky Way are at the
center of an extended dark matter halo. This halo
is predominately in the form of a hot spherical plasma, which is composed of an array of mirror particles:
$e', H', He', O', Fe',....$. These particles are continuously undergoing both dissipative and non-dissipative self interactions,
with the energy dissipated from the halo being replaced by energy produced from ordinary supernovae, made possible if kinetic
mixing with strength $\epsilon \sim 10^{-9}$ exists. 
The current mirror-star formation rate in such a plasma
is expected to be very low; the plasma cannot locally cool and condense into stars. The vast bulk of mirror
star formation is therefore expected to have occurred at very early times - in the first billion years or so.
As discussed above, this is the presumed origin of the mirror metal component of the halo plasma. 
Although very rare, mirror supernovae might still occur today. 
Observationally, a mirror supernova might appear to be something like a Gamma Ray Burst given
the assumed kinetic mixing interaction, and indeed it has been proposed as a candidate for
the central engine powering (at a class of) such objects \cite{sil,blinn}.

The halo dynamics described above requires
the kinetic mixing interaction to not only exist but have strength $\epsilon \sim 10^{-9}$ .
This interaction induces also an interaction between charged mirror particles and ordinary nuclei.
This enables halo mirror particles to thereby scatter off ordinary nuclei, essentially
a Rutherford-type (spin independent) elastic scattering process.
Hence, mirror particle interactions might potentially be seen in {\it direct} detection experiments searching
for halo dark matter
\footnote{
Given that mirror dark matter  arises from a particle-antiparticle asymmetry
in the Universe, signals from the annihilation of mirror baryons
with mirror antibaryons are not anticipated. There would be far too few mirror antibaryons in the halo for 
such annihilation to be detected.  
Thus, observable {\it indirect} detection 
signatures of mirror dark matter are expected to be very limited; possibly only an excess of positions produced via kinetic
mixing induced processes in mirror supernovae, should such supernovae occur at a sufficient rate \cite{sil}.  
}.

Consider a mirror nuclei of type $A'$ of atomic number $Z'$ (e.g. $A' = H', He', O', Fe',...$) that is moving
with velocity $v$. If this mirror nuclei passes close to an ordinary nucleus
$A$ of atomic number $Z$ (presumed at rest), then it can scatter leaving $A$ with a recoil energy $E_R$. 
The differential cross-section for this process has a characteristic $1/E_R^2$ dependence:
\begin{eqnarray}
{d\sigma \over dE_R} = {2\pi \epsilon^2 Z^2 Z'^2 \alpha^2 \over E_R^2 m_A v^2}\ F_A^2 F^2_{A'}
%\label{cs}
\end{eqnarray}
where $F_{A}$, $F_{A'}$ are the relevant form factors.
If this kinetic mixing induced interaction does indeed exist, then halo mirror dark matter can be probed
in direct detection experiments.
In fact, a kinetic mixing strength $\epsilon \sim 10^{-9}$ happens to be just the right magnitude for the current generation
of direct detection experiments to be sensitive to this interaction \cite{foot2003,footdd,footdd2,foot2,foot05}. 

The rates in such an experiment depend not just on the cross-section but also on the halo velocity distribution of the mirror particles.
The self interactions of the mirror particles in the halo plasma should help keep these particles in thermal equilibrium.
Their velocity distribution is therefore expected to be Maxwellian: 
\begin{eqnarray}
f_{A'} = exp(-E/T) = exp(-m_{A'}v^2/2T) = exp (-v^2/v_0^2)\ .
\end{eqnarray}
The quantity $v_0$, which characterizes the velocity dispersion, evidently depends 
on the mass $m_{A'}$, of the particular component: 
\begin{eqnarray}
v_0^2 = {2T \over m_{A'}} \approx {\bar m \over m_{A'}}\ v_{rot}^2 \ 
\label{11b} 
\end{eqnarray}
where Eq.(\ref{10}) has been used.
Observe that such a mass dependent velocity dispersion is very different from the distribution expected for collisionless cold dark matter,
where $v_0 \approx v_{rot}$ is anticipated \cite{smith}.
 
Clearly, mirror dark matter has a number of distinctive features:
It is (a) multi-component, with a spectrum of particles with known masses (b) interacts with ordinary matter via
kinetic mixing induced interactions, leading to Rutherford-type (spin independent) elastic scattering
and (c) heavy mirror particles, $m_{A'} > m_{He}$, have small velocity dispersion ($v_0^2 \ll v_{rot}^2$).
These features, it turns out, combine to provide a consistent explanation of the DAMA annual modulation signal \cite{dama,dama2,dama3}
and also the low energy excesses found by CoGeNT \cite{cogent}, CRESST-II \cite{cresst} and CDMS/Si \cite{cdms/si}.
In this interpretation, these experiments have detected the kinetic mixing induced interactions of 
halo mirror metal components, $A' \sim O' - Fe'$ \cite{foot05}. While these developments appear to be
very encouraging, the experimental situation is still not completely settled.
Significant tension with the null results of XENON100 \cite{xenon100} and LUX \cite{lux} exists.
Also, important and necessary checks have yet to be made such as an experiment located in the  southern hemisphere.
Such an experiment is important, not just as a check of the DAMA annual modulation signal, but also
to search for the expected diurnal modulation \cite{footdiurnal}.

\newpage
\section{The particle physics}

\vskip 0.2cm
The standard model of particle physics is a highly predictive gauge theory, based on the gauge symmetry:
$G_{SM} \equiv SU(3)_c \otimes SU(2)_L \otimes U(1)_Y$.
This theory has been extremely successful in accounting for the electromagnetic, weak, and strong 
interactions of the known particles \cite{glashow,qcd}. For a review, see for instance \cite{veltman}.
The electromagnetic and strong interactions are associated with unbroken gauge symmetries $U(1)_Q$ and $SU(3)_c$, 
while weak interactions arise from the spontaneous breaking of $SU(2)_L \otimes U(1)_Y \to U(1)_Q$.
The recent discovery of a Higgs-like resonance at the Large Hadron Collider (LHC) \cite{lhchiggsdiscovery1,lhchiggsdiscovery2} appears to confirm 
that this symmetry breaking is due to
the nonzero vacuum expectation value of an elementary Higgs doublet field, $\phi$ \cite{higgs}.
Indeed, the measured properties of the Higgs-like resonance are (currently) consistent with those expected for the standard model Higgs scalar
\cite{higgsspin,higgs5,higgs6}.

The standard model can be described by a renormalizable Lagrangian:
\begin{eqnarray}
{\cal L} = {\cal L}_{SM} (e, u, d, \gamma,...) 
\ .
\label{lag1sm}
\end{eqnarray}
This Lagrangian respects an array of symmetries including
proper orthochronous Lorentz transformations, space-time translations and gauge symmetries, as discussed above.
Notably, the standard model Lagrangian does not respect improper  
Lorentz symmetries, such as  parity and time reversal.
Parity, in particular, is violated maximally by the weak interactions as the $SU(2)_L$ gauge bosons $W$ couple only to the 
left-handed chiral fermion fields.
However improper  space-time symmetries, appropriately defined, can
be exact and unbroken symmetries of nature if a 
set of mirror particles exist. The simplest such model has been called the exact parity symmetric model \cite{flv}.

\vskip 0.6cm
\subsection{Exact parity symmetric model}

\vskip 0.4cm

Mirror particles are defined as follows. For every known particle, a mirror partner is hypothesized,
which we shall denote with a prime ($'$). The interactions of these duplicate set of particles are
described by a Lagrangian of exactly the same form as that of the standard model.  That is,
the ordinary particles and mirror particles are described by the Lagrangian:
\begin{eqnarray}
{\cal L} = {\cal L}_{SM} (e, u, d, \gamma,...) + {\cal L}_{SM} (e', u',
d', \gamma', ...) + {\cal L}_{mix}
\label{lag1}
\end{eqnarray}
where ${\cal L}_{mix}$ accounts for possible non-gravitational interactions connecting 
ordinary and mirror particles, which we set aside for the moment.
That is, the ordinary and mirror particles form parallel sectors, each
respecting independent gauge symmetries $G_{SM}$.  This means that the gauge symmetry of the full Lagrangian, ${\cal L}$, is $G_{SM} \otimes G_{SM}$.
As defined in Eq.(\ref{lag1}) above, the Lagrangian has a $Z_2$ discrete symmetry which swaps each ordinary particle 
with its partner. If we make a slight adjustment, and interchange left and right-handed chiral fields
in the mirror sector so that mirror weak interactions couple to right-handed chiral fermion fields (instead of left-handed
fields) then
the Lagrangian, ${\cal L}$, respects an exact parity symmetry, which we also refer to as mirror symmetry \cite{flv}: 
\footnote{
Technically, there are two possible theories depending on whether or not we flip the left 
and right chiralities in the mirror sector.
Although these two theories are formally distinct, they
are phenomenologically
almost indistinguishable. Certainly, 
for the applications to dark matter phenomena, this distinction is unimportant.
See also section 6 of \cite{footvolkas95} for further discussions about this dichotomy.
}
\begin{eqnarray}
& x, y, z, t  \to -x,-y,-z,t \nonumber \\
& 
G^{\mu} \leftrightarrow G'_{\mu},\ 
W^{\mu} \leftrightarrow W'_{\mu}, \ B^{\mu} \leftrightarrow B'_{\mu}, \ \phi \leftrightarrow \phi'
\nonumber \\
& \ell_{iL} \leftrightarrow \gamma_0 \ell'_{iR}, \
e_{iR} \leftrightarrow \gamma_0 e'_{iL}, \
q_{iL} \leftrightarrow \gamma_0 q'_{iR}, \
u_{iR} \leftrightarrow \gamma_0 u'_{iL}, \
d_{iR} \leftrightarrow \gamma_0 d'_{iL} \ . 
\label{msym}
\end{eqnarray}
\vskip 0.1cm
\noindent
Here $G^{\mu}, W^{\mu}, B^{\mu}$ are the
$SU(3)_c, \ SU(2)_L, \ U(1)_Y$ spin-one gauge bosons,
the fermion fields
$\ell_{iL} \equiv (\nu_i, e)_L, \ e_{iR}, \ q_{iL} \equiv (u_i, d_i)_L, \ u_{iR}, \ d_{iR}$ represent the
leptons and quarks, $i=1,2,3$ is the generation index and $\gamma_0$ is a Dirac gamma matrix. Also included is 
the Higgs doublet $\phi$ along with its mirror partner, $\phi'$. 
This review discusses the parameter region (to be defined in section 2.3) where $\langle \phi \rangle = \langle \phi' \rangle$,
so that the mirror symmetry
is {\it not} spontaneously
broken by the vacuum; mirror symmetry is an exact, unbroken
symmetry of the theory.

The parity transformation as given in Eq.(\ref{msym}), which we here define as ${\cal P}$, involves swapping ordinary particles
with mirror particles in addition to $(x, y, z, t) \to (-x, -y, -z, t)$.
Although this is non-standard, and is perhaps subtle, it is of course a perfectly acceptable
definition of space-time parity in the presence of degenerate partners \cite{flv,flv2,footvolkas95} (see also \cite{lee,salam} for early related discussions).
This theory also exhibits an exact time reversal symmetry ${\cal T}$, defined by ${\cal P} {\cal T} \equiv $ CPT 
where CPT is the conventionally defined CPT transformation (the CPT transformation is
an invariance of ${\cal L}_{SM}$ itself and so is also an invariance of ${\cal L}$).
The ${\cal P}$ and ${\cal T}$
transformations do not separately commute with proper Lorentz transformations
(which is, of course, a general property of space and time inversion transformations)
but together with space-time translations close to form the Poincar$\acute{e}$ 
group - the group of isometries of Minkowski space-time.
\vskip 0.6cm
%\centerline{\epsfig{file=revfig2.1.eps,angle=270,width=5.4cm,angle=90}}
\centerline{\epsfig{file=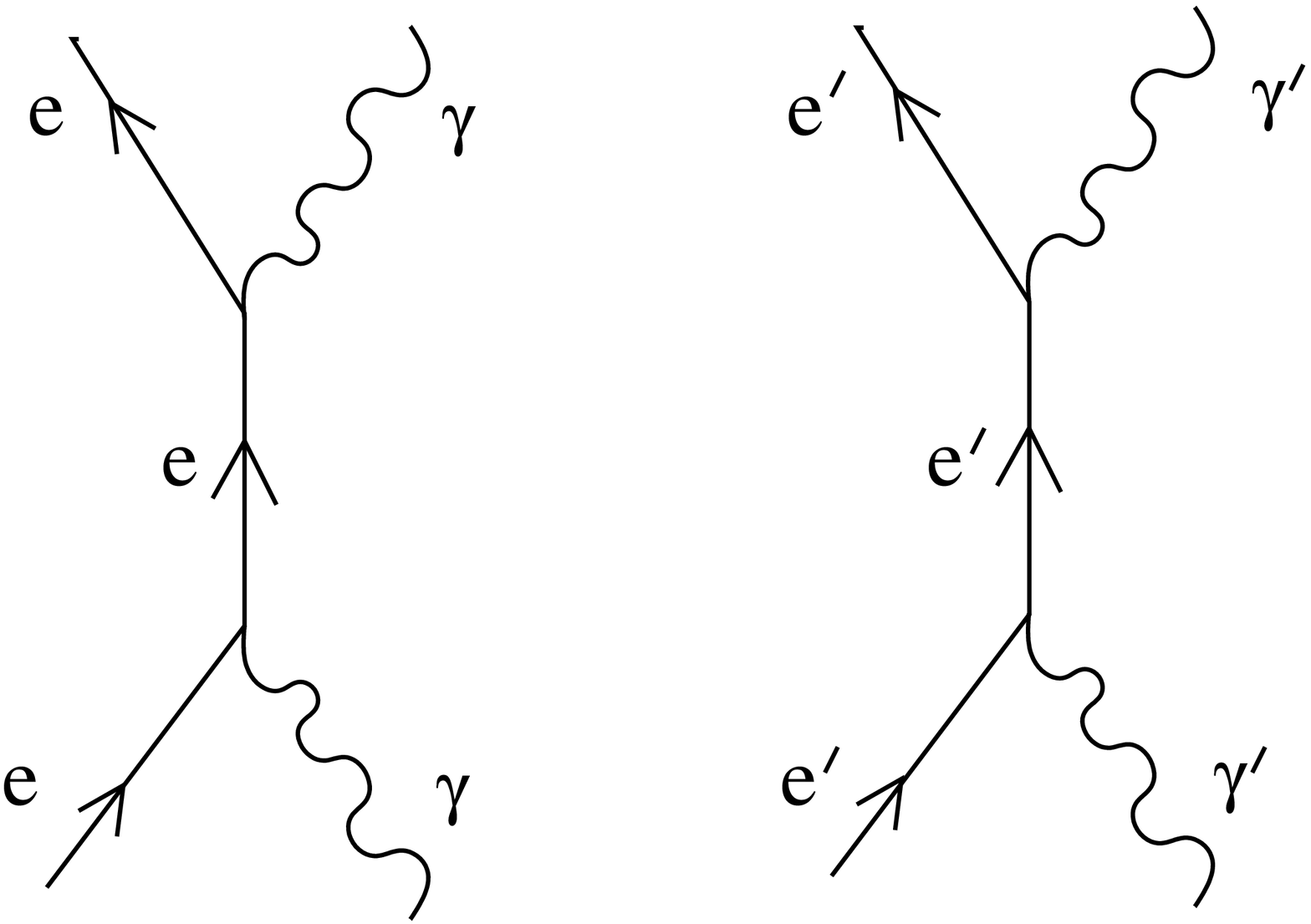,angle=270,width=5.4cm,angle=90}}
\vskip 0.3cm
\noindent
{\small
Figure 2.1: The process $e\gamma  \to e\gamma$ and the mirror particle analogue: $e'\gamma' \to e'\gamma'$.
Mirror symmetry implies that the
cross-section for both processes is exactly the same.
}
\vskip 1cm
\newpage

Mirror symmetry, so long as it is not spontaneously, or otherwise broken, ensures that the masses and 
couplings of the 
particles in the mirror sector are exactly identical to the 
corresponding ones in the ordinary sector.
The only new parameters are those in ${\cal L}_{mix}$, which by hypothesis conserve mirror symmetry.
An important, but trivial consequence of mirror symmetry is that every ordinary particle process has a mirror
particle analogue. Take $e\gamma \to e \gamma$ elastic scattering as an example (figure 2.1).
Mirror symmetry implies that $e'\gamma' \to e' \gamma'$ can also occur, and since the
symmetry is exact and unbroken, the cross-section for each process is exactly the same.
In the Thomson limit, for instance, the (Born) cross-section for both processes is $\sigma_T = {8\pi \alpha^2/3m_e^2}$.
The same thing happens, of course, for every other ordinary particle process.

Mirror symmetry does not exclude the possible existence of new interactions coupling ordinary and mirror particles together.
However, with the minimal particle content, the (mirror, gauge, Lorentz) symmetries of the theory restrict
such renormalizable interactions to just two terms \cite{flv}:
\begin{eqnarray}
{\cal L}_{mix} = - \frac{\epsilon}{2} F^{\mu \nu} F'_{\mu \nu} - 
\lambda \phi^{\dagger}\phi \phi'^{\dagger}\phi' \ ,
\label{mix}
\end{eqnarray}
where $F_{\mu \nu}$ ($F'_{\mu \nu}$) is the ordinary (mirror) $U(1)_Y$
gauge boson field strength tensor. The first interaction is a mixing of the kinetic terms for the $U(1)_Y$ and $U(1)'_Y$
gauge bosons, while the second interaction is a Higgs - mirror Higgs quartic coupling which forms part of the full Higgs
potential.  We now discuss each of these terms
in more detail.

\subsection{Photon - mirror photon kinetic mixing}

The $U(1)_Y - U(1)'_Y$ kinetic mixing term in Eq.(\ref{mix}) is 
gauge invariant, since $F_{\mu \nu} \equiv \partial_\mu B_\nu - \partial_\nu B_\mu$ itself is gauge invariant
under the $U(1)_Y$ gauge transformation, $B_\mu \to B_\mu + \partial \chi$.
Kinetic mixing respects mirror symmetry Eq.(\ref{msym}), and all the other known symmetries of the theory.
Furthermore, since kinetic mixing is a renormalizable interaction, $\epsilon$ can be viewed as a fundamental parameter of the 
theory \cite{fh}. 
%\footnote{
%Observe that $\epsilon \to 0$ is technically natural (in the absence of gravity) as there is increased symmetry in this limit. 
%For example, with $\epsilon = 0$, ${\cal L}$ 
%is invariant with respect to independent global Lorentz transformations in the ordinary and mirror sectors. For this
%reason all radiative corrections to $\epsilon$ are proportional to $\epsilon$ itself.}.

In standard electroweak theory, the $U(1)_Y$ gauge boson $B_\mu$, is a linear combination of the photon $A_\mu$
and the Z-boson $Z_\mu$:
\begin{eqnarray}
B_\mu = \cos\theta_w \ A_\mu + \sin \theta_w \ Z_\mu
\ .
\end{eqnarray}
It follows that there is both $\gamma - \gamma'$ and  $Z - Z'$
kinetic mixing. However, experiments and observations are much more 
sensitive to $\gamma - \gamma'$ kinetic mixing interaction so we need not discuss $Z-Z'$ mixing any further.

What is the physical effect of photon - mirror photon kinetic mixing?
Consider $U(1) \otimes U(1)'$ quantum electrodynamics of the electron $\psi_e$, and photon $A_\mu$, mirror electron
$\psi'_e$, and mirror photon $A'_\mu$:
\begin{eqnarray}
{\cal L} &=& -{1 \over 4}F_{\mu\nu}^2\ -\ {1 \over 4}F_{\mu\nu}^{'2}\ -\ {\epsilon \over 2} F_{\mu\nu}F'^{\mu\nu}
\nonumber \\
&+& \bar{\psi}_e(i\hat{\partial}-m)\psi_e\ +\ \bar{\psi}'_e (i\hat{\partial}-
m)\psi'_e
\ + \ e\bar{\psi}_e\hat A\psi_e \ + \ e\bar{\psi}'_e \hat A'\psi'_e 
\ 
\end{eqnarray}
where we have adopted the convenient notation: $\left( F_{\mu \nu} \right)^2 \equiv F_{\mu \nu} F^{\mu \nu}$,
$\hat{\partial} \equiv \gamma^\mu \partial_\mu$, $\hat A \equiv \gamma^\mu A_\mu$ and so on.
The kinetic mixing can be removed with a non-orthogonal transformation:  
$A_\mu \to \ \stackrel{\sim}{A}_\mu \ \equiv \ A_\mu + \epsilon A'_\mu$,
$A'_\mu \to \ \stackrel{\sim}{A}'_\mu \ \equiv \ A'_\mu \sqrt{1 - \epsilon^2}$. One then has two massless 
(i.e. degenerate) and kinetically unmixed states; any orthogonal
transformation of which will leave the kinetic terms invariant. One can transform to a basis where only
one of these states couples to electrons. 
The state coupling to the electrons is the {\it physical photon} $A_1$, appropriate for an ordinary matter 
dominated environment, such as the Earth \cite{holdom} (see also \cite{sashafv}). The orthogonal state we call the 
{\it sterile photon} $A_2$:
\begin{eqnarray}
A_1^\mu = A^\mu \sqrt{{1 - \epsilon^2}}, \  \  A_2^\mu = A'^\mu + \epsilon A^\mu
\ .
\end{eqnarray}
In this physical basis for an ordinary matter environment, the Lagrangian is (to leading order in $\epsilon$):
\begin{eqnarray}
{\cal L} &=& -{1 \over 4} \left(F_1^{\mu\nu}\right)^2 \ - \ {1 \over 4}\left(F_2^{\mu\nu}\right)^2 \nonumber \\
&+& \bar{\psi}_e (i\hat{\partial}-m)\psi_e \  + \ \bar{\psi}'_e(i\hat{\partial}-
m)\psi'_e
\ +  \ e\bar{\psi}_e \hat A_1 \psi_e \ + \ e\bar{\psi}'_e (\hat A_2 - \epsilon \hat A_1)\psi'_e
\ 
\end{eqnarray}
where $F_j^{\mu \nu} \equiv \partial^\mu A_j^\nu - \partial^\nu A_j^\mu$ ($j=1,2$). 
Evidently, the physical photon couples to mirror electrons with electric charge $\epsilon e$, while
the mirror photon doesn't couple to ordinary matter at all.
The mirror symmetry appears to be broken, but it is not of course; it is simply the result of a mirror
asymmetric environment consisting of ordinary matter.

For completeness, let us briefly digress to discuss the physical states appropriate for a mirror matter environment, such as a star
composed of mirror baryons. These are the {\it physical mirror photon} $A'_2$, and the {\it sterile mirror photon} $A'_1$:
\begin{eqnarray}
{A'}_1^\mu = A^\mu + \epsilon A'^\mu, \ \  {A'}_2^\mu = {A'}^\mu \sqrt{{1 - \epsilon^2}}
\ .
\end{eqnarray}
In this physical basis for a mirror matter environment, the Lagrangian is (to leading order in $\epsilon$):
\begin{eqnarray}
{\cal L} &=& -{1 \over 4} \left(F_1^{\mu\nu}\right)^2 \ - \ {1 \over 4}\left(F_2^{\mu\nu}\right)^2 \nonumber \\
&+& \bar{\psi}_e (i\hat{\partial}-m)\psi_e \  + \ \bar{\psi}'_e(i\hat{\partial}-
m)\psi'_e
\ +  \ e\bar{\psi}_e (\hat A'_1 - \epsilon A'_2) \psi_e \ + \ e\bar{\psi}'_e \hat A'_2 \psi'_e
\ .   
\end{eqnarray}
Evidently, a mirror star would emit the state $A'_2$. In terms of the ordinary matter physical states,
$A'_2 = A_2 - \epsilon A_1$ (to leading order in $\epsilon$). Thus, the flux of mirror photons detectable in
an ordinary matter telescope is reduced by a factor $\epsilon^2$. This makes such radiation undetectable with current
technology. Also note that such radiation would decohere into ordinary matter eigenstates on passing through
ordinary matter and thus could not even be detected in underground experiments
\footnote{For a mixed ordinary/mirror matter environment oscillations between ordinary and mirror photons are possible in principle.}.

Generalization of this $U(1) \otimes U(1)'$ quantum electrodynamics to the exact parity symmetric model is straightforward.
The physical photon $A_1$, is {\it the} photon. It couples to the known particles in the usual
way and additionally couples to mirror charged particles with coupling suppressed by $\epsilon$.
That is, the photon couples to mirror protons with ordinary electric charge $\epsilon e$, 
mirror electrons with ordinary electric charge $-\epsilon e$ etc. As discussed above,
the orthogonal state $A_2$ doesn't couple to
ordinary matter at all \footnote{
In principle, the sign of $\epsilon$ can be either positive or negative.
Although these two cases are physically inequivalent, this detail is unimportant for the 
kinetic mixing applications discussed in this review. For this reason, subsequent reference to the $\epsilon$ parameter
are statements about its magnitude only.}.
The small induced electric charge means that mirror particles can elastically scatter off ordinary
nuclei and can thereby be directly detected in experiments such as DAMA, CoGeNT, CDMS etc. 
Another consequence of the small induced electric charge is that mirror electron - mirror positron pairs 
can be produced from processes such as $e \bar e \to e' \bar e'$ in the core of ordinary supernovae and in the early
Universe. The cross-section for this process is proportional to $\epsilon^2$.

The magnitude of the kinetic mixing parameter of astrophysical interest and also of interest for dark matter direct
detection experiments turns out to be very small: $\epsilon \sim 10^{-9}$.
This is nearly two orders of magnitude smaller than the direct laboratory upper limit of 
$\epsilon < 1.55 \times 10^{-7}$ (90\% C.L.) which arises from
the orthopositronium system \cite{ortholimit}.
The kinetic mixing interaction induces orthopositronium - mirror orthopositronium mass mixing which leads to 
oscillations of orthopositronium into mirror orthopositronium \cite{glas1} (see also \cite{footsergei,demi}).
There are important proposals to improve the precision of orthopositronium experiments to directly explore the $\epsilon \sim 10^{-9}$
parameter region \cite{proposal}.

As a final comment, the approach taken here is to consider kinetic mixing as a fundamental interaction in the 
Lagrangian \cite{fh,flv}. An alternative possibility is that kinetic mixing is radiatively generated \cite{holdom}.
In particular, in Grand Unified models, such as those based on $SU(5)\otimes SU(5)'$ gauge
symmetry, the $U(1)_Y$ is embedded in a non-abelian gauge symmetry. This additional symmetry  
prevents $U(1)_Y-U(1)'_Y$ kinetic mixing from arising at tree-level (i.e. in the classical limit).
However if there exists particles $X_i$ that are charged under both ordinary and mirror electromagnetism, 
e.g. $X \sim (5,5)$ under $SU(5)\otimes SU(5)'$, then
kinetic mixing can be radiatively generated
at 1-loop level. Such induced kinetic mixing is typically around $\epsilon \sim 10^{-3}$ \cite{holdom}. However if kinetic mixing
cancels at 1-loop, as happens if the particles $X_i$ are degenerate in mass,
then it can be shown to cancel also at 2-loop level \cite{collie}.
At three loops, kinetic mixing might conceivably 
be of order $\epsilon \sim 10^{-9}$,
although this has yet to be demonstrated in an actual calculation.

The 
kinetic mixing interaction is the only term in ${\cal L}_{mix}$ [Eq.(\ref{mix})] which is used in the applications to the astrophysical and cosmological 
problems discussed in subsequent sections of this review.
Nevertheless,
for completeness we now briefly consider other possible 
non-gravitational interactions connecting ordinary and mirror particles discussed in 
the literature.

\subsection{Higgs portal coupling}

\vskip 0.2cm

In addition to kinetic mixing, there is only
one other renormalizable term (in the minimal model) which can couple the known particles with 
the mirror particles. This is the Higgs - mirror Higgs quartic interaction,
also called Higgs portal coupling:
\begin{eqnarray}
{\cal L} = - \lambda_2 \phi^{\dagger}\phi \phi'^{\dagger}\phi' \ .
\label{portal}
\end{eqnarray}
The possible effects of this interaction have been discussed in a number 
of papers \cite{flv,flv2,henry,chacko,hall,sashav,chin2,archilfv}
\footnote{The Higgs portal coupling and kinetic mixing interaction have also
been discussed in the context of more general hidden sector dark matter models, for a flavour of such work 
see for example \cite{mambrini,foothidden,foothidden2}.}.
We shall summarize  some of the main results here.

The complete Higgs potential, including the above portal coupling, is:
\begin{eqnarray}
V(\phi, \phi') & = & -\mu^2 \left( \phi^{\dagger} \phi + \phi'^{\dagger}\phi' \right) + 
\lambda_1 \left[(\phi^{\dagger} \phi)^2 + (\phi'^{\dagger} \phi')^2\right] \nonumber\\
& + & \lambda_2 \phi^{\dagger}\phi \phi'^{\dagger} \phi' \ .
\end{eqnarray}
This potential can be minimized to obtain the non-trivial vacuum:
\begin{eqnarray}
\langle \phi \rangle &=& \langle \phi' \rangle = 
\left(\begin{array}{c} 0 \\
{v \over \sqrt{2}}
\end{array}\right), \ \ {\rm for\ } \lambda_1 > 0, \  \ |\lambda_2| < 2\lambda_1 
\label{first}
\end{eqnarray}
where $v = \sqrt{2\mu^2 \over 2\lambda_1 + \lambda_2} \simeq 246$ GeV.
There is a second possible vacuum, one with $\langle \phi \rangle \neq 0, \ \langle \phi' \rangle = 0$, 
in which the mirror symmetry is spontaneously broken \footnote{
When QCD effects are taken into account $\langle \phi' \rangle$ is perturbed away
from zero, but is still very small: $\langle \phi' \rangle \sim \Lambda_{QCD}^3/m_{\phi'}^2$
where $\Lambda_{QCD} \sim 200$ MeV \cite{spon1}.}.
This broken phase occurs for
a distinct region of parameter space, namely $\lambda_1 > 0$, $\lambda_2 > 2\lambda_1$. 
The phenomenology of this second solution is clearly quite different,
and has been discussed in several papers \cite{spon1}. Next to minimal models, with additional singlet scalar(s) and/or soft 
breaking terms have also been considered in the literature. Such models can accommodate 
$\langle \phi' \rangle \gg \langle \phi \rangle$ \cite{spon2}, or 
$\langle \phi'\rangle \sim \langle \phi \rangle $ \cite{hall,chinese}.
The mirror dark matter discussed in this review refers to
the theoretically unique case where mirror symmetry is completely unbroken. 
As discussed 
above, this assumes the minimal scalar content with $|\lambda_2| < 2\lambda_1$ 
so that the parity conserving vacuum, Eq.(\ref{first}), results.

Expanding the potential around the parity conserving vacuum allows one to identify the two mass eigenstate Higgs fields: 
$H_1$ and $H_2$. These states are maximal combinations
of the weak eigenstates:
\begin{eqnarray}
H_1 = {\phi_0 + \phi'_0 \over \sqrt{2}}, \ 
H_2 = {\phi_0 - \phi'_0 \over \sqrt{2}}
\end{eqnarray}
where $\phi_0$ and $\phi'_0$ are the real parts of the neutral components of $\phi$ and $\phi'$, respectively.
The two states, $H_1$ and $H_2$, have definite exact mirror parity, with $H_1$ being even while $H_2$ is odd.
When the Lagrangian, Eq.(\ref{lag1}), is rewritten in terms of $H_1$ and $H_2$, one finds that $H_1$ and $H_2$ each
couple to ordinary fermions and gauge bosons similar to the standard model Higgs, but with coupling reduced by $1/\sqrt{2}$\ \cite{flv2}.
Whether or not this is observable  depends on the mass difference between $H_1$ and $H_2$. 
The masses of $H_{1,2}$ are:
\begin{eqnarray}
m_{H_1}^2 = 2v^2 \left(\lambda_1 + {\lambda_2 \over 2}\right), \ 
m_{H_2}^2 = 2v^2\left(\lambda_1 - {\lambda_2 \over 2}\right).
\end{eqnarray}
We see that the effect of the Higgs portal coupling [Eq.(\ref{portal})] is
to break the mass degeneracy. 
The mass difference $|m_{H_1} - m_{H_2}|$, is given by
\begin{eqnarray}
|m_{H_1} - m_{H_2}| &=& \sqrt{2} v \left| \sqrt{\lambda_1 + {\lambda_2 \over 2}} -
\sqrt{\lambda_1 - {\lambda_2 \over 2}}\right|,  \nonumber \\
&\simeq & {|\lambda_2| v \over \sqrt{2 \lambda_1}} \ \ {\rm for} \ \ |\lambda_2| \ll \lambda_1 \ .
\label{20a}
\end{eqnarray}
The rough consistency of the Higgs-like resonance discovered at the 
LHC \cite{lhchiggsdiscovery1,lhchiggsdiscovery2} with standard model expectations  already
puts restrictions on $|m_{H_1} - m_{H_2}|$. This mass difference must be less than the Higgs decay width otherwise
the two states will be produced incoherently 
\footnote{
For the Higgs mass difference
to be less than the Higgs decay width requires
small values of $\lambda_2$. Small values of $\lambda_2$ (and also kinetic mixing, $\epsilon$) 
are  technically natural as 
the limit $\lambda_2 \to 0$, $\epsilon \to 0$ corresponds to the decoupling of the ordinary and mirror
sectors. There is consequent increase in symmetry in this limit (cf. \cite{kristian}) as one can perform independent
Poincar$\acute{e}$ 
symmetry transformations on the ordinary sector and mirror sector separately.}.
Incoherent $H_1, \ H_2$ production 
leads to a large deviation from standard Higgs physics \cite{flv2,chin2,archilfv}, which
is already excluded. 

Coherent $H_1, \ H_2$ production occurs when
$|m_{H_1} - m_{H_2}| \stackrel{<}{\sim} \Gamma_{H}$, where $\Gamma_{H} \simeq 4$ MeV is the standard model Higgs decay width \cite{higgswidth}.
In this parameter region, the weak eigenstate, $\phi_0$, is produced and starts to oscillate
into the mirror state $\phi'_0$ (the discussion below closely follows the treatment of \cite{archilfv}).
The oscillation probability is then:
\begin{eqnarray}
P(\phi_0 \to \phi'_0) = \sin^2 \left( {t \over 2t_{\rm osc}} 
\right)
\end{eqnarray}
where $t_{\rm osc} = 1/|m_{H_1} - m_{H_2}|$ is the oscillation time in the non-relativistic 
limit.
The average oscillation probability to the mirror state, which determines the invisible decay width,
is given by
\begin{eqnarray}
\langle P(\phi_0 \to \phi'_0) \rangle 
&=& \Gamma_H \int_0^{\infty} e^{-\Gamma_H t} \ \sin^2 \left({t \over 2t_{\rm osc}}
\right)
\ dt \nonumber \\
&=& {1 \over 2} \left( {1 \over 1 + \Gamma_H^2 t_{\rm osc}^2}\right).
\label{23a}
\end{eqnarray}
Evidently in this coherent production regime the branching fraction to invisible channels is always less than $50\%$.
The oscillations also modify the cross-sections into visible channels. These cross-sections are reduced by the factor $f$, where
\begin{eqnarray}
f &=& 1 - \langle P(\phi_0 \to \phi'_0)\rangle \nonumber \\
&=& {1 \over 2} + {1 \over 2} \left( {\Gamma_H^2 t_{\rm osc}^2 \over 1 + \Gamma_H^2 t_{\rm osc}^2} \right).
\end{eqnarray}
Observe that the Higgs physics becomes indistinguishable from that of the standard model in the limit where $t_{\rm osc} \to \infty$.
This occurs when $|m_{H_1} - m_{H_2}| \to 0$, or 
equivalently when $\lambda_2 \to 0$.

What is the experimental limit on $\lambda_2$ from collider data?
Ref. \cite{higgsinv} studied the standard model Higgs augmented with invisible decay modes.
There, they found that LHC and Tevatron data implied a limit on the branching ratio: $Br(H \to {\rm invisible}) < 0.23$ at $95\%$ C.L.
Setting $\langle P(\phi_0 \to \phi'_0) \rangle < 0.23$, and using Eq.(\ref{23a}), it follows that: 
\begin{eqnarray}
{\Gamma_H \over |m_{H_1} - m_{H_2}|} > 1.08 \ \ {\rm at \ \ 95\% \ C.L.}
\ .
\end{eqnarray}
Massaging this expression, using Eq.(\ref{20a}), leads to the limit:
\begin{eqnarray}
\lambda_2 < 7.7\times 10^{-6} \ \ {\rm at \ \ 95\% \ C.L.}
\ .
\end{eqnarray}
This experimental limit can be compared with the cosmological bound $\lambda_2 \stackrel{<}{\sim} 10^{-8}$ \cite{henry,sashav}.
This bound  
arises by demanding that $\phi \phi \to \phi' \phi'$ scattering be small enough so that the mirror sector is not thermalized 
with the ordinary matter sector in the early Universe. Note however that the cosmological limit can be evaded
in inflationary scenarios with low reheating temperature, $T_{rh} \sim 100$ GeV \cite{sashav}.

%\vskip 0.2cm

\subsection{Neutrino - mirror neutrino mass mixing}

%\vskip 0.1cm

Neutrino oscillations have been observed in a variety of experiments which indicates that neutrinos have nonzero masses.
For a review see for example \cite{zuber}.
This means that the standard model will have to be extended in some way
to accommodate 
massive neutrinos. Although the neutrinos have mass, their overall mass scale is sub eV, which is much smaller than
the other fermions in the standard model.
If mirror symmetry is unbroken, then we expect a set of mirror neutrinos, also sub eV mass scale.
They need not be exactly degenerate with their ordinary matter counterparts if there is
mass mixing between ordinary and mirror neutrinos. Such mass mixing
is possible since it does not violate any of the  fundamental
unbroken symmetries of the theory such as $U(1)_{Q}$ of electromagnetism or mirror symmetry.  
Neutrino mass mixing, if it exists, would 
lead to oscillations between the ordinary and mirror neutrinos \cite{flv2,foot94,footvolkas95,silearly,berezhinsky}.

Whether or not neutrino - mirror neutrino mass mixing is expected to occur depends on the mechanism by 
which neutrinos gain their masses.
Here we consider the three simplest seesaw neutrino mass generating models.
These are now called type-I, type-II and type-III seesaw models \cite{ernest}.
In principle the following analysis could be repeated for any other model
generating nonzero neutrino masses.

\vskip 0.3cm
\noindent
{\bf Type-I seesaw}
\vskip 0.3cm
\noindent
In this model, three gauge singlet right-handed neutrinos, $\nu^i_R$, $i=1,..,3$, are added to the standard model \cite{typeI}.
The coupling of these neutrinos to $\nu_L$ is described by the following Lagrangian, restricting here to the first generation for simplicity:
\begin{eqnarray}
{\cal L}_{\nu} = \lambda \bar f_L \phi \nu_R + M \bar \nu_R (\nu_R)^c \ + \ H.c.
\end{eqnarray}
where $(\nu_R)^c$ is the standard CP transformation. [In the Dirac-Pauli representation of the $\gamma$ matrices,
$(\nu_R)^c \equiv \gamma_2 \gamma_0 \nu_R^*$.]
In the Lagrangian above, $\bar f_L = (\bar \nu_e, \bar e)_L$ and $\phi$ is the Higgs doublet field whose neutral component 
develops a non-zero vacuum expectation value, $\langle \phi \rangle$.
If $M \gg \lambda \langle \phi \rangle$, then diagonalization of the resulting $2\times 2$ neutrino mass matrix yields
two Majorana states: $\nu_{light} \sim \nu_L$ and $\nu_{heavy} \sim \nu_R$,
with masses, $m_{light} \simeq {(\lambda \langle \phi \rangle)^2 \over M}$ and 
$m_{heavy} \approx M$.
\vskip 0.3cm
\noindent
{\bf Type-II seesaw}
\vskip 0.3cm
\noindent
For the type-II seesaw model, an electroweak triplet scalar $\Delta$,
is introduced instead of the $\nu_R^i$ \cite{typeII}.
In this case a Yukawa term:
\begin{eqnarray}
{\cal L}_{\nu} = \lambda \bar f_L (f_L)^c \Delta \ + \ H.c.
\end{eqnarray}
generates a Majorana mass for $\nu_L$ when the neutral component of $\Delta$ gains a nonzero vacuum expectation value:
$\langle \Delta^0 \rangle \neq 0$.
\vskip 0.3cm
\noindent
{\bf Type-III seesaw}
\vskip 0.3cm
\noindent
In the type-III seesaw option three fermionic triplets $\Sigma_R^i$, are introduced (instead of $\nu_R^i$).
These states couple to $\nu_L$ in the following way, again restricting to one generation for simplicity \cite{typeIII}:
\begin{eqnarray}
{\cal L}_{\nu} = \lambda \bar f_L \phi \Sigma_R + M \bar \Sigma_R (\Sigma_R)^c\ + \ H.c .
\end{eqnarray}
The resulting neutrino mass matrix has the same form as for the type-I seesaw.

\vskip 0.4cm

For each of these three models we can easily add an isomorphic Lagrangian (with ordinary fields replaced by mirror fields) 
to the mirror sector. As before, 
there is an exact parity symmetry again swapping each ordinary particle with its mirror partner.
An important question arises: Are masses that mix ordinary and mirror neutrinos
allowed in any of these three models?
With the particle content described above, only the type-I seesaw model
can have mass mixing between ordinary and mirror neutrinos.
[For the type-II and type-III seesaw, mass mixing between ordinary and mirror neutrinos is forbidden by
the gauge symmetry of the Lagrangian.]
This arises through terms such as:
\begin{eqnarray}
{\cal L}_{int} = M_{mix} \bar \nu_R (\nu'_R)^c \ + \ H.c.
\label{mix999}
\end{eqnarray}
The above $M_{mix}$ term leads to off-diagonal contributions to the mass matrix describing
neutrinos and their mirror partners.

The effect of mass mixing is to induce oscillations between ordinary and mirror neutrinos.
At one time it was suggested that such ordinary - mirror neutrino oscillations might
be implicated in the atmospheric and solar neutrino observations \cite{flv2,foot94,footvolkas95}. Experiments have shown that this is
not the case;  the solar, atmospheric, and long-baseline neutrino experiments can
all be accounted for with just the three ordinary neutrinos (see for example the review \cite{neureview}).
Some anomalies remain, but it seems unlikely that they could be explained with mirror neutrino oscillations,
unless the mirror symmetry was broken in some way (see \cite{gu} for some
recent work in this direction). 
The conclusion is that on length scales probed by the solar, atmospheric, and long-baseline neutrino experiments, 
there is no convincing evidence for any oscillations into mirror neutrinos. 
Thus, either the mass mixing between ordinary and mirror neutrinos is zero, as occurs in e.g. the type-II
and type-III seesaw models, or it is small. Small mass mixing is possible in the type-I seesaw model, and the 
experiments could be used to place an upper limit on the parameters $M_{mix}$ in Eq.(\ref{mix999}). 
A more sensitive probe of neutrino - mirror neutrino mass mixing could come from measurements 
of energetic neutrinos of astrophysical origin by experiments such as IceCube and ANTARES.
Indeed, ref.\cite{josh} points out that such oscillations would modify the flavour ratios of the observed
neutrinos away from standard expectations of $\nu_e : \nu_\mu : \nu_\tau = 1 : 1 :1$.
Experiments could thereby find evidence for, or against, ordinary - mirror neutrino oscillations,
which might tell us something about the neutrino mass generation mechanism.

Oscillations of ordinary neutrinos into mirror neutrinos can also be important for cosmology,
modify BBN etc., 
and potentially have also astrophysical implications.
Some cosmological effects of neutrino oscillations were discussed in the context of the now disfavored solutions
to  the atmospheric and solar neutrino anomalies involving oscillations of ordinary neutrinos into 
mirror neutrinos \cite{fvcos}. Cosmological effects of oscillations of the 
heavy $\nu_R$ Majorana fermions were also 
considered in \cite{bellnicole} and  some astrophysical applications of neutrino oscillations into
mirror neutrinos were discussed in \cite{blinn}.

\vskip 0.6cm

\subsection{Higher dimensional effective operators in ${\cal L}_{mix}$?}

\vskip 0.3cm

So far, it must be said, the discussion has been conservative. We have only examined the consequences of 
the two renormalizable interactions
given in Eq.(\ref{mix}), and briefly considered also neutrino mass mixing.
It is possible that we might be lucky. 
If nature is more liberal there could be interesting TEV scale physics connecting
ordinary particles with their mirror counterparts. A common gauge interaction perhaps,
coupling equally to both ordinary and mirror particles.
The LHC signatures of this type of $Z'$ interaction has been discussed in the context
of more generic hidden sector dark matter models in \cite{mamb2}. 
Even if such interactions are not (yet) directly observable at the LHC, they might be probed indirectly
through effective interactions inducing mixing of some of the known neutral particles 
with their mirror counterparts.
We consider here two 
examples that have been discussed in the literature.

\vskip 0.8cm
\noindent
{\bf Neutron oscillations into mirror neutrons}
\vskip 0.4cm

%%%% last bit %%%%

Effective interactions leading to
neutron - mirror neutron ($n-n'$) mass mixing  and thus to $n-n'$ oscillations are an interesting 
possibility \cite{nber1}.
Such oscillations could have important implications for cosmic ray physics \cite{nber2} and big 
bang nucleosynthesis \cite{french}.
Neutron - mirror neutron mass mixing doesn't violate any of the mirror, gauge or Lorentz unbroken symmetries,
however it does require the generation of a dimension nine operator: \footnote{
The neutron - mirror neutron mass mixing operator [Eq.(\ref{n3})] 
does violate the separate global $U(1)$ symmetries generated by baryon number $B$, 
and mirror-baryon number $B'$,  but conserves a {\it diagonal} $U(1)$ subgroup.
Baryon number is usually considered to be an accidental symmetry of the standard model, so
its violation in models beyond the standard model is possible, and of no cause for concern. Naturally, one must check that the
underlying particle physics model which produces the effective operator in Eq.(\ref{n3}) is
consistent with constraints such as proton lifetime bounds.}
\begin{eqnarray}
{\cal L}_{n n'} = {1 \over \Lambda_n^5 } u d d u' d' d' \ + \ H.c.
\ 
\label{n3}
\end{eqnarray}
While free neutrons can oscillate into mirror neutrons,
bound neutrons in nuclei cannot oscillate because of the negative nuclear binding energy. 
Experiments searching for free neutron disappearance provide only weak limits on the scale $\Lambda_n$: 
$\Lambda_n \stackrel{>}{\sim} 10$ TeV \cite{neutronexp}.
At first sight such a higher dimensional operator might be unexpected, nevertheless
it has been demonstrated in \cite{nmoh} that
renormalizable 
models generating the effective operator in Eq.(\ref{n3}), with a value of $\Lambda_n \sim 10$ TeV, are possible.

\vskip 0.5cm
\noindent
{\bf Muonium oscillations into mirror muonium}
\vskip 0.5cm

Muonium, the short lived $e \bar \mu$ bound state, might offer another means to probe
exotic higher dimensional effective couplings between the ordinary and mirror particles \cite{serg}.
In particular, an effective operator of the form:
\begin{eqnarray}
{\cal L}_{M M'} =  {1 \over \Lambda^2_{MM'}}
\bar{\mu}\gamma^{\lambda}(1 + \gamma_5)e 
\bar{e}'\gamma_{\lambda}(1 - \gamma_5)\mu' \ + \ H.c.
\label{serg}
\end{eqnarray}  
will induce mass mixing between muonium and its mirror partner.
Such mass mixing fascilitates oscillations of muonium into mirror muonium -
which can lead to a potentially detectable invisible decay of muonium. 
Such experiments are feasible and could greatly improve over existing limits or possibly find new physics.

\vskip 0.8cm 

Novel ideas for interactions coupling ordinary and mirror particles have been discussed in 
the context of large extra dimensions in \cite{ED1,ED2,ED3}. 
Within such theories, it may be possible to interpret mirror particles as ordinary particles on a distinct slice of space-time
within a higher dimensional space. Details aside, such a framework can lead to
massive Kaluza-Klein (KK) states which couple ordinary particles to mirror particles. If such KK induced interactions are not too feeble,
they could potentially lead to interesting missing energy signatures at the LHC.

\vskip 0.5cm

\newpage
\subsection{Generalized mirror models}
\vskip 0.1cm

Generalized mirror models, where more than one additional
sector of particles is postulated, have been discussed in \cite{gen1,gen2}.
Consider the standard model plus $(N-1)$ {\it copies}, that is a total of $N$ sectors:
\begin{eqnarray} 
{\cal L} = {\cal L}_{SM}^{1} + {\cal L}_{SM}^{2} + ... + {\cal L}_{SM}^{N} 
\ .
\end{eqnarray}
The {\it copies} of the standard model can be either {\it mirror} copies where the chirality is flipped,
or non-mirror copies where the chirality is not flipped. In fact,
one could imagine having $p$ ordinary and $q$ mirror sectors.  In the special case where $p = q$ the full Lagrangian
is left-right symmetric and an exact parity symmetry ${\cal P}$, can be defined, analogous to Eq.(\ref{msym}).
[The exact parity symmetric model corresponds to $p = q = 1$.]
In the case where $p=q$, the total discrete symmetry of the model is:
\begin{eqnarray}
C_p \otimes C_p \otimes {\cal P}
\end{eqnarray}
where $C_p$ is the group of permutations of $p$ objects. As with the minimal $N=2$ case, one can show that there exists a
large range of parameters where this discrete symmetry is unbroken by the vacuum \cite{gen2}.

For $N$ not too large, say less than around 10 or so, the dark matter phenomenology is broadly similar to the simplest
$N=2$ case. As $N$ increases, the main effect is to make the dark matter less dissipative and less self interacting,
as it can be harder for particles of the same {\it copy} to find each other. Of course
the details will depend on the matter mass fraction of the Universe which is contributed by each sector \footnote{
If each sector has the same number of baryonic particles, i.e. contributes the same mass fraction to the
Universe's matter budget, then $N=6$ is suggested
from the inferred matter abundance (from e.g. CMB observations): $\Omega_{m}/\Omega_b \approx 6$. 
The $N=6$ case, though, is (perhaps) not as well motivated as first impressions might indicate.
It requires production of identical baryonic asymmetry in each sector in the early Universe, yet one
still needs temperature asymmetries to be generated. After all, the energy density during the BBN epoch is radiation dominated 
and strongly constrained.
Having sectors with identical baryonic asymmetries and asymmetric temperatures, although certainly conceivable, seems non-trivial to realize.}.
The end result though, is that
galactic halos need not require such a large heat source to stabilize them from collapse.  
Indeed, for $N$ very large, say $N \gg 10$, dark matter halos might not require any heat source at all.
The dark matter phenomenology of the very large $N$ case is therefore quite different.
See \cite{dvali}
for further discussions and novel motivations for considering the case of very large $N$.

\vskip 0.6cm

Besides mirror symmetry, there are a few other potential space-time symmetries which might have something to do with particle interactions.
In particular, scale invariance and supersymmetry have both been discussed in the literature. 
However, unlike mirror symmetry, both scale invariance and supersymmetry 
appear to be broken symmetries. Such theories do require the existence of additional
particles, but none of these are guaranteed to be stable
unless further assumptions are made.  Scale invariance or supersymmetry may therefore have no direct relevance for dark matter.
Such theories could be important in other ways, in particular they might help 
explain the stability of the weak scale ($\langle \phi \rangle = 174$ GeV) in the presence of a higher energy physical scale, $\Lambda$.
Although not all high energy scales destabilize the weak scale, some do, and supersymmetry or scale invariance may well be
important. Such ideas have motivated some work on combining scale invariance with mirror symmetry in \cite{scale9}
and recently,  mirror symmetry (albeit softly broken) with supersymmetry in \cite{susy9} (the latter makes use of an interesting $U(4)$ symmetry
limit of the Higgs potential identified in \cite{chacko}).
Of course, both types of theories would predict the same low energy physics if mirror symmetry were unbroken.

Finally, let us note here that there has been some work exploring possible connections between
mirror particles and certain algebraic constructions \cite{pav} and the extra degrees of freedom
in quaternion quantum mechanics \cite{quat}.
% ref order xxx need to fix

\vskip 0.2cm

\subsection{Generic hidden sector dark matter}
\vskip 0.1cm

We conclude this section with a few final observations.
As discussed in the introduction and in this section, mirror dark matter is a very special
hidden sector model. The hidden sector is an exact copy of the standard model sector, which enables
the symmetries of the theory to be extended to include exact improper space-time symmetries: ${\cal P},\ {\cal T}$.
These symmetries connect each ordinary particle with a mirror partner,
and since they are unbroken, all of the properties of the hidden sector are completely specified. There are no
free parameters describing the masses, lifetimes, or self interactions of the mirror particles.
The only unknown parameters are those in ${\cal L}_{mix}$ which couple ordinary particles to mirror particles, and of these,
only the $U(1)_Y-U(1)'_Y$ kinetic mixing interaction appears to be important for the astrophysics and cosmological
applications discussed in this review.

Naturally, it is possible that dark matter arises from a hidden sector where the mirror symmetry, ${\cal P}$, is not
exact, or from a hidden sector not associated with the concept of mirror symmetry. Such models, here labeled as
generic hidden sector dark matter, have been extensively studied in the literature, especially in recent times.
See  \cite{spon2,mamb2,mambrini,foothidden,foothidden2,randal,generic1,generic2,generic2b,generic3,generic4,das66,generic5,gosh,fan,generic6} 
for a partial list.

\vskip 0.2cm
\section{Early Universe Cosmology}

\vskip 0.1cm

Our starting point is Einstein's equation
\begin{eqnarray}
R_{\mu \nu} - {1 \over 2} g_{\mu \nu} R = - {8\pi G_N} T_{\mu \nu} 
\end{eqnarray}
where $R_{\mu \nu}$ is the Ricci tensor, $R \equiv g^{\mu \nu} R_{\mu \nu}$, $T_{\mu \nu}$ is the stress energy tensor, and
$G_N$ is Newton's constant
(in this notation the cosmological constant,  $\Lambda$, is considered as a contribution to $T_{\mu \nu}$).
The usual assumptions of homogeneity and isotropy lead to the Robertson-Walker metric:
\begin{eqnarray}
ds^2 = -dt^2 + a(t)^2 \left[ {dr^2 \over 1 - kr^2} + r^2 d\Omega^2 \right]
\end{eqnarray} 
where $a(t)$ is the {\it scale factor} and $k = 1, -1, 0$ for closed, open or spatially flat Universe.
Henceforth we restrict discussion to the spatially flat case, consistent with precision CMB and other measurements.
Einstein's equation implies that the scale factor satisfies the Friedmann equation, which for a spatially flat Universe is:
\begin{eqnarray}
\left( {\dot{a} \over a}\right)^2 = {8\pi G_N \rho_{c} \over 3} \ .
\label{friedmann}
\end{eqnarray} 
Here $\rho_{c}$ is the total energy density of the Universe, also called the {\it critical} density given our presumption
of spatial flatness.
With the Hubble parameter defined as $H \equiv \dot{a}/a$, the Friedmann equation implies that
\begin{eqnarray}
\rho_c = {3H^2 \over 8\pi G_N}
\ .
\end{eqnarray} 
In the mirror dark matter context, the total energy density of the Universe has
contributions from both ordinary and mirror particles.
At early times, $t \stackrel{<}{\sim} 1$ second (i.e. $T \stackrel{>}{\sim} 1 $ MeV) the energy density was dominated by the relativistic 
species: $\gamma, \nu_\alpha, \bar \nu_{\alpha}, e, \bar e$ and their mirror counterparts:
$\gamma', \nu'_\alpha, \bar \nu'_{\alpha}, e', \bar e'$, $\alpha = e, \mu, \tau$.
In general, the relevant Bose-Einstein/Fermi-Dirac distributions of these particles are described 
by the temperatures $T_\gamma, \ T_\nu$ for the ordinary
particles and $T'_{\gamma}, 
\ T'_{\nu}$ for the mirror particles \footnote{The distributions also depend on chemical potentials. We make the usual assumption
that they are all small and can therefore be neglected.}.
[Electromagnetic interactions and mirror electromagnetic interactions are frequent enough to set $T_\gamma = T_e = T_{\bar e}$
and $T'_{\gamma} = T'_{e} = T'_{\bar e}$ \ .]

Early Universe cosmology can be used to constrain $T'/T$. (Our notation is: $T$ [$T'$] without subscript is the photon [mirror-photon] temperature.)
Within the standard big bang cosmology, 
the observed light element abundances and CMB anisotropies require 
that the relativistic energy density contributed by particles 
beyond the standard model is less than that of around one ordinary neutrino.
Thus, one is led to consider initial conditions with $T' < T$. 
It was recognized some time ago that
if photon - mirror photon kinetic mixing exists, interactions such as $e \bar e \to e' \bar e'$ will occur which
can potentially thermally populate the mirror sector.
The relevant Feynman diagram is shown in figure 3.1.
A very rough bound
of $\epsilon \stackrel{<}{\sim} 3 \times 10^{-8}$ was derived by requiring that
the mirror sector did not come into thermal equilibrium with the
ordinary sector, prior to the epoch of BBN \cite{glas2}.
For smaller $\epsilon$ values
the kinetic mixing induced interactions will heat the mirror sector, but with
$T'<T$.  
Evaluating the evolution of $T'/T$ as a function of $\epsilon$ is clearly an essential step needed
to check the compatibility of the theory with BBN and
CMB/LSS \cite{paolo1,footplb2012}.

\vskip 0.6cm
%\centerline{\epsfig{file=rfigfeynman1.eps,angle=270,width=5.7cm,angle=90}}
\centerline{\epsfig{file=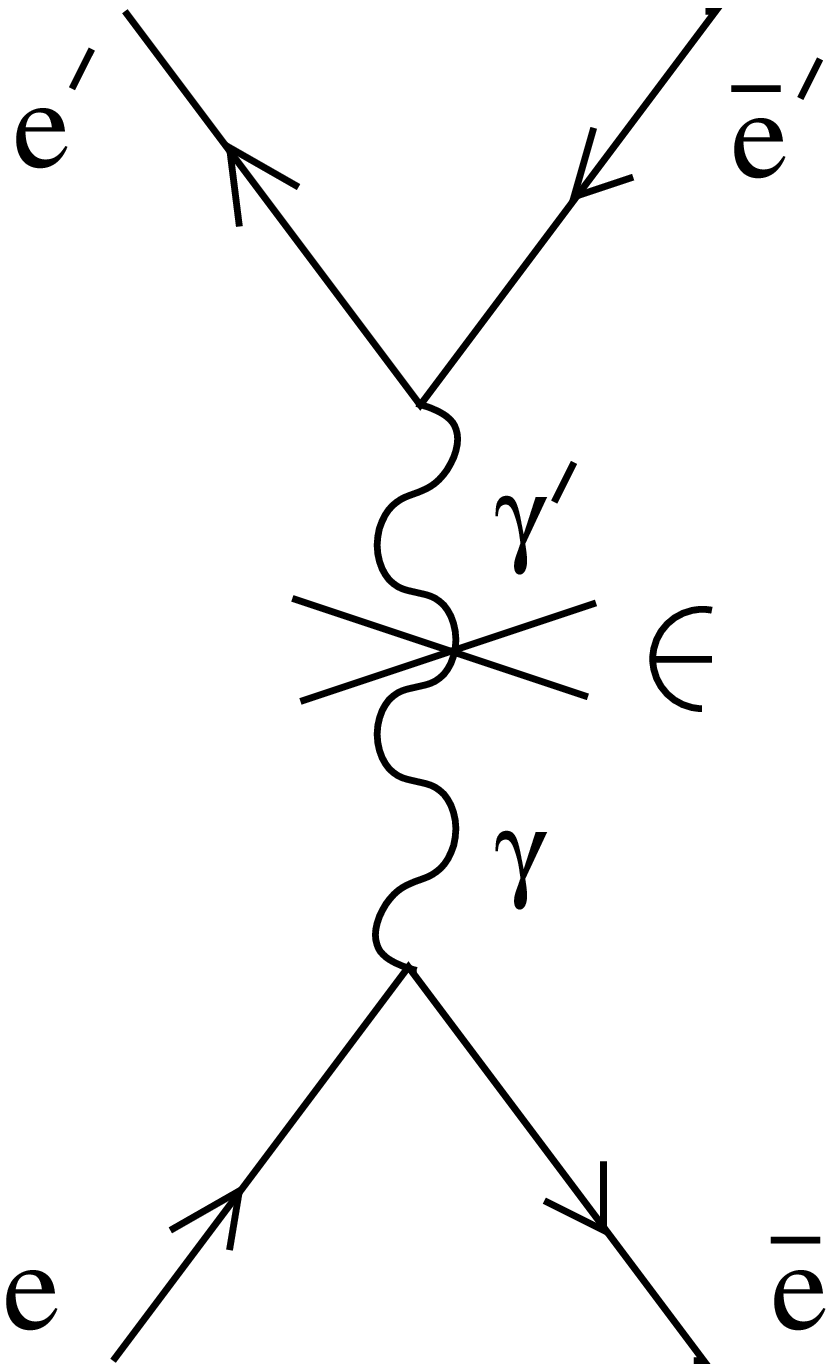,angle=270,width=5.4cm,angle=90}}
\vskip 0.3cm
\noindent
{\small
Figure 3.1: The process $\bar e e \to \bar e' e'$ induced via kinetic mixing, 
treated here as an interaction signified by the cross on the photon propagator.
}

\vskip 1.0cm

\subsection{Evolution of $T'/T$ as a function of $\epsilon$} 

\vskip 0.1cm

In the $\epsilon \to 0$ limit, the ordinary and mirror particles are completely decoupled from each other. 
This means they may have different temperatures: $T, \ T'$.
To proceed further, we define: $T_\gamma$ [$T'_\gamma$] for the temperature of the ordinary 
[mirror] photons and $T_\nu$ [$T'_\nu$] for the temperature of the ordinary [mirror] neutrinos.
We assume effective initial conditions $T'_\gamma , T'_{\nu}  \ll T_\gamma = T_\nu$ 
due to some physics at early times \footnote
{Asymmetric reheating within chaotic inflationary scenarios is one such candidate for this physics \cite{kolb,hodges,others}. 
In such models it is possible that only the ordinary matter
is reheated after inflation, leading to the initial condition: 
$T'_\gamma , T'_\nu \simeq 0$ and $T_\gamma = T_\nu = T_{RH}$.}.
With these initial conditions it is reasonably safe to neglect the  $\nu'$ contribution to the relativistic energy density 
since $T'_\nu \ll T'_\gamma$ in the period of
interest. 
This is because entropy generation in the mirror sector
occurs mainly in the low temperature region: $T \stackrel{<}{\sim}$ 10 MeV,
during which time
the mirror weak interaction rate is always 
much less than the expansion rate: $G_F^2 T'^5 \ll \sqrt{G_N} T^2$. \footnote{
Instead of assuming that $T' \ll T$ is the effective initial condition,
one could imagine having nonzero $T'/T$ initially. In this case
$T'_\nu$ need not be negligible for $T'/T \stackrel{>}{\sim} 0.2$ initially.
Such initial values of $T'/T$ 
would lead to more stringent limits on the kinetic mixing strength, $\epsilon$. 
To keep our analysis simple, though, we have not considered this possibility.
Note however, that the possible effects for early Universe 
cosmology of a nonzero $T'/T$ initial condition, in the absence of kinetic mixing, have
have been studied in the literature in \cite{ber1,ig1,lss,previous,lss4}. See also the
reviews \cite{ber69,creview}.}

For $T_\gamma < 100$ MeV, $\bar e e \to \bar e' e'$ is the dominant process which generates entropy in the 
mirror sector.
The (spin-averaged) cross-section for this process is:
\begin{eqnarray}
\sigma = {4\pi \over 3} \alpha^2 \epsilon^2 {1 \over s^3} (s + 2m_e^2)^2\ .
\end{eqnarray}
Here $\alpha \equiv e^2/4\pi$ is the fine structure constant, $m_e$
is the electron mass, and $s$ is the Lorentz invariant Mandelstam variable ($\sqrt{s}$ is the total combined energy of the $e$ and $\bar e$ 
in the center-of-mass frame).
Considering a comoving volume $R^3$, the
rate at which energy is transferred to the mirror sector is: \footnote{
The mirror particle number densities of $\bar e', e'$ are
always much less than that of $\bar e, e$ for the $\epsilon$ parameter space consistent with BBN and other 
observations. It follows that the correction to the energy transfer rate due to the 
back reaction: $\bar e' e' \to \bar e e$ is always very small, and for this reason it need not be included in this analysis.}
\begin{eqnarray}
{dQ \over  dt} =
R^3 n_{e} n_{\bar e} \langle \sigma v_{M\o l} {\cal E} \rangle \ 
\label{qyiy}
\end{eqnarray}
where the brackets $\langle ... \rangle$ denote the appropriate average over the momentum distributions of $e$ and $\bar e$.
Here ${\cal E}$ is the energy transferred in the process 
$\bar e e \to \bar e' e'$,
$v_{M\o l}$ is the M\o ller velocity (see \cite{gondolo} and references therein),
and $n_{e}$ 
[$n_{\bar e}$]  
is the total number density of electrons [positrons] (i.e. including both spin states): 
\begin{eqnarray}
n_{e} \simeq n_{\bar e} = {1 \over \pi^2} \int^{\infty}_{m_e} { \sqrt{E^2 - m_e^2} \ E
\over  1 + exp(E/T_\gamma) }  \ dE \ .
\label{bla76}
\end{eqnarray}
The quantity $\langle \sigma v_{M\o l} {\cal E} \rangle$ has been evaluated in \cite{paolo1} 
(with essential help from \cite{gondolo}):
\begin{eqnarray}
\langle \sigma v_{M\o l} {\cal E} \rangle =
{\omega \over 8m^4_e T_{\gamma}^2 [K_2 (z)]^2} \int_{4m_e^2}^{\infty} ds \ \sigma \ (s -
4m_e^2)
\sqrt{s} \int_{\sqrt{s}}^{\infty} dE_+ \ e^{-E_+/T_\gamma} E_+ \sqrt{{E_+^2 \over s}
- 1} 
\nonumber
\\
\
\label{bla25}
\end{eqnarray}
where $z \equiv m_e/T_\gamma$ and $K_2 (z)$ is the modified Bessel function of order two. The quantity $\omega \approx 0.8$
accounts for the effect of various approximations used, such as 
replacing the $\bar e, e$ Fermi-Dirac distribution with the simpler Maxwellian one \cite{paolo1}.

%\newpage

The ordinary particles form one system with temperature $T$ and mirror particles another with temperature $T' < T$ 
(where we have momentarily set aside the difference between $T_\gamma$ and $T_\nu$).
Heat is transferred from
the ordinary-particle system to the mirror-particle one.
Since the self-interaction rate in each of these systems, due to e.g. $\bar e e \to \bar e e$ or $\bar e' e' \to \bar e' e'$, is much larger than the 
transfer rate: $\bar e e \to \bar e' e'$, 
each system remains in equilibrium described by its own temperature.
In this situation, the second law of thermodynamics can be applied to each system. Considering a transfer of heat $dQ$ from
the ordinary-particle system to the mirror-particle one, the entropy change of the ordinary-particle system is $dS = -dQ/T$,
while the entropy change of the mirror particle one is $dS = dQ/T'$. The total entropy of the combined system increases: $dS = dQ(1/T' - 1/T) > 0$
given $T' < T$. The same equations apply to more familiar systems; a textbook example is the cooling of a hot stone in a glass of water \cite{feynman}.
%We now proceed to formalize this heuristic discussion.

To do things properly, one needs to consider separately the neutrinos and $e, \ \bar e, \ \gamma$ as two subsystems
because they are not in equilibrium with each other at low temperatures: $T \stackrel{<}{\sim} 3$ MeV.
The process $\bar e e \to \bar e' e'$  transfers entropy from the $e,\ \bar e,\ \gamma$ subsystem to the $e', \ \bar e', \ \gamma'$
system. The change in entropy of the $e, \ \bar e, \ \gamma$ subsystem is then: 
\begin{eqnarray}
dS = {-dQ \over T_\gamma} \ .
\label{syiy}
\end{eqnarray}
The entropy density of a species $i$, of density $\rho_i$ and pressure $p_i$, is given by \cite{kolbbook}:
\begin{eqnarray}
 s = {\rho_i + p_i \over T_i} \ .
\end{eqnarray}
Use of the above relation and also Eq.(\ref{qyiy}) allows  Eq.(\ref{syiy}) to be rewritten in the form: 
\begin{eqnarray}
{d \over dt} \left[ { (\rho_\gamma + p_\gamma + \rho_e + \rho_{\bar e} + p_e + p_{\bar e} )R^3 \over T_\gamma}\right] 
= - {n_{e} n_{\bar e} \langle \sigma v_{M\o l} {\cal E} \rangle R^3 \over T_\gamma}
\label{1yay}
\end{eqnarray}
where \cite{kolbbook}
\begin{eqnarray}
\rho_\gamma &=& {\pi^2 \over 15} T_{\gamma}^4 \nonumber \\
p_{\gamma} &=&  {\rho_\gamma \over 3} \nonumber \\
\rho_e &=&  \rho_{\bar e} = {T_\gamma^4 \over \pi^2} \int^{\infty}_{z} {(u^2 - z^2)^{1/2} u^2 
\over 1 + e^u} \ du \nonumber \\
p_e &=& p_{\bar e} = {T_\gamma^4 \over 3\pi^2} \int^{\infty}_{z} {(u^2 - z^2)^{3/2} 
\over 1 + e^u} \ du 
\ 
\end{eqnarray}
and recall $z = m_e/T_\gamma$.

In the above discussion, we have neglected the neutrino subsystem.
Actually, there is also a small effect due to
the transfer of heat between the $e, \bar e, \gamma$ subsystem and the neutrino subsystem. However one can check that this small
contribution is indeed negligible.  Furthermore, the second law of thermodynamics applied to the
neutrino subsystem then implies $dS \simeq 0$ for that system. It follows that  $R \propto 1/T_\nu$ is 
always a good approximation in the period of interest.

Similarly, the second law of thermodynamics can be applied to the mirror-particle system:
\begin{eqnarray}
dS' = {dQ \over T'_\gamma} \ .
\end{eqnarray}
That is,
\begin{eqnarray}
{d \over dt} \left[ {(\rho'_\gamma + p'_\gamma + \rho'_e + \rho'_{\bar e} + p'_e + p'_{\bar e} )R^3 \over T'_\gamma}\right] 
=  {n_{e} n_{\bar e} \langle \sigma v_{M\o l} {\cal E} \rangle R^3 
\over T'_\gamma}
\label{2yay}
\end{eqnarray}
where
\begin{eqnarray}
\rho'_\gamma &=& {\pi^2 \over 15} {T_{\gamma}'}^4 \nonumber \\
p'_{\gamma} &=& {{\rho'_\gamma} \over 3} \nonumber \\
\rho'_e &=& \rho'_{\bar e} = {{T_{\gamma}'}^4 \over \pi^2} \int^{\infty}_{z'} {(u^2 - {z'}^2)^{1/2} u^2 
\over 1 + e^u} \ du \nonumber \\
p'_e &=& p'_{\bar e} = {{T_{\gamma}'}^4 \over 3\pi^2} \int^{\infty}_{z'} {(u^2 - {z'}^2)^{3/2} 
\over 1 + e^u} \ du 
\end{eqnarray}
and $z' = m_e/T'_\gamma$.

There is one more equation needed, which is the 
Friedmann equation [Eq.(\ref{friedmann})]:
\begin{eqnarray}
\left({\dot{R} \over R}\right)^2 &=& {8\pi G_N \over 3}\left[ \rho_\gamma + \rho_e + \rho_{\bar e} + \rho_\nu +
\rho'_\gamma + \rho'_e + \rho'_{\bar e}\right] \ 
\label{3yay}
\end{eqnarray}
where $\rho_\nu = {7\pi^2 \over 40} T_\nu^4$ is the total neutrino energy density (i.e. including all three flavours and antineutrinos).

The three equations:
Eqs.(\ref{1yay}),\ (\ref{2yay}) and Eq.(\ref{3yay}) (along with $R \propto 1/T_\nu$) form a closed system which can be numerically solved to 
give the evolution
of $T_\gamma$, $T_\nu$ and $T'_\gamma$.  Let us first check the $\epsilon \to 0$ special case. In that limit, no entropy
is transferred to the mirror sector so that $T'_\gamma = 0$ at all times (assuming that the initial value of $T'_\gamma$ is zero).
Furthermore, in the $\epsilon = 0$ case, the above equations
reduce to the usual equations governing $T_\gamma, \ T_\nu$ evolution as given in e.g. \cite{Weinberg}. 
If $\epsilon \neq 0$ entropy will be transferred to the mirror sector and $T'/T$ will grow
with time. This is illustrated in figure 3.2 which shows the numerical solution 
of the equations for the example with $\epsilon = 10^{-9}$.
In this numerical work we have taken an initial condition $T'/T = 0$ at $T = T^{initial}_{\gamma} = 1$ GeV.
The figure shows the low temperature evolution, which is independent of $T^{initial}_{\gamma}$ so long as $T^{initial}_{\gamma} \gg 10$ MeV.

\vskip 1.0cm
\centerline{\epsfig{file=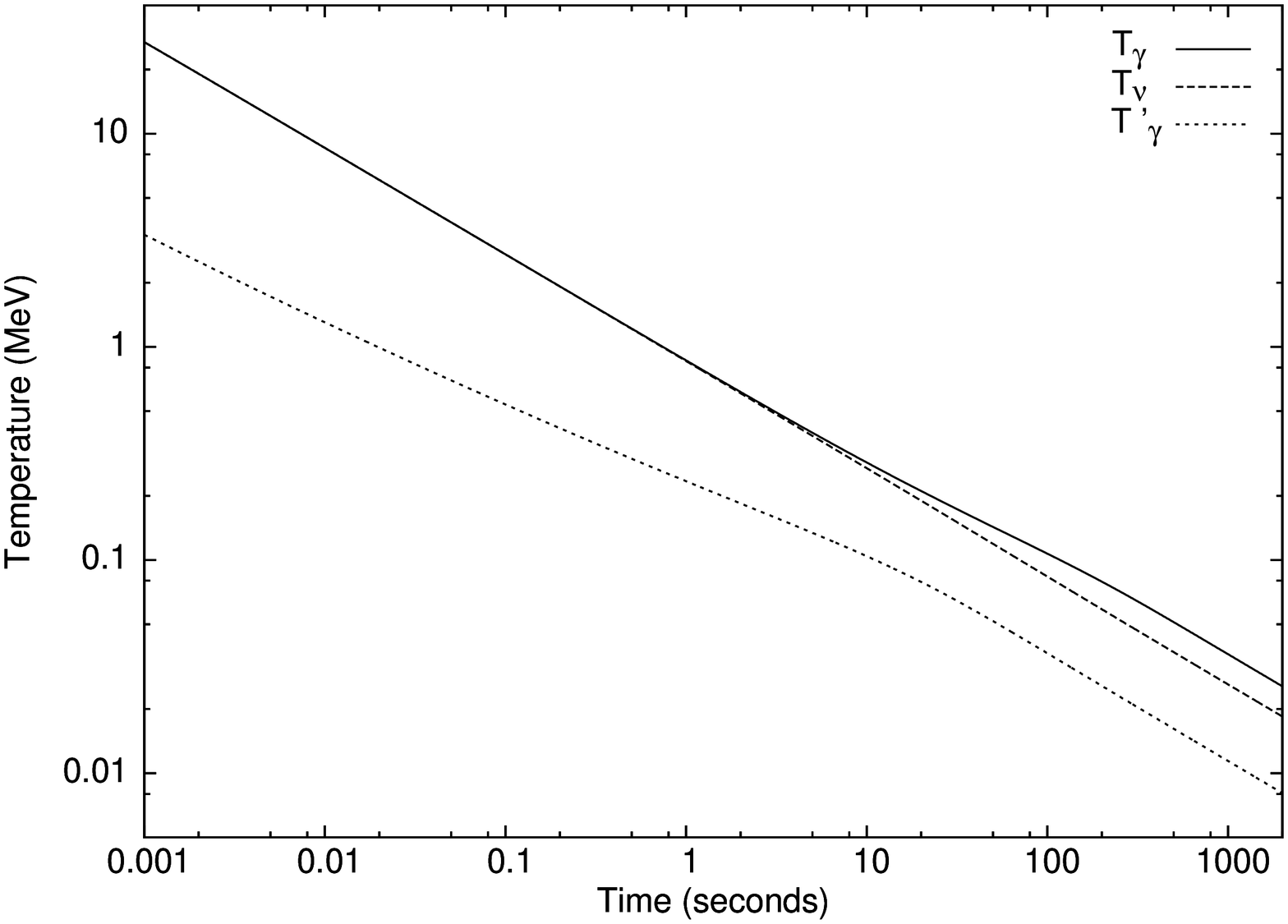,angle=270,width=12.4cm}}
\vskip 0.5cm
\noindent
{\small
Figure 3.2: Evolution of $T_\gamma$ (solid line), $T_\nu$ (dashed line) and $T'_\gamma$ (dotted line) for
$\epsilon = 10^{-9}$.
}

\vskip 1.6cm
\newpage

For this example with $\epsilon = 10^{-9}$, 
$T'_\gamma/T_\gamma$ evolves to $0.31$ as $t \to \infty$.
For more general $\epsilon$
values near $10^{-9}$, it is found numerically that $T'_\gamma/T_\gamma$ evolves to a constant which
satisfies \cite{footplb2012,paolo1}:
\begin{eqnarray}
{T'_\gamma \over T_\gamma} \simeq 0.31\left( {\epsilon \over 10^{-9}}\right)^{1/2} \ \ {\rm as}\ \ t \to \infty
\ .
\label{constant}
\end{eqnarray}
For $t \stackrel{<}{\sim} 1 $ sec, the ratio, $T'_\gamma/T_\gamma$ is slowly varying $\sim (1/T_{\gamma})^{1/4}$
which can be understood analytically, as we shall see in section 3.2.

\subsection{An analytic solution of $T'/T$ for $T > m_e$}

The evolution of $T'/T$, as outlined above, requires solution of three simultaneous differential equations: 
Eqs.(\ref{1yay}),\ (\ref{2yay}) and Eq.(\ref{3yay})  .
Some useful analytic results can be derived assuming (a) the 
massless electron limit, i.e. relevant for $T \gg m_e$ (in this limit $T_\gamma = T_\nu \equiv T$) and (b)
$\rho' \ll \rho$, which is generally expected to be roughly valid if we keep within the one additional effective neutrino energy density limit. 
%Here, $\rho \equiv gT^4_{\gamma} \pi^2/30$ is the total ordinary particle energy density and  
Here, $\rho$ is the total ordinary-particle energy density and  
%$\rho' \equiv g' T'^4_{\gamma} \pi^2/30$ that of the mirror particles. [The quantities: $g$, $g'$ count degrees of freedom,
$\rho'$ that of the mirror particles. 
The ordinary-particle energy density can
be approximated by summing over only those particles with $m_i \ll T$:
\begin{eqnarray}
\rho &=& \left( \sum_B g_B + {7 \over 8}\sum_F g_F \right) {\pi^2 T^4 \over 30}
\nonumber \\
     &\equiv & {g\pi^2 T^4 \over 30} 
\end{eqnarray} 
where $g_B$ ($g_F$) is the number of degrees of freedom 
of each boson (fermion) with $m_i \ll T$.
An analogous  relation defines the mirror-particle energy density: $\rho' \equiv g'T'^4_{\gamma} \pi^2/30$.
Consider evolution during a period of constant $g$ and $g'$.
During such times Eq.(\ref{2yay}) reduces to:
\begin{eqnarray}
{d(\rho'/\rho) \over dt} \simeq  
{n_{e} n_{\bar e} \langle \sigma v_{M\o l} {\cal E} \rangle
\over \rho } \ .
\label{3.2a}
\end{eqnarray}
For high temperatures where $m_e/T \to 0$, the quantities  $\langle \sigma v_{M\o l} {\cal E} \rangle$ and $n_e, \ n_{\bar e}$
have the analytic solution \cite{paolo1}:
\begin{equation}
\langle \sigma v_{M\o l} {\cal E} \rangle = {2\pi \omega \alpha^2 \epsilon^2
\over 3T}\ , \ \ n_e \simeq n_{\bar e} = {3\zeta (3) T^3 \over 2\pi^2}
\ 
\label{ms}
\end{equation}
where $\zeta$ is the Riemann zeta function with $\zeta (3) \simeq 1.202$.
Eq.(\ref{3yay}) (with $R \propto 1/T$) can be used to derive
an approximate time-temperature relationship. Assuming $\rho' \ll \rho$ (and $T \gg m_e$) we have 
$dT/dt = - \sqrt{8\pi G_N \rho/3}\ T$, and
Eq.(\ref{3.2a}) reduces to:
\begin{eqnarray}
{d(\rho'/\rho) \over dT} = {-A \over T^2}
\label{rho}
\end{eqnarray}
where
\begin{eqnarray}
A = {135 \sqrt{5}\ \zeta(3)^2 \ \omega \ \alpha^2 \ \epsilon^2 \over 2\sqrt{G_N}\ \pi^6 \sqrt{\pi}\ g\sqrt{g}  }\ .
\end{eqnarray}
Eq.(\ref{rho}), derived in \cite{paolo1} using a more heuristic line of reasoning, can be analytically solved once $g$ and $g'$ are specified.

Let us now determine $g,\ g'$ for the period: $1 \ {\rm MeV} < T_\gamma < 100$ MeV. 
Again, we can neglect the production of 
$\nu'_{e,\mu,\tau}$ [since $G_F^2T'^5_{\gamma} \ll \sqrt{G_N} T^2 $
in this low temperature region].
This means that,  to a good approximation, the radiation content of the mirror sector consists of just: 
$e', \bar e', \gamma'$,  
while that of the ordinary sector 
contains: $e, \bar e, \nu_\alpha, \bar \nu_\alpha, \gamma$ ($\alpha = e,\mu,\tau$). 
It follows that $g' = 11/2, \ g = 43/4$  and hence
$\rho'/\rho = (g'/g)(T'^4/T^4)$,
with $g'/g \approx 22/43$.
Eq.(\ref{rho}) then has the analytic solution:
\begin{eqnarray}
{T' \over T} &=& \left(\frac{g}{g'}A\right)^{1/4} \left[ {1 \over T} - {1 \over T_{initial}}\right]^{1/4}\ ,
\nonumber \\
 &\simeq & \frac{0.25}{(T/{\rm MeV})^{1/4}} \sqrt{\frac{\epsilon}{10^{-9}}} \ \ \ {\rm for} \ T \ll T_{initial}
\ 
\label{ana}
\end{eqnarray}
where the initial condition $T' = 0$ at $T = T_{initial}$ is assumed.
[$T_{initial}$ is model dependent, e.g. it might be the ordinary-particle reheating temperature in inflationary
scenarios with asymmetric reheating \cite{kolb,hodges,others}.]
In figure 3.3 we compare this analytic solution with the numerical solution of 
Eqs.(\ref{1yay}),\ (\ref{2yay}) and Eq.(\ref{3yay}) for the example with $\epsilon = 10^{-9}$.
This figure clearly illustrates the validity of the analytic solution for $T \stackrel{>}{\sim}$ 1 MeV.

\vskip 0.4cm
\centerline{\epsfig{file=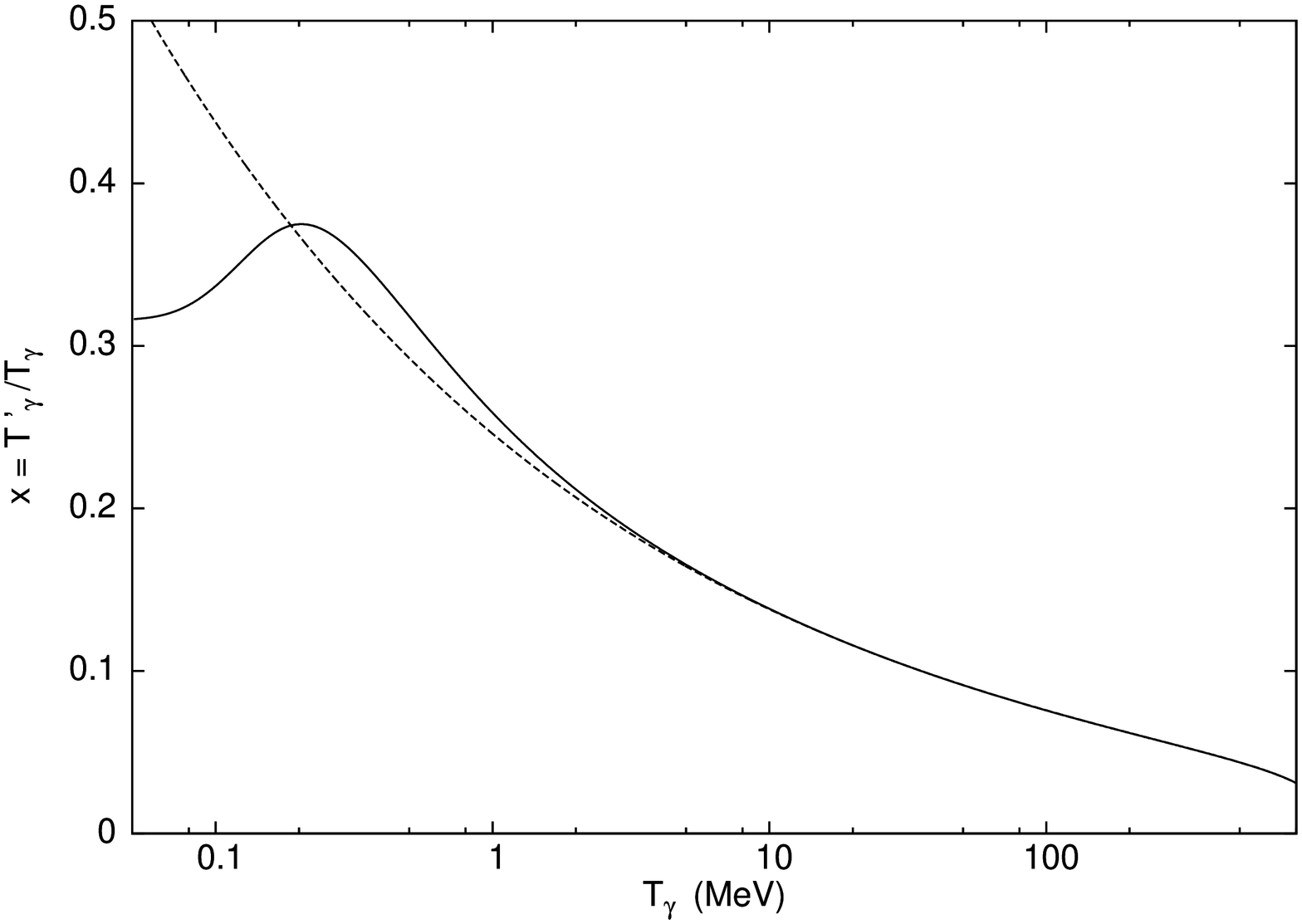,angle=270,width=12.0cm}}
\vskip 0.3cm
\noindent
{\small
Figure 3.3: Evolution of $x \equiv T'_\gamma/T_\gamma$ the (near) exact solution (solid line) in 
comparison with the analytic solution in Eq.(\ref{ana}) (dashed line).
This example assumes $\epsilon = 10^{-9}$. The figure demonstrates the validity of the analytic
solution for the temperature region: $T_\gamma \stackrel{>}{\sim} 1$ MeV.
}

\vskip 1.0cm
\noindent

\subsection{Mirror BBN: The $He'$ abundance}

\vskip 0.3cm

The equations describing the $T'_\gamma,\ T_\gamma, \ T_\nu$ evolution can be used in conjunction with $n' \leftrightarrow p'$
conversion rates to calculate the primordial value
for the mirror-helium mass fraction as a function of the kinetic mixing parameter, $\epsilon$ \cite{paolo2}.
The $He'$ mass fraction is an important quantity to know, e.g. it will be required if one is 
interested to study the formation and evolution of mirror stars,
or to understand the properties of mirror-particle plasmas.

The primordial mirror-helium mass fraction $Y'_{p}$, can be calculated in a similar way to the helium
mass fraction $Y_p$ (for a review of the latter, see for instance \cite{Weinberg}).
Recall that the ordinary helium mass fraction is set by two-body and three-body processes:
$n + \bar e \leftrightarrow p + \bar \nu_e, \ \ n + \nu_e \leftrightarrow p + e, \ \ n \leftrightarrow p + e + \bar \nu_e$.
At high temperature, $T \gg 1$ MeV, the rates of these reactions are much greater than
the expansion rate of the Universe and they drive the neutron to proton ratio to unity.  As
the temperature drops to around 1 MeV, the neutron - proton mass
difference leads to a larger proportion of protons. At a temperature of $T = T_{freeze} \sim 0.8$ MeV, 
the neutron/proton ratio is `frozen'  as
the two-body reactions become less frequent than the expansion rate of the Universe. 
Only neutron decays can further modify this ratio,
which occur until deuterium formation at
$T \sim 0.07$ MeV. The end result is that around 25\%  of the baryons are converted into helium,   
75\%  into hydrogen, with
trace amounts of other light elements. 

Mirror nucleosynthesis proceeds in a similar manner.
The main difference is that  
mirror BBN occurs somewhat earlier than ordinary BBN given
$T' < T$.  At earlier times,  the expansion rate is greater so that the mirror-neutron/mirror-proton ratio 
freezes-out at a higher temperature, $T'_{freeze} > T_{freeze}$. For this reason, 
and also because there is insufficient time
for mirror neutrons to decay, the mirror-neutron/mirror-proton ratio is expected to be much closer to 
unity \cite{ber1,previous}.

As discussed earlier, we may
assume $T'_{\nu} \ll T'_{\gamma}$, since the 
process: $\bar e e \to \bar e' e'$ is important only for temperatures where
the mirror weak interaction rate is always much less than the expansion rate: $G_F^2 T'^5 \ll \sqrt{G_N} T^2$.
It follows therefore that the only two-body reactions needed to compute $Y'_{p}$ are 
\begin{eqnarray}
n' + \bar e' \to p' + \bar \nu', \ \ \ p' + e' \to n' + \nu' ~ .
\label{reactions}
\end{eqnarray}
Also, we can neglect mirror neutron decay $n' \to p' + e' + \bar \nu'$, since mirror BBN occurs during the first
$\sim$ 10 seconds, i.e. on a much shorter time scale than the free $n'$ lifetime 
(which by mirror symmetry is identical to the free neutron lifetime: $\tau_n \approx 881$
seconds).
The reaction rates of the above processes (\ref{reactions}) can be adapted from standard relations given in 
\cite{Weinberg}, which can be further simplified by neglecting the Pauli blocking effect on neutrinos (as $T'_{\nu} \ll T'_\gamma$):
\begin{eqnarray}
&&\lambda_{n'\rightarrow p'} = \lambda(n'+ \bar e' \to p' + \bar \nu') 
  = B\int_0^\infty E_{\nu'}^2 p_{e'}^2 dp_{e'} [e^{E_{e'}/T'} + 1]^{-1} 
\nonumber \\
&&\lambda_{p'\rightarrow n'} = \lambda(p'+e' \to n' + \nu') 
%  = B\int_{(Q^2-m_e^2)^{1/2}}^\infty E_{\nu'}^2 p_{e'}^2 dp_{e'} 
  = B\int_{\sqrt{Q^2-m_e^2}}^\infty \ E_{\nu'}^2 p_{e'}^2 dp_{e'} 
    [e^{E_{e'}/T'} + 1]^{-1} 
\label{1}
\end{eqnarray}
where 
\begin{eqnarray}
B = \frac{G_{F}^2 (1+3g_A^2) \cos^2\theta_C}{2 \pi^3} \ .
\end{eqnarray}
Here
$G_{F} = 1.166 \times 10^{-5} \ {\rm GeV^{-2}}$ is the Fermi constant, $g_{\rm A} = 1.257$ is the axial
vector coupling relevant for these beta decay processes, and $\theta_{\rm C}$ is the Cabibbo angle ($\cos^2 \theta_C \simeq 0.95$).
For $n' + \bar e' \to p' + \bar \nu', \ 	Q = E_{\nu'} - E_{e'} $, and for $p' + e' \to n' + \nu', \ Q = E_{e'} - E_{\nu'}$, 
where $Q \equiv m_n - m_p = 1.293$ MeV.
The range of the integrations in Eqs.(\ref{1}) are fixed by considering that the integrals are taken over all 
kinematically allowed positive values of $p_{e'}$.

The ratio of mirror neutrons to mirror nucleons $X_{n'}$, is governed by the rate equation:
\begin{eqnarray}
{dX_{n'} \over dt} = \lambda_{p'\rightarrow n'} (1 - X_{n'}) 
                   - \lambda_{n'\rightarrow p'} X_{n'} ~~.
\label{eq:Xn}
\end{eqnarray}
Given that, to a good approximation, all available mirror neutrons go into forming $He'$ 
it follows that 
the primordial mirror-helium mass fraction satisfies: $Y'_{p} \simeq 2X_{n'} (\infty)$.
[The quantity
$X_{n'}(\infty)$ is the asymptotic value ($t \to \infty$)  
of $X_{n'}$.] 
Thus, to determine $Y'_p$ we simply need to solve Eq.(\ref{eq:Xn})
(along with Eqs.(\ref{1yay}),\ (\ref{2yay}) and Eq.(\ref{3yay}) of course)
to obtain $X_{n'}(\infty)$. The appropriate initial condition is $X_{n'}(0)=0.5$.
The result of this numerical work is shown in Figure 3.4.

\vskip 0.4cm

\vskip 0.4cm
\centerline{\epsfig{file=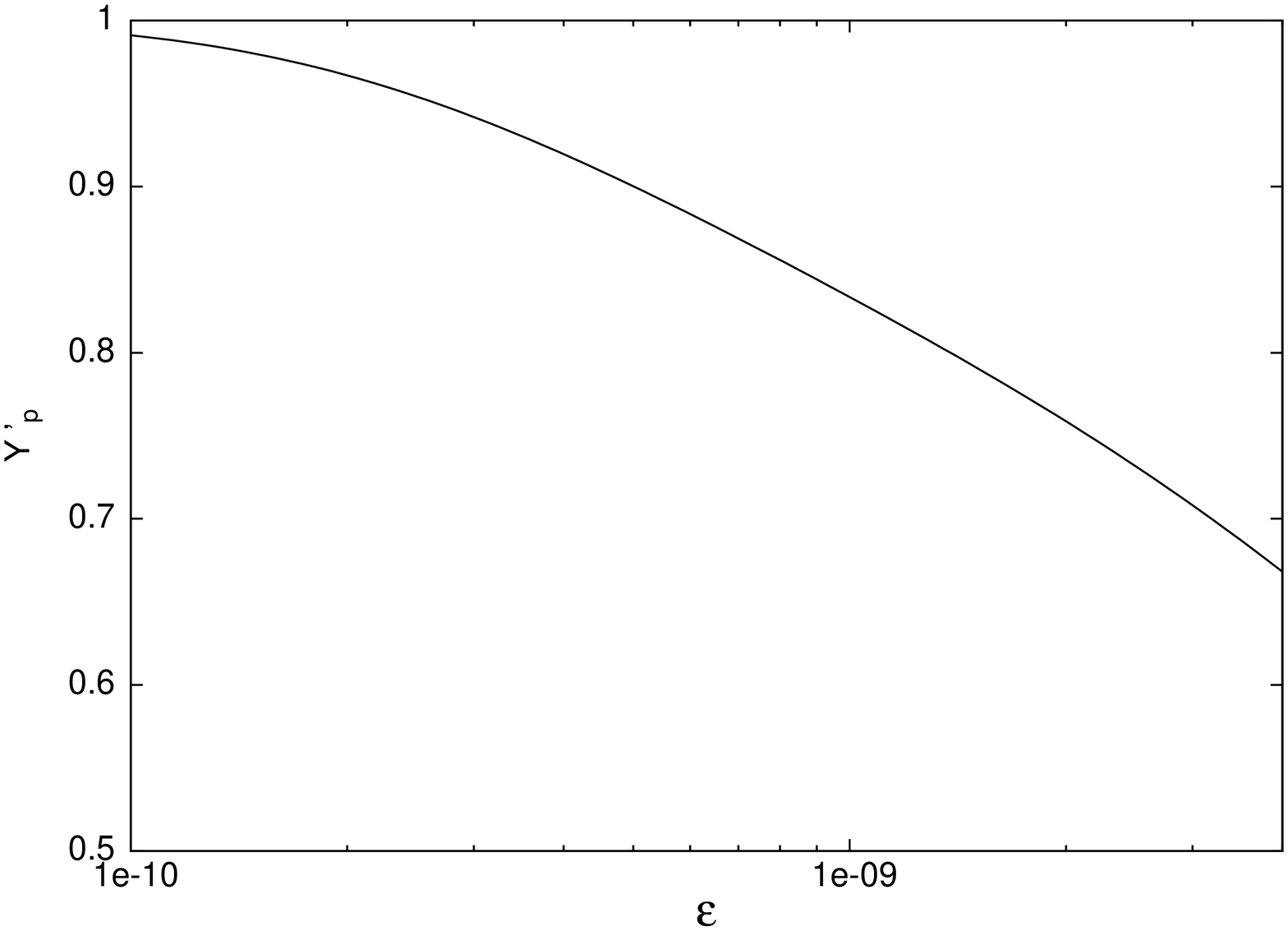,angle=270,width=12.4cm}}
\vskip 0.4cm
\noindent
{\small
Figure 3.4:
Primordial mirror-helium mass fraction ($Y'_{p}$) versus the kinetic mixing interaction strength, $\epsilon$.  }
\vskip 1.8cm

As with ordinary BBN, the early Universe 
production of mirror carbon and heavier mirror elements can 
also occur but are highly suppressed.
These are produced via three-body interactions such as the triple alpha process \cite{wag}:
\begin{eqnarray}
^4 He' \ + \ ^4He'\ +\ ^4He' \to \ ^{12}C' \ + \ \gamma'\ .
\end{eqnarray}
The total mass fraction of $C'$ and heavier elements
produced in the early Universe, here define by $X_{C'}$,  has been estimated to be small: 
$X_{C'} < 10^{-8}$, for $\epsilon \sim 10^{-9}$ \ \cite{paolo2}.
Mirror metals are therefore expected to be synthesized mainly in mirror stars;
this is analogous, of course, to the synthesis of ordinary metals, which occur in ordinary stars.

\vskip 0.4cm
\noindent
{\bf The mean particle mass: $\bar m$}
\vskip 0.4cm
\noindent
Knowledge of the primordial mirror-helium mass fraction allows the
mean mass ($\bar m$) of the particles in a hot mirror particle plasma to be estimated. This is an important parameter
in the equation governing
hydrostatic equilibrium. In fact $\bar m$ sets the temperature of the mirror particle halo of spiral galaxies [Eq.(\ref{10})],
as will be discussed in more detail in section 4.
As such, it also sets the scale of the velocity dispersion of the halo particles [Eq.(\ref{11b})] and thereby influences rates in direct
detection experiments (section 5).

For a homogeneous plasma consisting of fully ionized mirror-helium nuclei,  mirror-hydrogen nuclei, and mirror electrons, the mean mass
is given by:
\begin{eqnarray}
\bar m & \equiv & {\sum n_{A'} m_{A'} \over \sum n_{A'}} \nonumber \\
       & \simeq & {n_{He'}m_{He} + n_{H'} m_{p} \over n_{He'} + n_{H'} + n_{e'}}
\end{eqnarray}
where the mirror electron mass has been neglected relative to the mirror nucleon masses and $m_p$ is the proton mass.
Two more relations follow from the definition of $Y'_p$ and from the $U(1)_{Q'}$ electrical neutrality of the plasma:
\begin{eqnarray}
Y'_p &\simeq & {n_{He'} m_{He} \over n_{He'} m_{He} + n_{H'} m_{p}}\ , \nonumber \\
n_{e'} &=& n_{H'} + 2n_{He'}, \ \ \ {\rm Q' \ \ neutrality}\ .
\end{eqnarray}
Using $m_{He} = 4m_{p}$, the above equations can be solved to obtain:
\begin{eqnarray}
{\bar m \over m_p} \simeq {1 \over 2 - {5 \over 4} Y'_p}\ .
\label{mbar99}
\end{eqnarray}
Thus, for $\epsilon \sim 10^{-9}$ we see from figure 3.4 that $Y'_p \approx 0.9$ and $\bar m \simeq 1.1$ GeV.

\newpage

\subsection{Calculation of $N_{eff}[CMB]$ and $N_{eff}[BBN]$}

\vskip 0.3cm
\noindent 
{\bf $N_{eff}[CMB]$}
\vskip 0.4cm
\noindent
It has become standard to parameterize the
relativistic energy density at the hydrogen recombination epoch in terms of the
{\it effective number of neutrino species}, $N_{eff} [CMB]$, by:
\begin{eqnarray}
\rho_{rad} = \left( 1 + {7 \over 8} \left[ {4 \over 11}\right]^{4/3} N_{eff} [CMB] \right) \rho_\gamma
\label{5yeah}
\end{eqnarray} 
where $\rho_\gamma$ is the CMB photon energy density. 
The $(4/11)^{4/3}$ factor takes into account the heating of the photons due to $\bar e e$ annihilation
while the $7/8$ factor results from the Fermi-Dirac statistics of neutrinos cf. Bose-Einstein statistics of the photons.
The canonical value for $N_{eff}$ is
$N_{eff} \simeq 3.046$; it is marginally larger than three due to the slight heating of the
neutrinos from $\bar e e$ annihilation \cite{mango}. 

We wish to compute  how $N_{eff}$ at the CMB epoch (i.e. at recombination) changes in the presence of the 
kinetic mixing induced process:
$\bar e e \to \bar e' e'$ \cite{footplb2012}. 
One effect of this process is to slightly cool the $\bar e, e$ and thus also the photons. Most of this cooling
occurs after neutrinos have decoupled, and thus the net effect is to increase the energy density 
of the neutrinos relative to that of the photons. This increases $N_{eff} [CMB]$ by
\begin{eqnarray}
\delta N_{eff}^{a} [CMB] =  
3 \left( \left[ {T_\nu (\epsilon) \over T_\nu (\epsilon = 0)}\right]^4  \ - \ 1\right) \
\label{add1}
\end{eqnarray}
where the temperatures are evaluated at photon decoupling,
i.e. when $T_\gamma = T_{dec} = 0.26$ eV.
Additionally, there is also the mirror photon contribution to the energy density, which increases
$N_{eff} [CMB]$ by
\begin{eqnarray}
\delta N_{eff}^{b} [CMB] =   {8 \over 7}\left( {T'_\gamma (\epsilon) \over T_\nu (\epsilon =0)}\right)^4  
\ .
\label{add2}
\end{eqnarray}
Again, the temperatures are evaluated at the photon decoupling time.
The change in  $N_{eff}$ due to kinetic mixing induced interactions, 
$\delta N_{eff} \equiv N_{eff} (\epsilon) - N_{eff} (0)$, is the sum of these two contributions:
\begin{eqnarray}
\delta N_{eff} [CMB] =  
\delta N_{eff}^{a} [CMB] \ +  \ \delta N_{eff}^{b} [CMB]   
\ .
\label{79z}
\end{eqnarray}
The evolution of the various temperatures are described by 
Eqs.(\ref{1yay}),\ (\ref{2yay}) and Eq.(\ref{3yay}).
Numerical solution of these equations
allows the calculation of $\delta N_{eff} [CMB]$; the result is given in figure 3.5. 
Also shown in this figure is the
separate contributions $\delta N_{eff}^{a} [CMB]$ and $\delta N_{eff}^{b} [CMB]$.

\vskip 0.5cm
\centerline{\epsfig{file=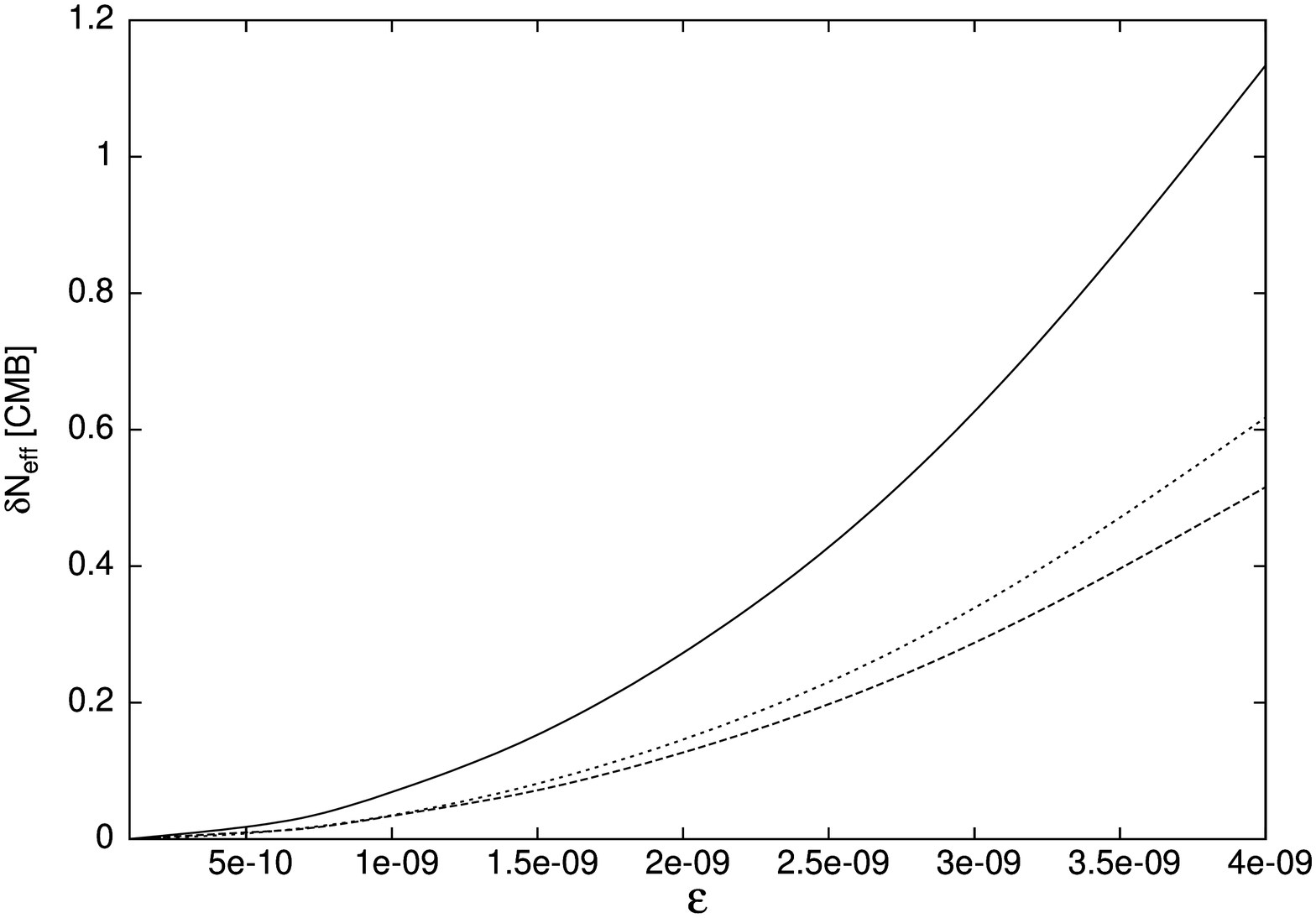,angle=270,width=12.0cm}}
\vskip 0.3cm
\noindent
{\small
Figure 3.5: $\delta N_{eff} [CMB]$ versus $\epsilon$ (solid line). 
The separate contributions, discussed in the text, $\delta N_{eff}^{a} [CMB] $ (dashed line)
and $\delta N_{eff}^{b} [CMB]$ (dotted line) are also shown.
}

\vskip 1.2cm
\noindent

An approximate analytic expression for the  $\delta N^b_{eff} [CMB]$ contribution is given by:
\begin{eqnarray}
\delta N_{eff}^b [CMB] \simeq {8x^4 \over 7} \left( {11\over 4}\right)^{4/3} \simeq 0.041\left({\epsilon \over 10^{-9}}\right)^2\ 
\end{eqnarray}
where $x$ is the asymptotic value for $T'_\gamma/T_\gamma$, obtained from Eq.(\ref{constant}).
Figure 3.5 indicates that $\delta N_{eff}^a [CMB] \approx 0.8\delta N_{eff}^b [CMB]$.

Observations indicate that $\delta N_{eff} [CMB]$ is small but might possibly be non-zero.
For instance, the {\it Planck} collaboration, using their measurements of the CMB combined with Baryon Acoustic oscillation surveys,
find that $N_{eff} = 3.30 \pm 0.27$ \cite{planck55}. Such results, if applicable to the mirror
dark matter model, would suggest a $2 \sigma$ upper limit
on $\epsilon$ of around $\epsilon \stackrel{<}{\sim} 3.5 \times 10^{-9}$ (from figure 3.5).
However, the implications for CMB are more complex than merely an increase in relativistic energy density.
Mirror baryons undergo acoustic oscillations prior to mirror-hydrogen recombination.
%(in much the same way that ordinary baryons undergo acoustic oscillations prior to ordinary hydrogen recombination).
It turns out that the acoustic oscillation effect is generally more important than the effects due to the increase
in relativistic energy density. We will see in section 3.5 that CMB and LSS observations suggest an upper bound on kinetic mixing
of around $\epsilon \stackrel{<}{\sim} 1-2 \times 10^{-9}\ . $

\vskip 1.0cm
\noindent 
{\bf $N_{eff} [BBN]$}
\vskip 0.5cm
\noindent 
The heating of the mirror sector via the process: $\bar e e \to \bar e' e'$ also affects BBN.
Recall that the primordial Helium abundance is determined by the evolution of the neutron and proton
number densities, which evolve as the Universe expands via 
the weak interactions:
\begin{eqnarray}
n + \bar e \leftrightarrow p + \bar \nu_e,\ n + \nu_e \leftrightarrow p + e, \
n  \leftrightarrow p + e + \bar \nu_e.
\label{rates88}
\end{eqnarray}
The rates for these processes depend on $T_\gamma$, $T_\nu$ and are given in standard texts \cite{Weinberg}.
The  primordial Helium abundance can be obtained
by evolving these rates  
down to the deuterium `bottle neck' temperature $T_\gamma = 0.07$ MeV, where   
Eqs.(\ref{1yay}),\ (\ref{2yay}) and Eq.(\ref{3yay}) are used to determine the $T_\gamma, \ T_\nu$ evolution.
This procedure allows the calculation of 
the helium mass fraction for a particular value of $\epsilon$, $Y_p (\epsilon)$ \cite{footplb2012}
\footnote{
The modifications of BBN  due to mirror dark matter with non-zero $T'/T$
arising from an assumed initial condition, instead of due to kinetic mixing, was studied in \cite{previous,creview}.}.
Of course, the standard value, $Y_p (0) \simeq 0.25$,
arises for $\epsilon = 0$. 
It is known that the energy density increase due to one extra neutrino species 
increases $Y_p$ by 0.013 \cite{bbn013}.
Thus, we can parameterize the effect on $Y_p$ by $\delta N_{eff} [BBN]$:
\begin{eqnarray}
\delta N_{eff} [BBN] = {Y_p (\epsilon) - Y_p (0) \over 0.013} \ .
\end{eqnarray}
In figure 3.6 the results for $\delta N_{eff} [BBN]$ versus
$\epsilon$ are given.  Comparison of figure 3.5 with figure 3.6 shows that
$\delta N_{eff} [CMB] > \delta N_{eff} [BBN]$. The diminished effect for BBN happens
because the process: $\bar e e \to \bar e' e'$ 
continues to occur for temperatures somewhat 
below the relevant freeze-out temperature for 2-body $n \leftrightarrow p$ processes: $T_w \approx 0.8$ MeV.

%%%% observations here  %%%%%

Observations currently do not strongly constrain $\delta N_{eff} [BBN]$.
For example, the analysis of \cite{mans} finds an upper limit of $\delta N_{eff} \le 1$
at $95\%$ C.L.
(see also \cite{bbn013} for further discussion).
It follows from figure 3.6 that BBN is currently not very sensitive to the predicted modifications 
due to kinetic mixing. 

%%% indirect evidence positron excess %%%%

\vskip 0.6cm
\centerline{\epsfig{file=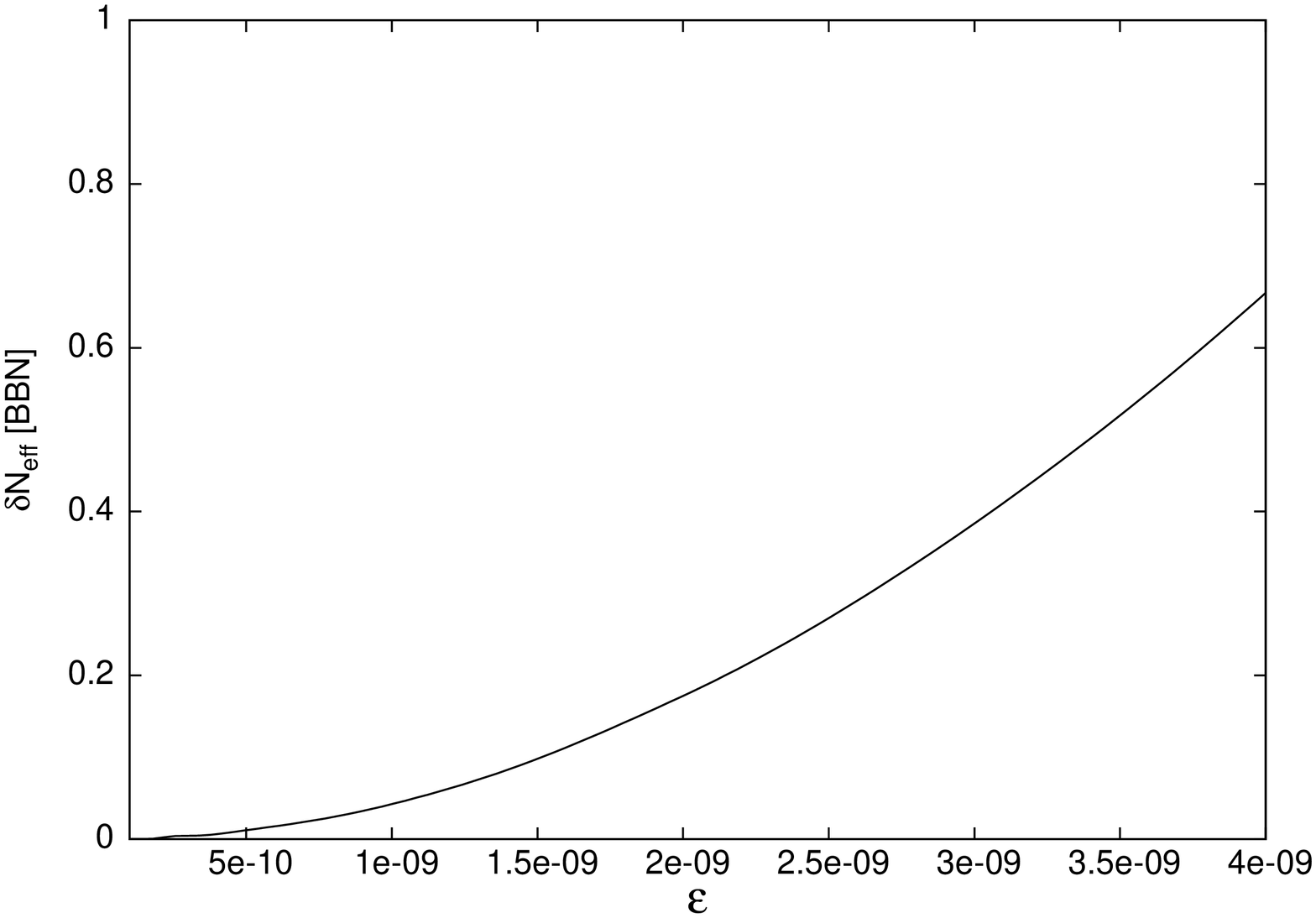,angle=270,width=12.0cm}}
\vskip 0.3cm
\noindent
{\small
Figure 3.6: $\delta N_{eff} [BBN]$ versus $\epsilon$.
}

\vskip 0.8cm

\subsection{Implications for CMB and LSS}
\vskip 0.6cm

The large-scale structure of the Universe has  been identified as an important probe of the basic 
constituents of the Universe and their properties. Structure in the Universe arose from
tiny perturbations in the density field which grew via gravitational instability.  
These density perturbations are defined in the obvious way:
\begin{eqnarray}
\delta ({\bf x}) \equiv {\rho({\bf x}) - \langle \rho \rangle \over \langle \rho \rangle}\ .
\end{eqnarray}
It is most convenient to work with the Fourier transformed quantities, $\delta ({\bf k})$.
In terms of these quantities, the power spectrum $P(k)$ is defined by:
\begin{eqnarray}
\langle \delta ({\bf k}) \delta ({\bf k}') \rangle = (2\pi)^3 P(k) \delta^3 ({\bf k} - {\bf k}')\ 
\end{eqnarray}
where the angular brackets denote the average over the whole distribution.

The anisotropies of the CMB provide another important cosmological probe.
They give information about the density perturbations around the time of last
scattering, i.e. at redshift $z \approx 1100$.
It is standard practice to express these anisotropies using a spherical harmonic expansion
of the photon temperature field:
\begin{eqnarray}
T(\theta,\phi) = \sum_{\ell m} \ a_{\ell m}\ Y_{\ell m} (\theta, \phi)
\ .
\end{eqnarray}
The anisotropy spectrum today can then be characterized in terms 
the variance of the
coefficients $a_{\ell m}$ in the above expansion, i.e.
\begin{eqnarray}
\langle a_{\ell m} a^*_{\ell' m'} \rangle = \delta_{\ell \ell'} \delta_{m m'} C_\ell\ .
\end{eqnarray}
Both CMB anisotropies and LSS of the Universe can be used to constrain mirror dark matter. 
%as we shall now explain.

We found in section 3.4 that the kinetic mixing interaction induces an additional contribution
to the energy density at the CMB epoch, $\delta N_{eff}[CMB]$.  
It is known that additional relativistic energy density can dampen the tail of the 
CMB anisotropy spectrum \cite{bsaj,dampen}.
However, the generation of $T'/T$ by kinetic mixing induced processes leads to another 
important effect for the CMB.
Prior to mirror-hydrogen recombination mirror particles formed a tightly coupled fluid consisting of
$e', H', He'$ and $\gamma'$. This tightly coupled fluid
experiences significant pressure, due to the mirror radiation ($\gamma'$) component.
Fourier modes which enter the horizon before mirror-hydrogen recombination epoch 
undergo acoustic oscillations due to the pressure of this tightly coupled fluid. 
If $T'_\gamma < T_\gamma$ then mirror-hydrogen recombination occurs prior to ordinary hydrogen recombination which
means that only the small-scale modes would be affected \footnote{
For $\epsilon = 10^{-9}$ we find that the mirror photons decouple from matter at a temperature of around 
$T'_\gamma \simeq 0.32$ eV. This is somewhat higher than the temperature at which
ordinary photons decouple from ordinary matter, $T_\gamma \simeq 0.26$ eV.
This difference arises due to the higher densities at earlier times which
enhances the rate: $e' + p' \to H' + \gamma'$ at the expense of: $H' + \gamma' \to e' + p'$.
Furthermore, we found in section 3.1 that $T'_\gamma/T_\gamma \simeq 0.31$ for $\epsilon = 10^{-9}$
[Eq.(\ref{constant})]. The net effect is that in this example, mirror photons decouple from mirror matter at a relatively early time:
$T_\gamma \approx 1$ eV that is, at redshift $z \approx 4400$.
}.
Their amplitudes would become suppressed.
In other words,  the effect of acoustic oscillations due to the pressure of the 
mirror baryon - mirror photon fluid is to suppress small-scale inhomogeneities in the mirror-matter density field.
Thus, one can anticipate a suppression of power on small scales when compared with 
collisionless cold dark matter in the linear regime.

The mirror dark matter model introduces only one additional parameter,
$x \equiv T'_\gamma/T_\gamma$. We found in section 3.1 that this parameter
is related to the fundamental Lagrangian kinetic mixing parameter $\epsilon$ via Eq.(\ref{constant}).
It is instructive to consider first the limit where 
$x \to 0$ (i.e. $\epsilon \to 0$). 
In this limit, the cosmological evolution of
mirror dark matter is indistinguishable from collisionless cold dark matter in
the linear regime.
This follows since mirror particles feel
negligible pressure after mirror-hydrogen recombination occurs at $t'_{dec}$, and $t'_{dec} \to 0$ as $x \to 0$.
For mirror dark matter with $x$ nonzero differences start to appear.  

The modifications to CMB and LSS which the kinetic mixing interaction induces, have been
computed in \cite{foot2013plb} \footnote{There are also studies examining related
effects for the physically distinct case where $\epsilon = 0$ and nonzero $T'/T$ is an initial
condition \cite{ber1,ig1,lss}.
Having $T'/T$ induced via kinetic mixing is not equivalent to having $T'/T$ as an initial condition imposed at $T \gg 1$ MeV.
This is because the kinetic mixing induced effects mainly occur after neutrino decoupling, $T \stackrel{<}{\sim}$ few MeV, and this
leads to two main differences. Having $T'/T$ imposed as an initial condition at $T \gg 1$ MeV would imply 
that there is a mirror neutrino
contribution to the energy density. Another difference is that, in the absence of kinetic mixing, 
there is no photon cooling contribution, $\delta N_{eff}^a [CMB]$. Numerically though, these two differences partially compensate each other.
Formally, the $x$ parameter referred to in this review assumes $T'/T \ll 1$ initially, with entropy generated in the mirror sector via kinetic
mixing induced processes at $T \stackrel{<}{\sim}$ few MeV.}.
%%% cite xxxx cw if mistake fixed xxxxx
This task is reasonably easy since the relevant equations, summarized in the appendix of \cite{foot2013plb},
are a straightforward generalization to the equations governing the perturbations of the ordinary 
baryons and photons (see e.g. \cite{cmbfast,sjak,hu} and references therein).
Standard initial conditions are assumed, i.e. Gaussian distributed adiabatic scalar perturbations, 
for a review see for instance \cite{dodelson}.

Since mirror dark matter reduces to collisionless cold dark matter in the limit $x \to 0$, we know that parameters near the $\Lambda CDM$ best fit
will also be a good fit for mirror dark matter for small $x$.
To study the effects of nonzero $x$ [or equivalently,  of nonzero $\epsilon$ given Eq.(\ref{constant})] for the CMB,
one could choose the $\Lambda CDM$ best fit parameters $\Omega_m h^2, \Omega_b h^2, h, ...$ 
and vary $x$. However doing this would modify the epoch of matter-radiation equality, 
$z_{EQ} +1 = \Omega_m/\Omega_{rad} \propto \Omega_m h^2/\rho_{rad}$,
given the extra contributions to $\rho_{rad}$ from Eq.(\ref{79z}).
The matter-radiation equality is fairly precisely determined by the data, so the parameter: $\Omega_m h^2$ can
be adjusted so that $z_{EQ}$ is fixed as $x$ is varied.
In fact, similar arguments apply to
$\Omega_b h^2$ and $\theta_s$ (the angular size of the sound
horizon at decoupling). We therefore hold $z_{EQ},\ \Omega_b h^2, \ \theta_s$ fixed as $x$ is varied.  
An analogous situation has been noted when
considering the effect of additional relativistic neutrino degrees of freedom \cite{dampen, bsaj}.
In this parameter space direction, the observable effects from
varying $x$ occur at small angular scales.

The choosen reference  parameters are:
$\Omega_m h^2 = 0.14$, $\Omega_b h^2 = 0.022$, $\Omega_{\Lambda} = 1 - \Omega_m$, $h = 0.70$
[$\Omega_m \equiv \Omega_b + \Omega_{b'}$]. 
These reference parameters are defined at $x = 0$. 
As discussed above, for nonzero values of $x$ these parameters are adjusted in such a way so that
$z_{EQ}$, $\Omega_b h^2$ and $\theta_s$ are held fixed.
A standard scale invariant 
(Harrison-Zeldovich-Peebles spectrum \cite{HZP}) 
initial perturbation spectrum is assumed, 
with normalization adjusted so that the height of the first peak is fixed. 
%is assumed and 
[Small effects due
to primordial tilt or reionization are not important to leading order since the point of this 
exercise is to compare
the effects of varying $x$ along the chosen parameter space direction.]

The CMB spectrum with these reference parameters, adjusted for nonzero $x$ as described above,
was computed in \cite{foot2013plb}. The results are given here in figure 3.7 and figure 3.8.
Also shown in figure 3.9 is a plot of $F_\ell (x) \equiv C_\ell (x)/C_\ell (x = 0)$ for several values of $x$.

\vskip 0.9cm
\centerline{\epsfig{file=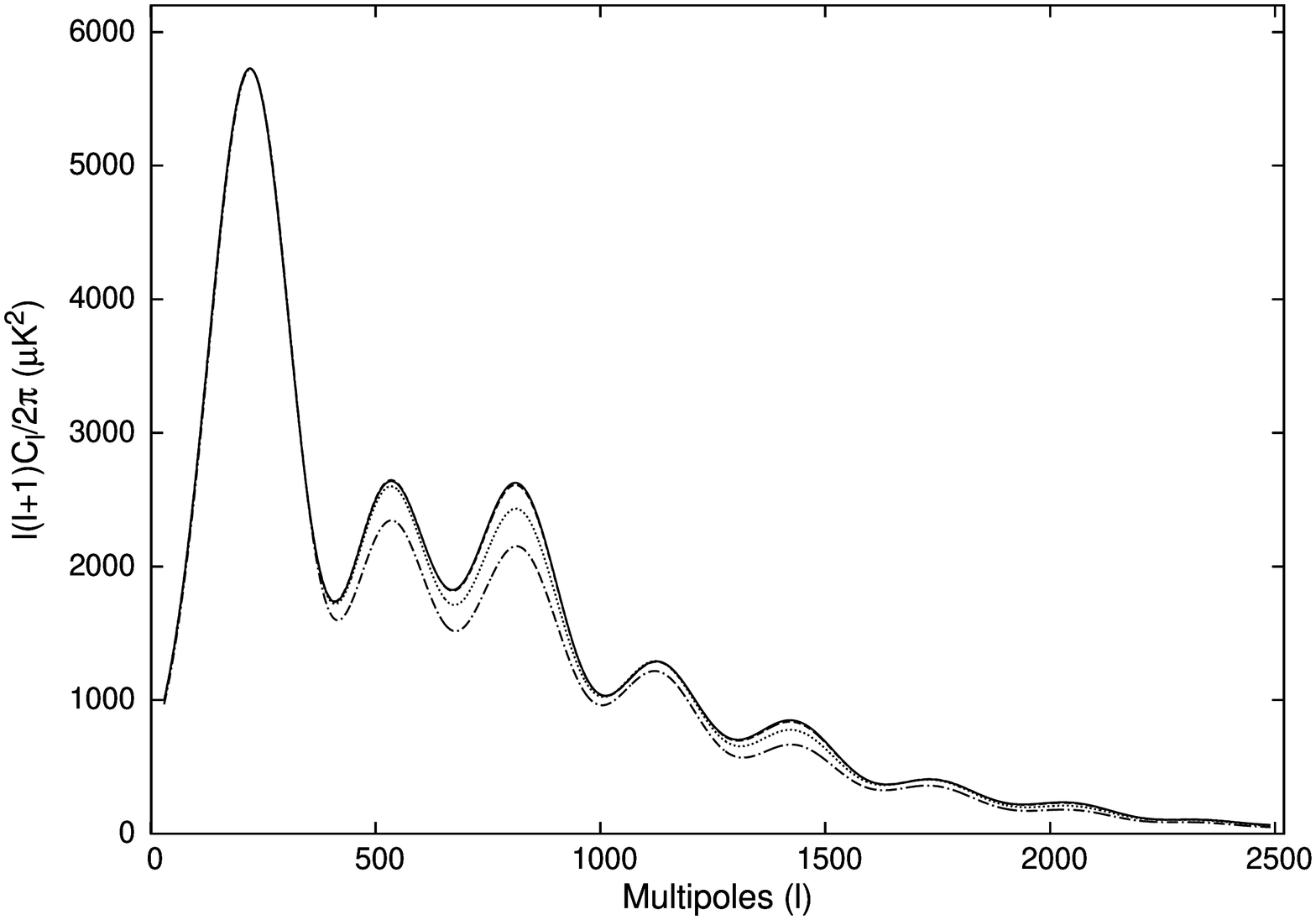,angle=270,width=12.4cm}}
\vskip 0.5cm
\noindent
{\small
Figure 3.7: The anisotropy spectrum for mirror dark matter. 
The solid line is  mirror dark matter with $x=0$ [i.e. $\epsilon = 0$]
with parameters described in the text.
This case is cosmologically equivalent to collisionless cold dark matter [$\Lambda CDM$]. 
Mirror dark matter with $x=0.3$ [$\epsilon = 10^{-9}$] (dashed line), $x = 0.5$ [$\epsilon = 2.6 \times 10^{-9}$] (dotted line) and $x=0.7$ 
[$\epsilon = 5.1 \times 10^{-9}$]
(dashed-dotted line) are also shown.  
}

\vskip 0.3cm

\centerline{\epsfig{file=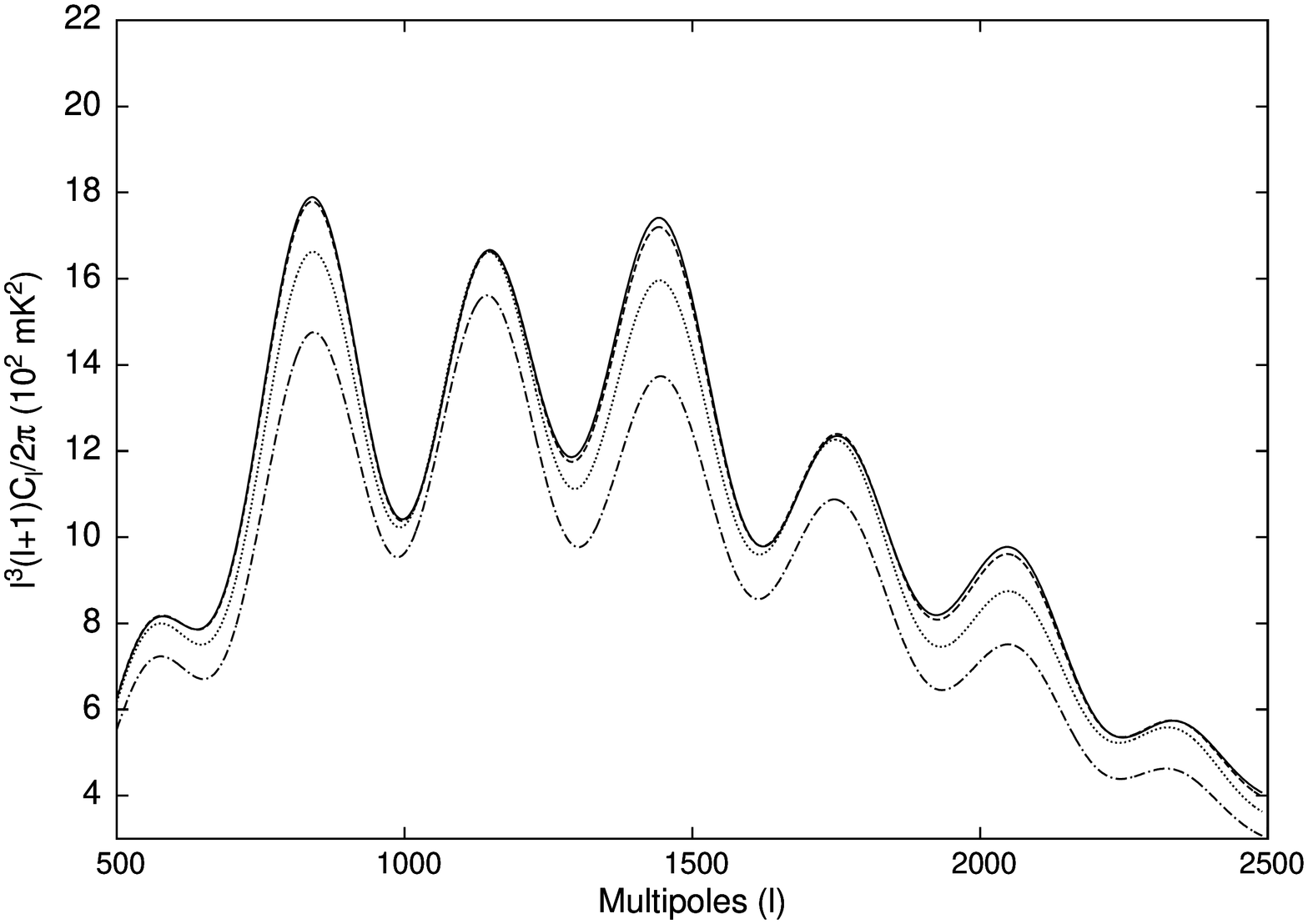,angle=270,width=12.4cm}}
\vskip 0.5cm
\noindent
{\small
Figure 3.8: The CMB tail. The curves correspond to the same parameters as figure 3.7.
}

\vskip 0.4cm

\vskip 0.4cm

\centerline{\epsfig{file=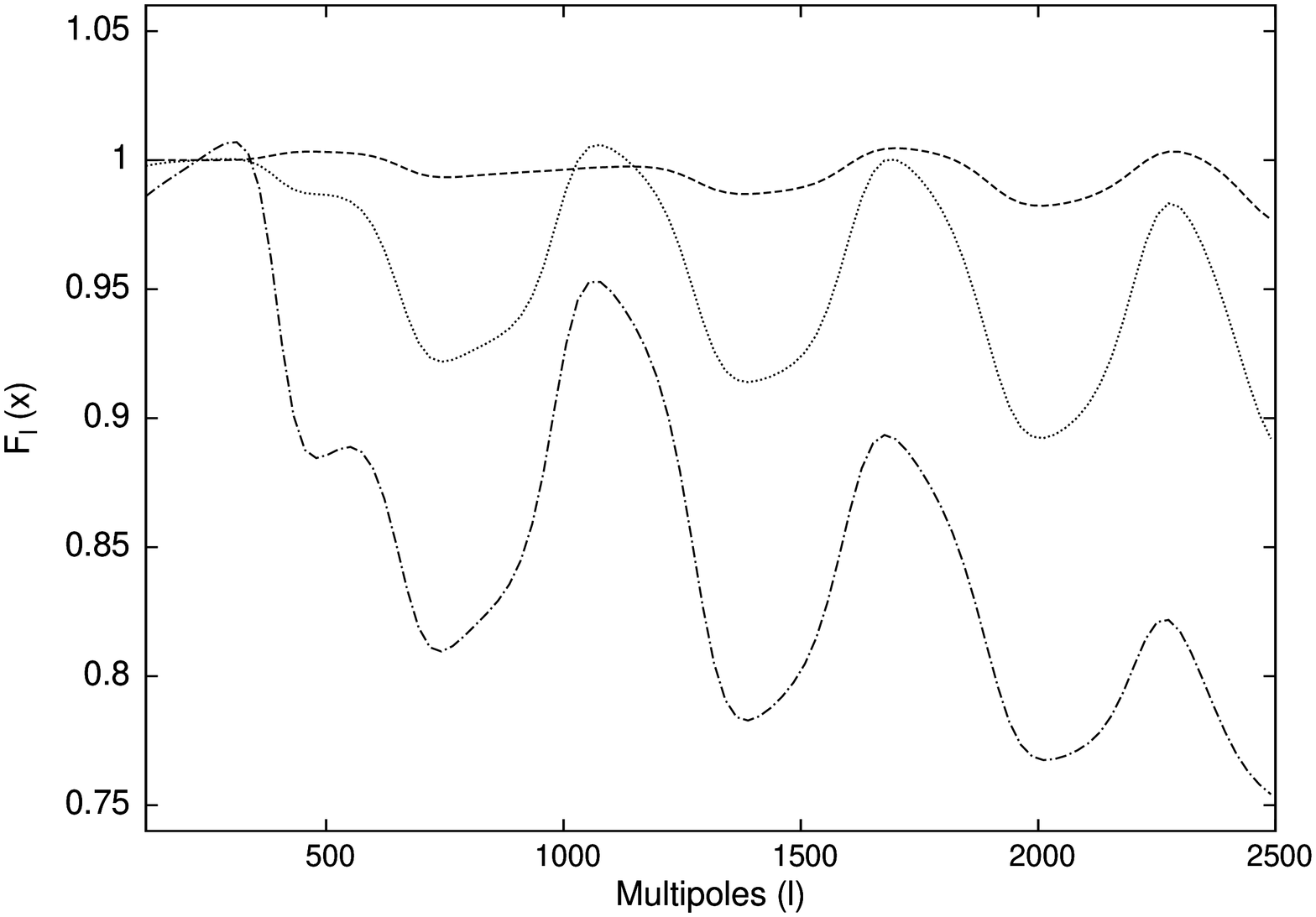,angle=270,width=12.4cm}}
\vskip 0.5cm
\noindent
{\small
Figure 3.9: $F_\ell (x) \equiv C_{\ell}(x)/C_{\ell} (x = 0)$ for $x = 0.3$ (dashed line) and
$x = 0.5$ (dotted line), and $x = 0.7$ (dash-dotted line) are shown.
}

\vskip 1.0cm
\newpage

Figures 3.7 - 3.9 show suppression of CMB anisotropies at small angular scales as $x$ increases. As discussed
above, this suppression is expected; it results primarily from the acoustic  oscillations 
occurring prior to mirror-hydrogen recombination.
Interestingly the suppression, which starts around the third peak, is larger for the higher odd peaks than the even ones.
This feature can be easily understood. The odd peaks arise from compressions of the baryon - photon
fluid while even peaks are due to rarefactions. 
The suppression of small-scale inhomogenities 
suppresses also the gravitational driving force on small scales. It is well known that this has a much greater effect
for compressions (the odd peaks) than for the rarefactions
as related effects occur when $\Omega_b h^2$ is reduced \cite{dodelson}. 
See also \cite{m45} for a more thorough discussion of essentially the same
physics in the context of more generic hidden sector dark matter models.

In addition to CMB anisotropies the matter power spectrum can also be used
to probe the kinetic mixing interaction. 
The power spectrum of matter is given by,
\begin{eqnarray}
P(k) = 2\pi^2 \delta_H^2 {k \over H_0^4} T^2(k)
\end{eqnarray}
where $H_0 = 100h\ {\rm km } \ {\rm sec}^{-1} \ {\rm Mpc}^{-1}$ characterizes the Hubble rate today and
$T(k)$ is the transfer function (see e.g. \cite{dodelson} for further details).
However since small scales $k \stackrel{>}{\sim} 0.1\ h\ {\rm Mpc}^{-1}$
have gone nonlinear today, linear perturbation theory can only reliably be used, at the current epoch, to 
calculate the matter power spectrum on  scales larger than this.

\vskip 0.5cm
\centerline{\epsfig{file=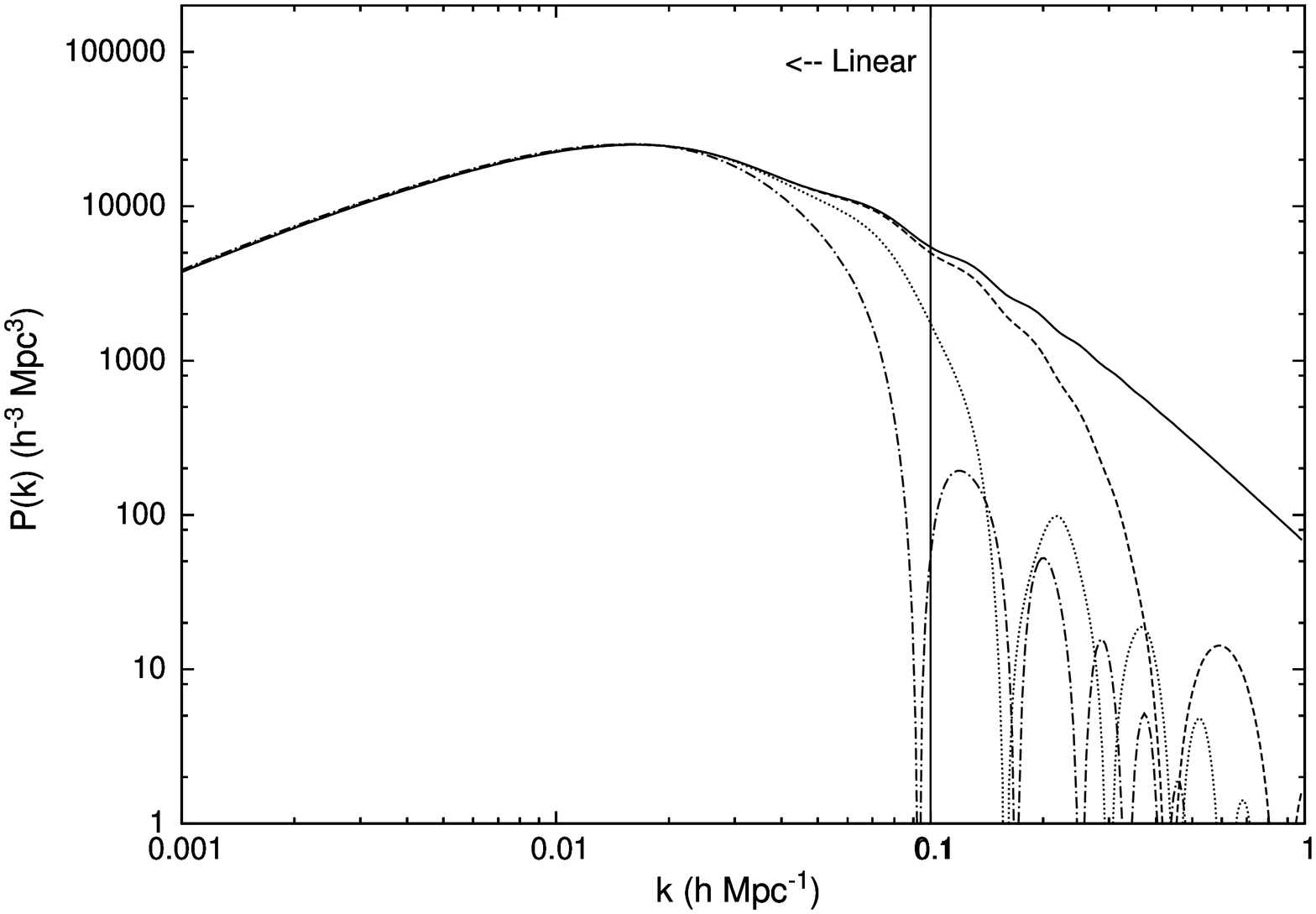,angle=270,width=12.0cm}}
\vskip 0.3cm
\noindent
{\small
Figure 3.10: Power spectrum of matter for the same reference ($\Omega_m, \ \Omega_b, \ \Omega_\Lambda, \ h$) parameters as figure 3.7. Again 
$x = 0$ (solid line), $x = 0.3$ (dashed line), $x=0.5$ (dotted line) and
$x = 0.7$ (dashed-dotted line).
The $x=0$ case is cosmologically equivalent to collisionless cold dark matter.
}

\vskip 0.7cm
\centerline{\epsfig{file=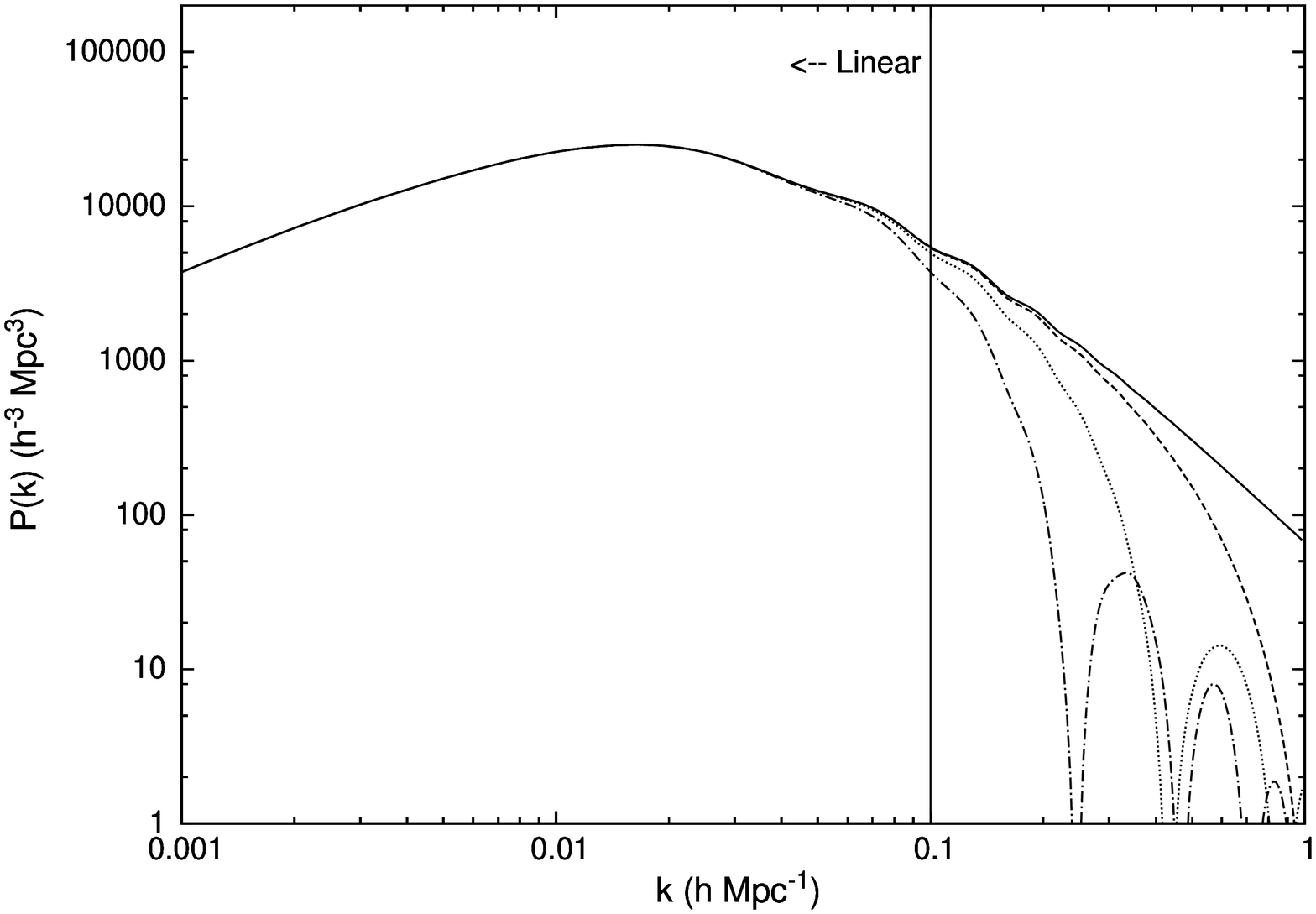,angle=270,width=12.4cm}}
\vskip 0.5cm
\noindent
{\small
Figure 3.11:  Power spectrum for 
$x = 0$ (solid line), $x = 0.2$ (dashed line), $x=0.3$ (dotted line) and
$x = 0.4$ (dashed-dotted line).
}

\vskip 1.4cm

In figure 3.10 the predicted matter power spectrum
is given for the considered $x$ values, for the same parameters used in figures 3.7 - 3.9,
while figure 3.11 considers just the low values of $x$ of most interest: $x = 0, \ 0.2, \ 0.3, \ 0.4$.
As $x$ increases from zero, deviations occur due (primarily) to acoustic oscillations experienced by
perturbations on small scales which enter the horizon before mirror-hydrogen recombination. 
These figures indicate that a rough upper bound of $x \stackrel{<}{\sim} 0.3-0.4$ could
be expected from galaxy surveys. See also \cite{lss,ber69,creview} for related discussions.

Collisionless cold dark matter provides a very successful explanation of the observed CMB anisotropy spectrum.
It also provides a good explanation of large-scale structure in the linear regime.  
We have seen here that
mirror dark matter closely resembles collisionless dark matter for the relevant observables provided
$x \stackrel{<}{\sim} 0.3-0.4$. Using Eq.(\ref{constant}), this translates to a rough upper limit on kinetic mixing
of: $\epsilon \stackrel{<}{\sim} 1-2 \times 10^{-9}$ (which we henceforth abbreviate to $10^{-9}$).
We conclude that mirror dark matter with $\epsilon \stackrel{<}{\sim} 10^{-9}$ 
is consistent with CMB observations and large-scale structure.

\vskip 0.6cm
\newpage

\section{Galaxy structure}

Structure arises from tiny perturbations in an otherwise homogeneous and isotropic Universe.
It is commonly believed that these perturbations are seeded in the very early Universe
by quantum fluctuations that were amplified by inflation, e.g. \cite{brau}. After the inflationary phase, one is
left with a set of nearly scale invariant (Gaussian distributed) perturbations.
These grow over time under the influence of gravity. Prior to mirror-hydrogen recombination, pressure can also play an important
role for mirror baryons.
As discussed in section 3.5, the pressure leads to acoustic oscillations which can suppress
inhomogenities 
on small scales. Exactly the same effect occurs for ordinary baryons until 
hydrogen recombination, which is much later if $T' \ll T$. For this reason and also
because the matter density is dominated by mirror baryons, mirror-particle density perturbations are expected
to form structure well in advance of ordinary baryons.

While the perturbations are small,
$\delta \rho \ll \langle \rho \rangle$, their evolution
is governed by a set of linear equations arising from the Boltzmann-Einstein equations \cite{dodelson}.
Eventually though, perturbations can grow to the point where $\delta \rho \sim \langle \rho \rangle$ after which their evolution
becomes mildly nonlinear. Around this time they can break away from the expansion and begin to collapse.
Up to this point, mirror dark matter closely resembles collisionless cold dark matter, except for 
very small scales where acoustic oscillations can be important at early times (as discussed above). 
The physics is also under control - the evolution of 
perturbations is well understood by linear perturbation theory.
The subsequent non-linear evolution, mergers, and the  
eventual formation of individual galaxies is much more complex and poorly understood.
For mirror dark matter, dissipative and non-dissipative interactions, heating and 
potentially other processes become relevant.
In this nonlinear regime mirror dark matter is expected to behave very differently from collisionless cold dark matter.
The relevant hydrodynamical simulations have not been done, and any discussion  
of galaxy formation and early evolution is, of course, preliminary.
With this caveat in mind, we briefly outline how structure might have evolved initially.

Structure in the mirror sector should form first, given the required initial conditions 
($T' \ll T$, \ $\Omega_{b'} \approx 5\Omega_b$).  One can further check (section 4.1) that the collapse is not
impeded by radiative cooling for typical galaxy-scale perturbations, i.e. radiative cooling typically occurs faster
than the free-fall time scale.
Mirror star formation can occur, either during the collapse or later in a disk, potentially producing also mirror supernovae.
With $\epsilon \sim 10^{-9}$ mirror
supernovae would be expected to provide a huge flux of ordinary X-ray photons: Kinetic mixing
induced processes in the hot core of mirror supernovae transfer core-collapse gravitational potential energy
into production of light ordinary particles ($e,\ \bar e,\ \gamma$) which can escape the collapsing mirror star; this energy
is ultimately radiated away as ordinary photons. Within this framework it is very natural to suppose that this
radiation might have been responsible for the reionization of ordinary matter 
at $6 < z < 20$  inferred to exist from CMB and other observations.
Once the ordinary matter is ionized, the plasma cannot absorb radiation very efficiently;
the Thomson scattering cross-section is simply too small and photoionization is ineffective due to the
low metal content at this early time.
One expects, therefore, that the ordinary baryons will ultimately collapse potentially forming a separate disk. 
Gravitational interactions between the two disks, assuming both form, could lead to their alignment cf. \cite{randal}.
Ordinary star formation and hence also ordinary supernovae would be expected to occur.
With $\epsilon \sim 10^{-9}$ these ordinary supernovae would provide a huge flux of X-ray mirror photons.
These mirror photons can potentially heat the mirror plasma component via photoionization if there is sufficient 
mirror metal enrichment of the plasma by this time.
If this does indeed happen, then this huge energy
input could expand the mirror gas out of the disk into a spherical distribution. 
This, it is alleged, is the origin of the roughly spherical dark matter
halos inferred to exist around spiral galaxies today.

In this picture, therefore, halos around spiral galaxies are currently composed predominately of a 
hot mirror metal enriched mirror-particle
plasma.
This plasma halo is kept `puffy' by non-trivial dynamics. 
The halo evolves to a quasi-equilibrium configuration where the energy radiated by dissipation is 
replaced by supernovae heating (via the kinetic mixing induced processes transferring the ordinary supernovae core collapse energy
into mirror photons, as described above). 
This dynamics allows many of the halos {\it current} properties, such as its radial mass distribution, 
to be determined, essentially independently of its past history. This {\it dynamical halo model}
was developed in a series of articles \cite{sph,r69,foot5,rf2013b,zurab2013} and will be
the subject of this section. 

The kinetic mixing induced supernova heating is applicable only to galaxies with appreciable
star formation occurring at the present time.
This includes the spiral and irregular galaxies.  
Elliptical galaxies, on the other hand, generally show little
current star formation activity. Their dark matter properties
are therefore expected to be very different. In the absence of any heating mechanism a substantial
mirror particle plasma halo could not exist.
In these galaxies the mirror dark matter presumably has cooled and condensed into mirror stars. 
For these galaxy types, therefore, the dark matter is likely to take the form of massive compact
objects rather than a diffuse plasma.
\footnote{Of course, it is possible that the halo of spiral galaxies might also have a substantial compact-object
component \cite{hodges,silearly,macho,macho2}.  
Indeed, some old mirror stars should exist as a remnant of the early
epoch of mirror star formation, expected to have occurred during the first billion years or so. 
On the observational front, initial searches for gravitational
microlensing \cite{pacman}, appeared to indicate the presence of large 
Massive Compact Halo Objects (MACHO) in the Milky Way
halo \cite{alcock}. More recent studies, though, have been somewhat less encouraging \cite{machorecent}.
The most recent observations suggest that the proportion of the halo's mass in the form of MACHO's is likely
to be less than around 30\%.
The totality of these observations, though, appear to be consistent with a Milky Way halo fraction of $\sim 10\%-20\%$ old mirror stars.
}

\subsection{Preliminaries}

\vskip 0.3cm
\noindent
{\bf The early history of Spiral Galaxies}
\vskip 0.3cm
\noindent
As discussed above, we expect the first structures to form from mirror-particle density perturbations.
Imagine one such perturbation, at the point in time where its evolution has become mildly nonlinear,
$\delta \rho \sim \langle \rho \rangle$. Around this time, the perturbation will break away from
the expansion and begin to collapse.

Consider a uniform collapsing spherical density perturbation of radius $R$ and mass density $\rho$ and
temperature $T$.  In the absence of pressure, such a perturbation would collapse to a point on
the free-fall time scale: \footnote{
The discussion here follows closely that of \cite{sph,zurab2013} with some equations adapted from
the standard treatment of collapsing baryonic structures, given in e.g. \cite{mo}.}
\begin{eqnarray}
t_{ff} = \sqrt{{3\pi \over 32 G_N \rho}}\ .
\end{eqnarray}
Since mirror dark 
matter is collisional the collapse can potentially be halted by pressure.
Whether or not this happens depends on the cooling time scale.
Let us assume for simplicity that the plasma is composed of fully ionized mirror helium so that
$n_{e'} = 2n_{He'} =  2n_T/3$, where $n_T$ is the total particle number density.
The cooling rate per unit volume due to thermal bremsstrahlung is then:
\begin{eqnarray}
\Gamma_{cool} = n_{e'}^2 \Lambda
\end{eqnarray}
where $\Lambda \sim 10^{-23}\ {\rm erg}\ {\rm cm^3}\ {\rm s^{-1}}$ for $T \sim 100$ eV.
[This $T \sim 100$ eV scale is a rough estimate from the virial theorem for a `typical' 
galactic halo, such as that which surrounds the Milky Way
\cite{sph}.]
The cooling time scale, $t_{cool}$, can be obtained from $\Gamma_{cool} t_{cool} \approx n_T (3/2) T$,
i.e.
\begin{eqnarray}
t_{cool} &\approx & {9T \over 4\Lambda n_{e'}} \nonumber \\
& \sim & 100 \ \left({T \over 100 \ {\rm eV}}\right) \left( {10^{-2}\ {\rm cm^{-3}} \over n_{e'}}\right) \ {\rm Myr} 
\ .
\label{cool6}
\end{eqnarray}
In the absence of any heat source, 
perturbations satisfying $t_{cool} < t_{ff}$ can collapse unimpeded.
Evaluating, $t_{cool}/t_{ff}$, using $\rho = n_{He'} m_{He}$, we have:
\begin{eqnarray}
{t_{cool} \over t_{ff}} \sim 0.3 \ \left( {T \over 100 \ {\rm eV}}\right)\left( {10^{-2}\ {\rm cm^{-3}} \over n_{e'}}\right)^{1/2}
\ .
\end{eqnarray}
Evidently there is no significant impediment for collapse of (typical) galaxy-sized perturbation due to pressure effects.

%For a Milky Way sized galaxy, with $T \stackrel{<}{\sim} T_{virial}$, $t_{cool} < t_{grav}$.
%This means that, in the absence of any heat source, galaxy scale mirror density perturbations can collapse unimpeded.

During an early epoch, say the first billion years or so, collapsing and merging mirror-particle halos would be occurring.
These mirror-particle structures, if they have collapsed, might be very dense structures. That is, their central densities 
can be orders of magnitude denser than the central densities of galactic halos at the current epoch.
Mirror-star formation is likely to be extremely rampant in these high density collapsed structures.
Once formed, mirror stars evolve very rapidly given the high primordial $He'$ abundance: $Y'_p \sim 0.90$
for $\epsilon \sim 10^{-9}$ (discussed in section 3.3).
[The study \cite{bersar} found that such mirror helium dominated stars evolve
more than an order of magnitude faster than ordinary stars which have $Y_p \approx 0.25$.]
One therefore expects that the interstellar gas will be quickly enriched with mirror metal components.
Meanwhile the ordinary baryons will accumulate within these structures, and ultimately ordinary star formation
and evolution will occur. 
In fact, the star formation rate (and hence also the supernova rate) can be very high
in these collapsed structures, given that star formation rates are observed to 
depend sensitively on the gas density: 
\begin{eqnarray}
\dot{\Sigma}_* \propto n_{gas}^N
\label{gas56}
\end{eqnarray}
where $N \sim 1-2$ \cite{s1959,ken}. This simple power-law behaviour, known as the Schmidt-Kennicutt law,
has been observed over a wide range of densities and environments, for a recent review see also \cite{kenrev}.

At this early time, the anticipated large rate of ordinary supernovae can
play a critical role. Kinetic mixing induced processes in the core of these early supernovae convert
a substantial proportion of the core-collapse energy into light mirror particles which ultimately provide a huge
energy source. This energy source, as we will see, might be  
able to efficiently heat the mirror plasma component, but only if the plasma is sufficiently enriched with mirror metals.
The heating may even be larger than the plasma's cooling rate due to dissipative processes.
If this happens the mirror plasma component will start expanding.
As it expands, the supernovae rate will reduce as the ordinary gas density reduces in response
to the weakening gravity. If, over time, the heating rate falls faster than the cooling rate, then 
the system will expand until it approaches a quasi-stable configuration.
This is illustrated schematically in figure 4.1. There may well be some damped oscillations (not shown in figure 4.1) as the
system first approaches and adjusts to the equilibrium configuration.

\vskip 1.6cm
%%%%%%%%%%%%%%%%%%%%%%%%%%%%%%%%%%%%%%% figure 4.1 %%%%%%%%%%%%%%%%%%%%%%%%%%%%%%%%%%%%%%%%%%%
\centerline{\epsfig{file=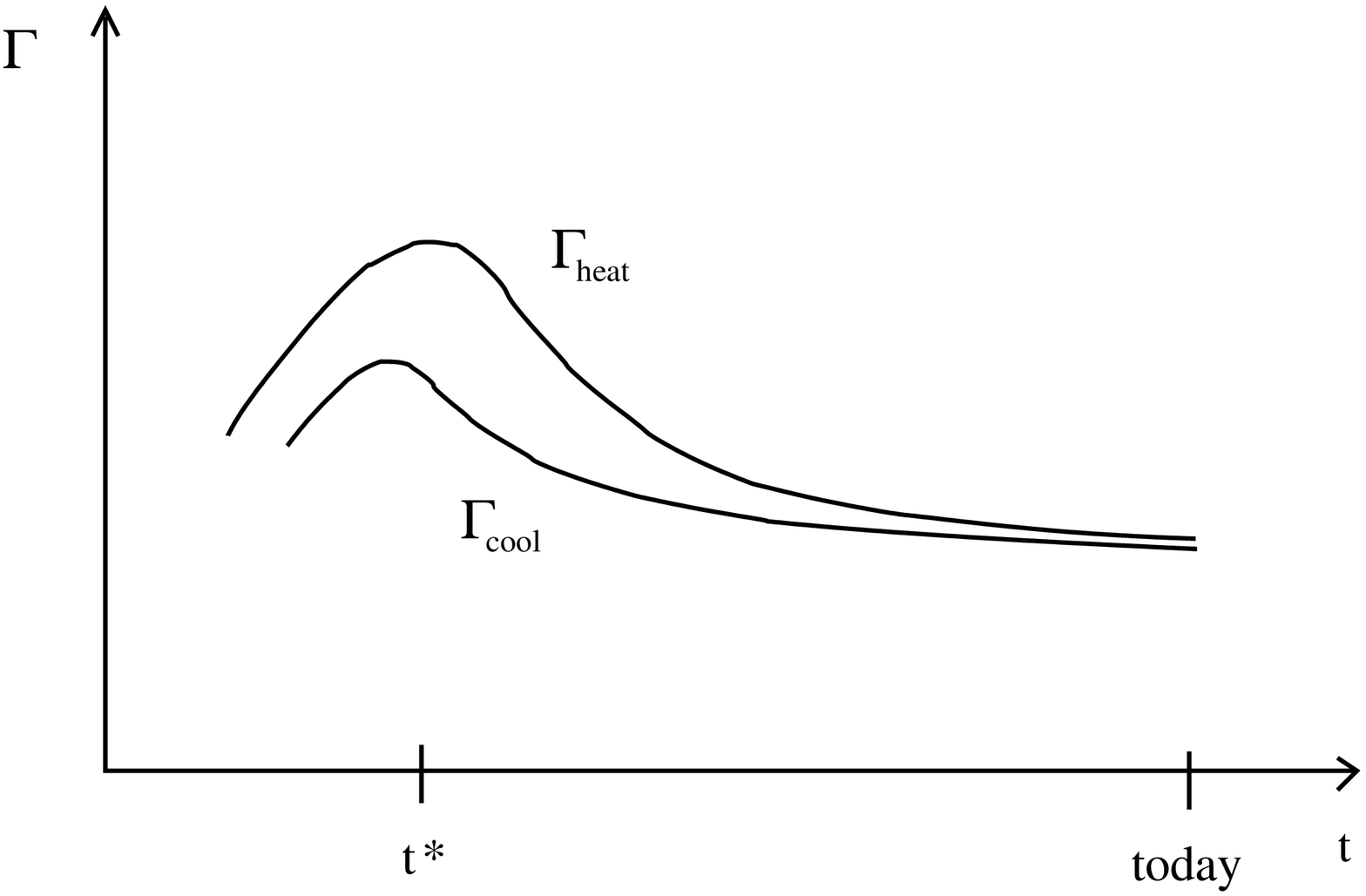, angle=0, width=9.8cm}}
\vskip 0.5cm
\noindent
{\small Figure 4.1: 
Schematic diagram of the heating and cooling rates of the galactic
mirror-particle halo, evolved from some early time to today. 
}

\vskip 1.0cm
\newpage

This reasoning suggests that the system will evolve until the halo heating and cooling rates 
approximately balance.
This motivates the dynamical requirement:
\begin{eqnarray}
\Gamma_{heat} = \Gamma_{cool}
\ .
\label{current}
\end{eqnarray}
In fact, this condition should be approximately valid for any point
in the halo for a stable configuration.
The above equation forms the basis of recent attempts to constrain the {\it current} properties 
of spiral galaxies in this framework.

The balancing of heating and cooling rates 
suggested by this dynamics [Eq.(\ref{current})] 
is consistent with the approximately constant star formation rate inferred for spiral galaxies
over the last few billion years (see e.g. \cite{galhun} and references therein for relevant
discussions). In fact, this dynamics provides a mechanism to regulate star formation rates. 
Curiously, the above line of reasoning suggests that 
the star formation rate in spiral galaxies should  have been substantially higher in the more distant past than it is at the present epoch.
Interestingly, there is ample observational evidence for this. 
The study \cite{noe},
for instance, found a gradual decline of the star formation rate of most galaxies since $z = 1.1$ and there are
studies suggesting a much higher star formation rate at earlier times e.g. \cite{beacom}
(see also   \cite{kenrev2,kenrev} and references therein for relevant discussions).
Also, the result that the $t_{cool}$ time scale [Eq.(\ref{cool6})] is larger for higher-mass
galaxies suggests that higher-mass galaxies should take more time to reach the $\Gamma_{heat} \simeq \Gamma_{cool}$ configuration,
and thus should have a relatively higher star formation rate (SFR) in the past cf. lower mass galaxies.
This is consistent with observations, which find that the SFR/$M_* \ \propto (M_*)^{-0.36}$,
that is SFR/$M_*$ has negative slope \cite{bush,kenrev}.

For the specific case of the Milk Way and Milky Way-like galaxies, 
there are recent studies which find evidence for a much higher star formation rate
in the distant past, at $z \stackrel{>}{\sim} 1.5$ (i.e. $\stackrel{>}{\sim} 9$ Gyr ago) \cite{mwa,mwb}.
These studies found that the star formation rate peaks at around $z \sim 2$, that is, $t^*$ in figure 4.1
is around $3$ Gyr.  The study \cite{mwb} also finds evidence that the intense period
of star formation at $z \stackrel{>}{\sim} 1.5$ is associated with the Milky Way's thick disk.
The thick disk has a much shorter scale length ($r_D \approx 2.0$ kpc) than the thin disk ($r_D \sim 3.6$ kpc) \cite{thick1}.
That is, there is evidence that the baryons in the Milky Way were much more compact at early times ($z \sim 2$) 
during this early period of intense star formation.
Incidently, the study \cite{mwb} also finds evidence for a dip in the Milky Way star formation rate at around $z \sim 1.1$ which
lasts for around 1 Gyr, after which the star formation rate is approximately constant. 

All these observations  appear to be qualitatively 
consistent with the picture advanced here. When the baryons assembled into the galaxy
producing stars and supernovae the mirror halo was extremely compact. 
The heating of the mirror-particle halo by ordinary supernovae made the halo expand. This in turn caused the
baryon gas density to reduce [and thus also the ordinary star formation rate, via Eq.(\ref{gas56})] in the weakening gravitational field. 
In this context,
it is tempting to interpret the observed dip in the star formation rate at $z \sim 1.1$
as the result of the first damped oscillation of the mirror particle plasma halo.
That is, the plasma halo has over-expanded as it overshoots the equilibrium configuration. In this over-expanded phase, the 
star formation rate is suppressed due to the weaker gravity and consequent reduction in $n_{gas}$ [Eq.(\ref{gas56})].
Further oscillations are presumably too damped to (currently) be resolved by observations.
Naturally, these observations suggest that it would be worthwhile to extend this qualitative discussion 
to a more quantitative one. That is, to try to
quantitatively estimate the evolution of the star formation rate using this assumed halo dynamics.
However, at present only the current properties of galaxies, utilizing Eq.(\ref{current}), have been
studied.

\vskip 0.3cm
\noindent
{\bf The structure of Spiral Galaxies today}
\vskip 0.3cm
\noindent
To summarize the picture so far, the dark matter halo of spiral galaxies
is predominately in the form of a mirror-particle plasma which is both self-interacting and dissipative. 
This plasma has evolved from a more collapsed state into the current
expanded distribution via heating from ordinary supernovae. The baryon distribution has also expanded
in response to the weakening gravity. This heating mechanism requires the kinetic mixing interaction to convert
the supernova core-collapse energy into light mirror particles, which we will elaborate on in sections 4.2 and 4.4. 
The halo will expand until it has evolved into a dynamically stable configuration where
the heating and cooling rates are approximately balanced. At this time, the halo is governed by Eq.(\ref{current}) and
also the equation of hydrostatic equilibrium, which we now proceed to describe.

Consider a mirror-particle plasma halo. 
%and for simplicity let us assume that the mass neglect any possible compact object 
%component (old mirror stars etc.). 
Such a plasma is influenced both by gravity and also pressure.
Hydrostatic equilibrium is simply the condition that these forces are balanced.
Under the assumption of spherical symmetry, hydrostatic equilibrium requires
the pressure $P$ to satisfy:
\begin{eqnarray}
{dP \over dr} &=& -\rho (r) g(r) 
= -\bar m n_T (r) {v_{rot}^2 \over r}
\ .
\label{p9u}
\end{eqnarray}
Here
$\bar m$ is the mean mass of the mirror particles ($e', H', He',...$) in the plasma, 
$n_T (r)$ is the total mirror  particle
number density
[so that  $\rho (r) = \bar m n_T (r)$ is the mirror-particle plasma mass density]  
and $g(r)$ is the local acceleration due to gravity.
For a spherically symmetric distribution, the local rotational velocity, $v_{rot}$, can be 
related to 
the total mass density, $\rho_{total}$, via Newton's law:
\begin{eqnarray}
g(r) = {v_{rot}^2 \over r} = {G_N \over r^2} \int_0^r  \rho_{total} \ dV
\ .
\label{88a}
\end{eqnarray}
The total mass density contains three components: The mirror particle plasma, mirror baryonic compact objects (e.g. old mirror stars),
and the baryonic component (ordinary stars and gas).
At sufficiently large distances from the galactic center,  various observations suggest that $\rho_{total}$ 
should be dominated by the mirror particle plasma component.
For the purposes of this preliminary (analytic) discussion we therefore keep only the 
mirror plasma contribution to $\rho_{total}$. 
The baryonic compact-object component will, however, be included when the relevant equations are solved 
numerically in section 4.4.

For a spherically symmetric halo, neglecting ordinary baryons and mirror-baryonic compact objects, these equations simplify to:
\begin{eqnarray}
T{dn_T \over dr} + n_T {dT \over dr} = - \bar m n_T {G_N \over r^2} \int^r_0 \bar m n_T 4\pi r'^2 dr'
\end{eqnarray}
where the ideal gas law has been used to relate the pressure to the local halo temperature: 
$P (r) = n_T (r) T(r)$. This should be reasonable as the mirror-particle self interactions should keep the halo in local thermodynamic equilibrium
(see section 4.7 for an estimate of the self-interaction collision rate in such mirror particle halos).
%% page number here xxxxx
If we further assume that the halo is isothermal, so that $dT/dr = 0$, then the above nonlinear equation
has the analytic solution \cite{sph}:
\begin{eqnarray}
n_T (r) = {T \over 2\pi G_N \bar m^2 r^2}
\ .
\label{t34}
\end{eqnarray}
Thus we have that $n_T \propto 1/r^2$, provided of course, that our assumptions are reasonable.
Using Eq.(\ref{88a}), together with the above solution,  allows the rotational velocity to be related to the temperature: 
\begin{eqnarray}
T = {1 \over 2} \bar m v_{rot}^2\ .
\end{eqnarray} 
Importantly, $v_{rot}$ is found to be independent of $r$. This resulting flat $v_{rot}$ could be connected with the
observed asymptotically flat rotation curves in spiral galaxies. Indeed, the two key assumptions of this analytic result,
that ordinary baryons can be neglected and that the halo is isothermal, might reasonably be approximately valid
at large distances (i.e. at distances much greater than the disk scale length).

Of course, the halo temperature is not expected to be completely isothermal. To figure out the temperature profile of the halo, 
one must understand the details of the heating and cooling processes.
How energy is transported to the halo at each location and how it is dissipated from the same location.
Eq.(\ref{current}) can then be used in conjunction with the hydrostatic equilibrium equation to determine the halo's physical
properties: the $n_T (r), \  T(r)$ radial profiles.

It is enlightening to first  consider a `toy' model, where the baryonic component is taken as a point source
whose energy output supports a spherical mirror dark matter halo (via kinetic mixing induced supernova heating as described above). 
That is, we have a mirror-photon luminosity, ${L'}_{SN}$, originating at $r=0$.
The energy being absorbed in a volume element, $dV = 4\pi r^2 dr$, assuming mirror radiation dominates the energy
transport, is
\begin{eqnarray}
d{\cal E}_{in} 
 =  {{L'}_{SN} \ e^{-\tau} \over 4\pi r^2} \ \sum_{A'} \sigma_{A'} n_{A'} \ dV  \ .
\end{eqnarray}
Here $\sigma_{A'}$ is 
the relevant cross-section, dominated by the photoionization of the halo mirror metal components, 
to be discussed in more detail in section 4.4.
The quantity
$\tau \equiv \sum_{A'} \int_0^r \sigma_{A'} n_{A'} dr $ is the optical depth.
The energy radiated out of the same volume element, due to thermal bremsstrahlung, is
\begin{eqnarray}
d{\cal E}_{out} = \Lambda (T) \ n^2_{e'} \ dV
\ .
\label{Eout}
\end{eqnarray}
Matching $d{\cal E}_{in} = d{\cal E}_{out}$ implies
\begin{eqnarray}
n_{e'} 
= {{L'}_{SN} \ e^{-\tau}  \over \Lambda (T) 4\pi r^2}\ {\sum_{A'} \sigma_{A'} n_{A'} \over n_{e'}}
\ .
\end{eqnarray}
If the ratio $n_{A'}/n_{e'}$ is approximately independent of $r$, reasonable given the neutrality of the plasma,
and if the halo is optically thin, so that $\tau \ll 1$, 
then $n_{e'} \propto 1/r^2$ follows.
This $n \propto 1/r^2$ behaviour is exactly the same as that derived from 
the hydrostatic equilibrium equation for a self-gravitating isothermal spherical distribution,
Eq.(\ref{t34}).
This suggests that the assumed isothermality of the halo is actually justified in this 
`toy' model.
This model thus appears to be self-consistent, 
except it is unphysical at $r = 0$.

Consider now the more realistic case, where the energy source is not taken as a point at $r=0$, but as a distribution extended
over a finite volume of radius $\sim$ $r_D$. In this case one expects $n \sim 1/r^2$ for $r \gg r_D$ and a softer behaviour 
for $r \stackrel{<}{\sim} few\ r_D$ (assuming, as before, that the halo is optically thin).
Specifically, if heating by ordinary  supernova sources is assumed, then this energy source will
be, on the average,  distributed (roughly) in the same manner as the mass of the galactic thin disk. 
This mass distribution, can be approximated by a Freeman disk with 
surface density \cite{freeman}:
\begin{eqnarray}
\Sigma (\stackrel{\sim}{r})= {m_D \over 2\pi r_D^2} \ e^{-\stackrel{\sim}{r}/r_D} \  
\end{eqnarray}
where $r_D$ is the disk scale length and $m_D$ is the total mass of the disk.
In cylindrical co-ordinates,
$(\stackrel{\sim}{r}, \stackrel{\sim}{\theta}, \stackrel{\sim}{z})$ with the disk at $\stackrel{\sim}{z}=0$,
the average flux at the point 
$P = (r_1, 0, z_1)$ in the optically thin limit ($\tau \ll 1$) is then
\begin{eqnarray}
F(r_1,z_1) = {{L'}_{SN} \over 4\pi M_D} \int \int 
{\Sigma (\stackrel{\sim}{r}) \over 
\stackrel{\sim}{r}^2 - 2\stackrel{\sim}{r} r_1 \cos \stackrel{\sim}{\theta} + r_1^2 + z_1^2 } \ 
\stackrel{\sim}{r} d\stackrel{\sim}{r} d\stackrel{\sim}{\theta}
\ .
\end{eqnarray}
The rate at which energy is absorbed in a volume element, $dV$, in the galaxy halo is then
 $d{\cal E}_{in} = F(r_1,z_1) \sum_{A'} \sigma_{A'} n_{A'} dV$ while $d{\cal E}_{out}$ is given
as before in Eq.(\ref{Eout}).
Again,
matching $d{\cal E}_{in} = d{\cal E}_{out}$, we find:
\begin{eqnarray}
n_{e'} = {F(r_1, z_1)
\over \Lambda (T)}
\ {\sum_{A'} \sigma_{A'} n_{A'} \over n_{e'}} 
\ .
\end{eqnarray}
One can indeed show that $F(r_1,z_1) \propto 1/r^2$ (where $r^2 = r_1^2 + z_1^2$) for $r \gg r_D$ and has 
a much softer behaviour for $r \stackrel{<}{\sim} few \ r_D$ with  $F(r_1,z_1) \sim log(r)$  as $r \to 0$.

The above analytic arguments suggest a quasi-isothermal distribution for the mirror-particle plasma density,
\begin{eqnarray}
\rho (r) = {\rho_0 r_0^2 \over r^2 + r_0^2}
\label{qI6}
\end{eqnarray}
where the core radius ($r_0$) scales with disk scale length:
$r_0 \propto r_D$.
As we will see in section 4.4,
these results are supported by more
detailed numerical analysis.
It seems therefore that the dark matter core exists because the supernova sources responsible
for heating the halo have a spatially extended distribution over $r \sim r_D$.
% \cite{foot5,rf2013b}.

Finally, let us conclude this preliminary discussion with a rough estimate of the halo's total heating
and cooling rate. The total cooling rate can be obtained by integrating Eq.(\ref{Eout}).
Assuming a mirror helium dominated halo, so that $n_{e'} (r) = 2\rho(r)/m_{He}$ and 
assuming also that $\rho (r) $ follows the quasi-isothermal distribution of Eq.(\ref{qI6}),
we obtain:
\begin{eqnarray}
{\cal E}_{out}  = \Lambda (T) \rho_0^2 r_0^3 (4/m_{He}^2) I
\end{eqnarray}
where
\begin{eqnarray}
I \equiv 4\pi \int_0^\infty {x^2 \over (1 +x^2)^2} \ dx 
\simeq 9.9
\ .
\end{eqnarray}
Putting some numbers in, we have:
\begin{eqnarray}
{\cal E}_{out} \approx \left({\Lambda (T) \over 10^{-23} \ {\rm erg \ cm}^3/{\rm s}} \right) 
\left( {\rho_0 r_0 \over 50 \ M_\odot/{\rm pc}^2}\right)^2
\left( {r_0 \over 5\ {\rm kpc}}\right) \ 3\times 10^{43} \ {\rm erg/s}
\ .
\end{eqnarray}
The total heating rate can be written as:
\begin{eqnarray}
{\cal E}_{in} &=& f_{SN} \langle E_{SN} \rangle R_{SN}
\nonumber \\
& \simeq & 
\left( {f_{SN} \over 0.1}\right)
\left({\langle E_{SN} \rangle \over 3\times 10^{53}\ {\rm erg}} \right) \left({R_{SN} \over 0.03 \ {\rm yr}^{-1}} \right) 
\ 3\times 10^{43} \ {\rm erg/s}
\label{fsn}
\end{eqnarray}
where $f_{SN}$ is the proportion of the supernova total energy, $E_{SN}$, absorbed by the halo, and $R_{SN}$
is the supernova frequency (the number of supernova per unit time) in the galaxy under consideration.
Equating ${\cal E}_{in}$  with ${\cal E}_{out}$ for the Milky Way galaxy
requires $f_{SN} \sim 0.1$ with an uncertainty potentially as large as an order of magnitude.
Evidently, a substantial fraction of supernova core-collapse energy needs to be converted into mirror photons
and ultimately absorbed by the halo.
We will see in section 4.4 that this will have important implications for the kinetic mixing 
strength, $\epsilon$ and also for the halo mirror-metal mass fraction.

Observe that  the condition: ${\cal E}_{in} \simeq {\cal E}_{out}$ should hold for any galaxy, not just the Milky Way.
This requirement
suggests a rough scaling relation:
\begin{eqnarray}
R_{SN} \propto \Lambda (T) \rho_0 r_0^2
\end{eqnarray}
where we have used $f_{SN} \propto \tau \propto \rho_0 r_0$, valid in the optically thin limit of the halo.
This galactic scaling relation, originally obtained in \cite{r69}, is in rough agreement with the dark matter properties 
inferred from observations of spiral galaxies. 
This analytic result prompted more detailed numerical studies \cite{foot5,rf2013b}, which we shall return to 
in section 4.4, and we 
postpone till then the comparison of predicted scaling relations with those obtained from observations.

\vskip 0.4cm

\subsection{The heating of the galactic halo}  

\vskip 0.2cm

Ordinary type II supernovae can potentially supply a substantial
amount of energy to the  
mirror particle halo, provided of course that kinetic mixing exists.
Can this energy possibly replace the energy lost from the halo
due to dissipation?
To answer this question let us examine this supernova heating mechanism
in more detail.

The kinetic mixing interaction gives the mirror electron and mirror positron a tiny ordinary electric charge 
of magnitude $\epsilon e$.  
This enables processes such as $e\bar e \to e' \bar e'$ and also plasmon decay into $e' \bar e'$
to occur in the hot dense core of ordinary supernovae, leading to the production of light mirror 
particles: $e', \ \bar e', \ \gamma'$ \cite{sil}.
Such particles interact weakly enough with ordinary matter so that they can escape from the supernova core
and also from the collapsing star \footnote{As these mirror particles pass out of the collapsing star,
a small proportion 
%($\sim 1\%$) 
% need to check proportion xxxxx
of the energy of these mirror particles
can be transferred to ordinary matter by processes such as photoionization of ordinary iron, $e'e \to e'e$ scattering etc.
This could be very important, as a major unsolved problem with type II supernova is the mechanism by
which energy is transferred to the shock wave which causes the star to explode.
It has been suggested in \cite{sil} that the energy transferred to ordinary matter via kinetic mixing
induced interactions with the escaping mirror particles might supply
enough energy to the shock wave to enable the star to explode. Naturally, more detailed studies are warranted to carefully
check this idea.}.
The supernova energy carried off by the light mirror  particles can be estimated from the work of \cite{raffelt}
which looked at the general case of energy loss due to light {\it minicharged} particles.
The energy loss rate per unit volume is given by:
\begin{eqnarray}
Q_P = {8 \zeta_3 \over 9 \pi^3}  \epsilon^2 \alpha^2 \left(
\mu^2_e + {\pi^2 T^2
\over 3}\right) T^3 Q_1
\label{raf1x}
\end{eqnarray}
where $Q_1$ is a factor of order unity, and $\mu_e$ is the electron chemical potential and 
$T \sim 30$ 
MeV is the temperature of the supernova core.
The observation of around a dozen neutrino events associated with SN1987A suggests that
$Q_P$ should not exceed the energy loss rate due to neutrino emission.
This indicates a rough upper limit on $\epsilon$ of
$\epsilon \stackrel{<}{\sim} 10^{-9}$ \cite{raffelt}.

For $\epsilon$ near this upper limit, a supernova can provide a huge energy source.
Initially this energy is distributed among the light mirror particles:
$e', \bar e', \gamma'$ (potentially also some fraction in $\nu'$).
These particles, injected into the region around ordinary supernova, would undergo a variety
of complex processes, shocks etc., with the bulk of this energy is expected to be (ultimately) converted into $\gamma'$ emission.
These mirror photons, with total energy up to around half the supernova core collapse
energy ($\sim 10^{53} \ {\rm erg}$ per supernova) can heat the halo.
The idea is that such mirror-photon heating, powered by ordinary supernovae, 
can replace the energy dissipated in
the halo due to bremsstrahlung and other processes.

The supernova energy needed to replace the energy dissipated by the halo 
is sizable. In section 4.1 it was estimated that at least  
a few percent of the total supernova energy needs to be absorbed by the halo. 
Thus, basic particle processes are required in the halo 
which can lead to the absorption of the $\gamma'$ radiation emitted from
the region around each ordinary supernova.
The $H'$ and $He'$ components of the mirror particle plasma are expected to be nearly fully ionized (section 4.3)
and thus one could consider 
Thomson scattering of $\gamma'$ off free $e'$ (figure 2.1).
However, there are two good reasons why such elastic scattering cannot transfer much energy to the halo.
Firstly, the kinematics of Thomson scattering is such that it transfers only a very small proportion of the $\gamma'$ energy to $e'$, and
secondly, the
Thomson cross-section ($\sigma_T = 8\pi\alpha^2/3m_e^2$) is several orders of magnitude too small \cite{sph,r69}. 
Clearly, another particle process
is needed. What about the scattering of mirror photons off the mirror electrons 
which are bound in mirror atoms? In particular, atomic K-shell states of mirror metals have sufficiently
high binding energy ($I > T$) so that these states are typically completely filled (section 4.3). Furthermore, 
it happens that the
scattering of $\gamma'$ off such bound atomic mirror electrons (photoionization) has a much
larger cross-section than the scattering off free mirror electrons.  
Photoionization can also efficiently transfer energy to the halo as the $\gamma'$ is completely
absorbed in this process. To be efficient enough, though, does require that the
halo contain mirror metal components since $H'$ and $He'$ are nearly fully ionized, as we will see. 

\vskip 0.5cm

\subsection{The ionization state of the halo}

\vskip 0.2cm

The halo of spiral galaxies is assumed to be in the form of a spherical plasma consisting of mirror particles: $e', H', He', Fe',...$
In principle, this plasma can be ionized via interactions with mirror photons (from the $\gamma'$ produced near ordinary
supernovae) or by mirror electron collisions.
We will see later that the halo temperature range of interest for typical spiral galaxies, 
is $0.01 \ {\rm keV} \stackrel{<}{\sim} T \stackrel{<}{\sim}$  few keV.
In this temperature range the ionization rate due to $e'$ collisions is generally much
more important than photoionization.
The relevant ionization processes for the mirror-helium components are:
\begin{eqnarray}
 e' + {He'}^{0} &\rightarrow & {He'}^{+} + e' + e' 
\nonumber \\
 e' + {He'}^{+} &\rightarrow & {He'}^{2+} + e' + e' 
\label{14x}
\end{eqnarray}
where ${He'}^0, {He'}^+, {He'}^{++}$ denote the neutral mirror-helium atom,
singly charged mirror-helium ion 
and doubly charged mirror-helium ion.
The cross-sections for the above two processes,  denoted respectively as $\sigma_I^a$ and $\sigma_I^b$,
are known to be well approximated by the Lotz formula \cite{Lotz}:
\begin{eqnarray}
\sigma_I = 4.5 \times 10^{-14} \ \left[ {ln(E/I) \over EI/{\rm eV^2}}\right] \ {\rm cm^2}
\end{eqnarray}
where $E \ge I$ is the energy of the incident $e'$ and $I$ is the relevant ionization potential.
[For the first process above, $I = 24.6$ eV, while for the second one, $I = 54.4$ eV.]

Opposing ionization are the $e'$ capture processes:
\begin{eqnarray}
e ' + {He'}^{+}  &\rightarrow & {He'}^{0}  + \gamma '  
\nonumber \\
e ' + {He'}^{2+}  &\rightarrow & {He'}^{+}  + \gamma '  
\ .
\label{go}
\end{eqnarray}
The cross-sections for these two process, denoted respectively by $\sigma_C^a$,  $\sigma_C^b$, can be approximated by a modified
Kramers formula \cite{kramers}:
\begin{eqnarray}
\sigma_C = \sum_n {8\pi \over 3\sqrt{3}} {\alpha^5 \over n^3} {Z_{eff}^4 \over E_{e'} E_{\gamma'}}
\label{cap}
\end{eqnarray}
where $E_{\gamma'} = E_{e'} + {Z^2_{eff}\alpha^2 m_e \over 2n^2}$.
For the  $He'$ processes in Eq.(\ref{go}), and also for analogous processes involving $e'$ capture by $H'$ and also $Fe'$ ions to be
considered here,
$Z_{eff} = (Z_C + Z_I)/2$, where $Z_C$ is the charge
of the nuclei and $Z_I$ is the ionic charge before $e'$ capture \cite{kramers}.
It follows that $Z_{eff} \approx 1.5$ for the first process in Eq.(\ref{go}), and $Z_{eff} = 2$ for the
second process. 

The above processes determine the number density of ${He'}^{2+}$ via:
\begin{eqnarray}
{dn_{{He'}^{2+}} \over dt} = 
n_{e'} n_{{He'}^{+}}
\langle
\sigma_I^b v_{e'} \rangle
- 
n_{e'} n_{{He'}^{2+}}
\langle \sigma_C^b  
v_{e'} \rangle
\ 
\end{eqnarray}
where the brackets $\langle ... \rangle$
denote the thermal average over the $e'$ energy distribution: 
\begin{eqnarray}
\langle \sigma^b_I v_{e'} \rangle &\equiv &  \sqrt{ {1 \over m_e \pi} } \left(
{2 \over T}\right)^{3/2} \ \int_I^{\infty}
\sigma^b_I
\ e^{-E_{e'}/T} 
\ E_{e'}\ dE_{e'}
\nonumber \\
\langle \sigma^b_C v_{e'} \rangle &\equiv &  \sqrt{{1 \over m_e \pi }} \left(
{2 \over T}\right)^{3/2} \ \int_0^{\infty}
\sigma^b_C
\ e^{-E_{e'}/T} 
\ E_{e'}\ dE_{e'}
\ .
\end{eqnarray}
In a steady state situation $dn_{{He'}^{2+}}/dt = 0$ and thus
\begin{eqnarray}
R_2^{He'} \equiv {n_{{He'}^{2+}}
\over n_{{He'}^{+}}} = 
{
\langle \sigma_I^b v_{e'}\rangle 
\over 
\langle
\sigma_C^b v_{e'} \rangle 
}
\ .
\end{eqnarray}
Similarly, we can derive additional equations by setting $dn_{{He'}^0}/dt = 0$, and for the 
corresponding processes for mirror hydrogen:
\begin{eqnarray}
R_1^{He'} \equiv {n_{{He'}^{+}}
\over n_{{He'}^{0}}} &=& 
{
\langle \sigma_I^a   
v_{e'} \rangle
\over 
\langle \sigma_C^a
v_{e'}\rangle
}
\nonumber \\
R_1^{H'} \equiv {n_{{H'}^{+}}
\over n_{{H'}^{0}}} &=& 
{ 
\langle \sigma_I v_{e'} \rangle
\over 
\langle \sigma_C v_{e'}\rangle}
\end{eqnarray}
where $\sigma_I$ and $\sigma_C$ are the relevant cross-sections for the mirror hydrogen 
processes. 

With the above definitions we can determine the number density of each component as a function of one of them,
taken to be $n_{He'} \equiv n_{{He'}^0} + n_{{He'}^+} + n_{{He'}^{2+}}$:
\begin{eqnarray}
n_{{He'}^{2+}} &=& \left( {R_1^{He'} R_2^{He'} \over 1 + R_1^{He'} + R_1^{He'} R_2^{He'}}\right) \ n_{He'} \nonumber \\
n_{{He'}^{+}} &=& \left( {R_1^{He'}  \over 1 + R_1^{He'} + R_1^{He'} R_2^{He'}} \right) \ n_{He'} \nonumber \\
n_{{He'}^{0}} &=& n_{He'} - n_{{He'}^{+}} - n_{{He'}^{2+}} \nonumber \\
n_{{H'}^{+}} &=& \left( {R_1^{H'} \over 1 + R_1^{H'}} \right) \ f n_{He'} \nonumber \\
n_{{H'}^{0}} &=& f n_{He'} - n_{{H'}^{+}} \nonumber \\
n_{e'} &=& 2n_{{He'}^{2+}} + n_{{He'}^{+}} + n_{{H'}^{+}} \nonumber \\
n_{T} &=& (1 + f)n_{{He'}} + n_{e'} 
\end{eqnarray}
where $f \equiv n_{H'}/n_{He'}$ (with $n_{H'} \equiv n_{H'^{+}} + n_{H'^0}$). 
The fraction, $f$, can be related to the $He'$ mass fraction:
\begin{eqnarray}
\xi_{He'}
= {1 \over 1 + f/4}
\ .
\label{f40}
\end{eqnarray}
We use the primordial mirror-helium mass fraction
as a rough estimate for the mirror-helium mass fraction
in galactic halos.
Figure 3.4 indicates that
$Y'_p \approx 0.9$ for $\epsilon \sim 10^{-9}$ and this motivates
the value:
$f \approx 0.4$\ .

It is now straightforward
to compute the $He'$ ionization fractions, which depend only on the temperature:
$F^{He'}_0 \equiv n_{{He'}^{0}}/n_{He'}$, 
$F^{He'}_1 \equiv n_{{He'}^{+}}/n_{He'}$, 
$F^{He'}_2 \equiv n_{{He'}^{2+}}/n_{He'}$ and
also the $H'$ ionization fractions: 
$F^{H'}_0 \equiv n_{{H'}^{0}}/n_{H'}$, 
$F^{H'}_1 \equiv n_{{H'}^{+}}/n_{H'}$. 
Figure 4.2a gives the results for $He'$ and figure 4.2b results for $H'$.
Figure 4.2a indicates that $He'$ is nearly fully ionized for $T \stackrel{>}{\sim}$ 10 eV, 
substantially below the, $I = 54.4$ eV ionization energy of ${He'}^{+}$.
This happens because for the $He'$ processes, the ionization cross-section is several orders of magnitude larger than
the $e'$ capture cross-section.
Qualitatively similar results occur also for $H'$.

%%%%%%%%%%%%%%%%%%%%%%%%%%%%%%%%%%%%%%% figure 4.2a %%%%%%%%%%%%%%%%%%%%%%%%%%%%%%%%%%%%%%%%%%%
\centerline{\epsfig{file=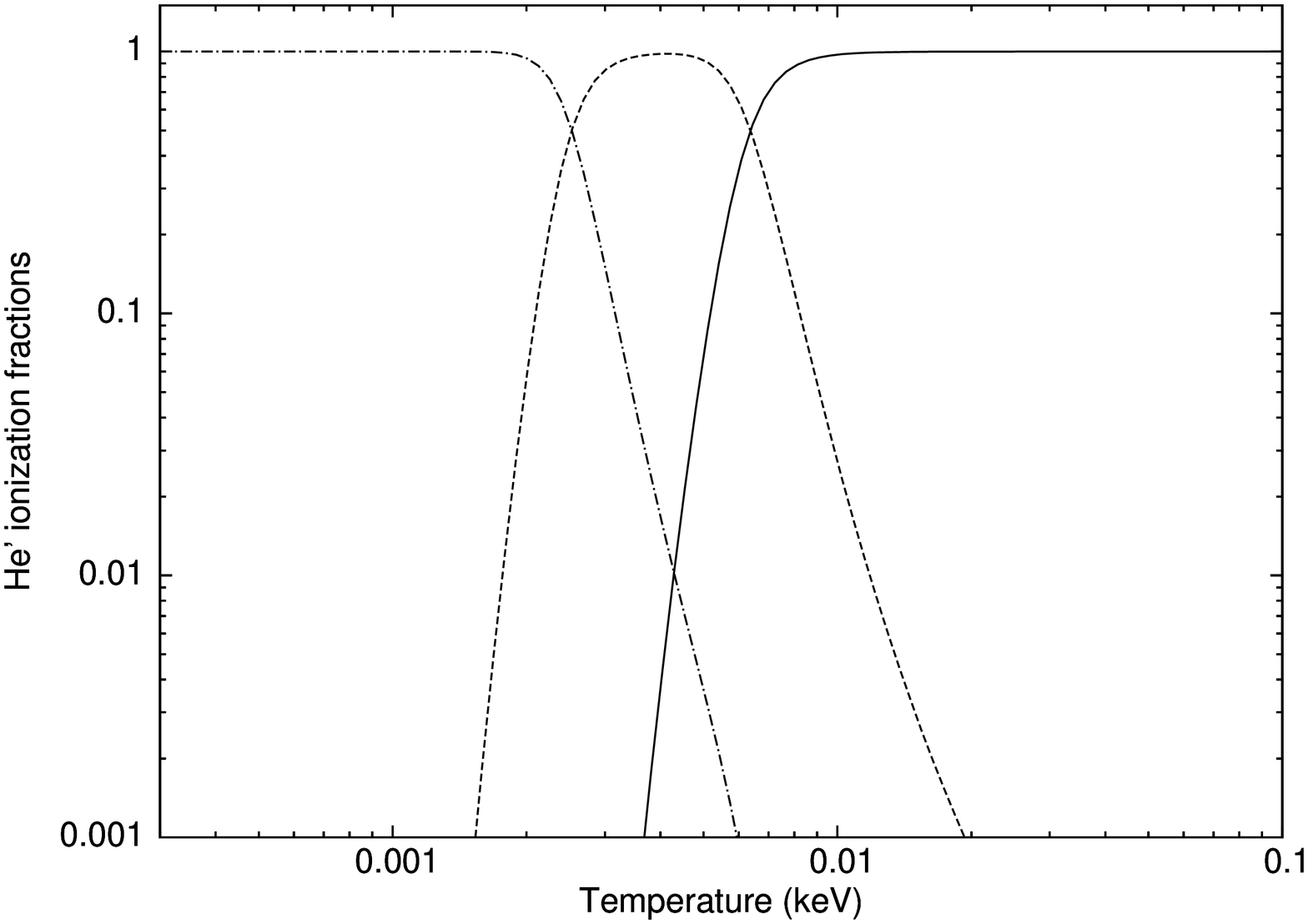, angle=270, width=12.0cm}}
\vskip 0.4cm
\noindent
{\small Figure 4.2a: The $He'$ ionization fractions,
as a function of the local halo temperature, $T$.
Shown are 
$F^{He'}_0 \equiv n_{{He'}^{0}}/n_{He'}$ (dashed-dotted line), 
$F^{He'}_1 \equiv n_{{He'}^{+}}/n_{He'}$ (dashed line) 
and $F^{He'}_2 \equiv n_{{He'}^{2+}}/n_{He'}$ (solid line). 
}

\vskip 0.2cm
%%%%%%%%%%%%%%%%%%%%%%%%%%%%%%%%%%%%%%%%%%%%%%%%%%%%%%%%%%%%%%%%%%%%%%%%%%%%%%%%%%%%%%%%%%%%%%%%%
%%%%%%%%%%%%%%%%%%%%%%%%%%%%%%%%%%%%%%% figure 4.2b %%%%%%%%%%%%%%%%%%%%%%%%%%%%%%%%%%%%%%%%%%%
\centerline{\epsfig{file=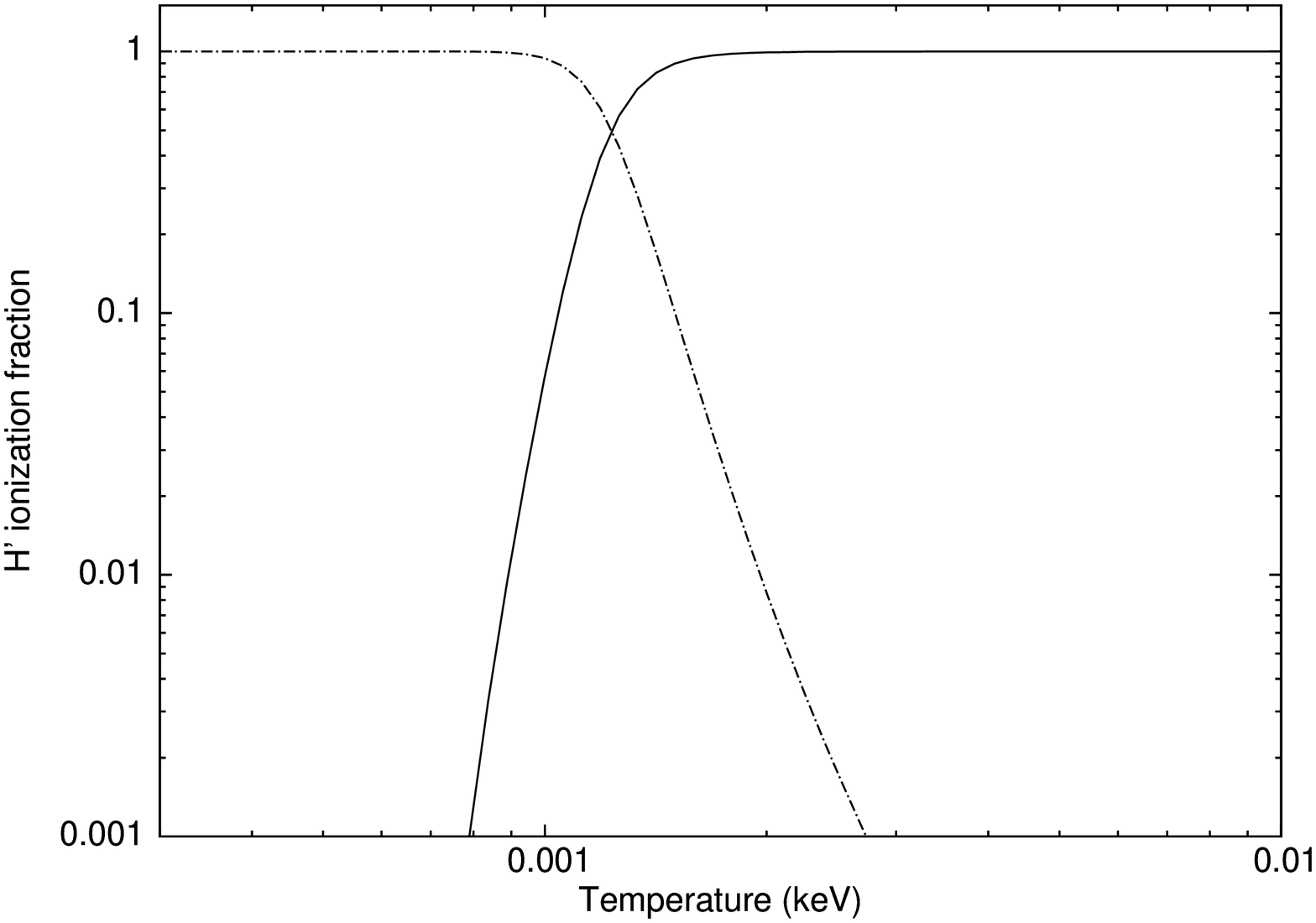, angle=270, width=12.0cm}}
\vskip 0.3cm
\noindent
{\small Figure 4.2b: The $H'$ ionization fractions,
as a function of the local halo temperature, $T$.
Shown are 
$F^{H'}_0 \equiv n_{{H'}^{0}}/n_{H'}$ (dashed-dotted line) 
and $F^{H'}_1 \equiv n_{{H'}^{+}}/n_{H'}$ (solid line). 
}

\vskip 1.0cm
%%%%%%%%%%%%%%%%%%%%%%%%%%%%%%%%%%%%%%%%%%%%%%%%%%%%%%%%%%%%%%%%%%%%%%%%%%%%%%%%%%%%%%%%%%%%%%%%%

%%%%%%%%%%%%%%%%%%%%%%%%%%%%%%%%%%%%%%% figure 4.3 %%%%%%%%%%%%%%%%%%%%%%%%%%%%%%%%%%%%%%%%%%%
\centerline{\epsfig{file=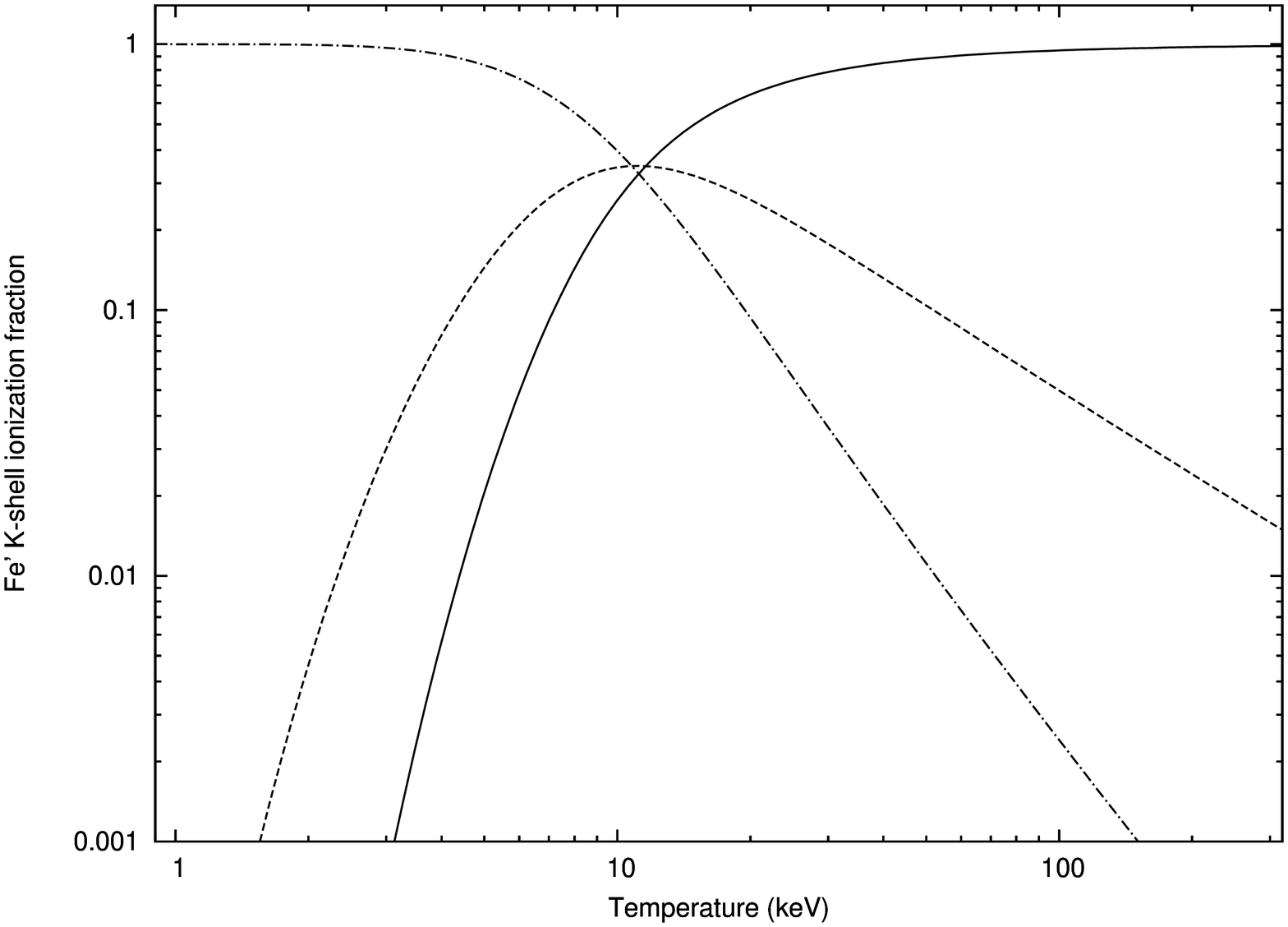, angle=270, width=12.0cm}}
\vskip 0.3cm
\noindent
{\small Figure 4.3: The $Fe'$ ionization fractions as a function of the local halo temperature, $T$.
$F^{Fe'}_1 \equiv n_{{Fe'}^{*}}/n_{Fe'}$ (dashed line) is the fraction of states with one K-shell state filled, 
$F^{He'}_2 \equiv n_{{Fe'}^{**}}/n_{Fe'}$ (solid line) is the fraction of states that are fully ionized.
The dashed-dotted line is the fraction of states with both K-shell states filled.
}

\vskip 1.4cm
%%%%%%%%%%%%%%%%%%%%%%%%%%%%%%%%%%%%%%%%%%%%%%%%%%%%%%%%%%%%%%%%%%%%%%%%%%%%%%%%%%%%%%%%%%%%%%%%%

In addition to the $H', He'$ components, we will consider a small metal component, 
taken as $Fe'$, with
total number density, $n_{Fe'}$.
We denote the number density of fully ionized $Fe'$ as $n_{{Fe'}^{**}}$ and  $Fe'$ with one K-shell $e'$ 
as $n_{{Fe'}^*}$.
%The ionization energy of the bound mirror electron in ${Fe'}^{*}$ is 9.3 keV
%and if both K-shell mirror electrons are present, the binding energy is 8.8 keV\cite{sugar}.
The relevant equations governing the K-shell ionization state of mirror iron 
are a straightforward
generalization of the equations for mirror helium.
In figure 4.3 the 
computed ionization fractions, $F^{Fe'}_1 \equiv n_{{Fe'}^*}/n_{Fe'}$ and
$F^{Fe'}_2 \equiv n_{{Fe'}^{**}}/n_{Fe'}$ are given.
Also shown is the fraction of states with both K-shell states filled: $1 - F^{Fe'}_1 - F^{Fe'}_2$.
For the temperature range of interest for spirals, 
$0.01 \ {\rm keV} \stackrel{<}{\sim} T \stackrel{<}{\sim}$  few keV,
this figure indicates that $Fe'$ generally has both K-shell states filled,
and figure 4.2 indicates that both $H'$ and $He'$ are typically nearly fully ionized.

\vskip 0.4cm

\subsection{The dynamical halo model}

\vskip 0.2cm

The dark matter halo surrounding spiral galaxies is assumed to be in the form of a spherically distributed mirror-particle
plasma. This halo has non-trivial dynamics.
As described in section 4.1, the halo (currently) is expected to be in hydrostatic
equilibrium where gravity is balanced by pressure. The halo is dissipative continuously losing
energy as mirror radiation escapes and this energy is replaced by heating. The candidate heat source is ordinary core-collapse
supernova which can convert a significant fraction of their gravitational potential energy into creation of energetic 
mirror particles via kinetic mixing
induced processes in the supernova core (section 4.2).
The end result is that halo dynamics is determined by three things:
(a) dissipation (b) heating and
(c) hydrostatic equilibrium.

The key idea is that the dynamics will drive the system to a configuration whereby dissipation exactly matches heating [Figure 4.1].
If this does indeed happen, then
independently of a galaxy's past history, 
the halo should have evolved to a state where
the energy being absorbed in each volume element is equal to the energy being
radiated from the same volume element.
Thus, we have the dynamical requirement:
\begin{eqnarray}
{d^2 E_{in} \over dt dV}= 
{d^2 E_{out} \over dt dV} 
\ .
\label{meal8}
\end{eqnarray}
Approximate equations for the left and right-hand sides
of the above condition have been derived in \cite{foot5}. These equations, and their derivation, are briefly
outlined below.

\vskip 0.5cm
\noindent
{\bf (a) dissipation}
\vskip 0.5cm
\noindent
It has been argued in \cite{foot5} that thermal bremsstrahlung is the most important dissipative process.
The rate at which energy is radiated via thermal bremsstrahlung per unit volume, 
at a particular point $P$ in the halo, is \cite{bookastro}:
\begin{eqnarray}
{d^2 W \over dt dV} = {16\alpha^3 \over 3m_e}\left({
2\pi T \over 3m_e}\right)^{1/2} 
\ \sum_j
\left[ Z_j^2 n_j n_{e'} \bar g_B  \right]
\label{john}
\end{eqnarray}
where the index $j$ runs
over the mirror ions in the plasma (of charge $Z_j$).
Also,
$\bar g_B$ is the frequency average of the velocity averaged
Gaunt factor for free-free emission.
We set $\bar g_B = 1.2$, which is known to be
accurate to within about 20\% \cite{bookastro} .

In principle not all energy radiated will escape from the halo. In fact, rough estimates
indicate that the halo is likely to be optically thick for mirror photons of energies greater
than around the (mean) halo temperature. The precise details will depend on
the chemical composition of the halo.  Thus, the energy radiated should be multiplied by
an efficiency factor $\epsilon_f$ to obtain the effective cooling rate:
\begin{eqnarray}
{d^2 E_{out} \over dt dV} = \epsilon_f {d^2 W \over dt dV}\ .
\label{epsilon41}
\end{eqnarray}
In our numerical work, though,
we set $\epsilon_f = 1$. See \cite{foot5} for further discussions.

\vskip 0.3cm
\noindent
{\bf (b) heating}
\vskip 0.3cm
\noindent
As discussed in section 4.2, kinetic mixing induced interactions lead to the production of
light mirror particles, $e', \bar e', \gamma'$ in the core of 
ordinary type II supernovae.
The energy carried away by mirror particles can be
comparable to that of neutrinos for $\epsilon \sim 10^{-9}$ \cite{raffelt,sil}.
The bulk of this energy is, ultimately, expected to be converted into mirror photons in the region around
each ordinary supernova. It is proposed that
these mirror photons, with total energy up to around half the supernova core collapse
energy ($\sim 10^{53} \ {\rm erg}$ per supernova), are responsible for replacing the energy lost in the halo due to dissipation.
Energy is transferred to the halo by the 
interactions (photoionization) of $\gamma'$ with heavy mirror
metal ions, which is possible because these components
retain their K-shell mirror electrons. 

To illustrate these ideas, we consider a halo composed of $H',\ He',\ e'$ and a small $Fe'$ component.
Considering a metal component consisting only of $Fe'$ can, in fact, be a reasonable approximation if 
the proportion of the supernova $\gamma'$ energy
contributed by $\gamma'$ with $E_{\gamma'}$
less than the $Fe'$ K-shell binding energy ($I \approx$ 9 keV)
is small. In fact, even if it is not small, inclusion of additional metal components, such as $O', Si'$,
would merely increase the energy absorbed by the halo. This effect can be accounted for, to a first approximation, by considering
just the $Fe'$ component with an exaggerated number density, which is anyway an unknown parameter
\footnote{Knowledge of the
detailed chemical composition of the halo metal component, although not absolutely essential
for the approximate computations discussed here, would be very useful to have.
This will depend on factors such as the relative rates of type II and type Ia mirror supernovae at
early times when the mirror metals were (presumably) synthesized.}.
A similar argument can be used to justify another simplifying assumption made:
In the subsequent analysis, we consider only the
photoionization of K-shell $e'$ atomic states in $Fe'$. 
In principle higher atomic shells will be filled and their inclusion will increase the total $Fe'$ photoionization 
cross-section, and hence the total energy
absorbed by the halo. Again, this effect can be accommodated, to a first approximation, by a moderate increase in the $Fe'$ number density.

The $Fe'$ number density, can be expressed in terms of the halo $Fe'$ mass fraction, $\xi_{Fe'}$, 
\begin{eqnarray}
n_{Fe'} = n_{He'} \left( {1 \over 1 + f/4}\right)  \left({m_{He} \over m_{Fe}}\right) \left({\xi_{Fe'} \over 1 -
\xi_{Fe'}}\right)
\ 
\end{eqnarray}
where $f \equiv n_{H'}/n_{He'}$ . 
As discussed after Eq.(\ref{f40}), early Universe cosmology suggests $f \approx 0.4$ for $\epsilon \sim 10^{-9}$.
The total $Fe'$ ($Z=26$) photoelectric cross-section
is given by (see e.g. \cite{bookastro}):
\begin{eqnarray}
\sigma_{PE}^g (E_{\gamma'}) = 
{g 16\sqrt{2} \pi \over 3m_e^2 } \alpha^6  Z^5 \left[ {m_e \over E_{\gamma'}
}\right]^{7/2} 
\ \ {\rm for} \ E_{\gamma'} \gg I
\ .
\label{pe4}
\end{eqnarray}
Here $g = 1$ or 2 counts the initial number
of K-shell mirror electrons present.
Evidently, the photoelectric cross-section scales with $\gamma'$ energy like $(E_{\gamma'})^{-7/2}$ for $E_{\gamma'}$ much greater
than the ionization energy, $I$. Near threshold, the cross-section has a marginally softer behaviour: $(E_{\gamma'})^{-3}$
and drops abruptly to zero at $E_{\gamma'} = I$ \cite{bookastro}.

The mirror photons produced in the regions around ordinary supernovae are responsible,
it is alleged, for heating the halo.
These mirror photons combine to yield the time averaged flux $F$, at a particular point $P$, in the halo.
The energy deposited per unit volume per unit
time at $P$ is then:
\begin{eqnarray}
{d^2 E_{in} \over dt dV}=  \int {dF  \over dE_{\gamma'}} \ \sum_{g=1,2} n_{Fe'}^g  (r) \ \sigma^g_{PE} 
\ dE_{\gamma'}
\ .
\label{meal}
\end{eqnarray}
where $n_{Fe'}^g$ is the number of $Fe'$ ions with $g = 1,2$ mirror-electrons bound in the atomic K-shell 
state (calculated as in section 4.3).
As a first approximation,
one expects that the supernova $\gamma'$ sources are distributed in proportion to the stellar disk
surface density, $\Sigma_*$, which can be modeled \cite{freeman}: 
\begin{eqnarray}
\Sigma_* = {m_D \over 2\pi r_D^2} \ e^{-r/r_D}\ 
\label{mum1}
\end{eqnarray} 
where $m_D$ is the total stellar mass of the disk 
and $r_D$ is the disk scale length.

To simplify the problem, we replace the disk with a spherically symmetric distribution, $\rho_D (r)$.
This distribution is
defined by requiring that the mass within a radius $r$ is the same as that of the 
disk, 
i.e. $\int^{r}_{0} \rho_D 4\pi r'^2 dr' \equiv \int^r_0 \Sigma_* \ 2 \pi r' dr' $, so that
\begin{eqnarray}
\rho_D (r) = {m_D \over 4\pi r_D^2 r} \ e^{-r/r_D} \ .
\label{mum2}
\end{eqnarray}
The supernova $\gamma'$ sources are then assumed to be
distributed in proportion to this spherically symmetric density, $\rho_D (r)$. It follows that
the average differential flux at a point $P$ in the halo is then:
\begin{eqnarray}
{dF(r) \over dE_{\gamma'}} = 
R_{SN} \ E_{\gamma'} {dN_{\gamma'} \over dE_{\gamma'}} 
%\kappa \left( E_{\gamma'}\right)^{c_1} R_{SN} 
\ 
\int^{\infty}_0 \int^1_{-1} \ {\rho_D \over m_D} \ {e^{-\tau} \ r'^2 \over 2{\rm d}^2}\ \ d\cos\theta \ dr'
\ 
\label{33a}
\end{eqnarray}
where $E_{\gamma'}dN_{\gamma'}/dE_{\gamma'}$ is the mirror photon energy-weighted spectrum resulting from
an average supernova (to be discussed in more detail in a moment) and
$R_{SN}$ is
the type-II supernova frequency in the galaxy under consideration. Also,
${\rm d} = \sqrt{ r^2 + r'^2 - 2rr' \cos\theta}$ is the distance between the supernova source at a point $Q$ and the point $P$ in the halo, while
$\tau$ is the  optical depth along this path: 
%$\tau = \int_0^{\rm d} n_{Fe'} \sigma_{PE} \ dy $.
%\end{eqnarray}
\begin{eqnarray}
\tau = \int_0^{\rm d} \sum_g n_{Fe'}^g (r_1) \sigma^g_{PE} \ dy
\ .
\label{str}
\end{eqnarray}

%%%%%%%%%%%%%%%%%%%%%%%%%%%%%%%%%%%%%%% figure 4.4 %%%%%%%%%%%%%%%%%%%%%%%%%%%%%%%%%%%%%%%%%%%
\vskip 1.0cm
\centerline{\epsfig{file=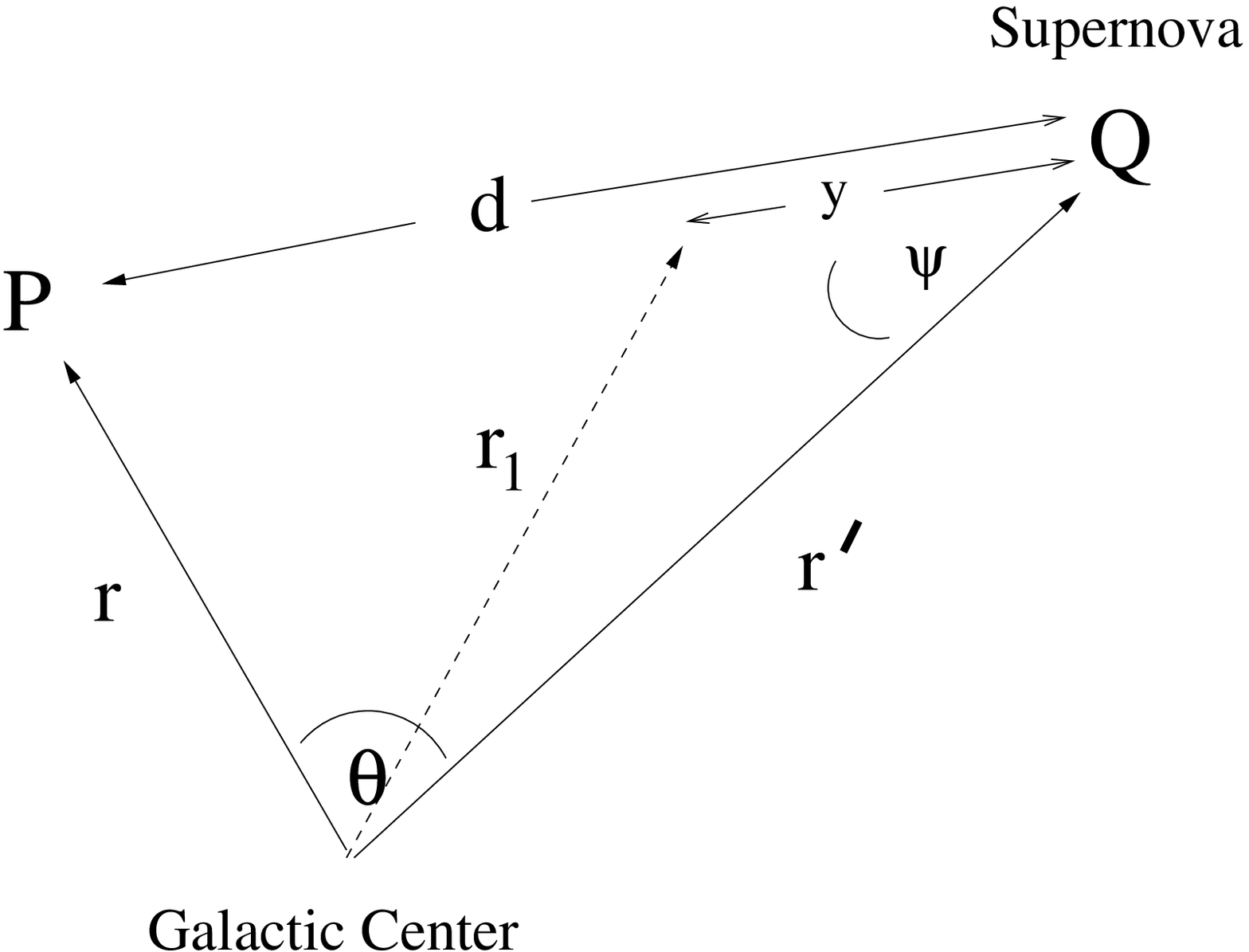, angle=0, width=8.3cm}}
\vskip 0.6cm
\noindent
{\small Figure 4.4: The geometry. Mirror photons travel a distance ${\rm d}$ from a supernova source at point $Q$, to heat the halo at a point $P$.
}
\vskip 1.9cm
%%%%%%%%%%%%%%%%%%%%%%%%%%%%%%%%%%%%%%%%%%%%%%%%%%%%%%%%%%%%%%%%%%%%%%%%%%%%%%%%%%%%%%%%%%%%%%%%%

\noindent
The relevant geometry is shown in figure 4.4, and we have
\begin{eqnarray}
r_1 &=& \sqrt{y^2 + r'^2 - 2r'y\cos\psi} \nonumber \\
\cos \psi &=& {{\rm d}^2 + r'^2 - r^2 \over 2r'{\rm d}} \ .
\end{eqnarray}
The integrals in Eq.(\ref{33a}) take into account all possible supernova source locations $Q$, appropriately weighted.

The rough analytic considerations of section 4.1 assumed an optically thin halo, but this may not be true for all $\gamma'$
energies. In the subsequent numerical analysis therefore, the effects due to the finite optical
depth, as encoded in the above equations, are computed.

The frequency at which supernovae occur in a given galaxy, $R_{SN}$, is an important ingredient.
The analysis of \cite{foot5} used the rough
baryonic scaling relation \cite{sal3} $m_D \propto (L_B)^{1.3}$ and $R_{SN}\propto (L_B)^{0.73}$
from supernova observations \cite{sn} to derive $R_{SN} \propto (m_D)^{0.56}$ for spiral galaxies.
The supernova study \cite{sn} also provides a measurement of $R_{SN}$ versus $\stackrel{\sim}{m}_D$,
where $\stackrel{\sim}{m}_D$ is the stellar mass
derived from photometry and spectral fitting,
finding $R_{SN} \propto (\stackrel{\sim}{m}_D)^{0.45}$.
These observations and studies motivate $R_{SN} \propto (m_D)^{0.5}$ which we adopt as a reference
value.
[A roughly similar scaling behaviour is found for the
overall star formation rate \cite{kenrev} in spiral galaxies; such
a correlation is of course not unexpected.]
Naturally, it is certainly possible that such a relation would be prone to some significant scatter
around the mean.

%%% new here
%%%% revised version below..remove above..keep below

The (average) mirror photon energy spectrum resulting from a single supernova is uncertain.
To proceed, let us assume that the peak of this (averaged) $\gamma'$ energy spectrum occurs at energies 
higher than the K-shell $e'$ atomic binding
energy of $Fe'$, $I \sim 9$ keV. 
[This is assumed for definiteness; it is not however expected to be an essential requirement.]
If this is the case then the $\sigma_{PE}^g \propto (E_{\gamma'})^{-7/2}$ behaviour indicates that 
the low energy part of the spectrum is mainly responsible for heating the halo.
We parameterize this unknown spectrum via a power law with index $c_1$, and cut-off $E_c$:
\begin{eqnarray}
E_{\gamma'} {dN_{\gamma'} \over dE_{\gamma'}} = {\cal E}_{SN}\ \left( {1 + c_1 \over E_c}\right)\left(
{E_{\gamma'}\over E_c}\right)^{c_1} \ 
%\nonumber \\
%\vspace 0.1cm
\equiv  \ \kappa \left(E_{\gamma'}\right)^{c_1}
\ .
\label{m3}
\end{eqnarray}
This spectrum has been normalized such that 
\begin{eqnarray}
\int^{E_c}_0 E_{\gamma'} {dN_{\gamma'} \over dE_{\gamma'}} \ dE_{\gamma'} = {\cal E}_{SN}
\ .
\end{eqnarray}
In the equations above, ${\cal E}_{SN}$ is the amount of energy ultimately converted into 
the creation of mirror photons
from an average supernova
(${\cal E}_{SN} \sim 10^{53}$ erg for $\epsilon \sim 10^{-9}$).

The total $\gamma'$ energy produced (on average) by ordinary supernovae per unit time in a given galaxy
is given by ${L'}_{SN} = R_{SN} \ {\cal E}_{SN}$.
With the above definitions, and the adopted scaling: $R_{SN} \propto (m_D)^{0.5}$, it follows that:
\begin{eqnarray}
\kappa R_{SN} = {1 + c_1 \over (E_c)^{1+c_1}}
\left( {m_D \over m_D^{MW}}\right)^{0.5}   
%{L_B \over L_B^{MW}}\right)^{0.73}
\ {L'}_{SN}^{MW}\ .
% \ 1.5 \times 10^{44} \ {\rm erg/s}
\label{kap2}
\end{eqnarray}
Here ${L'}_{SN}^{MW}$ is the total $\gamma'$ luminosity resulting, on the average, from ordinary supernovae for a
reference $\sim $ Milky Way sized spiral galaxy of stellar mass $m_D^{MW} = 5\times 10^{10} \ m_\odot$.
[${L'}_{SN}^{MW} = R_{SN}^{MW} {\cal E}_{SN} \sim 10^{44} \ {\rm erg/s}$ for $\epsilon \sim 10^{-9}$.]

Although the $\gamma'$ spectrum could not be a simple power law for energies sufficiently high, such details may not be so
important since the halo is optically thin for energies: $E_{\gamma'} \stackrel{>}{\sim} 30$ keV \cite{foot5}.
Therefore, replacing the entire spectrum by the extrapolation of the
low energy spectrum, i.e. by a power law and cut-off, can be a simple yet potentially accurate parameterization.
In the numerical analysis to follow, we examine the range of values for the index, $c_1$:
$1 \le c_1 \le 3$. [This range is centered around $c_1 = 2$ which corresponds to the low energy part of a thermal spectrum.]
It turns out that the numerical results, as we will see shortly, have practically no essential dependence on the precise value of the index $c_1$;
our poor knowledge of the mirror photon energy spectrum is, therefore, not a serious obstacle in deriving halo properties from
this dynamics. 

Regarding the cut-off parameter, $E_c$, its precise value simply
determines the proportion of the supernova energy contributed by mirror photons with energies below $\sim 30$ keV.
This proportion is unknown, however it cannot be too small otherwise there will be insufficient energy to power the halo.
In the numerical work we fix $E_c = 50$ keV.
Although this may seem arbitrary, it is really not so: 
Changing $E_c$ has approximately the same effect on the dynamics as changing ${L'}_{SN}^{MW}$  since both quantities simply
scale the total amount of energy supplied to the halo.
Thus, the $E_c$ parameter dependence is simply transferred to the parameter ${L'}_{SN}^{MW}$.

\vskip 0.8cm
\newpage

To summarize the halo heating, ordinary supernovae convert a substantial fraction of their core collapse energy into the production
of light mirror particles: $e',\ \bar e',\ \gamma'$, given that photon - mirror photon kinetic mixing of strength $\epsilon \sim 10^{-9}$ exists. 
In the region around each supernova this energy is converted into a flux
of X-ray $\gamma'$ radiation, $dF/dE_{\gamma'}$, which transports this energy to the halo. The energy absorbed at a particular point, $P$, is given 
in terms of this mirror-photon flux by
Eq.(\ref{meal}). This flux
can be approximated as:
\begin{eqnarray}
{dF(r) \over dE_{\gamma'}} = 
\kappa R_{SN} (E_{\gamma'})^{c_1}  
\ 
\int^{\infty}_0 \int^1_{-1} \ {\rho_D \over m_D} \ {e^{-\tau} \ r'^2 \over 2[r^2 + r'^2 - 2rr'\cos\theta]}\ \ d\cos\theta \ dr'
\ 
\label{sum44}
\end{eqnarray}
with $\kappa R_{SN}$ given in Eq.(\ref{kap2}), $\rho_D$ in Eq.(\ref{mum2}) and $\tau$ in Eq.(\ref{str}).

\vskip 0.3cm
%\newpage
\noindent
{\bf (c) hydrostatic equilibrium}
\vskip 0.3cm
\noindent
The mirror-particle plasma component of the halo consists of the mirror particles, $e', H', He',...$, 
interacting via mirror electromagnetic interactions.
To proceed we will need to know something about the {\it current} chemical composition of the halo.
We found in section 3.3 that early Universe cosmology yields a 
primordial mirror-helium mass fraction of $Y'_p \approx 0.9$ for $\epsilon \sim 10^{-9}$,
with negligible primordial production of mirror metal components.
As discussed earlier, a substantial mirror metal component can be produced in mirror stars at a very early time,
and is in fact needed for the halo to absorb enough energy.

The mirror particles, $e', H', He', Fe',...$
are presumed to be approximately spherically distributed, interacting with each other via Coulomb scattering.
Eventually the halo should end up in hydrostatic equilibrium, where gravity is balanced by the pressure gradient.
Spherical symmetry implies
that the pressure $P$, density $\rho$, and gravitational acceleration $g$, 
depend only on the radial distance $r$.
They are related by the hydrostatic equilibrium condition, Eq.(\ref{p9u}).
Introducing the total particle number density, $n_T(r)$, and the mean mass, $\bar m(r)$:
\begin{eqnarray}
n_T (r) = \sum_{A'} n_{A'}, \ \ \ \bar m(r) \equiv {\rho(r) \over n_T(r)} = {\sum_{A'} n_{A'}(r) \ m_{A'}\over n_T(r)
}
\ .
\end{eqnarray}
The hydrostatic
equilibrium equation can be rewritten in the form:
\begin{eqnarray}
n_T \ {dT \over dr} + T \ {dn_T \over dr} = -\bar m(r) \ n_T(r) \ g(r)
\ 
\label{p9}
\end{eqnarray}
where
the ideal gas law has been used to relate $P(r) = n_T (r)T(r)$ 
\footnote{
In principle a plasma need not be described by a single temperature $T(r)$. The mirror-nuclei (heavy) and  
mirror-electron (light) components could have different temperatures because energy is transferred inefficiently between
the heavy and light components.
However, if the heating of the halo
is due to the photoionization process
then the energy is transmitted initially to the mirror-electron component rather than the mirror nuclei.
The ejected $e'$ will interact with the plasma (via Coulomb interactions etc.)
mainly heating the $e'$ component.  Dissipative processes also
primarily cool the $e'$ component rather than the mirror nuclei. Thus, to a first approximation, both heating and cooling
processes act on the mirror-electron component only. It follows that the simplistic treatment of using a single local halo
temperature to describe both components should be reasonable. 
Also, note that local fluctuations in the halo temperature will be smoothed by thermal conduction.
}.
In section 3.3 we estimated that $\bar m \approx 1.1$ GeV [Eq.(\ref{mbar99})] assuming
a fully ionized plasma with mirror-helium mass fraction: $Y'_p \approx 0.9$ as suggested by mirror BBN (for $\epsilon \sim 10^{-9}$).
This initial $\bar m$ value can be adjusted to take into account the actual ionization state
of the halo at a given radius, $r$,
as determined by solution of the equations in section 4.3.

With the assumption of spherical symmetry, Newton's law can be used to relate
the acceleration $g(r)$ with 
the total mass density, $\rho_{total}$:
\begin{eqnarray}
g(r) = {G_N\over r^2} \int_0^r  \rho_{total}\ dV
\ .
\label{g77}
\end{eqnarray}
%where $G_N$ is Newton's constant.
The total mass density has a baryonic and mirror baryonic component.
In spiral galaxies the baryonic component can be further separated into a stellar component and a gas
component. The stellar component can be approximated by
a Freeman disk, Eq.(\ref{mum1}). [In principle, there can also be a stellar bulge component, which
we have chosen to disregard in this analysis.] 
As before, we replace this with the spherically symmetric distribution,
$\rho_D (r)$, Eq.(\ref{mum2}).
Although the baryonic density has both a stellar and gas component,
to a first approximation one can consider the stellar component only.
For typical spirals, $10^9 \ m_\odot \stackrel{<}{\sim} m_D \stackrel{<}{\sim} 10^{12}\ m_\odot$,
the mass of the gas component is generally smaller than that of the stellar
component and importantly its distribution is more (radially) extended.
Therefore, at any given radius the gas contribution to the gravitational acceleration, $g(r)$, is much
smaller than either the stellar contribution or the dark matter contribution.
Thus a simple, but also reasonable, first approximation is to take
\begin{eqnarray}
\rho_{total}(r) = \rho_D (r) + \rho (r) 
\label{neg53}
\end{eqnarray}
where $\rho_D (r)$ is the baryonic 
contribution (the subscript `D' stands for `Disk' not `Dark'!). It is given in 
the spherically symmetric approximation by Eq.(\ref{mum2}), 
and $\rho (r)$, of course, is the mirror-particle plasma component which we
seek to derive by solving the dynamical equations.

The total mass density, $\rho_{total}(r)$, also contains a compact-object component composed of old mirror stars, 
$\rho'_D (r)$, which we have negligently excluded in Eq.(\ref{neg53}).
One reason for this negligence is that the mass and distribution of this compact-object component is quite uncertain.
One might suspect that old mirror stars would be distributed in a `dark' disk, possibly aligned with the baryonic disk cf. \cite{randal}, 
even so, its scale length need not be the same as that of the thin baryonic disk.
We make the convenient assumption that this component is subdominant and can therefore be approximately neglected, to a first approximation. 
Naturally, the mirror star component cannot be completely negligible as mirror
star evolution is required to generate the needed mirror metal components of the plasma halo.
The neglect of $\rho'_D(r)$ is therefore a source of a potentially significant systematic uncertainty in the
subsequent analysis.

A boundary condition is needed to solve the
hydrostatic equilibrium equation. 
A natural choice, adopted in the numerical analysis (to follow) is to require $dT/dr \to 0$ at large galactic radius, $R_{gal}$, taken to be
50$r_D$.  The numerical results are
independent, to a very good approximation, of the particular value of $R_{gal}$ chosen
provided, of course, that $R_{gal} \gg r_D$.
Importantly, 
the physical quantities of interest, the $\rho(r),\ T(r)$ evolution in the inner
region: $r \stackrel{<}{\sim} 10 r_D$,
are also not very sensitive to the particular boundary condition used.
For example, one can check that the alternative boundary condition:
$T = 0$ at $r=R_{gal}$,
gives the same $\rho(r), \ T(r)$ evolution in the inner halo region of interest 
%$r \stackrel{<}{\sim} 10 r_D$.
\footnote{
The full extent
of the halo, i.e. the physical scale $R_{gal}$, is certainly of great interest, but is presently unknown. At large
distances, $r \gg r_D$, there can be a number of complications: Mirror particle outflows from the halo,
inflows from the surrounding medium (e.g. from the cluster's mirror particle halo), tidal interactions etc.
Additionally, for $r$ large enough,
the density inevitably becomes so low as to invalidate the assumption of hydrostatic equilibrium.   
}.

\vskip 0.6cm
\noindent
{\bf Baryonic properties of galaxies}
\vskip 0.4cm
\noindent
The halo dynamics, as portrayed above, intricately relates the dark matter properties of galaxies
to their baryonic properties. The heating, cooling and hydrostatic equilibrium
equations depend either directly or indirectly on the baryonic properties.
It is useful therefore,
to pause and clearly state the baryonic properties that we will use.
As already mentioned, around Eq.(\ref{kap2}), we assume that $R_{SN} \propto (M_D)^{0.5}$.
Additionally,
we utilize the empirical 
correlation between the baryonic disk mass $m_D$ and the disk radius, $r_D$ \cite{lapi,PSS}:
\begin{eqnarray}
\log \left( {r_D \over {\rm kpc}}\right) = 0.633 + 0.379
\log \left( {m_D \over 10^{11}\ m_{\odot}}\right) 
+ 0.069 \left[\log \left( {m_D \over 10^{11}\ m_{\odot}}\right)\right]^2
\ .
\label{rd}
\end{eqnarray} 
With these relations, the baryonic parameters of spiral galaxies are (roughly) specified by a single
parameter which can be taken as either  $m_D$ or  $r_D$.

In the discussion to follow, we apply the assumed halo dynamics to derive the dark matter properties 
for a generic galaxy. It can be a spiral or an irregular galaxy. So long as there is active star formation 
its morphology does not seem to be of critical importance in relation to its 
dark matter properties that we will derive. 
%parameterized by $m_D, \ r_D$.
We consider the stellar mass range:
\begin{eqnarray}
10^{9} \ m_\odot \stackrel{<}{\sim} m_D \stackrel{<}{\sim} 10^{12} \ m_\odot
\label{mdrange}
\end{eqnarray}
which is typical for spirals.
For each $m_D$ the baryons are assumed to be distributed 
as in Eq.(\ref{mum2}) with $r_D$ obtained from
Eq.(\ref{rd}).

\vskip 0.6cm
\noindent
{\bf Solving the equations}
\vskip 0.4cm
\noindent
The equations governing the dark matter distribution, Eqs.(\ref{meal8}), (\ref{p9}), 
are coupled integro-differential equations. Numerically solving these equations
is non-trivial. If we are able to `guess' the dark matter distribution, $\rho (r)$,
we can then use the hydrostatic equilibrium condition, Eq.(\ref{p9}), to determine the corresponding $T(r)$.
For each $r$ value, we can then determine the ionization state of the halo by solving the equations given in
section 4.3. Once
the ionization state is known, we can
evaluate $\bar m (r)$ and check whether it is significantly different
from the starting value, assumed to be $\bar m = 1.1$ GeV (independent of $r$).
[Recall from section 3.3 that $\bar m = 1.1$ GeV assumed the halo was fully ionized.]
The ionization state is also needed to evaluate the cooling [Eq.(\ref{john})] and 
heating rates [Eqs.(\ref{meal}), (\ref{sum44})].
This gives us enough information to check whether our guess for $\rho (r)$
satisfies Eq.(\ref{meal8}). A more sophisticated adaption of this procedure, well suited to a numerical code,
is to replace the `guess' for the dark matter distribution, $\rho (r)$, with a parameterization with
a set of adjustable parameters, $X$ (we use $\rho_0,\ r_0,\ \beta$ below). We can then repeat the procedure, scanning over the parameters, $X$, finding the
particular values of the parameters for which  
Eq.(\ref{meal8}) holds most accurately.
This methodology is summarized by the flowchart in figure 4.5.

The analytic arguments of section 4.1 provide a good starting point.
Recall, these arguments suggest $\rho (r) \propto 1/r^2$ at large radii and that $\rho (r) \to constant$ in an
inner region.
%The conditions, Eq.(\ref{meal8},\ref{p9}), can be solved numerically. 
This motivates a parameterization of the dark matter distribution along the lines of
Eq.(\ref{qI6}), which has two parameters: $\rho_0,\ r_0$. 
However, in an attempt to improve accuracy and provide a useful check we consider the more general profile with an
additional parameter, $\beta$:
\begin{eqnarray}
\rho (r) =  \rho_0 \left[ {r_0^2  \over r^2 + r_0^2
}\right]^\beta
\ .
\label{beta}
\end{eqnarray}
Starting with the above parameterization, we determine $T(r)$ by numerically solving
Eq.(\ref{p9}) for a particular choice of $X = \{\rho_0, \ r_0,\ \beta\}$. 
We can then determine (at each $r$ value) the cooling rate $d^2 E_{out}/dt dV$, and
heating rate $d^2 E_{in}/dt dV$. To quantify how well Eq.(\ref{meal8}) is satisfied,
we introduce the function,
$\Delta (r_0, \rho_0, \beta)$:
\begin{eqnarray}
\Delta (r_0, \rho_0, \beta) \equiv {1 \over 10 r_D} \int^{11r_D}_{r_D}
\left| 1 - {{d^2 E_{in} \over dt dV} \over {d^2 E_{out} \over dt dV}}\right| \ dr
\ . 
\label{delta}
\end{eqnarray}
We can now scan over the parameters, 
$r_0, \ \rho_0, \ \beta$ to find the ones for which $\Delta$ is minimized.
Numerically it is found that the parameterization, Eq(\ref{beta}), gives a near exact 
solution to the Eq.(\ref{meal8}):
The resulting $\Delta$ minimum is less than 0.01.
That is, the left and right-hand sides of Eq.(\ref{meal8}) agree to better than
1\%
\footnote{ 
One can also check that
$\Delta_{min} \stackrel{<}{\sim} 0.01$ occurs even when the
range of integration of the integral in Eq.(\ref{delta}) is extended from $\{r_D \le r \le 11 r_D\}$ to $\{0.2r_D \le
r \le  15 r_D\}$.
Extending this integration range has negligible effect on the derived values for $\beta, r_0$ and $\rho_0$.
}.
This means that the profile, Eq.(\ref{beta}), and the values of
$r_0, \ \rho_0$ and $\beta$
derived by minimizing $\Delta$, should provide an accurate representation
of the dark matter properties expected from this dynamics.

\vskip 0.9cm
\centerline{\epsfig{file=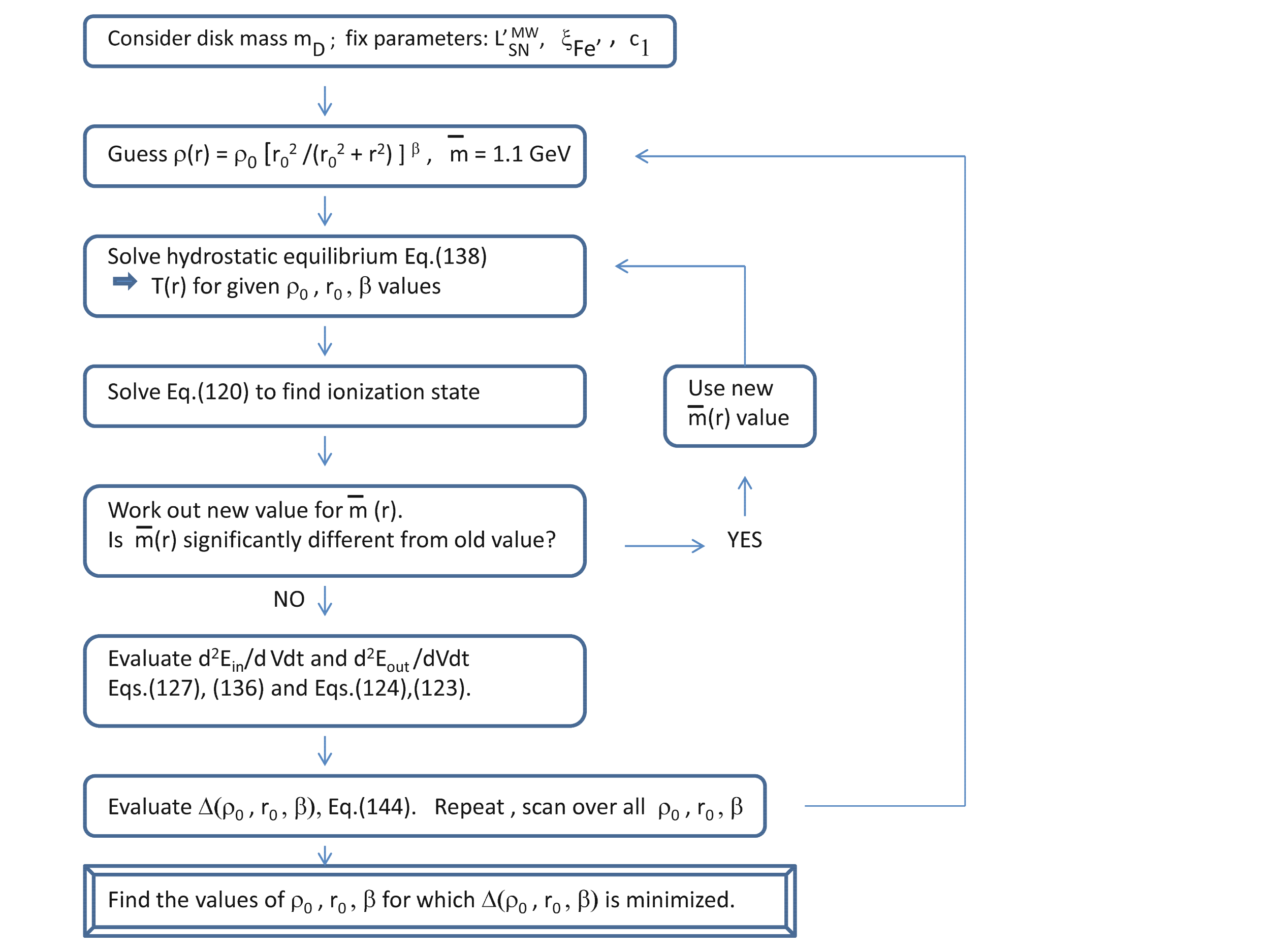, angle=0, width=17.9cm}}
\vskip 0.5cm
\noindent
{\small Figure 4.5:
Flowchart of the steps taken to solve the equations. 
}
\vskip 1.2cm

%xxxxx check eq in fig

The halo dynamics discussed earlier depends on three parameters: 
${L'}_{SN}^{MW}, \ \xi_{Fe'}, \ c_1$. [Technically, four if $E_c$ is included, although 
variation of $E_c$ is equivalent to variation of ${L'}_{SN}^{MW}$ and thus need not be considered.]
Recall ${L'}_{SN}^{MW}$
is the total (time averaged) mirror photon luminosity resulting from 
ordinary supernovae for a reference $\sim$ Milky Way sized spiral galaxy of stellar mass $M_D^{MW} = 5 \times 10^{10}\ m_\odot$,
$c_1$ parameterizes the low energy part
of the energy spectrum of these mirror photons [Eq.(\ref{m3})] and
$\xi_{Fe'}$ is the mirror-iron  mass fraction in the halo. 
These parameters 
determine (to a large extent) the amount
of heat energy absorbed by the halo via Eq.(\ref{meal}). 
Since the dynamics is mainly determined by this heating,  
there is significant parameter degeneracy of the derived halo properties. 

We consider the parameter range:
%%% implicitly assumed these parameters are independent of m_D....maybe possible \xi (M_D) dependence xxxxxxxxxxxxx
\begin{eqnarray}
0.3\times 10^{45} \ {\rm erg/s} < {L'}_{SN}^{MW} < 3\times 10^{45} \ {\rm erg/s}, 
\ \  
0.006 <  \xi_{Fe'} < 0.06,   \ \  1 \le  c_1 \le 3 
\label{range5}
\ .
\nonumber \\
\end{eqnarray}
The central value of $c_1$, i.e. $c_1 = 2$, corresponds  to the low energy part of a thermal 
spectrum while the
central values of the parameters, ${L'}_{SN}^{MW}$ and $\xi_{Fe'}$ are chosen
such that the derived value of $\rho_0 r_0$ is roughly consistent with observations (i.e. from fits
of rotation curves)
\footnote{
The central value of ${L'}_{SN}^{MW}$ is around an order of magnitude larger than the estimated value 
given in e.g. \cite{sn}. At this stage this is not a cause of concern given the various approximations and uncertainties. 
For example, the cooling
rate is likely to be an overestimate since we assumed that all of the bremsstrahlung
radiation escaped the halo [i.e. set $\epsilon_f = 1$ in Eq.(\ref{epsilon41})].
The effective cut-off scale $E_c$, which we set to 50 keV might be lower, in which case a greater
proportion of supernovae energy will be absorbed by the halo. Also, the metal
content might be higher than the $\xi_{Fe'} = 0.02$ central value, again increasing the proportion
of supernovae energy absorbed by the halo.
}.

For a given choice of parameters:
${L'}_{SN}^{MW}, \  \xi_{Fe'},\ c_1$, 
the quantity $\Delta (\rho_0, r_0, \beta)$ [Eq.(\ref{delta})] can be evaluated. 
Minimizing $\Delta (\rho_0, r_0, \beta)$ with respect to variations in $\rho_0, \ r_0,\ \beta$ defines
values of $\rho_0, \ r_0$ and $\beta$ for the chosen values of ${L'}_{SN}^{MW}, \  \xi_{Fe'}, \ c_1$.
That is, the dynamics completely determines the mirror particle density profile, Eq.(\ref{beta}).
This procedure 
can be repeated for each $m_D$ value, over the considered range Eq.(\ref{mdrange}).
The result of performing this numerical task is that $\Delta_{min} \stackrel{<}{\sim} 0.01$
(independently of $m_D$) and
\begin{eqnarray}
\beta & \approx & 1.0 \nonumber \\
r_0 & \approx & 1.4 r_D    \nonumber \\
\rho_0 r_0 &\approx & \left[ {\xi_{Fe'} \over 0.02}\right]^{0.8} \ 
\left[ { {L'}_{SN}^{MW} \over 10^{45} \ {\rm erg/s}} \right]^{0.8}
\left[ {2 \over c_1}\right] 
\ 50 \ m_\odot/{\rm pc}^2
\ .
\label{40}
\end{eqnarray}
Evidently, this dissipative halo dynamics yields a quasi-isothermal dark matter
density profile - a result consistent with the analytic arguments of section 4.1.

There are several noteworthy aspects of the above numerical solution.
Firstly, notice that $r_0$ depends on the baryonic properties through the $r_0 \propto r_D$ scaling.
Recall that such a result was suggested by the analytic considerations of section 4.1.
Secondly, the two other 
derived halo parameters, $\beta$ and the product $\rho_0r_0$, are (roughly)
independent of the baryonic properties ($m_D$) over the entire mass range: Eq.(\ref{mdrange}). 
This covers three orders of magnitude in $m_D$.
Thirdly, the derived values for $\beta, \ r_0$ (in Eq.(\ref{40}) above) hold even when the parameters 
(${L'}_{SN}^{MW}, \ \xi_{Fe'}, \ c_1$) 
are varied within the range, Eq.(\ref{range5}).
This is illustrated in Figure 4.6a,b,c, which show the effects of an order-of-magnitude variation in ${L'}_{SN}^{MW}$ 
on the derived values of  $\beta, \ r_0, \ \rho_0 r_0$ (with 
$c_1 = 2$, $\xi_{Fe'} = 0.02$, $E_c = 50$ keV held fixed).
This is probably not surprising  given the analytic results of section 4.1 where the quasi-isothermal halo could be motivated
from the basic dynamics of the model, i.e. independent of the precise value of parameters.
The central density, $\rho_0$, does depend on the parameters 
(${L'}_{SN}^{MW}, \ \xi_{Fe'}, \ c_1$) 
but there is an
approximate parameter degeneracy. This degeneracy can be understood from the dynamics:
The central density depends
(mainly) only on the total amount of energy transmitted to the halo. Thus, for example,
increasing the energy produced, ${L'}_{SN}^{MW}$, 
should have approximately the same effect as increasing the proportion of energy absorbed, $\xi_{Fe'}$.
Similarly, the variation with respect to $c_1$, given the adopted parameterization of the $\gamma'$ flux, can also be understood.

\vskip 1.2cm
%%%%%%%%%%%%%%%%%%%%%%%%%%%%%%%%%%%%%%% figure 4.6a %%%%%%%%%%%%%%%%%%%%%%%%%%%%%%%%%%%%%%%%%%%
%
\centerline{\epsfig{file=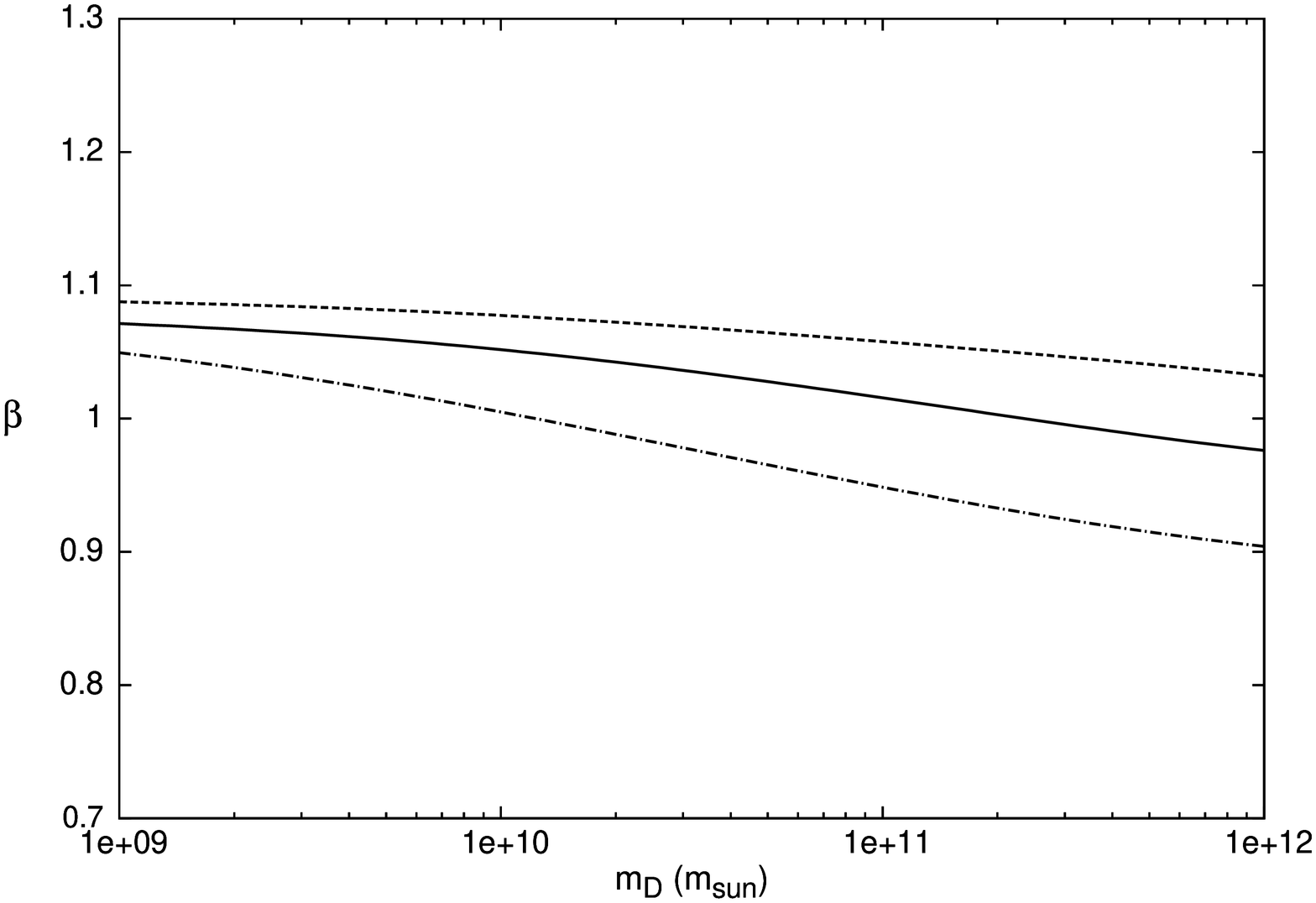, angle=270, width=12.2cm}}
\vskip 0.4cm
\noindent
{\small Figure 4.6a:
The dark matter density, slope, $\beta$ [Eq.(\ref{beta})], versus $m_D \ [m_\odot]$.
The dashed-dotted, solid and dotted lines correspond to
${L'}_{SN}^{MW} = 0.3\times 10^{45}$ erg/s,
${L'}_{SN}^{MW} = 1.0\times 10^{45}$ erg/s and
${L'}_{SN}^{MW} = 3.0\times 10^{45}$ erg/s respectively.
%The other parameters held fixed at the reference values:
%$c_1 = 2$ and $\xi_{Fe'} = 0.02$, $E_c = 50$ keV. 
}

\vskip 0.7cm
%%%%%%%%%%%%%%%%%%%%%%%%%%%%%%%%%%%%%%% figure 4.6b %%%%%%%%%%%%%%%%%%%%%%%%%%%%%%%%%%%%%%%%%%%
%%
\centerline{\epsfig{file=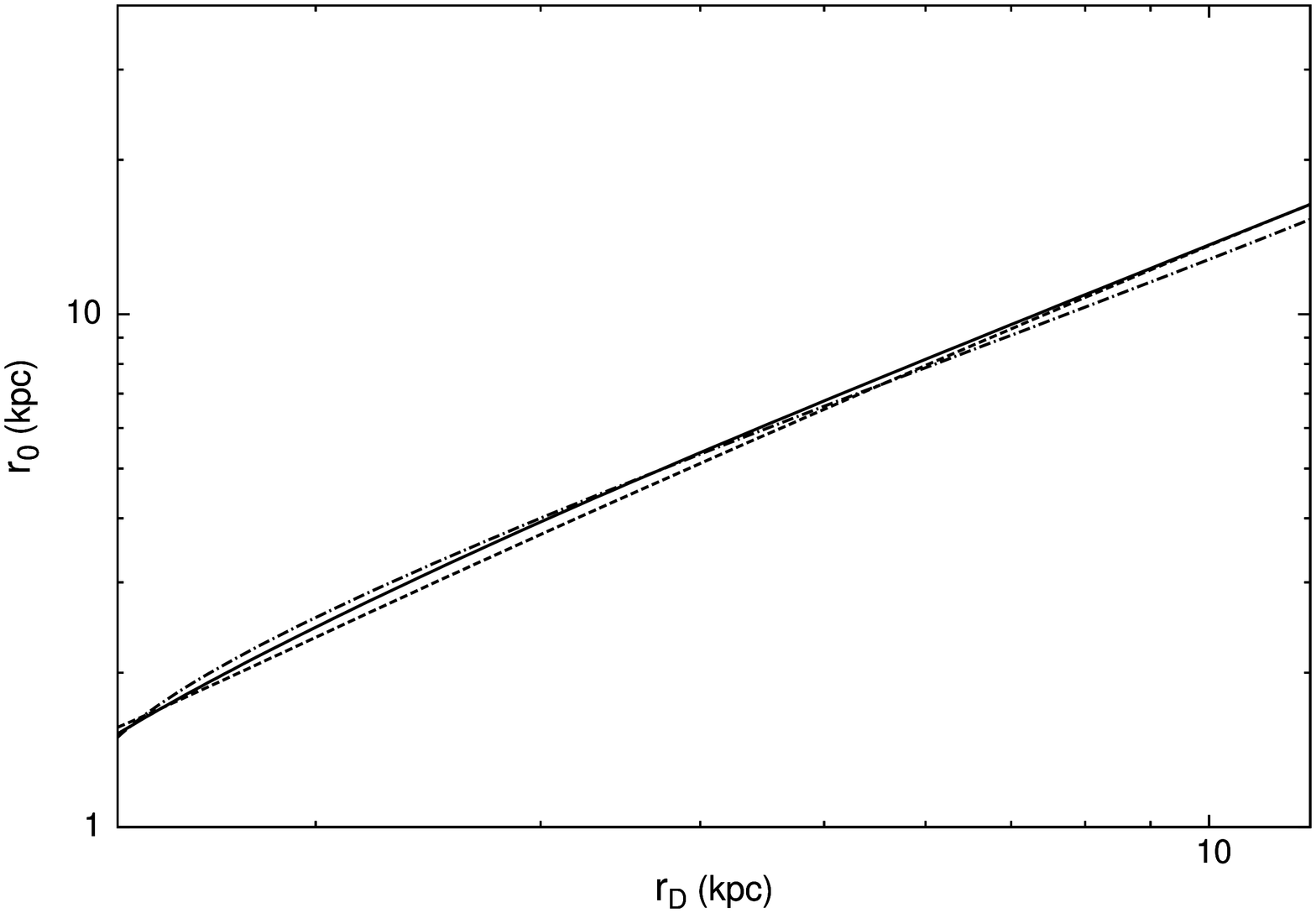, angle=270, width=11.8cm}}
\vskip 0.2cm
\noindent
{\small Figure 4.6b:
Dark matter core radius, $r_0$, versus
disk scale length, $r_D$.
Parameters as per figure 4.6a.
}
\vskip 0.4cm

%%%%%%%%%%%%%%%%%%%%%%%%%%%%%%%%%%%%%%% figure 4.6c %%%%%%%%%%%%%%%%%%%%%%%%%%%%%%%%%%%%%%%%%%%
%%
\centerline{\epsfig{file=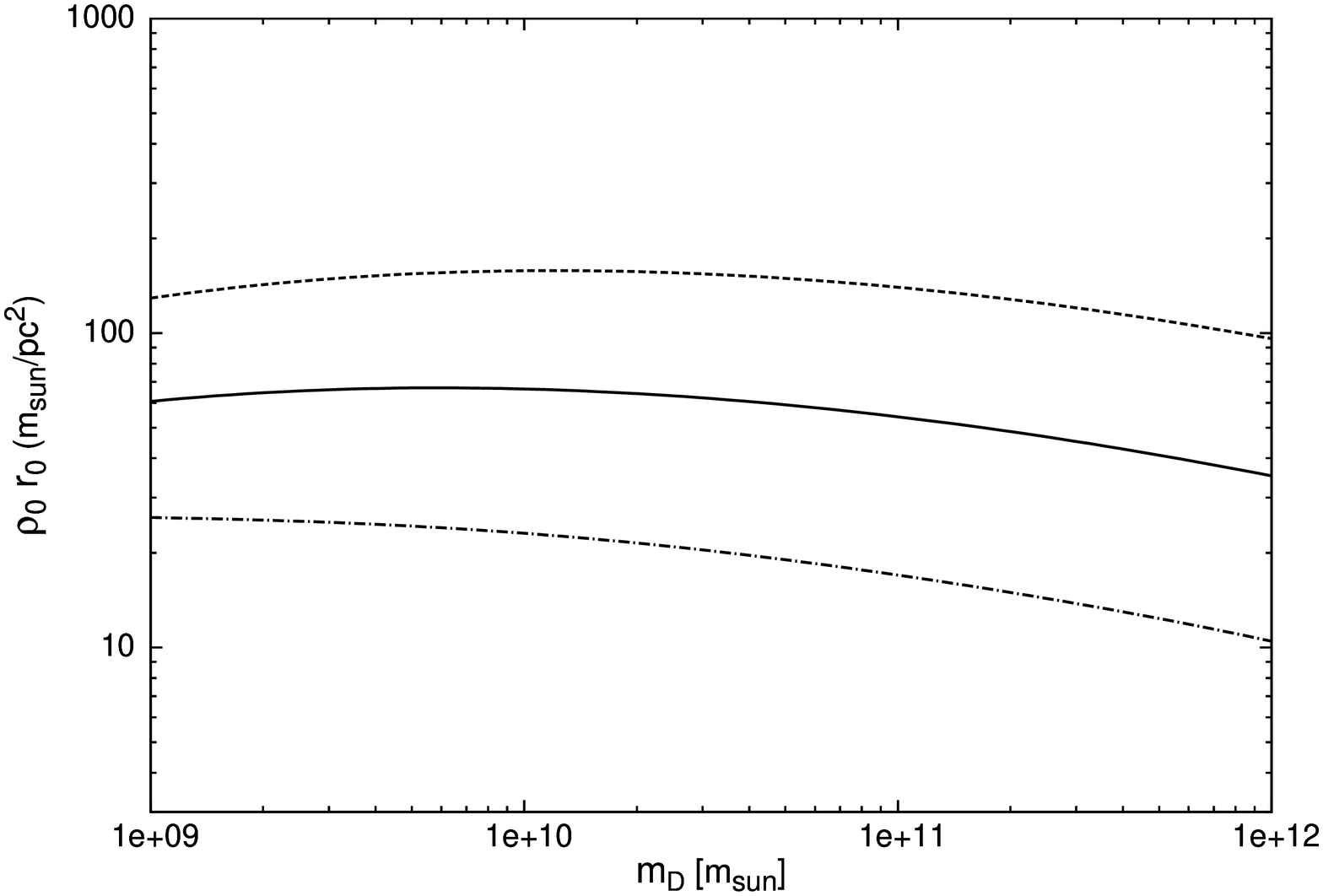, angle=270, width=11.7cm}}
\vskip 0.2cm
\noindent
{\small Figure 4.6c:
$\rho_0 r_0$ versus $m_D$.
Parameters as per figure 4.6a.
}

\vskip 0.9cm

The dynamics discussed here implies that the dark matter distribution
associated with spiral galaxies is closely linked with the galaxy's baryonic properties.
Observations, for many years, have also suggested this.
Galactic rotation curves, in particular, show surprising regularities.
Indeed, the observations indicate that rotation curves are (roughly) 
fixed once the galaxy's luminosity, $L$, is specified 
\cite{rubin2,PS2,PSS}.
This feature
can be explained, at least in a large part,
by the non-trivial dynamics
discussed here. The dynamics does not allow just any dark matter
density profile, but the highly constrained quasi-isothermal distribution,
whose properties are tightly linked with the baryonic properties of galaxies.
Most importantly, this particular 
distribution is consistent with the dark matter density profile
obtained from phenomenological studies of the rotation curves of actual spiral galaxies 
e.g. \cite{kent,begeman,burkert,bs,gentile,free,dsds,others2,others3,salucci,alabama}.
Such studies have long favored a cored dark matter distribution; the quasi-isothermal \cite{kent,begeman} and
Burkert profiles \cite{burkert} both able to provide a reasonable fit to the data.
Moreover, ref.\cite{dsds} found that measurements  of high resolution rotation curves 
in a sample of 
galaxies implied that the core radius of the quasi-isothermal 
distribution scales linearly with disk scale length:
\begin{eqnarray}
log(r_0) \ = \ (1.05 \pm 0.11) \ log(r_D) \ + \ (0.33\pm 0.04)
\ .
\end{eqnarray}
This $r_0 \approx 2r_D$ result, is broadly consistent the second relation in Eq.(\ref{40}).
The third relation, of Eq.(\ref{40}) is also very interesting. Although it is found that $\rho_0 r_0$ does depend on the parameters
(${L'}^{MW}_{SN},\ \xi_{Fe'},\ c_1$),
it is plausible that these parameters are roughly independent
of galaxy size and morphology. Thus a rough $\rho_0 r_0 \propto constant$ scaling
is anticipated. Again such a scaling relation has been derived 
from actual observations of galaxies \cite{free,others2,others3,salucci}.
The observed value of the $\rho_0 r_0$ `constant' 
is around $\rho_0 r_0 \approx 100 \ m_\odot/{\rm pc^2}$.

%%%%%%

%%%%%%%%%%%%%%%%

Before further discussing the properties of the solution, let us stop to briefly estimate the required 
($\epsilon, \ \xi_{Fe'}$) parameter range where the halo can plausibly be heated by ordinary supernovae.
Firstly, Eq.(\ref{raf1x}) shows that the fraction of the core-collapse energy of an ordinary supernova that is converted into mirror particles
is proportional to $\epsilon^2$. \footnote{This is only strictly true for $\epsilon \stackrel{<}{\sim} 10^{-9}$.
For $\epsilon \stackrel{>}{\sim} 10^{-9}$ the fraction of supernova energy converted into mirror particles
saturates to a constant. This constant might not necessarily be unity because mirror particle self interactions
can lead to trapping  of mirror particles and prevent their rapid escape from the supernova core. In any case,
this appears to be a mute point given that
early Universe cosmology, as discussed in section 3, suggests that $\epsilon \stackrel{<}{\sim} 10^{-9}$.}
Thus, the average $\gamma'$ luminosity in galaxies, powered by kinetic mixing
induced processes in ordinary supernovae, scales as: ${L'}_{SN} \propto \epsilon^2$.
Eq.(\ref{40}) then suggests that 
\begin{eqnarray}
\rho_0 r_0 \sim \left({\epsilon \sqrt{\xi_{Fe'}}\over 4\times 10^{-10}}\right)^{1.6} \ 50\ m_\odot/{\rm pc}^2\ 
\end{eqnarray}
where we have used $c_1 = 2$ and ${L'}^{MW}_{SN}  \sim (\epsilon/10^{-9})^2 \ 10^{44} \ {\rm erg/s}$ 
(valid for $\epsilon \stackrel{<}{\sim} 10^{-9}$).

The above estimate is subject to
various sizable uncertainties, but perhaps the largest is the fraction of ${L'}_{SN}$ contributed by
mirror photons with $E_{\gamma'} \stackrel{<}{\sim} 30$ keV. With the chosen parameterization, Eq.(\ref{m3}), 
this is controlled by the uncertain cut-off parameter: $E_c$. In the numerical work leading to Eq.(\ref{40}), and hence also the above equation, 
this parameter was fixed to $E_c = 50$ keV.
For lower (higher) $E_c$ values, a greater (smaller) proportion of supernova energy can be absorbed by the halo.
Naturally, this proportion cannot be larger than $100\%$, and thus a rough lower limit on $\epsilon \sqrt{\xi_{Fe'}}$
can be derived. A rough upper limit on this parameter can also be obtained.
Recall from section 3 that early Universe cosmology suggests an upper limit
on $\epsilon$ of around $\epsilon \stackrel{<}{\sim} 10^{-9}$, and of course, $\xi_{Fe'} \le 1$.
These considerations thereby lead to the rough estimate:
\begin{eqnarray}
\epsilon \sqrt{\xi_{Fe'}} \ \sim \ 10^{-10} - 10^{-9}\ .
\label{678}
\end{eqnarray}
This result has assumed the simplified model of a halo mirror metal component consisting of just $Fe'$.
It is possible, of course, that instead, the $Fe'$ component is negligible, and lighter mirror metals 
(e.g. $A' \sim O', \ Si'$) populate
the halo. In this case, these lighter metal components can play the same role as $Fe'$ in the halo dynamics, 
i.e. their photoionization provides the mechanism of transferring the energy from the $\gamma'$ to the halo.
Effectively little has changed, except the relevant parameter is then $\epsilon \sqrt{\xi_{O'}}$ (taking
here mirror oxygen, $O'$ as the chosen replacement for $Fe'$). Note, though, that the parameter range for $\epsilon \sqrt{\xi_{O'}}$ 
can be a little wider than that for $\epsilon \sqrt{\xi_{Fe'}}$ [given above in Eq.(\ref{678})] as its lower limit can be reduced because of 
the larger photoionization cross-section for $O'$ near threshold.

\subsection{Properties of the solution}

The considered dynamics appears to determine, to a large extent, the physical properties of the halo. Not only is the dark matter
density profile constrained but so to is its temperature. We now proceed to investigate this implied
temperature profile, $T(r)$, and then consider the galactic rotation curves resulting from this dynamics.

\vskip 0.3cm
\noindent
{\bf Halo temperature profile}
\vskip 0.3cm
\noindent
The numerical solution of the equations yields not just the halo mass density, $\rho (r)$, but also its temperature
profile: $T(r)$.
In figure 4.7 we plot the evaluated temperature versus radial distance for examples with
$m_D = 10^9 \ m_\odot, \ m_D = 10^{10} \ m_\odot, \ m_D = 10^{11} \ m_\odot, \
m_D = 10^{12} \ m_\odot$.
For each of these examples the temperature profile is obtained from the equation governing hydrostatic
equilibrium, Eq.(\ref{p9}), with the density, Eq.(\ref{beta}), parameterized by the values of
$\beta, \ r_0$ and $\rho_0$ obtained
by minimizing $\Delta$, Eq.(\ref{delta}) [with assumed reference parameters:
$c_1 = 2, \ \xi_{Fe'} = 0.02$, ${L'}_{SN}^{MW} = 10^{45}$ erg/s,
$E_c = 50$ keV].

Figure 4.7 shows that the plasma temperature is approximately constant 
in the outer part of the halo and rises in the inner region - 
a result not unexpected given the analytic considerations of section 4.1.
The halo mirror particle plasma temperature range of the outer region ($r \sim 5r_D$) for the $m_D$ range
$10^{9} \ m_\odot \stackrel{<}{\sim} m_D \stackrel{<}{\sim} 10^{12} \ m_\odot$ is
approximately
$20\ {\rm eV} \stackrel{<}{\sim} \ T \ \stackrel{<}{\sim} 0.6\ {\rm keV} $.
%See figure 2 for some examples. 
In this temperature range, figures 4.2a, 4.2b and 4.3 indicate that 
$H'$, $He'$ are fully ionized and $Fe'$ has both K-shell states filled.
This consistency of the halo's ionization state, over the entire mass range of
interest:
$10^{9} \ m_\odot \stackrel{<}{\sim} m_D \stackrel{<}{\sim} 10^{12} \ m_\odot$,
results in the smooth behaviour of the derived relations, Eq.(\ref{40}).
The consistency of the derived relations, Eq.(\ref{40}), with observations is
therefore a non-trivial test of this mirror halo dynamics.
It is perhaps an indication that mirror dark matter is favored over a more generic
dissipative hidden sector model. 

%This would appear to be an important constraint if one were to think about replacing
%mirror dark matter with a more generic dissipative hidden sector model.

\vskip 0.8cm
%%%%%%%%%%%%%%%%%%%%%%%%%%%%%%%%%%%%%%% figure 4.7 %%%%%%%%%%%%%%%%%%%%%%%%%%%%%%%%%%%%%%%%%%%
\centerline{\epsfig{file=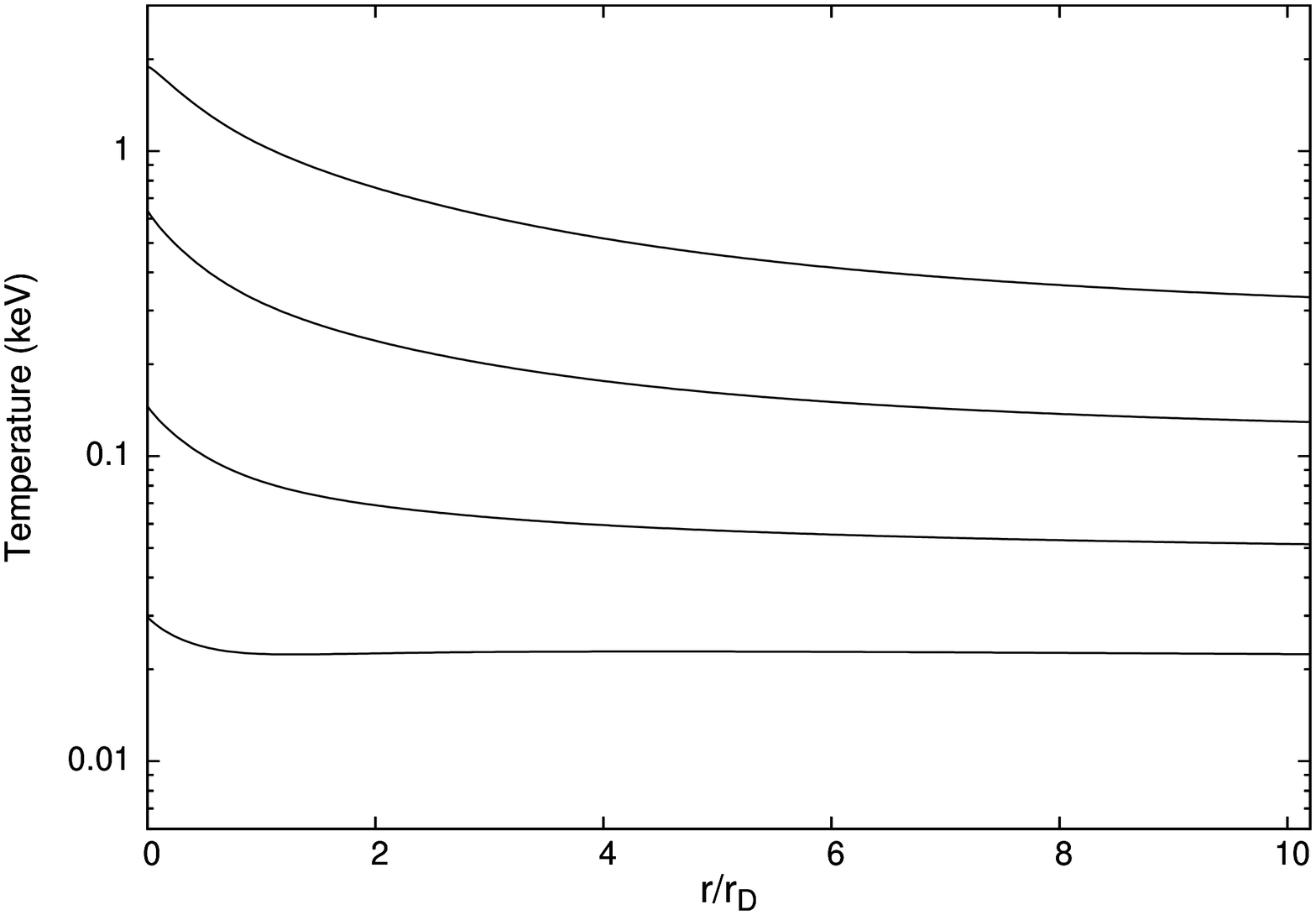, angle=270, width=11.9cm}}
\vskip 0.2cm
\noindent
{\small Figure 4.7: Halo mirror plasma temperature versus $r/r_D$ for
the examples with
(from bottom to top curves): $m_D = 10^9 \ m_\odot, \ m_D = 10^{10} \ m_\odot, \ m_D = 10^{11} \ m_\odot, \
m_D = 10^{12} \ m_\odot$.
%Standard reference parameters: 
%$c_1 = 2, \ \xi_{Fe'} = 0.02$, ${L'}_{SN}^{MW} = 10^{45}$ erg/s, 
%$E_c = 50$ keV are assumed.  
}

\vskip 0.4cm

\vskip 0.1cm
\noindent
{\bf Rotation curves}
\vskip 0.6cm
\noindent
Rotation curves of spiral galaxies have historically been extremely important in establishing the case for
non-baryonic dark matter in the Universe \cite{sr}.
They provide a direct probe of the mass distribution of galaxies assuming only that Newton's law is valid.
The nontrivial halo dynamics considered here strongly constrains the dark matter mass distribution.
It allows the rotation curves of spiral galaxies to be predicted (subject of course to uncertainties in the baryonic distribution), 
up to one unknown parameter, which
can be taken to be ${L'}^{MW}_{SN}$, given the parameter degeneracy.

The rotational velocity is given by $v_{rot}^2/r = g(r)$, with $g(r)$ obtained from Newton's Law, Eq.(\ref{g77}).
With the assumption of spherical symmetry we have
\begin{eqnarray}
v_{rot} (r) = \left[ {G_N \over r} \int_0^r  \rho_{total} \ dV \right]^{1/2}\ .
\end{eqnarray}
Recall, that we have approximated $\rho_{total}$ by just two components: 
$\rho_{total} = \rho_D (r) + \rho (r)$, where $\rho_D (r)$ is the baryonic contribution, given in spherically
symmetric approximation by, Eq.(\ref{mum2}) and $\rho (r)$ is the derived mirror particle plasma
distribution, given by Eq.(\ref{beta}).
Although inclusion of just these two components would not be expected to provide an adequate description 
of all spiral galaxies, it appears to be sufficient in many cases.

Consider first a specific example: the galaxy NGC3198.
This galaxy has stellar mass 
%$L_B = 2 \times 10^{10} L_\odot, \ 
$m_D = 3.0 \times 10^{10}\ m_\odot$ \cite{blok} and from Eq.(\ref{rd}) we find $r_D = 2.8$ kpc.
Measurements of the rotation curve for NGC3198 are given in \cite{ngc3198} which is consistent
with other measurements such as the one in \cite{blok}.
%NGC 3198 and in \cite{ngc3198}  a measurement of its rotation curve is provided.
In figure 4.8 we derive the rotation curve for NGC3198, determining the dark matter parameters, $\beta,\  r_0, \
\rho_0$
by minimizing $\Delta$ inputting the above baryonic parameters ($m_D, \ r_D$) for NGC3198.
It is found that the data can be fit with:
$c_1 = 2, \ \xi_{Fe'} = 0.02$, ${L'}_{SN}^{MW} = 2.2\times 10^{45}$ erg/s and
$E_c = 50$ keV - 
values close to the reference parameters considered earlier.
We emphasise that this is actually a one-parameter fit due to the parameter degeneracy.
%Including a small gas baryonic component would alleviate the slight discrepancy between the theoretical curve and
%data.

\vskip 0.6cm

%%%%%%%%%%%%%%%%%%%%%%%%%%%%%%%%%%%%%%% figure 4.8 %%%%%%%%%%%%%%%%%%%%%%%%%%%%%%%%%%%%%%%%%%%
\centerline{\epsfig{file=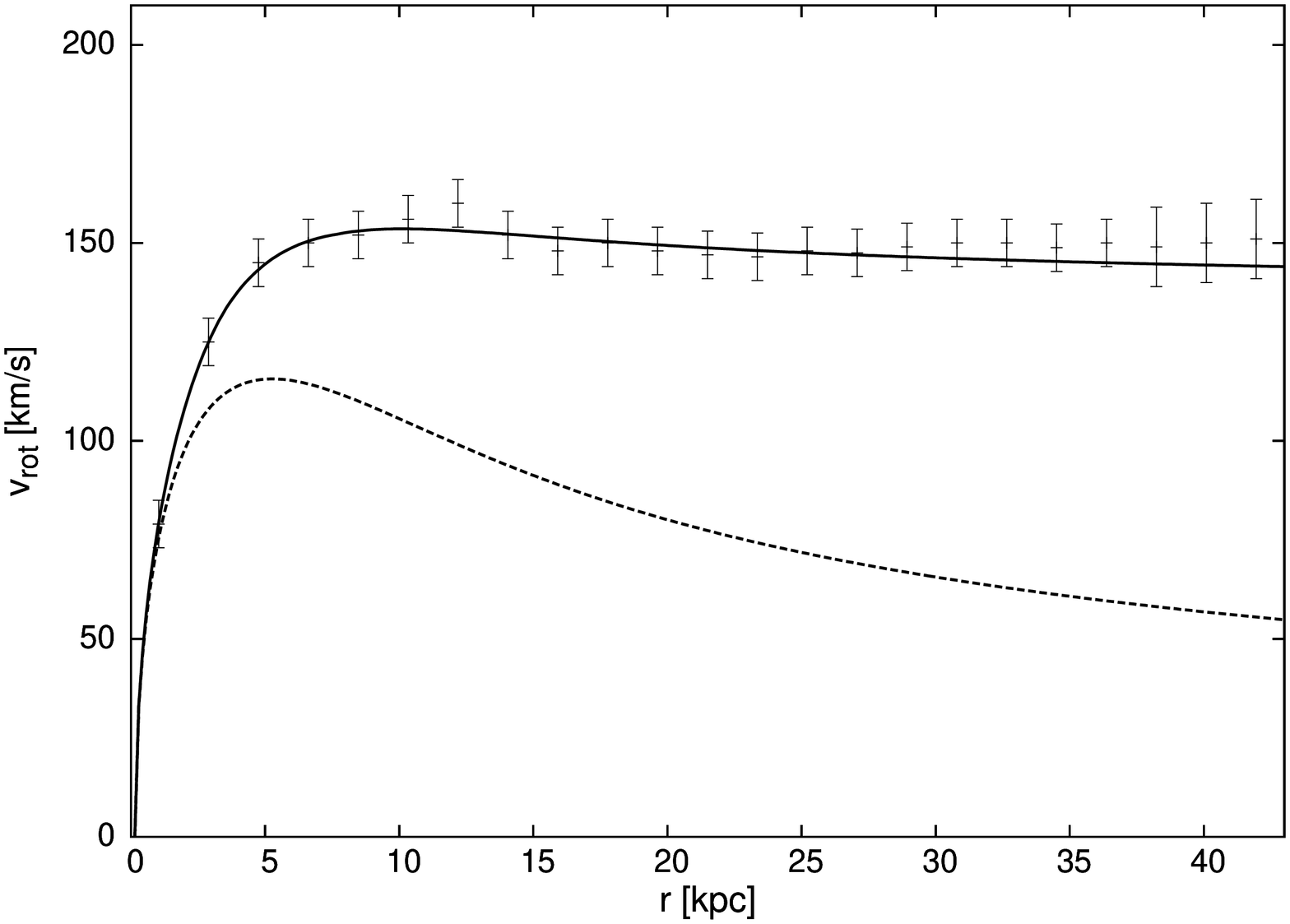, angle=270, width=12.0cm}}
\vskip 0.3cm
\noindent
{\small Figure 4.8: Rotation curve for NGC3198. The solid line is the rotation curve derived
from the assumed halo dynamics.  Also shown (dashed curve) is the baryonic contribution.
The data is obtained from \cite{ngc3198}.
}
\vskip 0.7cm

%%%%%%%%%%%%%%%%%%%%%%%%%%%%%%%%%%%%%%% figure 4.9 %%%%%%%%%%%%%%%%%%%%%%%%%%%%%%%%%%%%%%%%%%%
\centerline{\epsfig{file=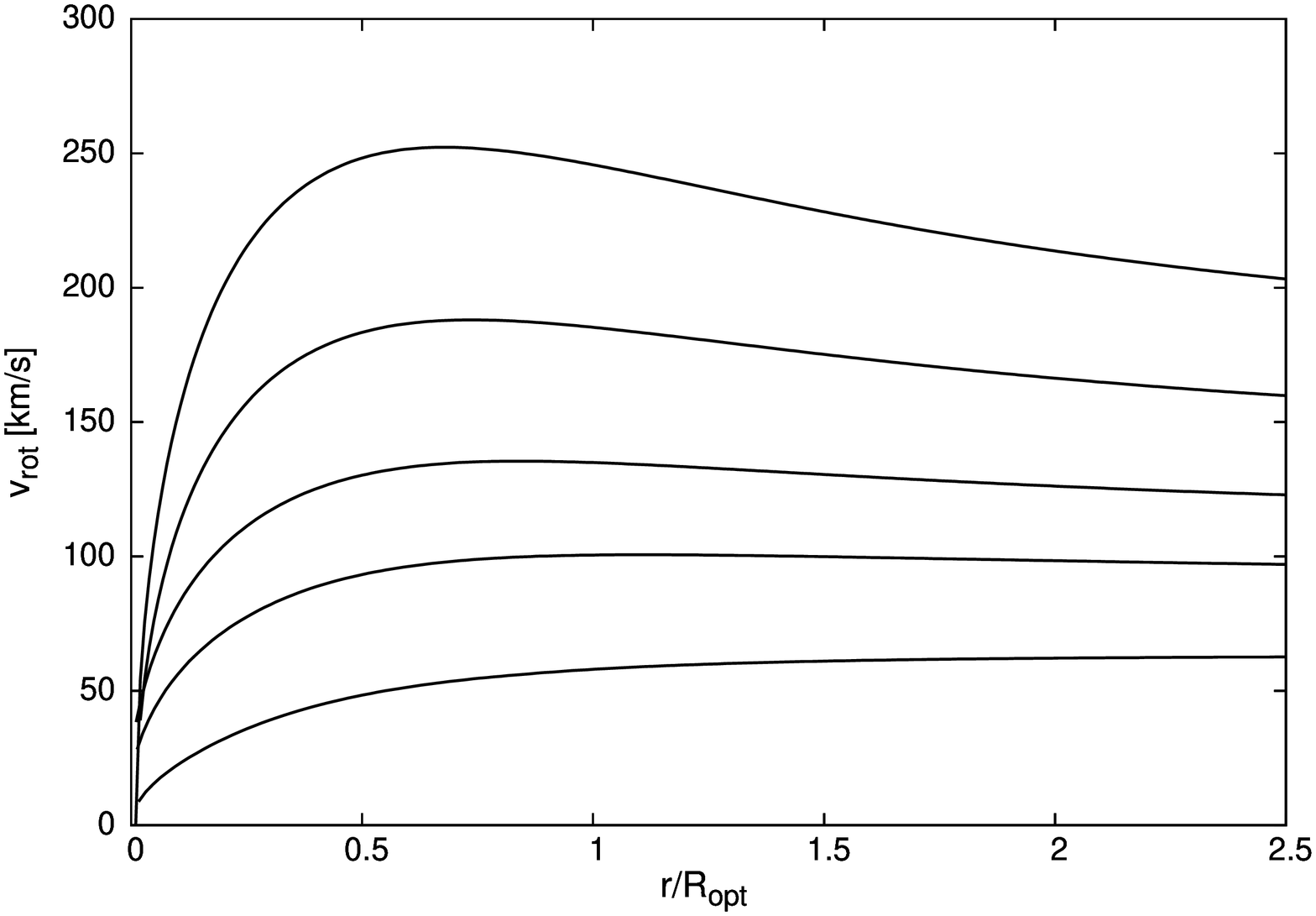, angle=270, width=11.8cm}}
\vskip 0.2cm
\noindent
{\small Figure 4.9: Derived rotation curves for examples with (from bottom to top)
$m_D = 10^9 \ m_\odot, \ 
m_D = 10^{10}  \ m_\odot, \ \ m_D = 3\times 10^{10} \ m_\odot, \
m_D = 10^{11}  \ m_\odot, \ \ m_D = 3\times 10^{11} \ m_\odot$.
%and $m_D = 10^{12} m_\odot$.
The rotational velocity, $v_{rot}$ [km/s] is plotted against $r/R_{opt}$,
where $R_{opt} = 3.2 r_D$. 
The reference  parameters $\{{L'}^{MW}_{SN},\ \xi_{Fe'}, \ c_1\}$ 
chosen are the same as per the solid line in figure 4.6.
}
\vskip 0.9cm

Consider next generic  examples of spiral galaxies.
Shown in figure 4.9 is the derived rotation curves for some representative examples with
$m_D = 10^9 \ m_\odot, \ 
m_D = 10^{10}  \ m_\odot, \ \ m_D = 3\times 10^{10} \ m_\odot,\
m_D = 10^{11}  \ m_\odot, \ \ m_D = 3\times 10^{11} \ m_\odot$.
%and $m_D = 10^{12} m_\odot$.
The reference  parameters $\{{L'}^{MW}_{SN},\ \xi_{Fe'}, \ c_1\}$ 
chosen are the same as per the solid line in figure 4.6.

One can also check the derived rotation curves against the Universal
Rotation Curve (URC) obtained in \cite{PSS}.
There they studied a large sample
of spiral galaxies ($\sim 1100$) and found that the rotation curves 
had an approximately universal profile, one that was completely
specified in terms of the galaxy's luminosity, $L$.
In Eqs.(8-10) of that reference an analytic
approximation for their URC is given.
The rotation curves theoretically derived here from nontrivial halo dynamics
are also (approximately) specified
by a single parameter, which can be $m_D, \ r_D$ or $L$.
[Of course this is true only after
the quantities 
${L'}^{MW}_{SN},\ \xi_{Fe'}, \ c_1$ 
are fixed, which
anyway are assumed to be approximately independent of galaxy size.]
Therefore a one-to-one comparison can be made between the empirical
URC and the rotation curve derived from the assumed halo dynamics.
As for figure 4.8 we adjust one of the unknown parameters
in ${L'}^{MW}_{SN},\ \xi_{Fe'}, \ c_1$, taken to be
${L'}^{MW}_{SN}$ to fit roughly the normalization of one of the URC's
given in \cite{PSS}, which gives:
$c_1 = 2, \ \xi_{Fe'} = 0.02$, ${L'}_{SN}^{MW} = 3\times 10^{45}$ erg/s.
Incidently, this is exactly the same
parameter set as for the dotted lines of figure 4.6.
In figure 4.10 we show the comparison between three representative URC's
of \cite{PSS} and the predicted curves from this halo dynamics \footnote{
We obtain the stellar disk mass estimate from the scaling relation $M_B = -0.38 + 0.92M_I$ 
given in \cite{PSS} and used the 
$L_B = (2.3 \times 10^{10} L_{B,\odot}) (m_D/5\times 10^{10} m_\odot)^{0.5/0.73}$
relation from footnote 38.
% footnote ref correct xxxxx
}

\vskip 0.2cm
\centerline{\epsfig{file=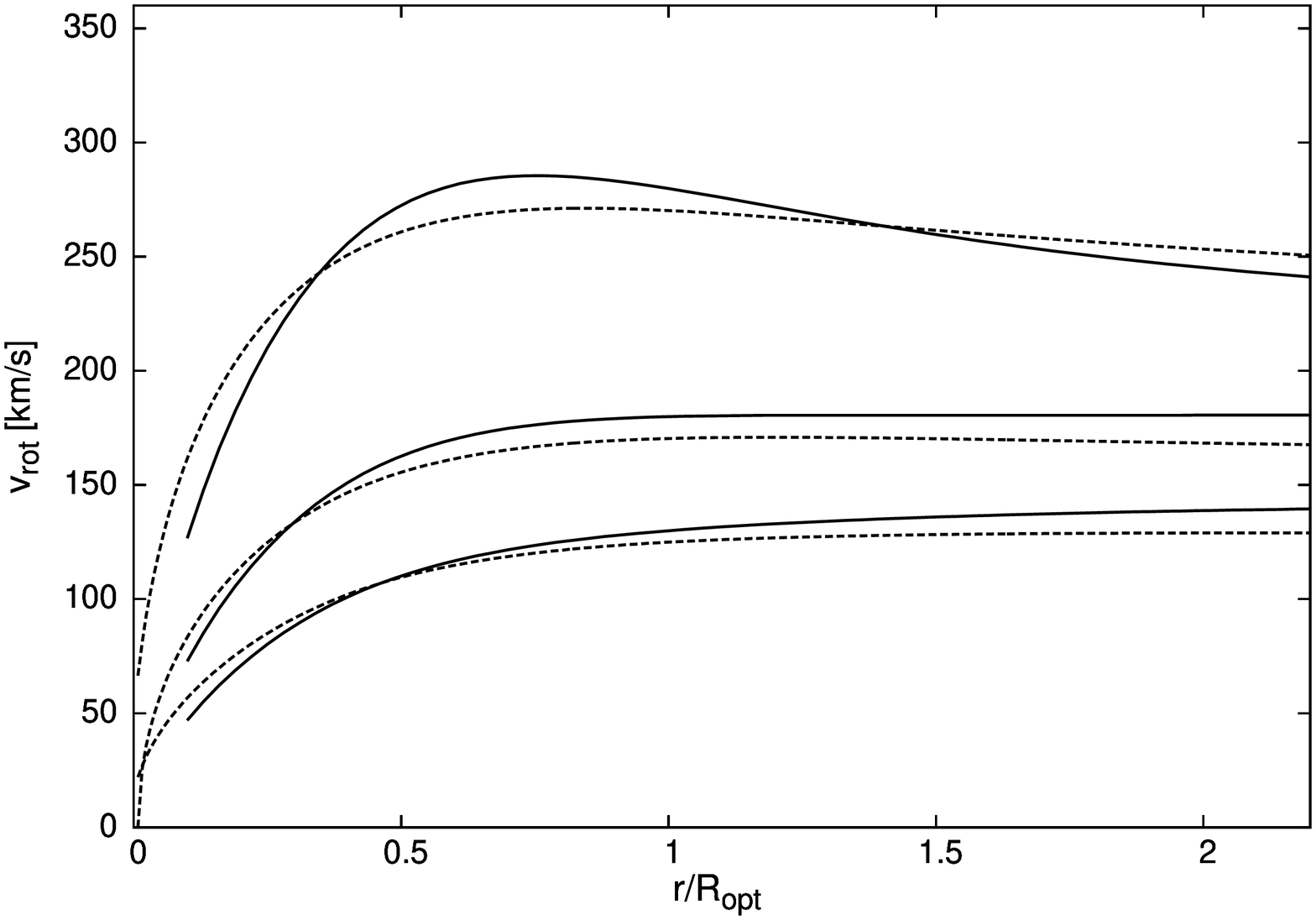, angle=270, width=11.6cm}}
\vskip 0.2cm
\noindent
{\small Figure 4.10: 
Comparison of the Universal Rotation Curve obtained from observations \cite{PSS}
(solid line)
with the rotation curve obtained from the assumed halo dynamics (dotted line).
The three examples shown are, from bottom to top: $M_I = -20.5$ ($m_D \approx 0.94\times 10^{10} \ m_\odot$),
$M_I = -21.6$ ($m_D \approx 3.7 \times 10^{10} \ m_\odot$)
$M_I = -23.2$ ($m_D \approx 2.7 \times 10^{11} \ m_\odot$).
[$M_I$ is the I-band absolute magnitude.]
}

\vskip 1.9cm

Figure 4.10 shows reasonable agreement between the derived rotation curves
and the empirically based URC. The agreement could probably be improved if the baryonic gas component 
was included which can increase $v_{rot}$ by $\sim 5\%$ for smaller spirals $m_D \stackrel{<}{\sim} 10^{10}\ m_\odot$.
However, the observational uncertainties of the URC are of order $5-10\%$, and there are, of course, a number
of other uncertainties [e.g. in the relation Eq.(\ref{rd})] and omissions [e.g. mirror stellar component].
Nevertheless, the results given in figures 4.9-4.10 are still very encouraging and provide interesting evidence 
that this type of halo dynamics might be on the right track in explaining galaxy structure.

\vskip 1.2cm
\noindent
{\bf Tully-Fisher relation}
\vskip 0.5cm
\noindent
In 1977 Tully and Fisher discovered that
the luminosity of a spiral galaxy is correlated
with the maximum value
of its rotational velocity \cite{tf}. 
Modern studies, e.g. \cite{bell,t1,mnrs}, find that: $L_B \propto (v_{max})^{\alpha_1}$, with
$\alpha_1 \approx 3.0-3.5$ ($L_B$ is the B-band luminosity). A similar relation, known as the baryonic Tully-Fisher
relation, relates $m_D$ and $v_{max}$ via $m_D \propto (v_{max})^{\alpha_2}$
with $\alpha_2 \approx 4.0-4.5$.
These relations have been examined in the context of the dynamical halo model in \cite{rf2013b}.
The B-band luminosity, $L_B$, can be extracted from the observed scaling, 
$R_{SN} \propto (L_B)^{0.73}$ \cite{sn}.
As before, the dynamical equations are solved by minimizing the function $\Delta$, Eq(\ref{delta}), 
assuming reference parameters:
%${L'}_{SN}^{MW} = 10^{45}$ ergs/s,
$c_1 = 2$, $\xi_{Fe'} = 0.02$, $E_c = 50$ keV.
To check  the sensitivity
of the results to the uncertain parameters,
${L'}^{MW}_{SN}$ is varied over an order 
of magnitude covering the range given in Eq.(\ref{range5}).
[Varying just ${L'}^{MW}_{SN}$ should be sufficient, due to the approximate parameter degeneracy.]
The result of this numerical work is shown in figure 4.11 for 
the B-band absolute magnitude, $M_B$, versus $v_{max}$ and
figure 4.12 for $m_{D}$ versus $v_{max}$ \footnote{
The absolute magnitude is given in terms of luminosity via
$M_B = M_B^{sun} - 2.5 log (L_B/L_{B,\odot})$, where $M_B^{sun} = 5.5$.
Also, with $R_{SN} \propto (L_B)^{0.73}$ and the adopted $R_{SN} \propto (m_D)^{0.5}$,
it follows that $L_B \propto (m_D)^{0.5/0.73}$.
The proportionality constant was fixed by setting
$L_B = 2.3\times 10^{10} L_{B,\odot}$ for $m_D = 5 \times 10^{10}\ m_{\odot}$.
}.
Our numerical results can be very well approximated by a power law:
$L_B \propto (v_{max})^{\alpha_1}$ and $m_{D} \propto (v_{max})^{\alpha_2}$,
with $\alpha_1 \approx 2.9$, $\alpha_2 \approx 4.1$.
The results are in reasonable agreement with the observations
which is not unexpected given the  
earlier results for galactic rotation curves.

\vskip 0.9cm

%%%%%%%%%%%%%%%%%%%%%%%%%%%%%%%%%%%%%%% figure 4.11 %%%%%%%%%%%%%%%%%%%%%%%%%%%%%%%%%%%%%%%%%%%
%%
\centerline{\epsfig{file=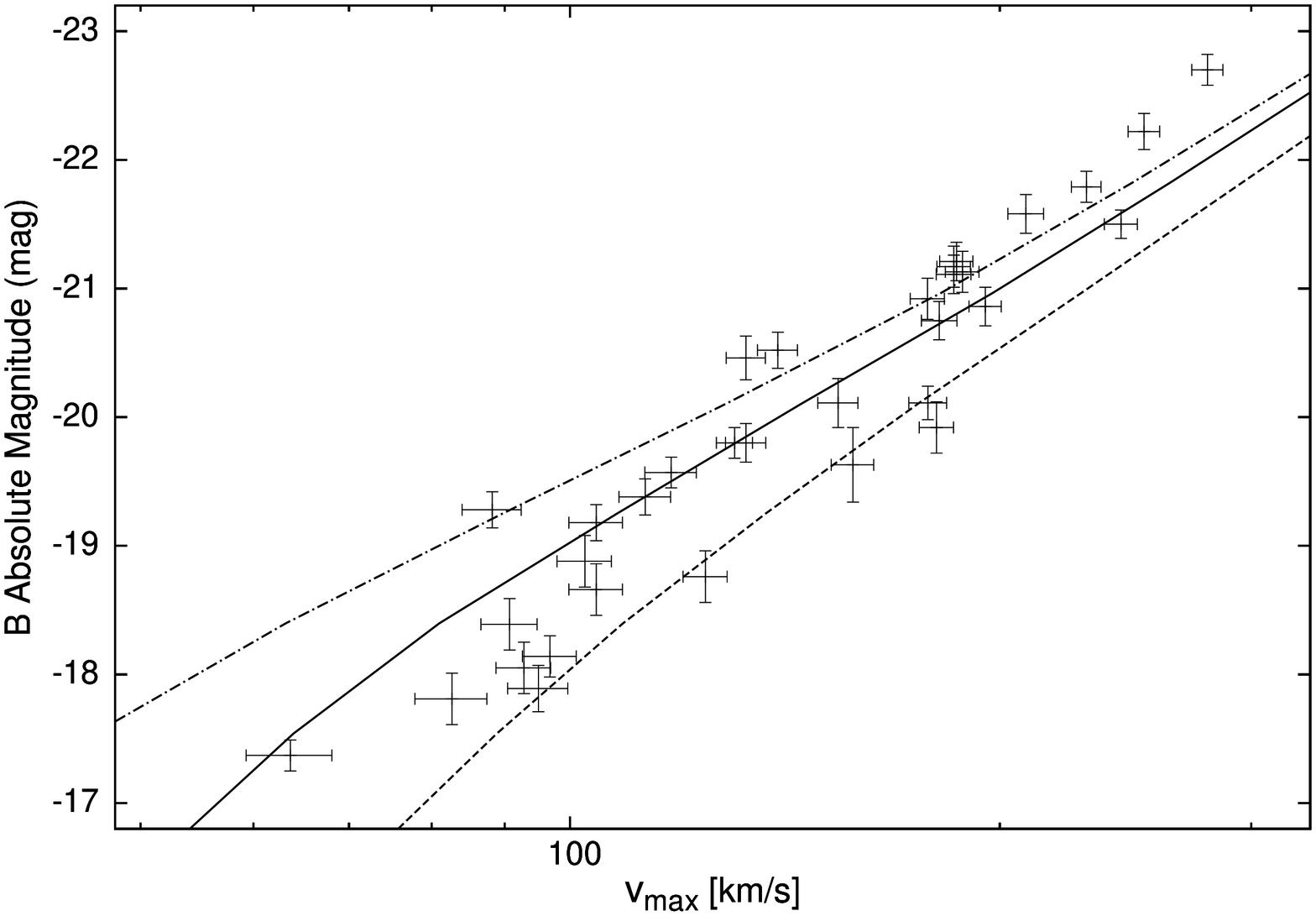, angle=270, width=11.9cm}}
\vskip 0.2cm
\noindent
{\small Figure 4.11: B-Band absolute magnitude, $M_B$, versus $v_{max}$, with halo profile derived
from the assumed dynamics. 
The dashed-dotted, solid and dotted lines correspond to
${L'}_{SN}^{MW} = 0.3\times 10^{45}$ erg/s,
${L'}_{SN}^{MW} = 1.0\times 10^{45}$ erg/s and
${L'}_{SN}^{MW} = 3.0\times 10^{45}$ erg/s respectively.
The other parameters are held fixed at the reference values:
$c_1 = 2$ and $\xi_{Fe'} = 0.02$, $E_c = 50$ keV. 
%%%% xxxxxxxxxx need to change y-axis to -M
Also shown is the data obtained from figure 12 of \cite{mnrs}. 
}
\vskip 0.9cm
%%%%%%%%%%%%%%%%%%%%%%%%%%%%%%%%%%%%%%% figure 4.12 %%%%%%%%%%%%%%%%%%%%%%%%%%%%%%%%%%%%%%%%%%%
%%%%
\centerline{\epsfig{file=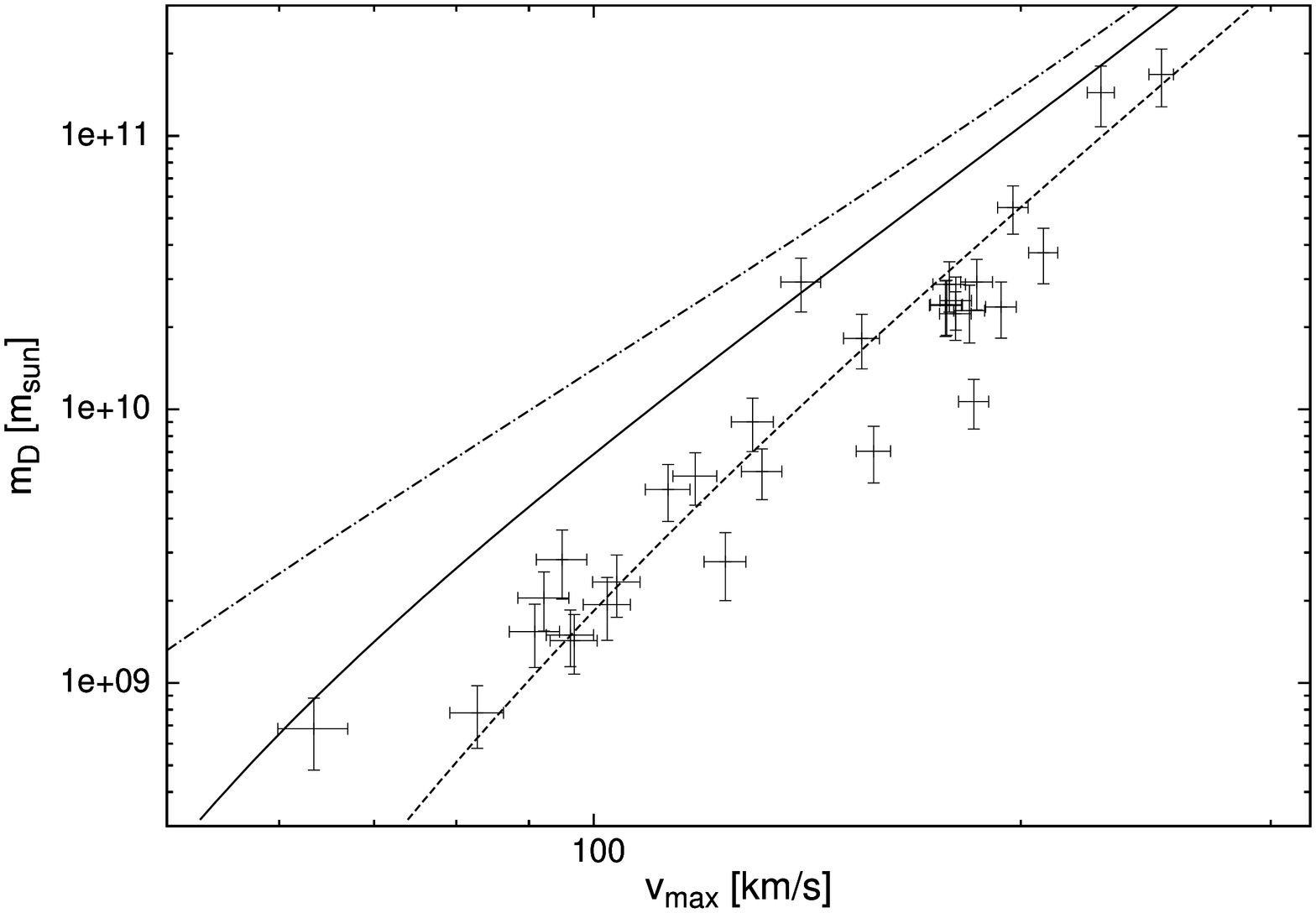, angle=270, width=11.9cm}}
\vskip 0.2cm
\noindent
{\small Figure 4.12: Stellar disk mass $m_{D}$ versus $v_{max}$, with halo profile derived
from the assumed dynamics. The parameters are the same as per figure 4.11.
Also shown is the data obtained from figure 13 (B stellar mass) of \cite{mnrs}.
}

\vskip 1.2cm

%%%%%%%%%%%%%%% starting to fix again here %%%%%%%%%%%%%%%%%%%%

\subsection{Dwarf Spheroidals, Ellipticals and the Bullet Cluster}

\vskip 0.2cm

The above analysis has focused on spiral galaxies. 
Here we briefly consider Dwarf spheroidal and 
elliptical galaxies. For reasons we shall shortly explain, the dark matter properties 
of these galaxies may be very different to that of spirals.
We then comment on the Bullet Cluster, which can potentially constrain dark matter properties in
clusters.

\vskip 0.6cm
\noindent
{\bf Dwarf Spheroidals and Ellipticals}
\vskip 0.5cm
\noindent
Dwarf spheroidal galaxies are small galaxies of luminosity $L \sim 10^6 \ L_\odot$, 
but feature a large dark matter proportion.
These galaxies are observed to have a relatively small ordinary gas
component and currently do not exhibit significant star formation \cite{grebel}.
Interestingly, they are small enough
so that the halo temperature could be very low, $T \sim 1$ eV, and still maintain hydrostatic equilibrium. 
For these temperatures the cooling rate is highly suppressed because 
the mirror hydrogen and mirror helium are mainly in the form of neutral
atoms (figures 4.2a, 4.2b).  
It is unclear if the relatively low amount of heat required in these small galaxies
could be obtained from the suppressed ordinary supernovae rate. Some other heating mechanism
might be possible in these systems. Alternatively the mirror particles might have condensed into
mirror stars in these galaxies.

\vskip 0.4cm

Another important class of galaxies in the Universe are elliptical galaxies. Like dwarf spheroidal galaxies, 
these galaxies are also observed to have only a very small proportion of baryonic gas and their 
current rate of star formation is very low.
Unlike dwarf spheroidals, ellipticals are typically large galaxies. Given the quenched star formation,
such galaxies could not have a substantial plasma halo supported by supernovae energy.
Thus, for ellipticals any significant mirror plasma halo component would presumably 
have collapsed or undergoing collapse. [For large ellipticals the collapse time scale can be quite long $>$ Gyr,
and so some large ellipticals might still be in the collapse phase.]
This means that the dark matter in such galaxies might be predominately in the form
of compact objects: mirror stars, mirror white dwarfs etc.\footnote{It 
is possible that mirror supernovae might have played an important role at some
earlier stage in the life of ellipticals  and possibly dwarf spheroidals as well.
In particular, the large X-ray flux of ordinary photons emitted by  mirror supernovae - expected
if kinetic mixing with $\epsilon \sim 10^{-9}$ exists - might have been responsible for 
heating and expelling the ordinary baryonic gas from these galaxies.}
It is not completely clear how these compact objects are distributed. 
The dark matter distribution might also
depend on the mechanism by which ellipticals are formed, whether in isolation or by major merger(s).
Presumably if mirror stars formed quickly during the collapse phase then this might explain the 
observed ellipticity of the mass distribution in elliptical galaxies, e.g. \cite{chand}.
In any case, the dark matter properties of ellipticals and also dwarf spheroidals are quite uncertain 
in the mirror dark matter framework (as presently understood).
Naturally, with more work and insight it might be possible to make more definitive statements about these galaxies.

We conclude this discussion with a final speculation.
Observations reveal that larger spiral galaxies have a relatively low baryonic gas mass fraction.
At some point in the life of a spiral galaxy, this gas fraction may reduce to the point where
the  supernova formation rate is insufficient to provide
an adequate heat source. If this happens  the mirror particle halo should begin to collapse. 
If the baryonic gas fraction is low enough, the star formation rate cannot increase fast
enough in this collapsing halo phase, and the collapse can become catastrophic. That is,
as the halo collapses, the cooling rate can become even greater than the heating rate, and the collapse accelerates.
It is certainly tempting to speculate that
at least some elliptical galaxies might be the end product of such a violent catastrophic collapse process.
This line of speculation suggests that all spiral galaxies will eventually undergo a complete metamorphosis; they will
be transformed at some point into elliptical galaxies.
%This might explain why the proportion of elliptical galaxies is increasing over time. xxxx

\vskip 0.5cm
\noindent
{\bf The Bullet Cluster}
\vskip 0.5cm
\noindent
So far in this section we have considered only galaxy structure. In larger structures, dark matter 
is of course very important. Information about the nature of dark matter within galaxy clusters
can be obtained from the observations of colliding clusters. Although such observations 
have been used to imply stringent constraints on the dark matter self-interaction cross-section \cite{clowe1},
we will argue below that these might instead imply constraints on the distribution of dark
matter within the cluster. 

%One expects that on very large scales, the properties of mirror
%dark matter should closely resemble collisionless cold dark matter, as such structures are still evolving linearly today
%(figure 3.10). However, the distribution of dark matter within moderately sized

The most well studied example of a colliding cluster is the Bullet cluster \cite{clowe1,clowe0,clowe2}.
The Bullet Cluster is a system in which 
two clusters have apparently collided. Each cluster consists of three
components, the galaxies, hot intergalactic baryonic gas and then there is the dark matter.
Some of the dark matter is bound into galaxy-scale halos and some can be in a diffuse intergalactic mirror-baryonic gas.
When two clusters collide, the hot X-ray emitting baryonic gas associated 
with each cluster appears to be slowed, but not stopped, by
interactions. On the other hand, the galaxies and the bulk of the dark matter appears to pass through
unimpeded. These observations pose a conundrum: Why
doesn't the mirror dark matter associated with each cluster slow down due to interactions? 

The answer might be very simple: It could be that the bulk of the dark matter is bound into 
galaxy-scale halos or smaller `compact' systems.
Indeed, in the extreme case where all the mirror dark matter is confined into galactic-scale halos, 
one would expect nearly all of the mirror dark matter associated with each cluster to pass 
through each other unimpeded \cite{sil99}. In fact there is some recent observational 
evidence supporting this possibility \cite{bacall}. Evidently, such a picture might really be possible, 
although it is constrained by strong lensing measurements which indicate
that dark matter, in at least some clusters, is smoothly distributed on scales $\sim$ 10-30 kpc, e.g. \cite{tyson}.

Of course, if the bulk of the mirror particles within the bullet cluster or main cluster are bound into 
halos of galactic scale or smaller, than this is very unlike the distribution of ordinary baryons in these clusters,
as the bulk of ordinary baryons appears to be in the form of a hot intergalactic gas.
As discussed already in section 1.2 and section 4.1, the early evolution of the ordinary and mirror particle components are very different.
Mirror baryons are expected to have collapsed first, at a very early epoch, 
and this collapse might have been very efficient on the scales of small clusters.
The early mirror supernovae occurring in these 
collapsed objects heated and ionized the ordinary baryons and thereby delayed galaxy formation.
Presumably, by the time the  galaxies began forming, only a fraction of the baryonic gas cooled and coalesced into galaxies.

How much diffuse intergalactic mirror-baryonic cluster gas is allowed by the Bullet cluster observations?
To answer this question, let us introduce the `plasma' surface mass density, $\Sigma^P_m$. This is the
dark matter surface mass density that a typical particle in the subcluster experienced after passing
through the main cluster. That is, it is a measure of the diffuse intergalactic plasma component.
If all the dark matter were in the form of 
a diffuse intergalactic cluster gas then  
it is estimated that $\Sigma^P_m = \Sigma_m \sim 0.3 \ {\rm g/cm^2}$ from weak lensing measurements \cite{clowe0}.
In general, though, $\Sigma^P_m \le \Sigma_m$. Our task now is to find the limits on $\Sigma^P_m$.

Limits on the plasma density, $\Sigma^P_m$ can be adapted from the work of \cite{clowe1,generic2b,sarkar}.
In \cite{generic2b} they argue that the most stringent limits on dark matter properties come from the survival
of the subcluster after having travelled through the main cluster.  
We assume that the cluster's mirror particle plasma component 
consists of fully ionized mirror helium. 
Consider a particular collision, involving a $He'$ particle from the main cluster colliding with a $He'$ particle from the subcluster.
A reasonable approximation is to neglect the thermal velocities of the particles since these are much
less than the velocity ($\sim 4800$ km/s) of the subcluster relative to the main cluster.
%% figure might be useful for kinematics xxxxx
In the reference frame of the subcluster, a net particle loss occurs when the post-collision  
velocities of both particles are larger than the escape velocity of the subcluster, $v_{esc} \approx 1200$ km/s.
This is equivalent to demanding that the recoil energy of the subcluster particle be in the range:
$E_{esc} < E_R < E_i - E_{esc}$, where $E_{esc} = m_{He} v_{esc}^2/2$, $E_{i} = m_{He}v_{1}^2/2$
and $v_1 \approx 4800$ km/s is the velocity of the incident main-cluster particle.
Assuming that a typical particle in the subcluster, after traveling through the main cluster experiences a $He'$ surface
number density of $\Sigma^P_m/m_{He}$,
the fraction of the subcluster's plasma halo which has evaporated is:
\begin{eqnarray}
f = 1 - exp\left(-{\Sigma^P_m \over m_{He}} \ \int^{E_i - E_{esc}}_{E_{esc}} \ {d\sigma \over dE_R} \ dE_R \right)\
.
\label{pool} 
\end{eqnarray}
Here, $d\sigma/dE_R$ is
the differential cross-section for Rutherford scattering of $He'$ off $He'$, and is given by:
\begin{eqnarray}
{d\sigma \over dE_R} = {32\pi \alpha^2 \over m_{He} E_R^2 v^2_1}\ .
\end{eqnarray}
Note that the nuclear form factors have been omitted which is a 
valid approximation given the magnitude of the momentum transfer
in this scattering process.

In the literature, bounds on the self-interaction cross-section have been obtained by 
fixing $\Sigma^P_m = \Sigma_m \sim 0.3 \ {\rm g/cm^2}$
and demanding that $f \stackrel{<}{\sim} 0.5$ \cite{clowe1,generic2b,sarkar}.
The bound $f \stackrel{<}{\sim} 0.5$ means 
that more than half of the dark matter mass of the subcluster does not evaporate after
passing through the main cluster.
For mirror dark matter, the cross-section is fixed, but instead the observations can be used to bound $\Sigma^P_m$.
Demanding that
greater than half of the dark matter mass of the subcluster survives
approximately unimpeded after passing through the main cluster is
roughly equivalent to requiring the bound:
\begin{eqnarray}
\Sigma^P_m \stackrel{<}{\sim} 0.5\Sigma_m \ .
\end{eqnarray}
This bound assumes that all of the intergalactic subcluster gas evaporates after passing through the main cluster.
One can check that for $\Sigma^P_m/\Sigma_m \stackrel{>}{\sim} 0.5$  essentially all of the subcluster evaporates
[numerically we find $f \stackrel{>}{\sim} 0.8$ from Eq.(\ref{pool})]. 
% need to check 0.9 xxxxx
We conclude that the Bullet cluster observations limit the mass fraction of mirror dark matter in the form 
of a hot diffuse gas
unbound to galactic-scale dark halos to be less than around: $\sim 50\%$ for that system.

There are several other known examples of colliding clusters, including MACS J0025.4-1222 \cite{bradac}, 
DLSCL J0916.2+295 \cite{dawson} and Abell 520 \cite{hints}.
Of these, Abell 520 appears to show qualitatively different behaviour, with the bulk of the mass of the colliding cluster appearing to
have slowed down and merged. This could mean that in that cluster there is a higher proportion of mirror dark matter
in the form of a diffuse intergalactic cluster gas. However, other interpretations are possible,
including that this cluster may have formed at the crossing of
three filaments of large scale structure \cite{hints2}.

%%% remove below ref hints xxxxxxxxxxxxxxx
%The mass distribution should lag behind the galaxy component.
%In fact, there are some interesting hints in this direction \cite{hints}.

% \vskip 0.5cm
%\noindent
%\newpage

\vskip 0.4cm

\subsection{
Appendix - Mirror dark matter self-interaction scale}

\vskip 0.5cm

The physical properties (density and temperature) of galaxy halos are such
that collisions are frequent and the mirror particles
can be described by a Maxwell-Boltzmann gas. How frequent are the collisions?
Here we make a rough estimate of the mean distance between mirror particle 
collisions in galactic halos.

%%%%%%%%%%%%%%%%%%%%%%%%%%%%%%%%%%%%%%% figure 4.12 %%%%%%%%%%%%%%%%%%%%%%%%%%%%%%%%%%%%%%%%%%%
%%
\vskip 1.2cm
\centerline{\epsfig{file=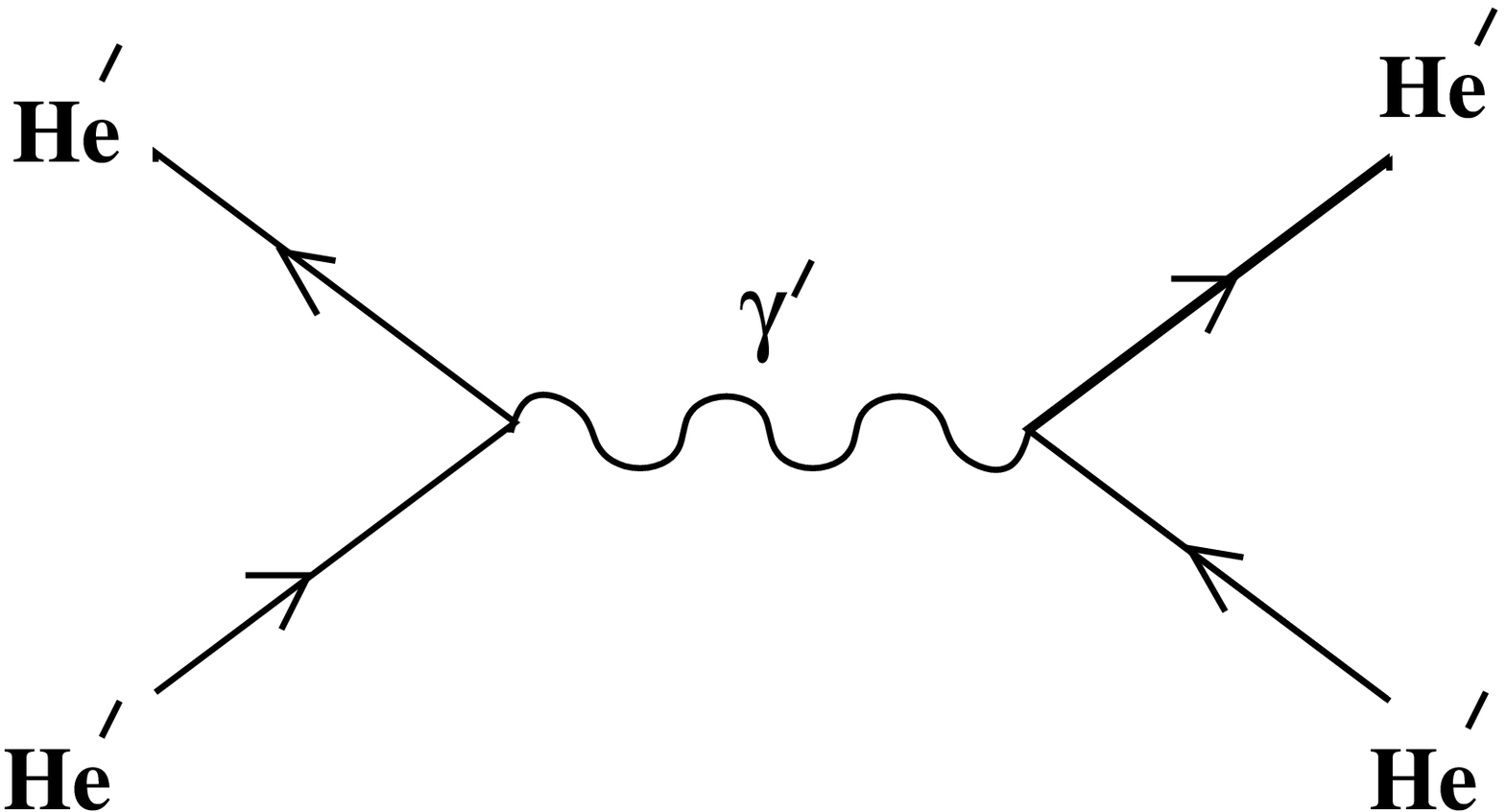, angle=0, width=6.8cm}}
\vskip 0.5cm
\noindent
{\small Figure 4.13: 
Feynman diagram for $He'-He'$ elastic scattering.
}
\vskip 1.7cm

\newpage
Consider a fully ionized mirror-helium plasma, so that $n_{e'} = 2n_{He'} = 2n_T/3$.
The cross-section for $He'-He'$ elastic (Rutherford) scattering,
in the center-of-mass frame is:
\begin{eqnarray}
{d \sigma \over d\Omega} = {\alpha^2  \over m_{He}^2 v_{cm}^4 \sin^4 {\theta \over 2} } \ ,
\end{eqnarray}
where $v_{cm}$ is the magnitude of the $He'$ velocity in the center of mass frame (see figure 4.13).
The total cross-section, $\sigma$ between two isolated $He'$ nuclei is divergent due
to the long range nature of the interaction. In practice, there is a
minimum angle, $\theta_{min}$,
for which elastic scattering can occur due to the shielding of  
the charge by neighboring particles in the plasma. This angle is given by:
\begin{eqnarray}
\theta_{min} \sim {1 \over \lambda_D p_{A'}}
\end{eqnarray}
where $\lambda_D = \sqrt{T/(4\pi \alpha n_{e'})}$ is the Debye length. 

Let us now estimate the distance scale between hard $He'-He'$ collisions, which we here define as those with $\theta > \pi/2$.
The cross-section for such hard collisions is: 
\begin{eqnarray}
\sigma^* = { 4\alpha^2 \pi \over  m^2_{He'} v_{cm}^4}\ .
\end{eqnarray}
Using, $m_{He} v_{cm}^2/2 \sim 3T/2$, the mean distance between hard collisions is estimated to be:
\begin{eqnarray}
{\ell^*} &=& {1 \over \sigma^* n_{He'}} \nonumber \\
& \approx &   1.12 \ {\rm kpc} \ \left[ {10^{-2} \ {\rm cm^{-3}} \over n_{He'}}\right]\ \left[{T \over {\rm keV}}\right]^2\ 
\end{eqnarray}
and the time scale between hard collisions is:
\begin{eqnarray}
t^* &\sim & \ell^*/v_{cm}\nonumber \\
& \approx &   4.1 \ {\rm Myr}\ \left[ {10^{-2} \ {\rm cm^{-3}} \over n_{He'}}\right]\ \left[{T \over {\rm keV}}\right]^{3/2}\ .
\end{eqnarray}
In galactic halos of spiral galaxies where $T \stackrel{<}{\sim}$ keV 
and $n_{He'} \stackrel{>}{\sim} 10^{-2} \ {\rm cm^{-3}}$
for the inner halo region of interest, $r \stackrel{<}{\sim} 6r_D$,  
it follows that ${\ell^*}$ is typically smaller than a kpc,
and $t^*$  smaller than 4 Myr.
For further discussion see also \cite{mfp}.

In much larger structures, such as large galaxy clusters and superclusters, the collision rate can become very small as 
the temperature is typically much larger than a keV and densities much smaller than $10^{-2} \ {\rm cm^{-3}}$.
In such structures $t^*$ can be greater than the Hubble time 
and treating mirror dark matter as collisionless can become a reasonable approximation.

%%% e' xxxxx influenced by E and B fields..etc.. xxxx

%%%%%%%%%%%%%%%%%%%%%%%%%%%%%%%%%%%%%%%%%%%%%%%%%%%%%%%%%%%%

%\newpage

\section{Direct detection experiments}

The dynamical halo model discussed in section 4, appears to provide an adequate description of the (current)
physical properties of dark matter in galaxies. 
In a spiral galaxy, such as the Milky Way, the dark matter halo is 
composed of mirror particles in a pressure supported, spherical, multi-component plasma
containing: $e'$, $H'$, $He'$, $O'$, $Fe'$,....  
Such a plasma dissipates energy
due to thermal bremsstrahlung and other processes, and if this energy were not replaced,
would collapse on a fairly short time scale ($<$ 1 Gyr). 
Ordinary supernovae can supply the needed energy provided two conditions are met.
Firstly, photon - mirror photon kinetic mixing has strength $\epsilon \sim 10^{-9}$ 
so that enough $\gamma'$ energy is produced by ordinary supernovae.
Secondly, the halo should contain a substantial mirror metal component, at least 1\% by mass, so that this energy
can be absorbed. That is, if $\xi_{A'}$ is the halo's mirror metal mass fraction (nominally taken
to be $A' = Fe'$), then 
$\xi_{A'} \stackrel{>}{\sim} 0.01$.
These conditions imply that
%In section 3 we found the constraint: $\epsilon \stackrel{<}{\sim} 10^{-9}$ from early Universe cosmology.
the parameter $\epsilon \sqrt{\xi_{A'}}$ is expected to be in the range:
\begin{eqnarray}
\epsilon \sqrt{\xi_{A'}} \sim 10^{-9} - 10^{-10}\ .
\label{ranged}
\end{eqnarray}
The rate, $R$, at which $A'$ particles scatter off ordinary
nuclei is proportional to the product of cross-section
and $A'$ number density.
The cross-section is proportional to $\epsilon^2$ and the $A'$ number density
is proportional to $\xi_{A'}$.
It follows therefore, that the rate $R$ is proportional to $(\epsilon \sqrt{\xi_{A'}})^2$.
Importantly, for $\epsilon \sqrt{\xi_{A'}}$ in the above range, Eq.(\ref{ranged}), the $A'$ interaction rate is large
enough for these particles to be observed in conventional direct detection experiments \footnote{Another type of direct
detection experiments have been proposed \cite{mitra}. The idea is that if mirror particles have a tiny ordinary electric charge
induced by kinetic mixing, 
they can be influenced by strong ordinary electromagnetic fields.
Mirror dark matter can thereby become polarized and give rise to potentially observable electric fields in cryogenic cavities.}.
Moreover $A'$ interactions can explain
the existing data:
The DAMA annual modulation signal \cite{dama,dama2,dama3} along with the other observations from CoGeNT \cite{cogent}, CRESST-II \cite{cresst}
and CDMS/Si \cite{cdms/si}.

The idea that mirror dark matter could be observed in direct detection experiments dates back to 2003 \cite{foot2003}
and has been explored in some detail over the last decade \cite{footdd,footdd2,foot2,foot05,foothidden,footold6,footelec,footdiurnal,tough}.
We now proceed to discuss the status of this particular  
interpretation of the direct detection experiments.

\subsection{The cross-section and halo distribution}

The mirror particle halo plasma consists of both mirror electrons
($e'$)
and mirror nuclei ($A'$).
In principle both of these components can scatter in an ordinary matter  target. 
Mirror electrons can scatter off loosely bound ordinary electrons while
mirror nuclei can scatter off ordinary nuclei.
[The kinematics is such that only $e' -e$ scattering and $A'-A$ scattering 
is expected to be important.]
For the present, we consider nuclear scattering
only, and defer further discussion of mirror electron scattering to section 5.3. 

The interaction rate in a direct detection experiment depends on the cross-section and halo velocity distribution
of the dark matter particles.  Both of these things depend on the particle physics underlying dark matter. This
is especially true for hidden sector dark matter in general, and for the specific mirror dark matter case.
Let us first discuss the cross-section.

\vskip 0.3cm
\noindent
{\bf The cross-section}
\vskip 0.3cm
\noindent
The cross-section for $A' - A$ scattering arises from the kinetic mixing induced interaction,
represented by the Feynman diagram in figure 5.1.
The photon - mirror photon kinetic mixing enables a mirror nucleus of speed $v$ 
[with mass and atomic numbers $A',\ Z'$] to (spin-independently) elastically scatter
off an ordinary nucleus presumed at rest [with mass and atomic numbers
$A,\ Z$].
Since kinetic mixing effectively provides the mirror nuclei, $A'$, with an ordinary electric 
charge of $\epsilon Z' e$,
the cross-section is just of the standard Rutherford form
corresponding
to a particle of electric charge $Ze$ scattering off a particle of
electric charge
$\epsilon Z'e$. 
The differential cross-section can be written in terms of the 
recoil energy of the ordinary nucleus, $E_R$ :
\begin{eqnarray}
{d\sigma \over dE_R} = {\lambda \over E_R^2 v^2}\ ,
\label{cs}
\end{eqnarray}
where 
\begin{eqnarray}
\lambda \equiv {2\pi \epsilon^2 Z^2 Z'^2 \alpha^2 \over m_A} F^2_A
(qr_A) F^2_{A'} (qr_{A'}) \ .
\end{eqnarray}
Here, the form factors, $F_X (qr_X)$ ($X = A, A'$),
take into account the finite size of the nuclei and mirror nuclei,
where $q = (2m_A E_R)^{1/2}$ is the magnitude of the momentum transferred and $r_X$
is the effective nuclear radius.
A simple analytic expression for
the form factor has been given by Helm \cite{helm,smith}:
\begin{eqnarray}
F_X (qr_X) = 3{j_1 (qr_X) \over qr_X}\ e^{-(qs)^2/2}\ ,
\end{eqnarray}
with $r_X \ = \ 1.14 X^{1/3}$ fm, $s = 0.9$ fm and $j_1$ is the spherical
Bessel function of index one.
\vskip 1.1cm
%\centerline{\epsfig{file=rfigfeynman2.eps,angle=270,width=3.8cm,angle=90}}
\centerline{\epsfig{file=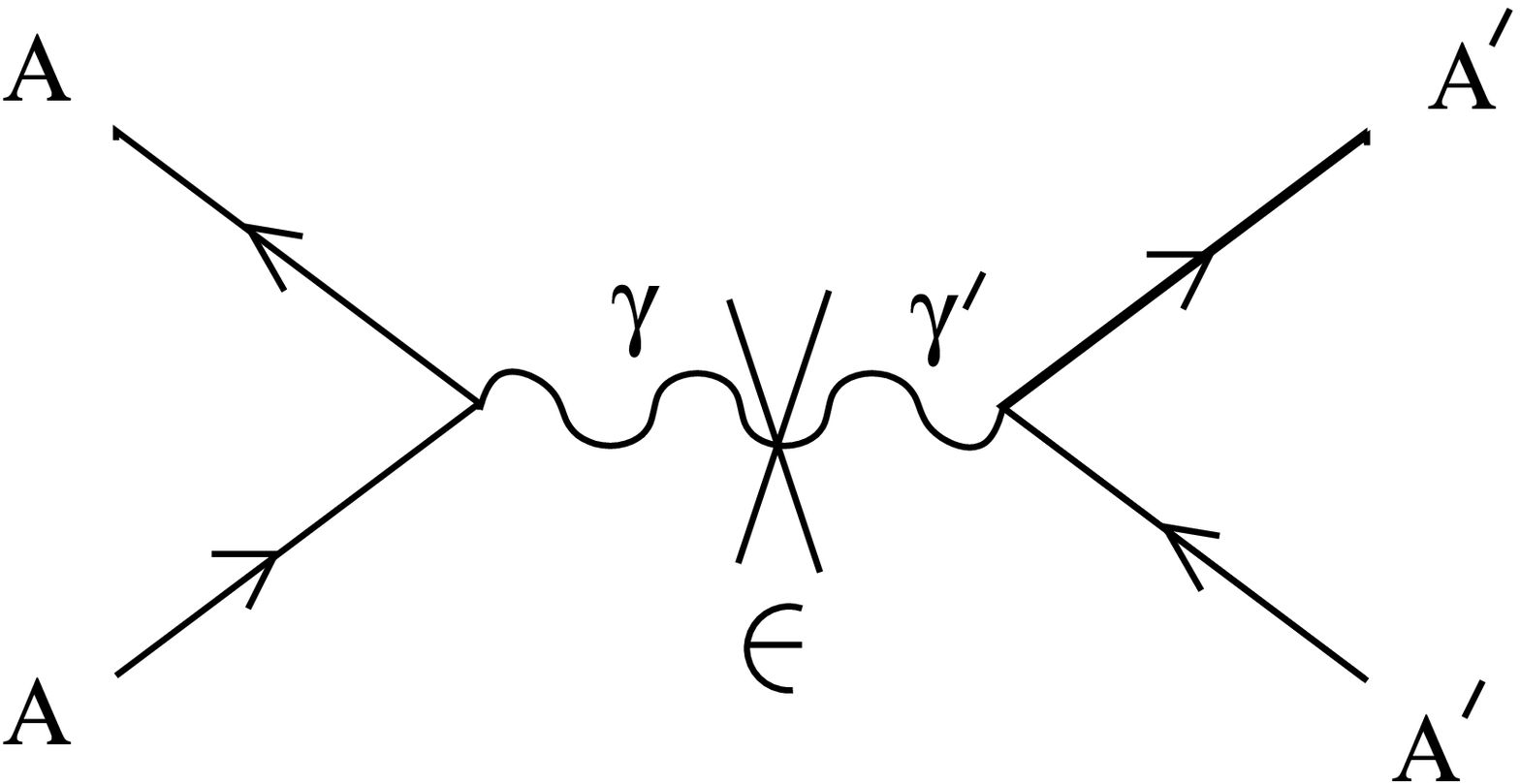,angle=270,width=3.6cm,angle=90}}
\vskip 0.6cm
\noindent
{\small
Figure 5.1: The elastic scattering process $AA' \to AA'$ induced via kinetic mixing, 
treated here as an interaction signified by the cross on the photon propagator.
}

\vskip 1.6cm

\noindent
{\bf The velocity distribution}
\vskip 0.5cm
\noindent
The mirror particles in the halo form a self-interacting plasma at a local temperature $T(r)$.
We saw in section 4 that the temperature profile of this plasma was roughly isothermal and that
the temperature of the plasma satisfied:
\begin{eqnarray}
T \approx  {1 \over 2} \bar m v_{rot}^2 \ .
\label{4}
\end{eqnarray}
Recall $\bar m = \sum n_i m_i/\sum n_i$ ($i = e', H', He', O', Fe',...$) refers to the mean mass of  
the particles in the plasma and $v_{rot}$ is the
galactic rotational velocity of the Milky Way.
In our numerical work we set
$\bar m \approx 1.1 $ GeV which is suggested by
mirror BBN computations
for $\epsilon \sim 10^{-9}$, discussed in section 3.3.  
Estimates of the rotational velocity of the Milk Way at the Sun's location ($r \approx 8.0$ kpc)
are typically in the range $\sim$ 200 - 280 km/s \cite{rot1,rot2,rot3}.

In a reference frame where there is no bulk halo motion,
the halo velocity distribution should be Maxwellian and thus $f_{A'} = e^{-E/T}$.
The halo
particles are nonrelativistic, so that $E = m_{A'} |{\textbf{u}}|^2/2$. It follows that 
the halo velocity distribution has the 
general form: 
\begin{eqnarray}
f_{A'} = e^ {-|{\textbf{u}}|^2/v_0^2
}
\end{eqnarray}
where $v_0^2 \equiv 2T/m_{A'}$ [$A'$ denotes the particle type e.g. $A'=H', He', O', Fe',...$].

A more useful reference frame, for
a direct detection experiment situated on the Earth, is one moving with the 
Earth's velocity through the halo, ${\textbf{v}}_E$
\footnote{ 
Assuming that there is no bulk halo 
rotation with respect to an observer located at the galactic center
then $\langle |{\textbf{v}}_E| \rangle \approx v_{rot} \ + \ 12$  
km/s. [The $12$ km/s offset is due to the Sun's peculiar velocity.]
Of course a small bulk 
halo rotation is possible but its size is essentially unknown. The effect of a small bulk halo motion 
can be incorporated, at least to a first approximation, by adjusting
$v_{rot} \to v_{rot}-v_{bulk}$ (where $v_{bulk}$ is the projection of ${\textbf{v}}_{bulk}$ in the ${\textbf{v}}_E$ direction).
The possibility of bulk halo rotation can thus be accommodated by
considering a liberal uncertainty on $v_{rot}$. }.
With respect to this reference frame, the halo particles have velocity, ${\textbf{v}} = {\textbf{u}} - {\textbf{v}}_E$,
and distribution:
\begin{eqnarray}
f_{A'}({\textbf{v}},{\textbf{v}}_E) = e^{-|{\textbf{v}} + {\textbf{v}}_E|^2/v_0^2}\ .
\label{997}
\end{eqnarray}
%%%% here xxxx

Consider the scattering of halo nuclei, $A'$ off target nuclei, $A$.
[In principle the scattering of mirror electrons off loosely bound ordinary electrons
in the detector target can also be important, but is more difficult to estimate
due to various complications (as will be discussed in section 5.3).]
For an Earth based detector, 
the rate at which $A'$ scatter on target nuclei, $A$, of total number $N_T$, is given by:
\begin{eqnarray}
{dR \over dE_R} = N_T n_{A'} 
\int^{\infty}_{|{\bf{v}}| > v_{min}}\ 
{d\sigma \over dE_R}
{f_{A'}({\textbf{v}},{\textbf{v}}_E) 
\over 
v_0^3 \ \pi^{3/2}
} \ |{\textbf{v}}|\ d^3 {\textbf{v}} 
\label{55}
\end{eqnarray}
where the integration limit is
\begin{eqnarray}
v_{min} \ = \ \sqrt{ {(m_{A} + m_{A'})^2 E_R \over 2 m_{A} m^2_{A'} }}\ .
\label{v}
\end{eqnarray}
In Eq.(\ref{55}), $n_{A'} = \rho_{dm} \xi_{A'}/m_{A'}$ 
is the number density of the halo $A'$ particles. 
We adopt the standard reference value for the dark matter mass density: 
$\rho_{dm} = 0.3 \  {\rm GeV/cm}^3$ and $\xi_{A'}$ is the  mass fraction
of species $A'$ in the halo \footnote{
The uncertainty in $\rho_{dm}$ is around a factor of two or so. Note that
the dynamical halo model, as discussed in section 4, could be used to estimate this density.
For the Milky Way, the dark matter density should be quasi-isothermal, Eq.(\ref{qI6}), with
$r_0 \approx 1.4r_D \approx 5$ kpc from Eqs.(\ref{40}),(\ref{rd}).
At $r = 8.0$ kpc (the sun's radial distance) this gives
$\rho_{dm} \approx 0.22 \ {\rm GeV/cm^3}$ using $\rho_0 r_0 = 100 \ m_{\odot}/{\rm pc^2}$.}.

The velocity integral in Eq.(\ref{55}) can be simplified, since the cross-section depends only on $|{\textbf{v}}|$.
Introducing the speed distribution:
\begin{eqnarray}
{d{\cal N}_{A'} \over d|{\textbf{v}}|} 
&\equiv & \int_{-1}^{1} 
{f_{A'}({\textbf{v}},{\textbf{v}}_E) 
\over 
v_0^3 \ \pi^{3/2}
}
\ |{\textbf{v}}|^2 2\pi \ \ d\cos\theta
\nonumber \\
& = & { |{\textbf{v}}| \over v_0 \sqrt{\pi}  \ |{\textbf{v}}_E| }
\
\left[ \ e^{-[(|{\textbf{v}}| - |{\textbf{v}}_E|)/v_0]^2} - e^{-[(|{\textbf{v}}| + |{\textbf{v}}_E|)/v_0]^2} \ \right]
\ .  
\label{dndv}
\end{eqnarray}
Then the rate, Eq.(\ref{55}) becomes:
\begin{eqnarray}
{dR \over dE_R} = N_T n_{A'} 
\int^{\infty}_{|{\bf{v}}| > v_{min}}\ 
{d\sigma \over dE_R}
\ 
{d{\cal N}_{A'} \over d|{\textbf{v}}|} 
\
|{\textbf{v}}|
\
d|{\textbf{v}}|
\ .
\label{55q}
\end{eqnarray}

The interaction rate, as defined above, has the same general form as for collisionless dark matter particles (also called `WIMPs'
in the literature).
See e.g. \cite{smith} for a pedagogical review.  However there are several important features distinguishing
the interactions of mirror dark matter particles from their collisionless peers.
Firstly, the mirror particle distribution has  no high velocity cutoff in the
velocity integral. Recall collisionless dark matter particles with velocities greater than around 600 km/s
have enough energy to escape from  our galaxy \cite{smith}.
High velocity mirror particles cannot escape from the halo (at our location) because their mean free path is
much shorter than galaxy scales (section 4.7). Another important difference is in the quantity $v_0$, which
characterizes the velocity dispersion.
In the collisionless particle case, $v_0 = v_{rot}$ \cite{smith}, while
for mirror dark matter, $v_0$ 
depends on the mass of the particular particle species, $m_{A'}$:
\begin{eqnarray}
v_0^2 [A'] = {2T \over m_{A'}} \simeq \frac{\overline{m}}{m_{A'}} \ v_{rot}^2 \ .
\label{dis}
\end{eqnarray}
Observe that $m_{A'} \gg \overline{m}$ implies $v_0^2[A'] \ll v_{rot}^2$. 
In figure 5.2 the speed distribution is shown for four representative halo components: 
$A' = H',\ He', \ O', \ Fe'$.
\vskip 0.9cm
\centerline{\epsfig{file=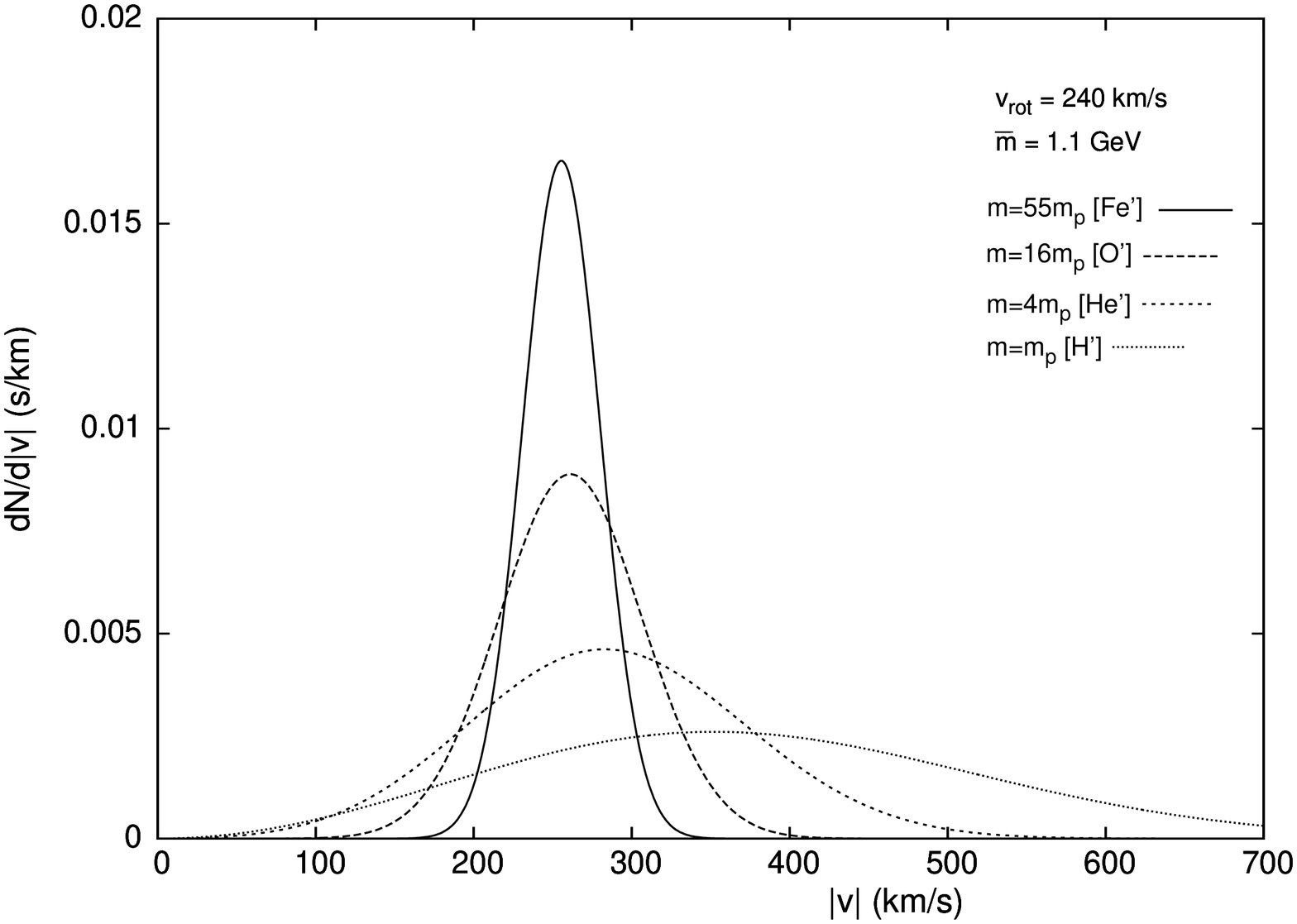,angle=270,width=12.3cm,angle=0}}
\vskip 0.5cm
\noindent
{\small
Figure 5.2: Speed distribution in the Sun's reference frame, $d{\cal N}_{A'}/d|{\textbf{v}}|$, 
versus particle speed 
%($|{\textbf{v}}|$),
for $A' = H'$ (thin dotted line), $He'$ (dotted line),  $O'$ (dashed line),  $Fe'$ (solid line). 
The parameters: $v_{rot} = 240$ km/s,  $\bar m = 1.1$ GeV
are assumed.
}
\vskip 1.0cm

%Another important feature of mirror dark matter is the form of the cross-section, which we now discuss.

%\newpage

\subsection{The interaction rate}

%\vskip 0.2cm

The differential interaction rate of dark matter interactions is given by Eq.(\ref{55}).
Inputting the specific form of the cross-section, Eq.(\ref{cs}), we have:
\begin{eqnarray}
{dR \over dE_R} 
= N_T n_{A'}
{\lambda \over E_R^2 } \int^{\infty}_{|{\bf{v}}| > v_{min}} \ 
{f_{A'}({\bf{v}},{\bf{v}}_E) \over v_0^3 \ \pi^{3/2} |{\bf{v}}|} \ 
d^3 {\textbf{v}} 
\ .
\label{552}
\end{eqnarray}
The velocity integral in Eq.(\ref{552}) is:
\begin{eqnarray}
I &\equiv& 
\int^{\infty}_{|{\bf{v}}| > v_{min}} \ {f_{A'} ({\bf{v}},{\bf{v}}_E) 
\over v_0^3 \ \pi^{3/2}|{\bf{v}}|} \ d^3 {\textbf{v}} 
  \ =   
\ \int^{\infty}_{|{\bf{v}}| > v_{min}} \ {d{\cal N}_{A'} \over 
d|{\textbf{v}}|} \ {1 \over |{\textbf{v}}|} \ d|{\textbf{v}}|
\ .
\label{55x}
\end{eqnarray}
With $d{\cal N}_{A'}/d|{\textbf{v}}|$ given by Eq.(\ref{dndv}),
$I$ can be evaluated in terms of error functions: 
\begin{eqnarray}
I = {1 \over 2yv_0} [erf(x+y) - erf(x-y)]\ ,
\end{eqnarray}
where
\begin{eqnarray}
x \equiv {v_{min}(E_R) \over v_0}, \ y \equiv {|{\textbf{v}}_E| \over v_0} \
\ .
\end{eqnarray}

The interaction rate, Eq.(\ref{552}), depends on $|{\textbf{v}}_E|$,
which varies in time due to the Earth's motion around the Sun:
\begin{eqnarray}
|{\textbf{v}}_E (t)| &=& v_{\odot} + v_{\oplus} \cos\gamma \cos \omega (t - t_0)
\nonumber \\
&=& v_{\odot} + \Delta v_E \cos \omega (t - t_0)
\end{eqnarray}
where $v_{\odot} = v_{rot} + 12 \ {\rm km/s} \sim 230 \ {\rm km/s}$ is the Sun's speed
with respect to the galactic halo
and $v_{\oplus} \simeq 30$ km/s is the Earth's orbital
speed around the Sun. The relevant phase, $t_0$, is $t_0 = 152.5$ days and 
$\omega = 2\pi/T$, with $T = 1$ year.
The inclination of the Earth's orbital plane relative
to the galactic plane is $\gamma \simeq 60^o$, which implies that 
$\Delta v_E \simeq 15$ km/s.

The differential interaction rate, Eq.(\ref{552}), can be expanded
in a Taylor series yielding a time
independent part 
and time dependent modulated component:
\begin{eqnarray}
{dR \over dE_R} \simeq {dR^0 \over dE_R} + {dR^1 \over dE_R} \cos \omega
(t-t_0)\ ,
\label{mod9}
\end{eqnarray}
with
\begin{eqnarray}
{dR^0 \over dE_R} &=& {N_T n_{A'} \lambda I(E_R, y_0) \over E_R^2}
\nonumber \\
{dR^1 \over dE_R} &=& {N_T n_{A'} \lambda \Delta y \over E_R^2} \left(
{\partial I \over \partial y}\right)_{y=y_0}\ .
\label{theoryrate}
\end{eqnarray}
Here $y_0 = v_{\odot}/v_0[A'], \ \Delta y = \Delta v_E/v_0[A']$  and
\begin{eqnarray}
\left( {\partial I \over \partial y}\right)_{y=y_0} 
= - {I(E_R,y_0) \over y_0} + { 1 \over \sqrt{\pi} y_0 v_0 [A']} \left[
e^{-(x-y_0)^2} + e^{-(x+y_0)^2} \right]\ .
\end{eqnarray}

The time dependence in Eq.(\ref{mod9})
is the dominant contribution. It arises due to the annual modulation of the Earth's velocity with
respect to the galactic halo \cite{spergel}.
There are secondary effects, the largest appears to be due to the gravitational
influence of the Sun. This effect, called {\it gravitational focussing} in \cite{peter}, 
leads to an enhancement of the dark matter flux when the Earth is behind the Sun
(for an observer looking in the same direction as ${\textbf{v}}_E$).
This means that the phase of the signal due to gravitational focussing is around March $2^{nd}$,
that is, 1/4 year earlier than the phase due to the Earth's velocity
modulation
in Eq.(\ref{mod9}).
In the following analysis,
the gravitational focussing 
effect is not included mainly because it has a relatively minor influence on the overall rates, allowed regions etc.
However, its most noteworthy aspect is that it can shift the overall phase $t_0$ by around 1-3 weeks 
earlier (depending on mass of the dark matter particle and other details) \cite{peter}. 
That is, the maximum signal is expected to occur around May 20 ($t_0 \approx 140$ days), which
is in very good agreement with the DAMA measured value of $t_0 = 144 \pm 7$ days (over the recoil energy range 2-6 keV) \cite{dama3}.

Interaction rates measured in experiments are smeared by the detector resolution.
Therefore, resolution effects must be modeled if we wish to
make a comparison of theory with data. This is usually done by convolving 
the rate with a Gaussian:
\begin{eqnarray}
{dR \over dE_R^m} = {1 \over \sqrt{2\pi}\sigma_{res}
} 
\int {dR \over dE_R} \ e^{-(E_R - E_R^m)^2/2\sigma^2_{res}} \ dE_R 
\label{556}
\end{eqnarray}
where $E_R^m$ is the measured energy and $\sigma_{res}$ describes the
resolution. Another complication is that the
measured energy of each event is typically in {\it electron equivalent} or keVee units.
This is the recorded energy of each event, usually
ionization or scintillation energy.  However, for
nuclear recoils only a portion of the energy is observed as ionization/scintillation.
This portion is described by the {\it quenching factor}, $q$: keVee = keV$_{NR}$/$q$. 
Naturally $q < 1$, although a class of events, known as {\it channeled} events, where scattered target
atoms travel down crystal
axis and planes, can have $q \simeq 1$ \footnote{
The idea that channeling could be important for the interpretation of direct detection experiments
was raised in \cite{russian} and supported by an initial study \cite{damachan}. 
However further theoretically modeling of this effect has found that it is probably
small \cite{ggchan}. There has also been some experimental work reaching
similar conclusions \cite{collarchan}. In view of these developments, effects
of channeling are not taken into account here, although one should be aware that 
this is a possible source of systematic uncertainty in any analysis.}.

%%%%% extra figure  %%%%%%

%\newpage

\vskip 0.5cm

\subsection{Scattering rates of the halo components}
\vskip 0.2cm

Mirror dark matter  has multiple halo components.  
There is the very light mirror electron component ($e'$) and then there are the
heavier mirror nuclei components.
It is reasonable to suppose that the lightest of these nuclei: $H', He'$  
dominates the mass density of the halo.
Additionally the heavier mirror metals form a subcomponent
with a spectrum spanning from mirror oxygen to mirror iron: $m_{O} \le m_{A'} \le m_{Fe}$.
Let us first briefly remark on the mirror electron component 
and then turn our attention to the mirror nuclei.
\vskip 0.5cm
%\newpage
\noindent
{\bf Electron recoils}
\vskip 0.4cm
\noindent
Halo mirror electrons can scatter off loosely bound atomic electrons in the target
providing them with $\sim keV$  recoils.
These recoils can potentially contribute to the $dR^0/dE_R$ rate in experiments such as 
CoGeNT and DAMA, as these experiments don't discriminate against electron
recoils.
Initial estimates indicated that the mirror electron contribution to the 
average event rate could be
comparable to the nuclear recoil rate at low energies \cite{footelec,foot2}.

However there is a serious complication. 
Mirror electrons being so light ($511$ MeV) can be strongly influenced
by mirror electric and mirror magnetic fields.
Although these effects are very difficult to estimate,
they are bound to be important (as we will explain in the following paragraph).
In the absence of mirror electric (${\bf E'}$) or magnetic fields (${\bf B'}$),
the mirror electron flux arriving at the Earth is:
\begin{eqnarray}
F_{e'} \sim n_{e'} \ v_0 (e') \ \ \  {\rm for} \ \ {\bf E'} = {\bf B'} = 0
\label{fluxe}
\end{eqnarray}
where we have approximated $\langle v_{e'} \rangle \sim v_0 (e')$ , which is roughly 
valid for mirror electrons as their velocity dispersion is much greater than the
Earth's speed through the halo. 

The mirror-electron velocity dispersion can be estimated from Eq.(\ref{dis}) to be 
around 10,000 km/s for $\bar m = 1.1$ GeV.
Because of the large velocity dispersion the mirror-electron flux, Eq.(\ref{fluxe}), 
is much greater than the flux of mirror nuclei arriving at the Earth.
In reality this could not be the case.
A larger mirror-electron flux would lead to a greater mirror-electron capture rate in
the Earth cf. the capture of mirror nuclei.  This would lead to an increasing 
mirror electric charge within the Earth, $Q'_E$.
Very quickly mirror electric and magnetic fields would be generated such that
the flux of mirror electrons is reduced until it approximately 
matches the flux of mirror nuclei hitting the Earth's surface.
With this effect taken into account the rate of electron recoils is significantly reduced,
leaving nuclear recoils as the dominant contribution to the average rate. 
Naturally, subtle effects 
are certainly possible, and further study of this
interesting physics is warranted.
\vskip 0.4cm
\noindent
{\bf Nuclear recoils}
\vskip 0.4cm
\noindent
Mirror nuclei can scatter off target nuclei potentially producing an observable nuclear recoil.
The rates have been given in Eq.(\ref{theoryrate}) of section 5.2. Here, we provide an illustrative example.
In figure 5.3a,b
the predicted event rates: $dR^0/dE_R$ and $dR^1/dE_R$ are given for a germanium target assuming a halo spectrum: 
$\xi_{He'} = 0.9$, $\xi_{O'} = 0.1$ and $\xi_{Fe'} = 0.001$, for $v_{rot} = 240$ km/s, $\bar m = 1.1$ GeV,
and $\epsilon = 2\times 10^{-9}$. 
We have assumed perfect energy resolution, and the recoil energy shown is the actual nuclear recoil energy
(not electron equivalent).
Considering first the time average rate, 
figure 5.3a clearly shows the $dR^0/dE_R \propto 1/E_R^2$ dependence which arises from the same dependence
in the Rutherford scattering cross-section, Eq.(\ref{cs}). As $E_R$ increases, the various kinematic thresholds 
are crossed, with dramatic reduction in event rate. 

The energy associated with each kinematic threshold can be easily estimated.
Given that the typical velocities in the Earth's reference frame are $|{\textbf{v}}| \sim v_{rot}$
and the narrow velocity dispersion 
(figure 5.2), the threshold occurs at energies where $v_{min}(E_R) \approx v_{rot}$. 
From Eq.(\ref{v}), this implies that 
\begin{eqnarray}
E^{threshold}_{A'} \approx {2 v^2_{rot} m_{A} m_{A'}^2 \over (m_A + m_{A'})^2} 
\ .
\label{aa5}
\end{eqnarray}
For a fixed $A'$ the threshold energy is maximized when $m_{A} = m_{A'}$.
The width of the threshold region, $\Delta E_R$ is determined by the velocity dispersion.
Roughly, 
\begin{eqnarray}
{\Delta E_R \over E^{threshold}_{A'}}\approx 4 {v_0 [A']
\over v_{rot}}
\approx 4 \sqrt{{\bar m \over m_{A'}}} 
\label{aa6}
\end{eqnarray}
where Eqs.(\ref{v}), (\ref{dis}), (\ref{aa5}) have been used.

In figure 5.3b, the annual modulation amplitude is given. It is possible to show via simple analytic arguments
that for a given component, $A'$, this amplitude is postive at sufficiently high recoil energy
with a maximum at $E_R \approx E^{threshold}_{A'}$ \cite{footdd2}. The annual modulation amplitude changes
sign at low recoil energies.
For the chosen abundances in this example, figure 5.3b indicates that the $O'$ contribution
dominates the annual modulation amplitude in the $E_R > 1$ keV region.

Of course, figure 5.3 is just an example; 
the precise chemical composition of the metal components is unknown, it could happen, for instance, that $\xi_{Fe'}$ is much
larger than 0.001, or even that $\xi_{Fe'} \sim \xi_{O'}$.
The contribution to the rate due to each component scales linearly with $\xi_{A'}$,
so the relative contributions of, say, $O'$ and $Fe'$ can change if different abundances are assumed.
There are several other important uncertainties.
As indicated in Eq.(\ref{aa5}),
the energy of the various kinematic thresholds 
changes if $v_{rot}$ is varied.
Increasing (decreasing) $v_{rot}$ moves the thresholds to higher (lower)
energies. Also, as indicated in Eq.(\ref{aa6}), the width of the threshold region is controlled by the parameter, $\bar m$. 
Increasing (decreasing) $\bar m$ broadens (sharpens) 
the threshold region.

Unfortunately, none
of the existing experiments have low enough thresholds and/or low enough backgrounds to be sensitive to the light $He'$ component. 
The CoGeNT experiment, for instance, has a nuclear energy threshold of around 2.5 keV.
The early CRESST-I experiment  had a low energy threshold of around $0.6$ keV 
and a target containing the light element oxygen \cite{cresstI}.  
The sensitivity of the CRESST-I experiment, though, was unable to constrain the $He'$ component.
A limit of around $\epsilon \sqrt{\xi_{He'}} \stackrel{<}{\sim} 3 \times 10^{-9}$ can be
obtained from the analysis in \cite{foot2003}. More recent very low energy threshold experiments, including Texono \cite{tex}
and CDMSlite \cite{cdmslite}, also have insufficient sensitivity to probe the light $He'$ component.
However, existing experiments can potentially detect the mirror metal component(s) as we shall now discuss.

\vskip 0.4cm
\centerline{\epsfig{file=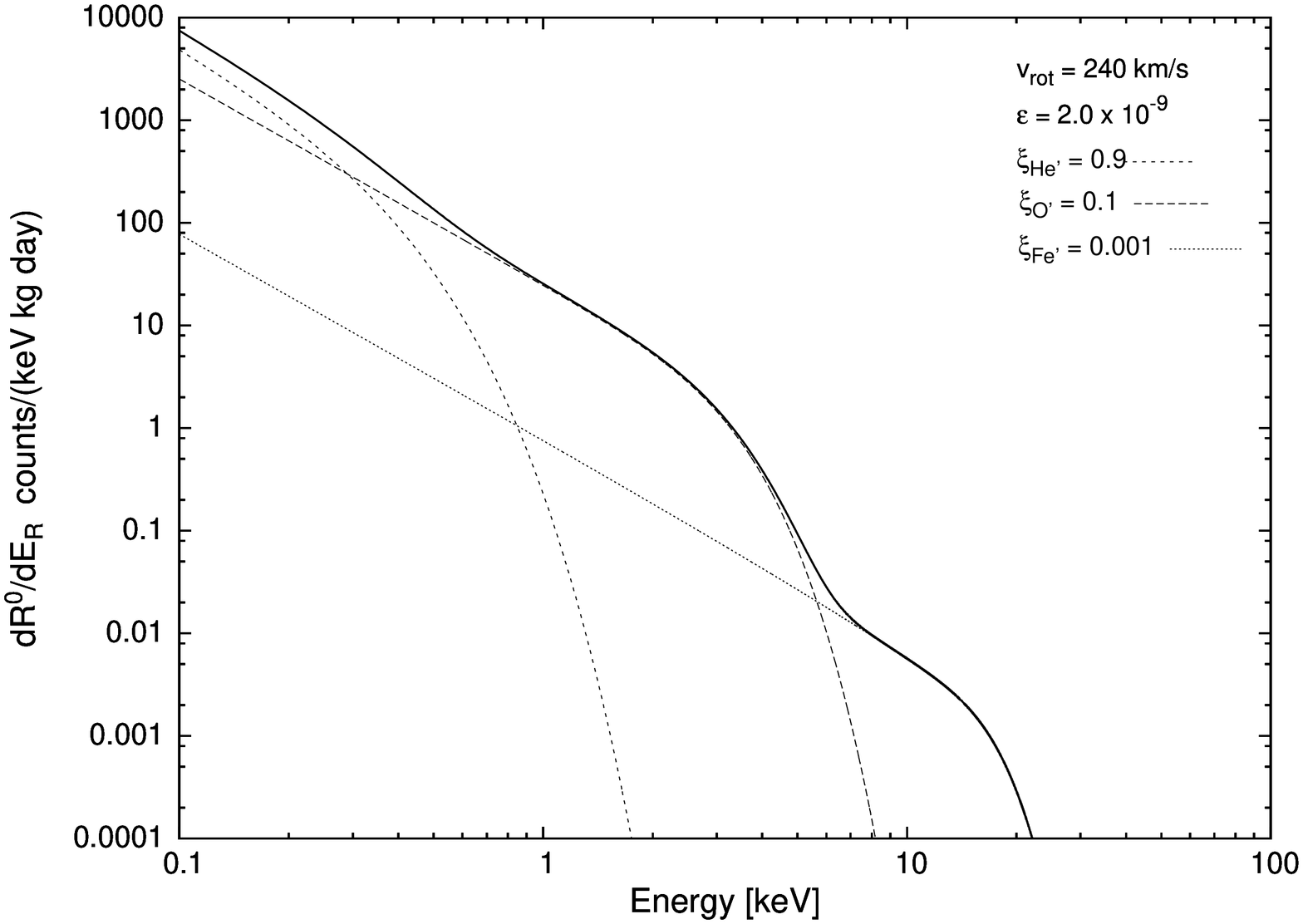,angle=270,width=12.0cm}}
\vskip 0.3cm
\noindent
{\small
Figure 5.3a: Predicted event rate: $dR^0/dE_R$ (solid line) on a germanium target for halo dark matter with composition:
$\xi_{He'} = 0.9$, $\xi_{O'} = 0.1$ and $\xi_{Fe'} = 0.001$, for $v_{rot} = 240$ km/s and $\bar m = 1.1$ GeV.
The contributions from the various components are also shown.
}
\vskip 0.4cm
\centerline{\epsfig{file=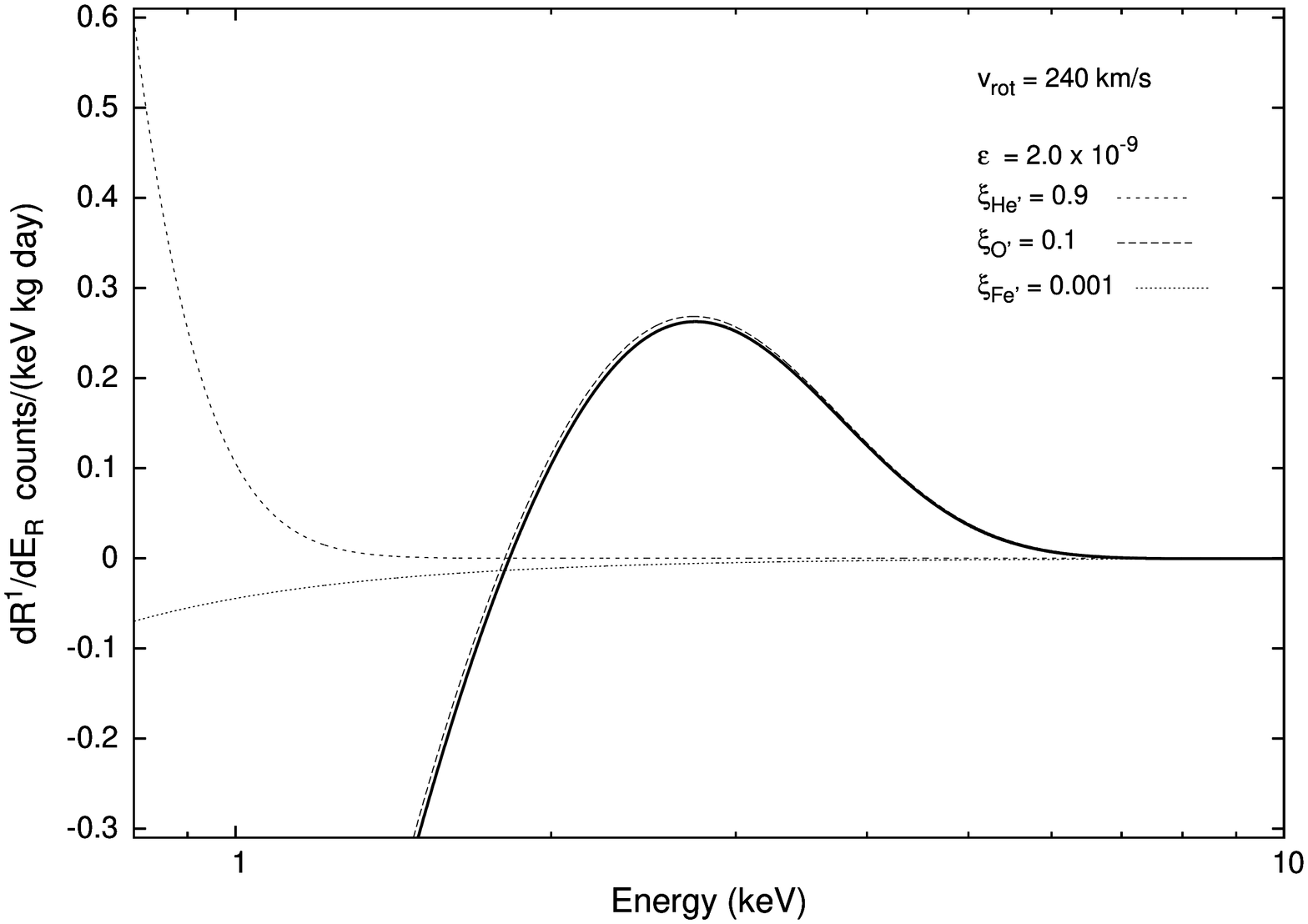,angle=270,width=12.0cm}}
\vskip 0.3cm
\noindent
{\small
Figure 5.3b: Predicted annual modulation amplitude: $dR^1/dE_R$ (solid line) on a germanium target, 
for the same parameters as per figure 5.3a. The contribution from the various components are also shown.
}
\vskip 1.4cm

%%%%%%%%%%% New subsection %%%%%%%%%%%%%%%%

\subsection{Analysis of the experiments}

\vskip 0.2cm

During the last decade or so, progress has been made in experimental efforts to
directly detect dark matter.
The DAMA/NaI \cite{dama} and DAMA/Libra \cite{dama2,dama3}
experiments, in particular, have obtained the first evidence for dark matter direct detection. 
These experiments have observed a modulation 
in the {\it single hit} event rate with a period and phase consistent with
expectations 
from dark matter interactions \cite{spergel}. Background rates are not expected to modulate,
with the possible exception of muon induced backgrounds. However it has
been known
for some time that muons cannot mimic the dark matter annual
modulation
signature \cite{damamuon}. The DAMA experiments thus provide a very strong case
that dark matter interactions have been detected, via the modulated component, $R^1$.

More recently, CoGeNT \cite{cogent}, CRESST-II \cite{cresst} and CDMS/Si \cite{cdms/si} experiments have obtained 
evidence for dark matter interactions.
These experiments aim to reduce backgrounds so that the unmodulated rate, $R^0$, can be revealed.
They have insufficient data, at present, to confirm an annual modulated component, $R_1$ (although
there is some tentative evidence for an annual modulation in the CoGeNT signal \cite{cogentmod}).
Our task now is to examine the data from each of these experiments
in the mirror dark matter framework.

As figure 5.3 illustrates,
none of the current direct detection experiments are sensitive to 
the $H', \ He'$ halo components; the kinetic energy of these particles (in the lab reference frame)
is simply too small to produce recoils energetic enough to be above the experimental
energy thresholds.
The current direct detection experiments are sensitive to the mirror metal components.
Of course, mirror dark matter predicts a spectrum of such particles, ranging in mass from 
mirror oxygen ($m_{O'} = 16m_p \simeq 15.0$ GeV) to
mirror iron ($m_{Fe'} = 55.8 m_p \simeq 52.5$ GeV).
To proceed we make the simple assumption that
the signal in each experiment is dominated by the interactions 
of a single such metal component, $A'$. 
Of course, this is only an approximation, however it can 
be a reasonable one given the fairly narrow energy range probed in each of the
experiments [the signal regions are mainly:
2-4 keVee (DAMA), 0.5-1 keVee (CoGeNT),  12-14 keV (CRESST-II) and 7-13 keV (CDMS/Si)].
With this assumption, the
interaction rate depends on the parameters $m_{A'},\ \epsilon \sqrt{\xi_{A'}}$
and also $v_{rot}$. 
We now consider each of the four experiments, DAMA, CoGeNT, CRESST-II and CDMS/Si in turn.  

\vskip 0.3cm
\noindent
{\bf DAMA }
\vskip 0.3cm
\noindent
The DAMA collaboration have operated an array of low radioactivity scintillating thallium doped 
sodium iodide NaI [Tl] crystals at the Gran Sasso underground laboratory \cite{damarev}. 
From 1996-2002 the total fiducial mass available was $\sim$ 100 kg, and was upgraded
to $\sim$ 250 kg from 2003-present.
A total of 1.33 ton-years of exposure has been collected  which allows a
sensitive probe of dark matter via the annual modulation signature.
Analysis of the data has revealed an annually modulated low energy component, at around $9\sigma$ C.L.
with phase consistent with dark matter interactions \cite{dama,dama2,dama3}. 

We consider this annual modulation signal over the relevant low energy range: 2 keVee - 8 keVee. 
We divide this energy range into 12 bins of width 0.5 keVee.
The theoretical annual modulation
signal, $dR^1/dE_R$, is evaluated as a function of $m_{A'},\ \epsilon \sqrt{\xi_{A'}}$, taking into account
detector resolution effects. A  $\chi^2$ function is defined by:
\begin{eqnarray}
\chi^2 (m_{A'}, \epsilon \sqrt{\xi_{A'}}) = \sum_{i=1}^{12} 
\left[ {R_i - data_i \over \delta data_i}\right]^2
\ .
\label{chi2yyy}
\end{eqnarray}
The $\chi^2$ function is minimized over quenching factor uncertainty which we take as: $q_{Na} = 0.28 \pm 0.08$ and
$q_I = 0.12 \pm 0.08$. \footnote{
There are some indications that the DAMA quenching factors could be smaller than the considered 
range \cite{collarchan}, 
and other indications
that the DAMA quenching factors could be larger \cite{tret}. Additionally, 
a few percent channeling fraction for iodine (and also sodium if there are lighter more abundant halo
components) can be important which can significantly lower the DAMA favored region. In view of these unknowns,
the DAMA favored region should be viewed as a rough guide only.}

\vskip 0.3cm
\noindent
{\bf CoGeNT }
\vskip 0.3cm
\noindent
The CoGeNT collaboration have been searching for dark matter with a low energy threshold 
P-type Point Contact germanium detector operating
in the Soudan Underground laboratory.
They have observed a low energy excess of events which cannot be explained by known backgrounds \cite{cogent}.
This excess can tentatively be interpreted as dark matter interactions.

We consider here the  most recent
data obtained from  $0.33$ kg $\times$ 807 days of exposure.
This data, stripped of known background components, and corrected for surface event contamination
and overall detection efficiency \cite{cogent}, is divided into
15 bins of width 0.1 keVee over the energy range 0.5 - 2 keVee. 
The theoretical rate, $dR^0/dE_R$, is obtained from Eq.(\ref{theoryrate}) taking into account 
the detector resolution, Eq.(\ref{556}).
The resulting 
$\chi^2$, defined as in Eq.(\ref{chi2yyy}), is minimized over the quenching factor uncertainty: $q_{Ge} = 0.21 \pm 0.04$
and a constant background component.

\vskip 0.3cm
\noindent
{\bf CRESST-II}
\vskip 0.3cm
\noindent
The CRESST-II collaboration have announced results for
their dark matter search with 730 kg-days of net exposure in a $CaWO_4$
target. The detector consists of eight modules with
energy thresholds (keV) of 10.2, 12.1, 12.3, 12.9, 15.0, 15.5, 16.2, 19.0.
Again a low energy event excess over known backgrounds is observed.

To facilitate a $\chi^2$ analysis,
the CRESST-II the data is divided into five bins with keV energy ranges of 10.2-13, 13-16, 16-19, 19-25, 25-40.
This data is summarized in table 1. This table
also indicates the expected background rate estimated from all known
sources of 
background \cite{cresst}.
\begin{table}
\centering
\begin{tabular}{c c c}
\hline\hline
Bin / keV & Total events & Estimated background  \\
\hline
10.2 -- 13.0 & 9 & 3.2  \\
13 -- 16 & 15 & 6.1 \\
16 -- 19 & 11 & 7.0  \\
19 -- 25 & 12 & 11.5  \\
25 -- 40 & 20 & 20.1 \\
\hline\hline
\end{tabular}
\caption{CRESST-II data: total number of events and estimated
background.}
\end{table}
\vskip 0.9cm
%No energy calibration uncertainty is considered for CRESST-II.
The rate in each energy bin, $R_i^0$,  can be calculated as per Eq.(\ref{theoryrate}) taking into account
the detector resolution, Eq.(\ref{556}). 
The exposure time in the appropriate step function (in energy) which
takes into account the various thresholds of the 8 detector modules.
The CRESST-II $\chi^2$ function is then defined by:
\begin{eqnarray}
\chi^2 (m_{A'}, \epsilon \sqrt{\xi_{A'}}) = 
\sum_{i=1}^{5}  \left[ {R_i^0 + B_i - data_i \over \delta data_i}\right]^2
\ 
\label{chi2bla}
\end{eqnarray}
where $B_i$ is the estimated
background in the $i^{th}$ energy bin. 
%We evaluate $\chi^2$ as per
%Eq.(\ref{chi2bla}),
%with the constraint $m_{A'} \le m_{Fe} \simeq 55.8m_p$.
No energy scale uncertainty is considered for CRESST-II.

For DAMA, CoGeNT and CRESST-II the 95\% C.L.
favored region is given by $\chi^2 \le \chi^2_{min} + \Delta \chi^2$ with $\Delta \chi^2 = 5.99$.

\vskip 0.3cm
\noindent
{\bf CDMS/Si}
\vskip 0.3cm
\noindent
The CDMS/Si experiment, utilizing an array of silcon detectors with 140.2 kg-days of exposure, has observed three dark matter
candidate events \cite{cdms/si}.
These three events have nominal recoil energies of 8.2 keV, 9.5 keV and 12.3 keV.
A $\chi^2$ analysis should not be used due to the small number of events.
Instead, the extended maximum likelihood formalism \cite{barlow} can be used to construct
the likelihood function:
\begin{eqnarray}
{\cal L}({\bf p}) = \left[ \Pi_{i} {dn (E_R^i) \over dE_R} \right] exp[-{\cal N({\bf p})}]
\end{eqnarray}
where the vector ${\bf p}$ denotes the unknown parameters.
Here, $dn (E_R^i)/dE_R$ 
is the interaction rate (defined more precisely in a moment) at the recoil energy for each of the three observed events, $i=1,...,3,$ while
$\cal N({\bf p})$ is the total number of events expected in the
acceptance recoil energy region:
\begin{eqnarray}
{\cal N({\bf p})} = \int {dn \over dE_R} \ dE_R
\ .
\end{eqnarray}
The event rate, $dn/dE_R$, is computed from the rate, $dR^0/dE_R$, by including
resolution effects \footnote{Resolution effects are taken into account
by convolving the rate $dR/dE_R$ 
with a Gaussian.
In the absence of resolution measurements, we use $\sigma_{res} = 0.1$ keV.} and 
detection efficiency, obtained from 
figure 1 of \cite{cdms/si}.

The CDMS collaboration \cite{cdms/si}, point out that the recoil energy calibration is 
likely around 10\% higher than nominally used, with some uncertainty. In view of this we  have scaled up the energies
by a factor: $f=1.1$ and adopted an energy calibration uncertainty of $\pm 10\%$, i.e. $f = 1.1 \pm 0.1$. 
For each value of the parameters: $m_{A'}, \ \epsilon \sqrt{\xi_{A'}}$ we have maximized ${\cal L}$ over this
range of $f$, to give profile likelihood function, ${\cal L}_P$  
The favored region for the parameters: $m_{A'},\ \epsilon \sqrt{\xi_{A'}}$
is then determined by  
\begin{eqnarray}
ln \ {\cal L}_P \ge ln \ {\cal L}_{P max} - \Delta \ ln \ {\cal L}_P
\ .
\end{eqnarray}
We set $2\Delta \ ln \ {\cal L}_p = 5.99 $  corresponding to 95\% C.L. for 2 parameters \cite{bbn013}. 
Since the estimated background rate in the energy region of interest, $E_{threshold} \le E_R \le 20$ keV, is much less than 1 event 
for the CDMS/Si exposure \cite{cdms/si},
we can simplify the analysis by neglecting any background contribution.

\vskip 0.8cm

The data set from each experiment can now be compared with the theoretical rate in the mirror dark matter framework.
The parameter space is scanned subject to the mild theoretical constraint: $m_{A'} \le m_{Fe'} \simeq 52.5$ GeV.
We also allow $A', \ Z'$ to have non-integer values, with $Z' = A'/2$, except
when we specifically consider $A' = Fe'$, where we use $Z' = 26$, $A' = 55.8$.
The best fit parameter values are given in table 2 for three representative values for $v_{rot}$.
In figure 5.4 we plot 
the $95\%$ C.L. favored region of parameter space for each experiment, for these same $v_{rot}$ values
%For each data set, we
%give $95\%$ C.L. favoured regions [$\chi^2 \le \chi^2_{min} + \Delta \chi^2$, with
%$\Delta \chi^2 = 5.99$]

\begin{table}
\centering
\begin{tabular}{c c c c c}
\hline\hline
$v_{rot}$ [km/s]   & CDMS/Si & CoGeNT & DAMA & CRESST-II\\
               &  &   
$\chi^2$ (min)/d.o.f.  &  
$\chi^2$ (min)/d.o.f.  & 
$\chi^2$ (min)/d.o.f.  \\
& best fit param. &  best fit param. &
 best fit param. &  best fit param. \\
\hline
&  & & & \\
200 &      & 9.7/12 & 6.1/10 & 2.6/3 \\
    & ${m_{A'} \over m_p}$ = 55.8 & ${m_{A'} \over m_p}$ = 39.0 & ${m_{A'} \over m_p}$ = 55.8 & 
    ${m_{A'} \over m_p}$ = 55.8 \\ 
    & $\frac{\epsilon \sqrt{\xi_{A'}}}{10^{-10}} = 0.93$ & ${\epsilon \sqrt{\xi_{A'}} \over 10^{-10}} = 2.5$ & 
    ${\epsilon \sqrt{\xi_{A'}} \over 10^{-10}} = 2.5$ & ${\epsilon \sqrt{\xi_{A'}} \over 10^{-10}} = 2.7$ \\
&  & & & \\
\hline
&  & & & \\
240 &      & 9.9/12 & 5.5/10 & 0.3/3 \\
    & ${m_{A'} \over m_p}$ = 37.0 & ${m_{A'} \over m_p}$ = 31.0 & ${m_{A'} \over m_p}$ = 45.2 & 
    ${m_{A'} \over m_p}$ = 55.8 \\ 
    & ${\epsilon \sqrt{\xi_{A'}} \over 10^{-10}} = 1.2$ & ${\epsilon \sqrt{\xi_{A'}} \over 10^{-10}} = 3.1$ & 
    ${\epsilon \sqrt{\xi_{A'}} \over 10^{-10}} = 3.8$ & ${\epsilon \sqrt{\xi_{A'}} \over 10^{-10}} = 1.7$  
\\
&  & & & \\
\hline
&  & & & \\
280 &      & 10.1/12 & 4.9/10 & 0.2/3 \\
    & ${m_{A'} \over m_p}$ = 25.5 & ${m_{A'} \over m_p}$ = 25.0 & ${m_{A'} \over m_p}$ = 37.1 & 
    ${m_{A'} \over m_p}$ = 36.0 \\ 
    & ${\epsilon \sqrt{\xi_{A'}} \over 10^{-10}} = 1.6$ & ${\epsilon \sqrt{\xi_{A'}} \over 10^{-10}} = 3.6$ & 
    ${\epsilon \sqrt{\xi_{A'}} \over 10^{-10}} = 4.7$ & ${\epsilon \sqrt{\xi_{A'}} \over 10^{-10}} = 2.4$  
\\
&  & & & \\
\hline\hline
\end{tabular}
\caption{Summary of $\chi^2 (min)$ and best fit parameters for the relevant data sets from the
CDMS/Si, CoGeNT, DAMA and CRESST-II experiments.}
\end{table}

\vskip 0.4cm
\centerline{\epsfig{file=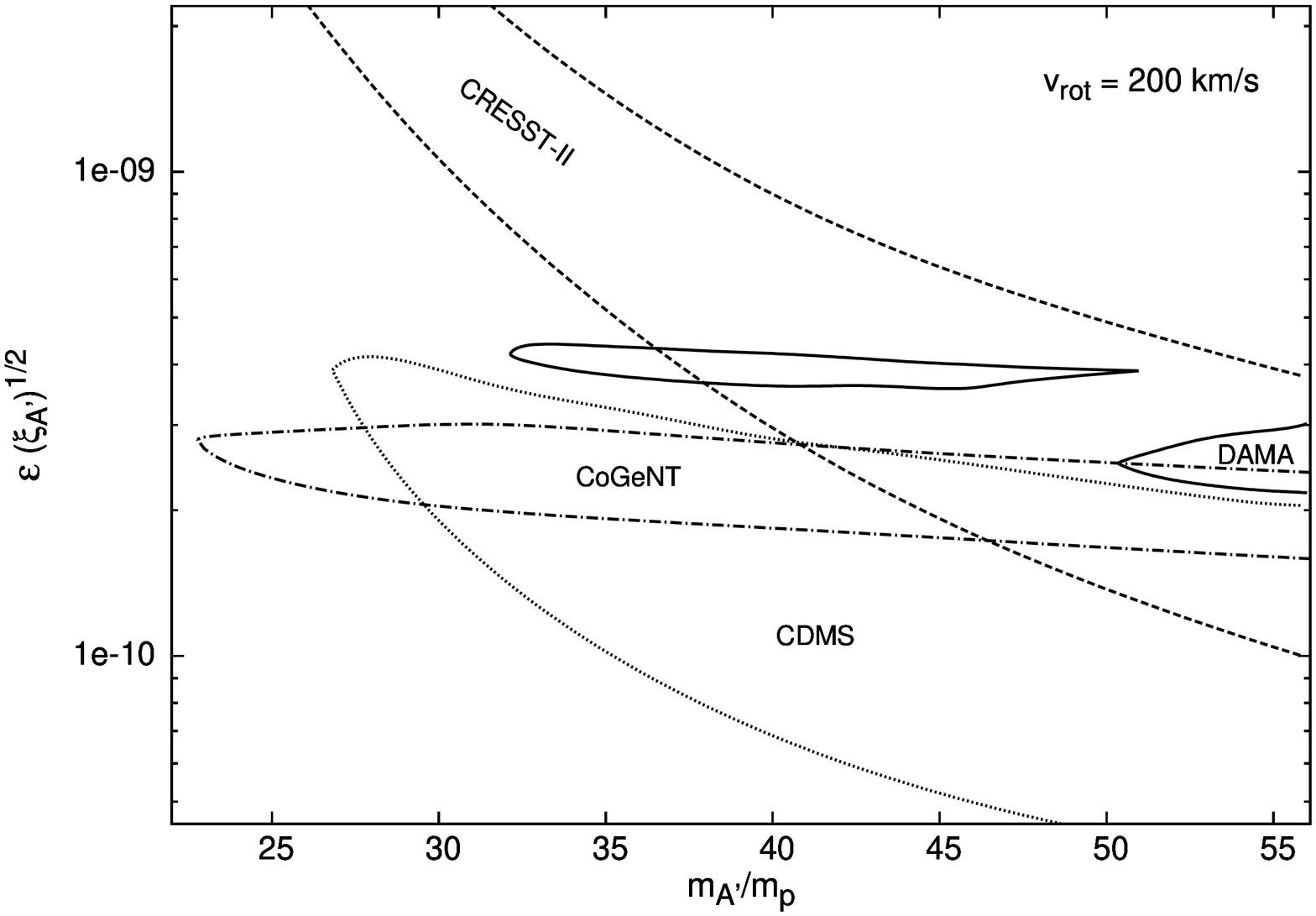,angle=270,width=12.0cm}}
\vskip 0.3cm
\noindent
{\small
Figure 5.4a: DAMA (solid lines), CoGeNT (dashed-dotted lines), 
CRESST-II (dashed lines) and CDMS/Si (dotted lines) favored regions of 
parameter space [95\% C.L.] in the mirror dark matter model for
$v_{rot} = 200$ km/s. As shown, the DAMA favored region consists of two parts: the lower-mass
region results from $A'-Na$ scattering while the higher-mass region
results from $A'-I$ scattering.
}
\vskip 0.5cm
\centerline{\epsfig{file=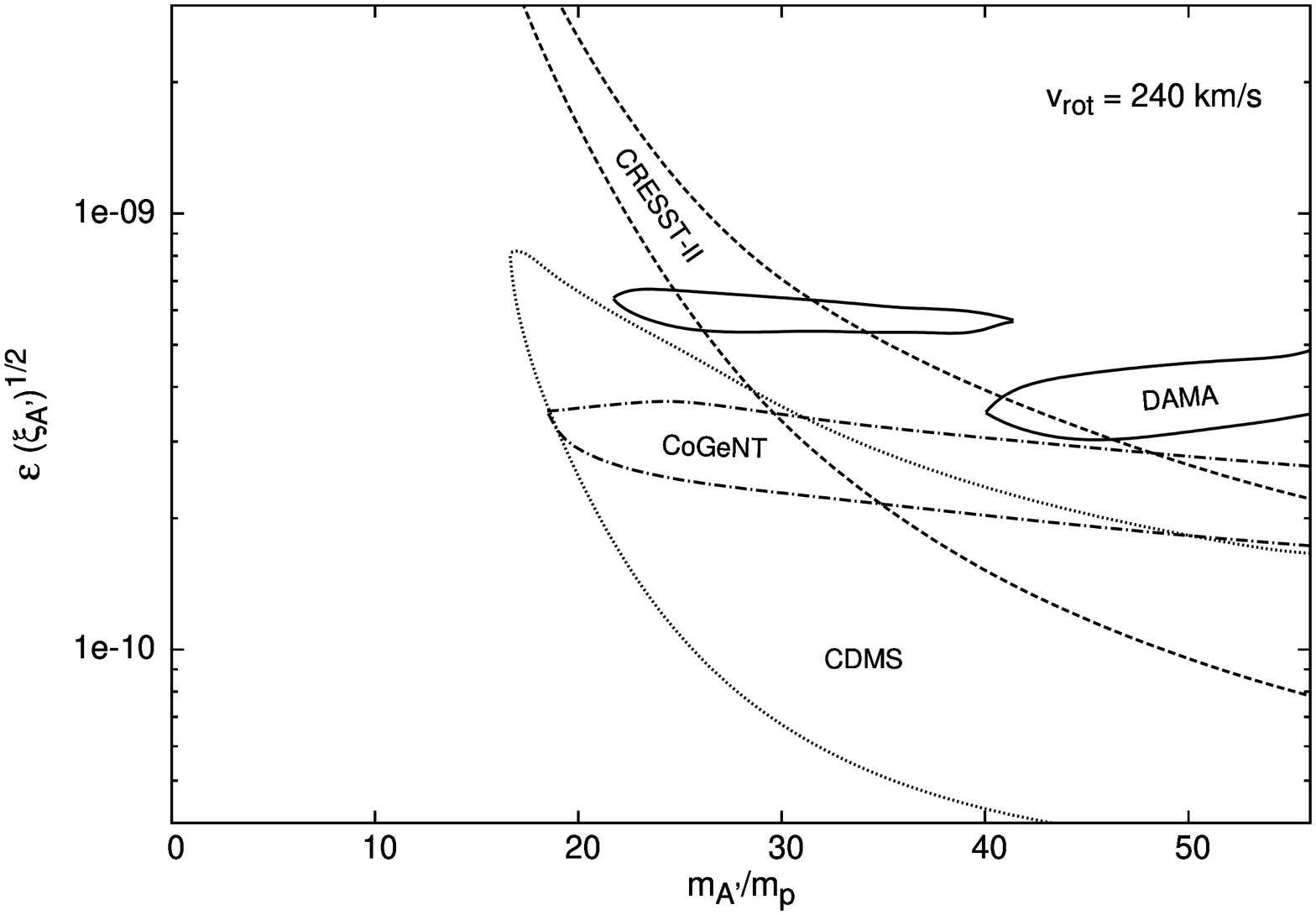,angle=270,width=12.0cm}}
\vskip 0.2cm
\noindent
{\small
Figure 5.4b: Same as figure 5.4a, except 
$v_{rot} = 240$ km/s. 
}
\vskip 0.5cm
\centerline{\epsfig{file=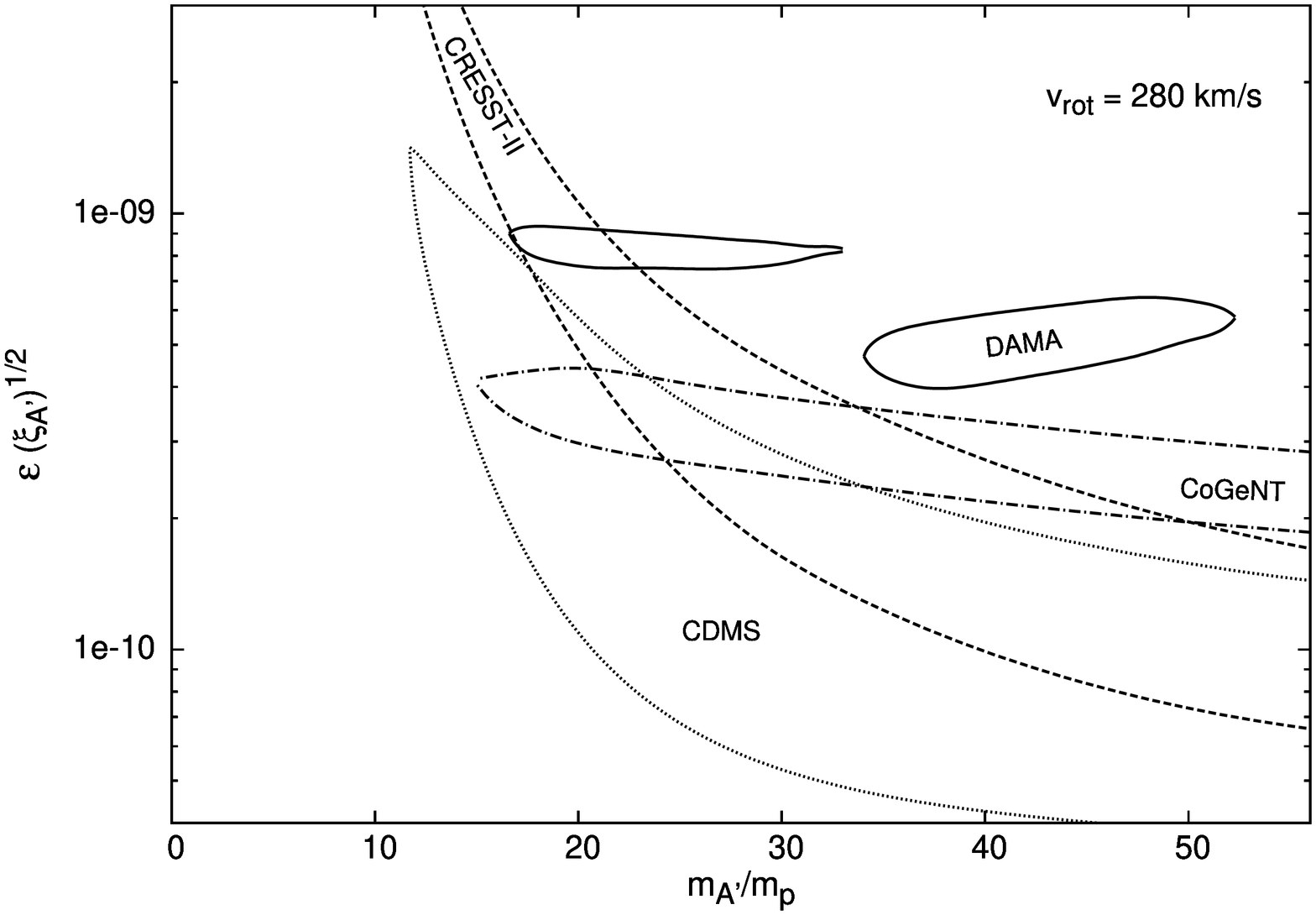,angle=270,width=12.0cm}}
\vskip 0.3cm
\noindent
{\small
Figure 5.4c: Same as figure 5.4a, except 
$v_{rot} = 280$ km/s.
}
\vskip 0.9cm

Figure 5.4 demonstrates that there is a substantial region of parameter space where
the data from all four experiments can be simultaneously explained within this theoretical framework.
The figures suggests a slight preference for
$A' \sim Fe'$, $v_{rot} \approx 200$ km/s, although of course,
the potential uncertainties cannot exclude other parameter space, with lighter $A'$ components and
higher $v_{rot}$ values.

\begin{table}
\centering
\begin{tabular}{c c c c c}
\hline\hline
$m_{A'}/m_p$   & CDMS/Si & CoGeNT & DAMA & CRESST-II\\
$Z'$ &  &   
$\chi^2$ (min)/d.o.f.  &  
$\chi^2$ (min)/d.o.f.  & 
$\chi^2$ (min)/d.o.f.  \\
& best fit param. &  best fit param. &
 best fit param. &  best fit param. \\
\hline
&  & & & \\
55.8 &      & 9.3/12 & 5.8/10 & 0.3/3 \\
26 & $v_{rot}$ = 205 km/s & 
$v_{rot}$ = 150 km/s & 
$v_{rot}$ = 210 km/s & 
$v_{rot}$ = 250 km/s  
\\
    & $\frac{\epsilon \sqrt{\xi_{A'}}}{10^{-10}} = 0.96$ & ${\epsilon \sqrt{\xi_{A'}} \over 10^{-10}} = 1.9$ & 
    ${\epsilon \sqrt{\xi_{A'}} \over 10^{-10}} = 3.1$ & ${\epsilon \sqrt{\xi_{A'}} \over 10^{-10}} = 1.7$ \\
&  & & & \\
\hline
&  & & & \\
28.1 &      & 9.8/12 & 10.2/10 & 0.3/3 \\
14 & $v_{rot}$ = 270 km/s & 
$v_{rot}$ = 210 km/s & 
$v_{rot}$ = 280 km/s & 
$v_{rot}$ = 300 km/s  
\\
    & $\frac{\epsilon \sqrt{\xi_{A'}}}{10^{-10}} = 1.5$ & ${\epsilon \sqrt{\xi_{A'}} \over 10^{-10}} = 2.5$ & 
    ${\epsilon \sqrt{\xi_{A'}} \over 10^{-10}} = 8.1$ & ${\epsilon \sqrt{\xi_{A'}} \over 10^{-10}} = 3.1$ \\
&  & & & \\
\hline
&  & & & \\
16.0 &      & 11.8/12 & 8.1/10 & 3.1/3 \\
8 & $v_{rot}$ = 300 km/s & 
$v_{rot}$ = 300 km/s & 
$v_{rot}$ = 300 km/s & 
$v_{rot}$ = 300 km/s  
\\
    & $\frac{\epsilon \sqrt{\xi_{A'}}}{10^{-10}} = 3.5$ & ${\epsilon \sqrt{\xi_{A'}} \over 10^{-10}} = 3.9$ & 
    ${\epsilon \sqrt{\xi_{A'}} \over 10^{-10}} = 10.1$ & ${\epsilon \sqrt{\xi_{A'}} \over 10^{-10}} = 11.0$ \\
&  & & & \\
\hline\hline
\end{tabular}
\caption{Summary of $\chi^2 (min)$ and best fit parameters for the relevant data sets from the
CDMS/Si, CoGeNT, DAMA and CRESST-II experiments.}
\end{table}

Instead of having a fixed value for $v_{rot}$ and varying $m_{A'},\ \epsilon \sqrt{\xi_{A'}}$
we also consider a fixed $A'$ element and treat $v_{rot}, \ \epsilon \sqrt{\xi_{A'}}$
as free parameters (subject to the mild constraint, $150 \le v_{rot} [{\rm km/s}] \le 300$).
We examine three representative $A'$ choices, $A' = Fe', \ Si', \ O'$. 
Table 3  summarizes the $\chi^2$ minimum and best fit points while
figure 5.5 provides the favored parameter region in each case.

\vskip 0.5cm
\centerline{\epsfig{file=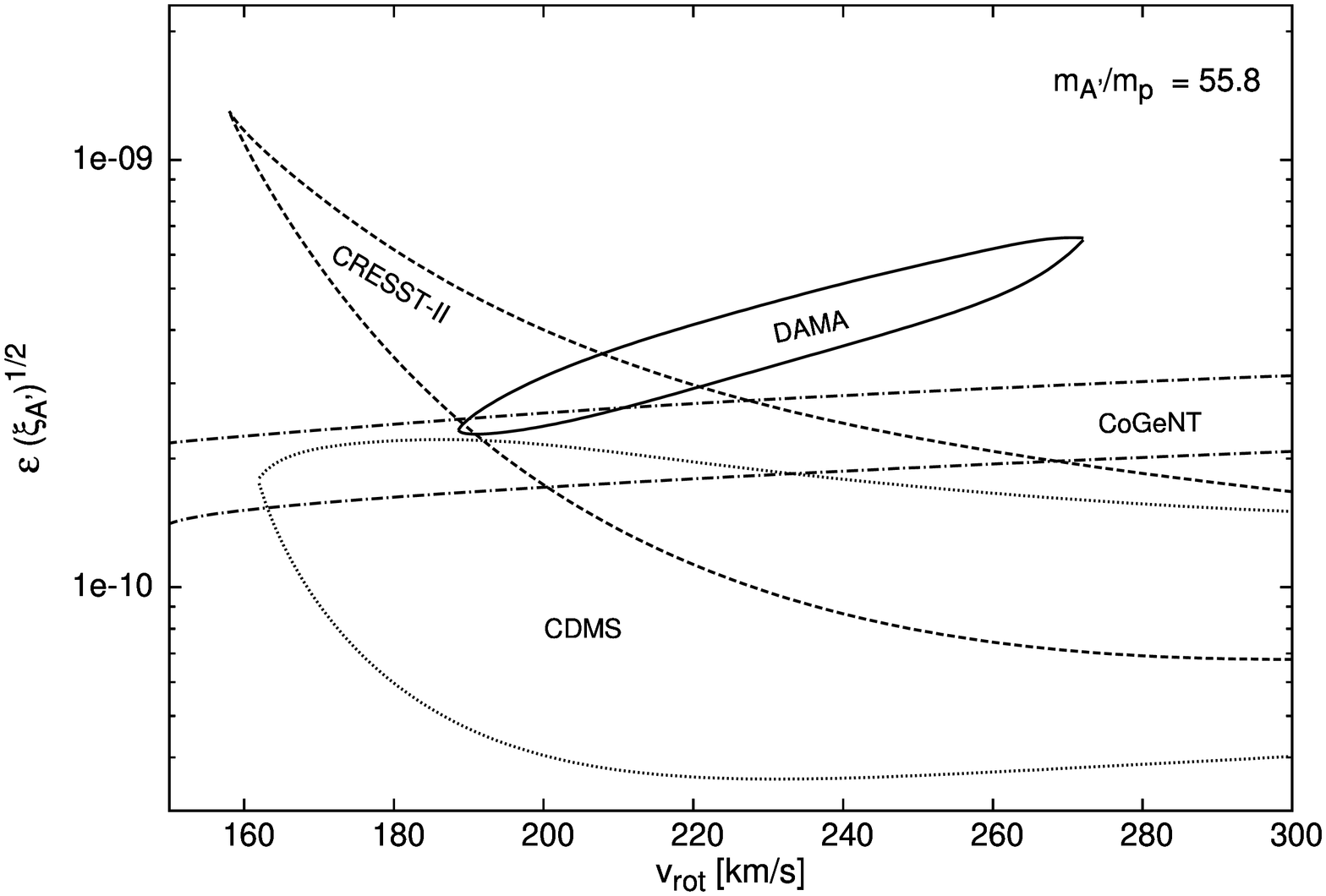,angle=270,width=12.0cm}}
\vskip 0.3cm
\noindent
{\small
Figure 5.5a: DAMA (solid lines) , CoGeNT (dashed-dotted lines), CRESST-II (dashed lines) and 
CDMS/Si (dotted lines) favored regions
of parameter space [95\% C.L.] for $A' = Fe'$.
}
\vskip 0.5cm
\centerline{\epsfig{file=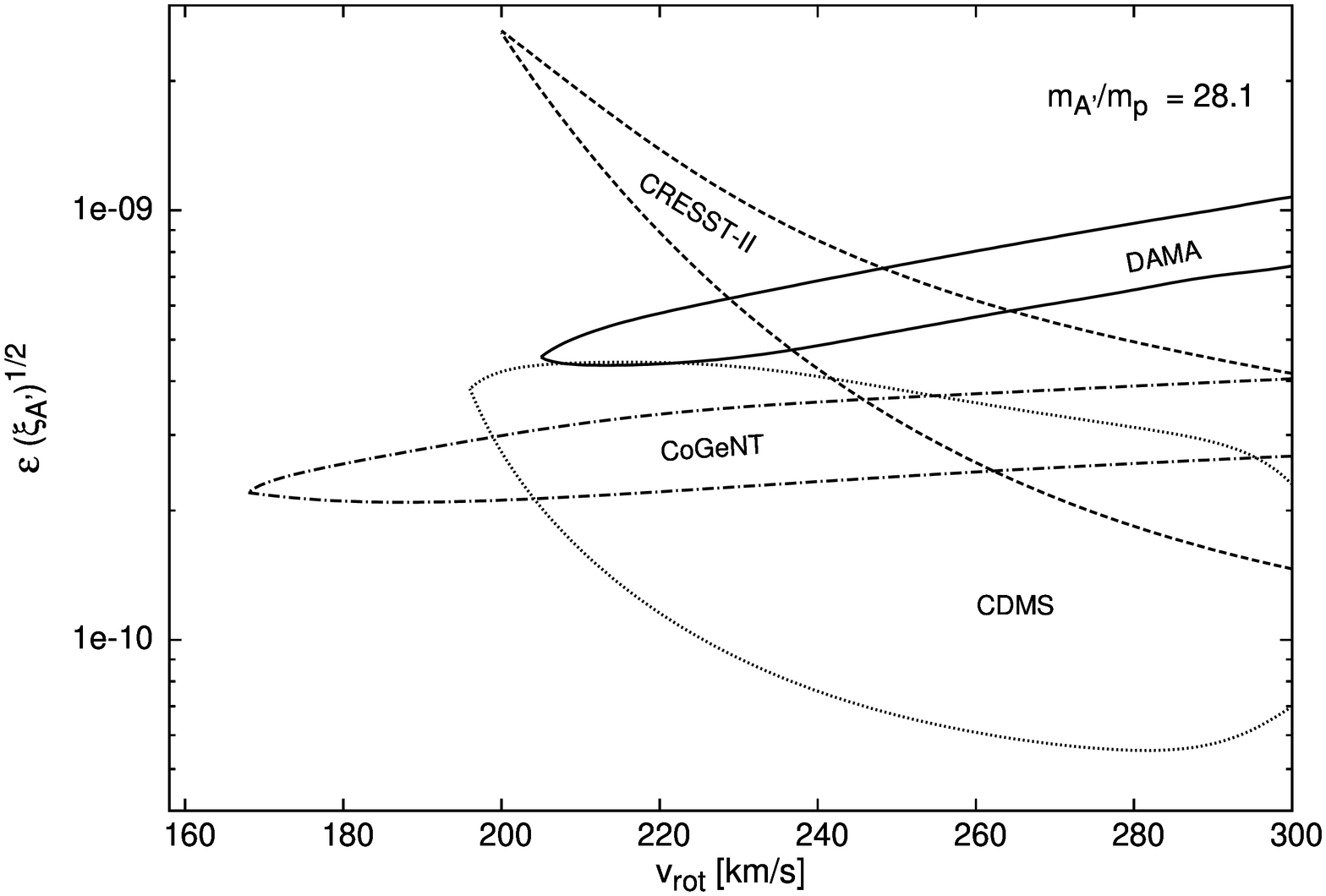,angle=270,width=12.4cm}}
\vskip 0.5cm
\noindent
{\small
Figure 5.5b: Same as figure 5.5a except for $A' = Si'$.
}
\vskip 0.7cm
\centerline{\epsfig{file=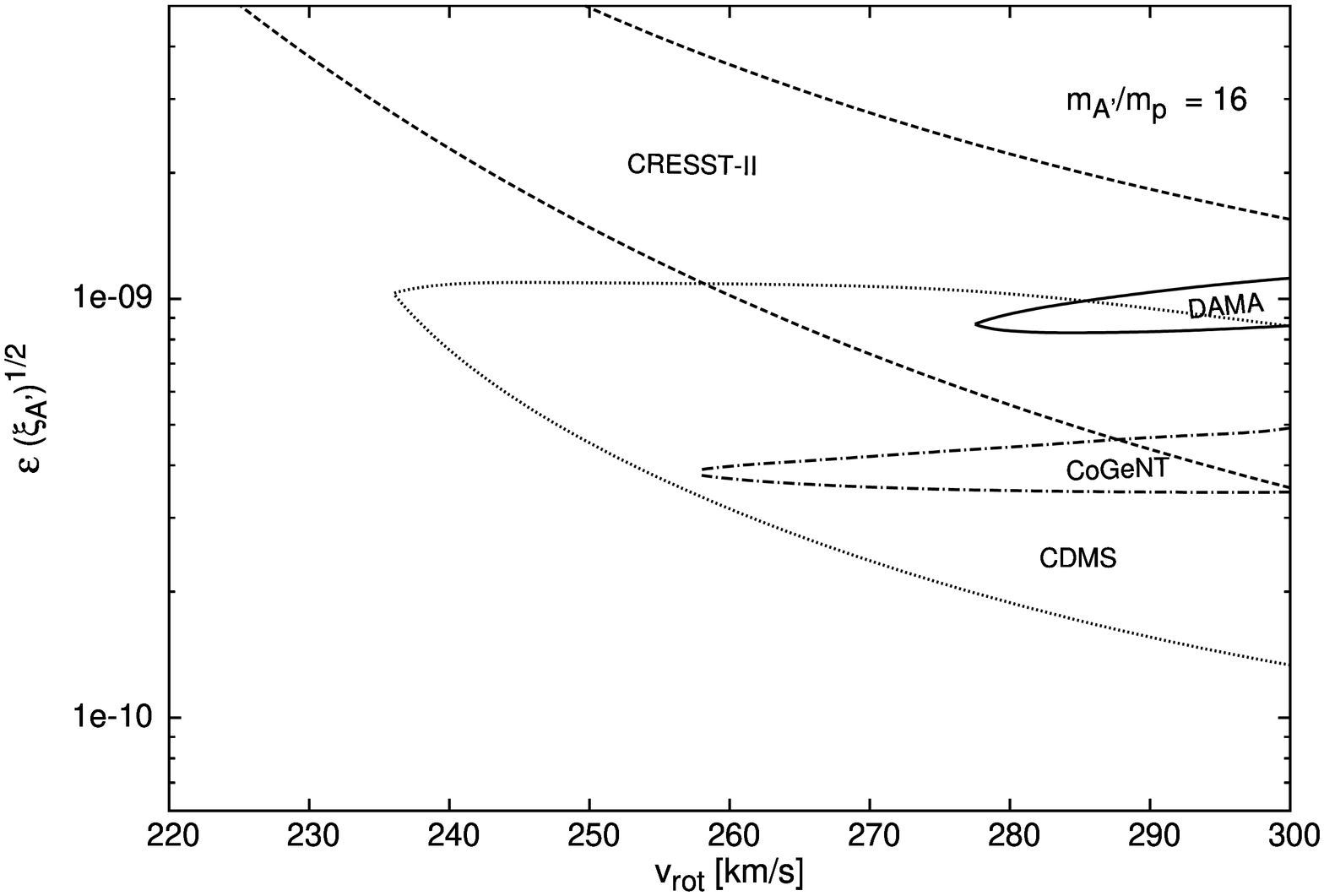,angle=270,width=12.4cm}}
\vskip 0.5cm
\noindent
{\small
Figure 5.5c: Same as figure 5.5a except for $A' = O'$.
}
\vskip 0.8cm
\newpage

The figures demonstrate that there is a substantial region of parameter space where each experiment can be
explained within this mirror dark matter framework. Furthermore, there is significant overlapping parameter space
between the allowed regions of the four experiments.  Examples are:
(a) from figures 5.4a and 5.5a: $A' = Fe'$ ($m_{A'}/m_p = 56$), 
$\epsilon \sqrt{\xi_{Fe'}} \simeq 2.5 \times 10^{-10}$ for $v_{rot} \approx 200$ km/s,
(b) from figure 5.4b:
$A' = Ca'$ ($m_{A'}/m_p = 40$), \ $\epsilon \sqrt{\xi_{Ca'}} \simeq  3\times 10^{-10}$ for $v_{rot} = 240$ km/s 
and figures 5.4c, 5.5c:
$A' = O'$ ($m_{A'}/m_p = 16$), \ $\epsilon \sqrt{\xi_{O'}} \simeq  6\times 10^{-10}$ for $v_{rot} = 280$ km/s.
In  section 5.5 we will examine two of these points in more detail, but before we do this, let us briefly
conclude here with a few more comments.

The DAMA target consists of Iodine and Sodium. For DAMA, 
the annual modulation signal is dominated by $A' - Na$ scattering if $m_{A'} \stackrel{<}{\sim} 40$ GeV, 
while for $m_{A'} \stackrel{>}{\sim} 40$ GeV then
both $A' - Na$ and $A' - I$ scattering conspire to produce the signal. 
Another noteworthy feature is that the signal is due to the scattering of 
$A'$ nuclei off target nuclei, where the phase space of $A'$ typically 
comes from the body of its 
Maxwellian velocity distribution (rather than, say, the tail).

While the DAMA signal is extracted from the annual modulation component, $dR^1/dE_R$, 
the low energy excesses seen in the other experiments are in the time averaged rate: $dR^0/dE_R$ [Eq.(\ref{theoryrate})].
For CoGeNT, the measured spectrum is consistent with the $dR^0/dE_R \propto 1/E_R^2$ behaviour predicted from the
energy dependence of the Rutherford cross-section. This explains why the CoGeNT spectrum is reproduced for 
a large $A'$ mass range. In other words, the shape of the predicted CoGeNT spectrum results from the dynamics 
rather than the kinematics. 
The data from the CRESST-II and CDMS/Si experiments, is also consistent with the predicted falling recoil energy spectrum, 
albeit with much larger experimental uncertainties.
Finally, note that the CRESST-II target consists of three components: Oxygen, Calcium and Tungsten, with
the dominant signal contribution arising from $A' - Ca$ and/or $A'- O$ scattering depending on the
mass of $A'$.  

\vskip 1.0cm

Important checks of these positive signals are expected in the near future. 
New results from the DAMA collaboration are anticipated this year from their upgrade in 
2010 which should result in a lower energy threshold.
Larger germanium experiments including CDEX \cite{cdex}, C-4 \cite{c4} should be able to more sensitively 
scrutinize CoGeNT's tentative
$dR^0/dE_R \propto 1/E_R^2$ spectrum. This should provide a useful means \cite{tough} of differentiating 
mirror dark matter from alternative explanations of the experiments, such as those invoking 
`light' WIMPs of mass $\sim 8$ GeV e.g. \cite{lightdm,lightdm2}.
The CRESST-II collaboration have upgraded their experiment to reduce backgrounds and are now collecting data
in the new configuration.
New results from CDMS/Ge and many other experiments are also awaited.

%xxxxxxxxxxxxxxxx

\newpage

\subsection{Some benchmark points}

\vskip 0.2cm

Figures 5.4 and 5.5 indicate that there is a substantial region of parameter space where
all three experiments can be explained within the mirror dark matter framework.
It is perhaps instructive to consider in detail a couple of example points.
The first one is 
near the combined best fit of the DAMA, CoGeNT, CRESST-II
and CDMS data for $v_{rot} = 200$ km/s:
\begin{eqnarray}
A' &=& {\rm Fe}' \ (m_{Fe'} = 55.8m_p),\ v_{rot} = 200\ {\rm
km/s}, \ 
\epsilon \sqrt{\xi_{Fe'}} = 2.5\times 10^{-10}\   \  \ {\rm [P1]} \ .
\label{p1}
\end{eqnarray}
As our second example point, we take
\begin{eqnarray}
A' &=& {\rm O}' \ \ (m_{O'} = 16m_p),\ \ v_{rot} = 280\ {\rm
km/s}, \ \
\epsilon \sqrt{\xi_{O'}} = 6.0 \times 10^{-10}\   \ \ \  {\rm [P2]}\ .
\label{p2}
\end{eqnarray}
The fit of the DAMA, CoGeNT and CRESST-II data for these example points are shown in figures 5.6a,b,c. 
These figures demonstrate that the data from all these experiments can
be reasonably well described by mirror dark matter with the above parameters.
(Note though that for the P2 parameter point, only a few dark matter events are 
expected for the CRESST-II exposure, so an alternative explanation may be needed
for that experiment, e.g. \cite{kuz}.) 
\vskip 0.7cm
\centerline{\epsfig{file=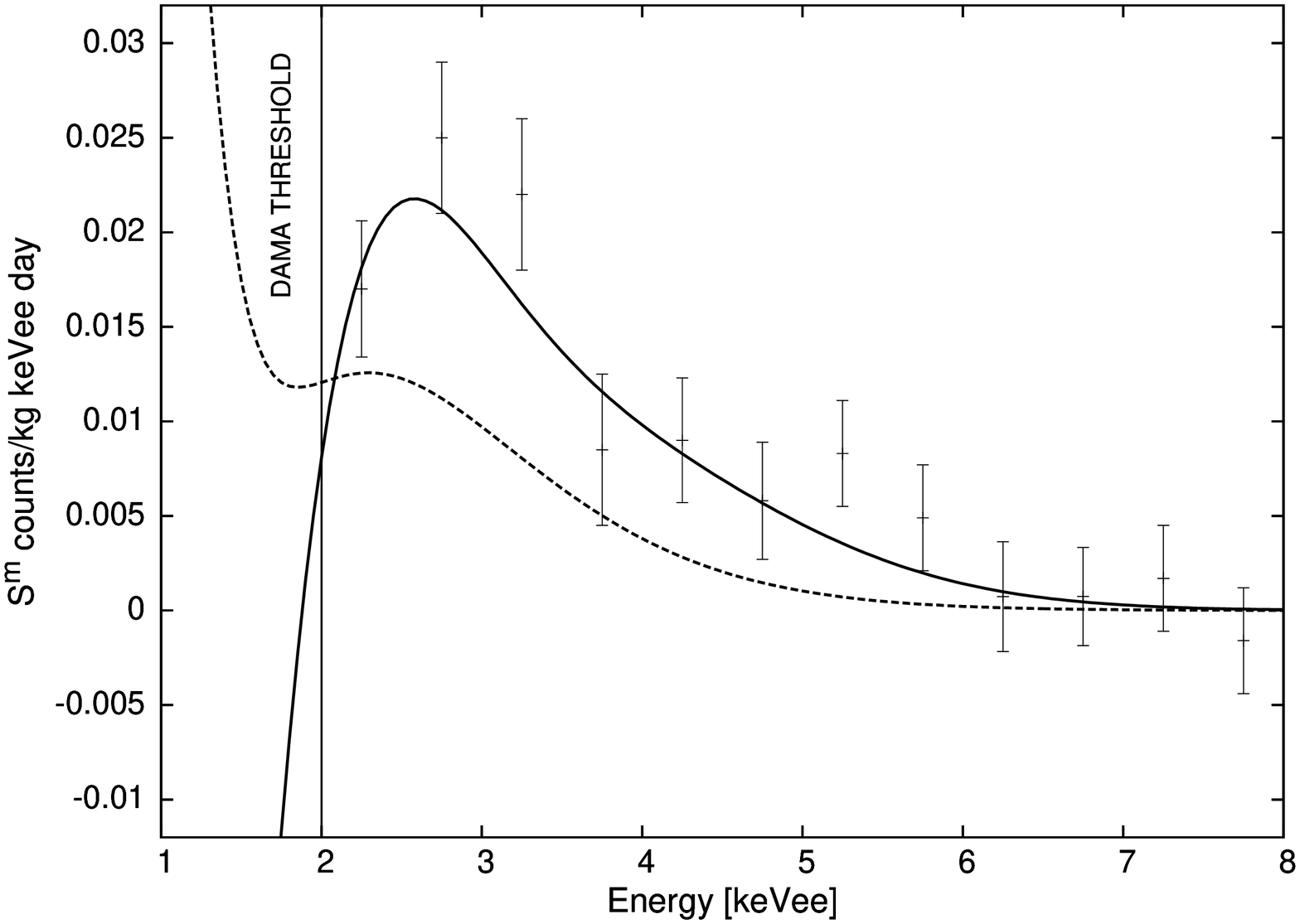,angle=270,width=12.3cm}}
\vskip 0.5cm
\noindent
{\small
Figure 5.6a: DAMA annual modulation spectrum for mirror dark matter
with P1 parameter choice
[Eq.(\ref{p1})]
(solid line) and P2 parameter choice [Eq.(\ref{p2})] (dashed line).
In these examples $q_{Na} = 0.36, \ q_I = 0.20$.
}
\vskip 0.5cm
\centerline{\epsfig{file=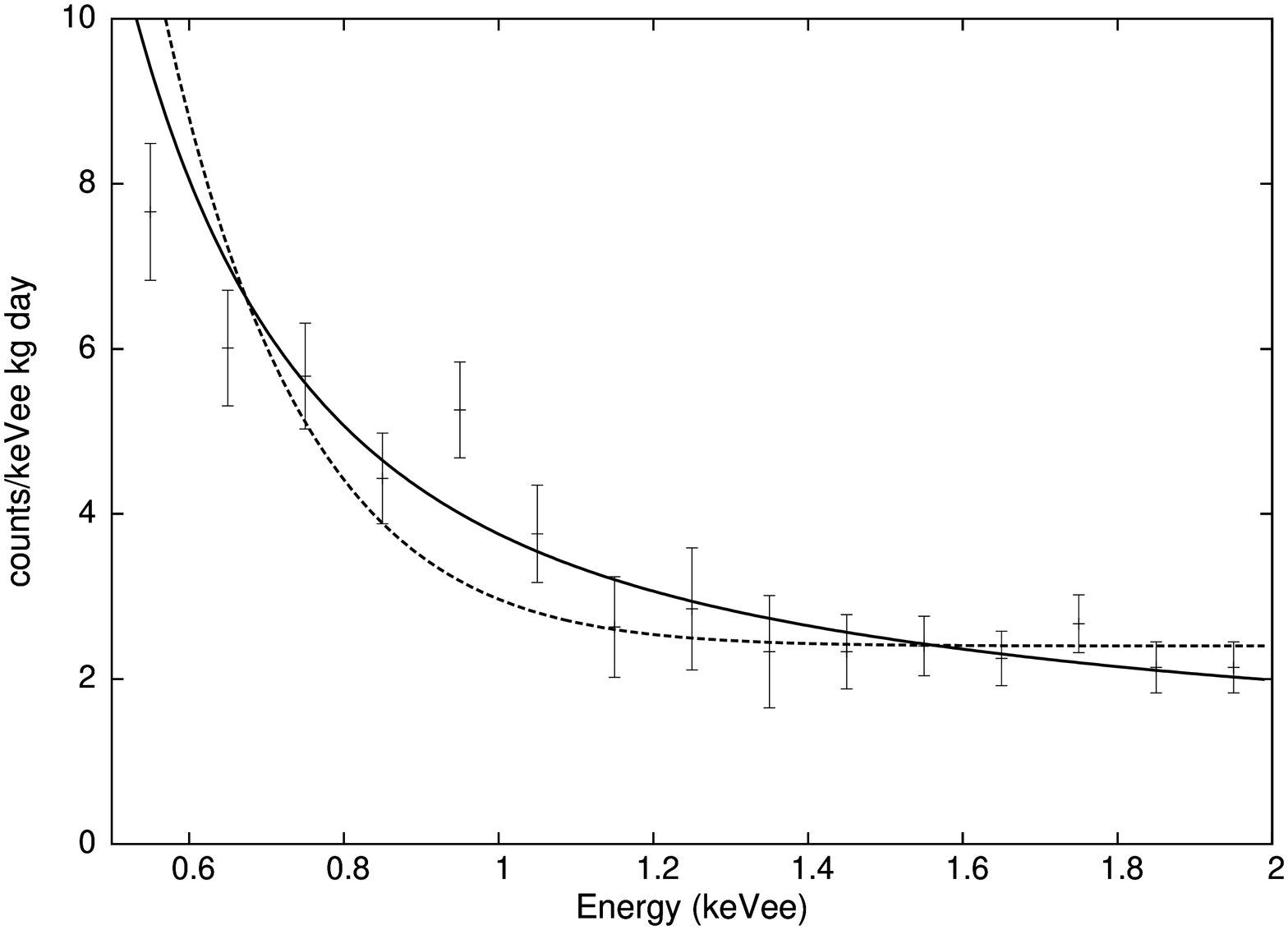,angle=270,width=12.4cm}}
\vskip 0.5cm
\noindent
{\small 
Figure 5.6b: CoGeNT spectrum for mirror dark matter with
the same parameters as figure 5.6a.
In these examples $q_{Ge} = 0.17$.}
\vskip 0.5cm
\centerline{\epsfig{file=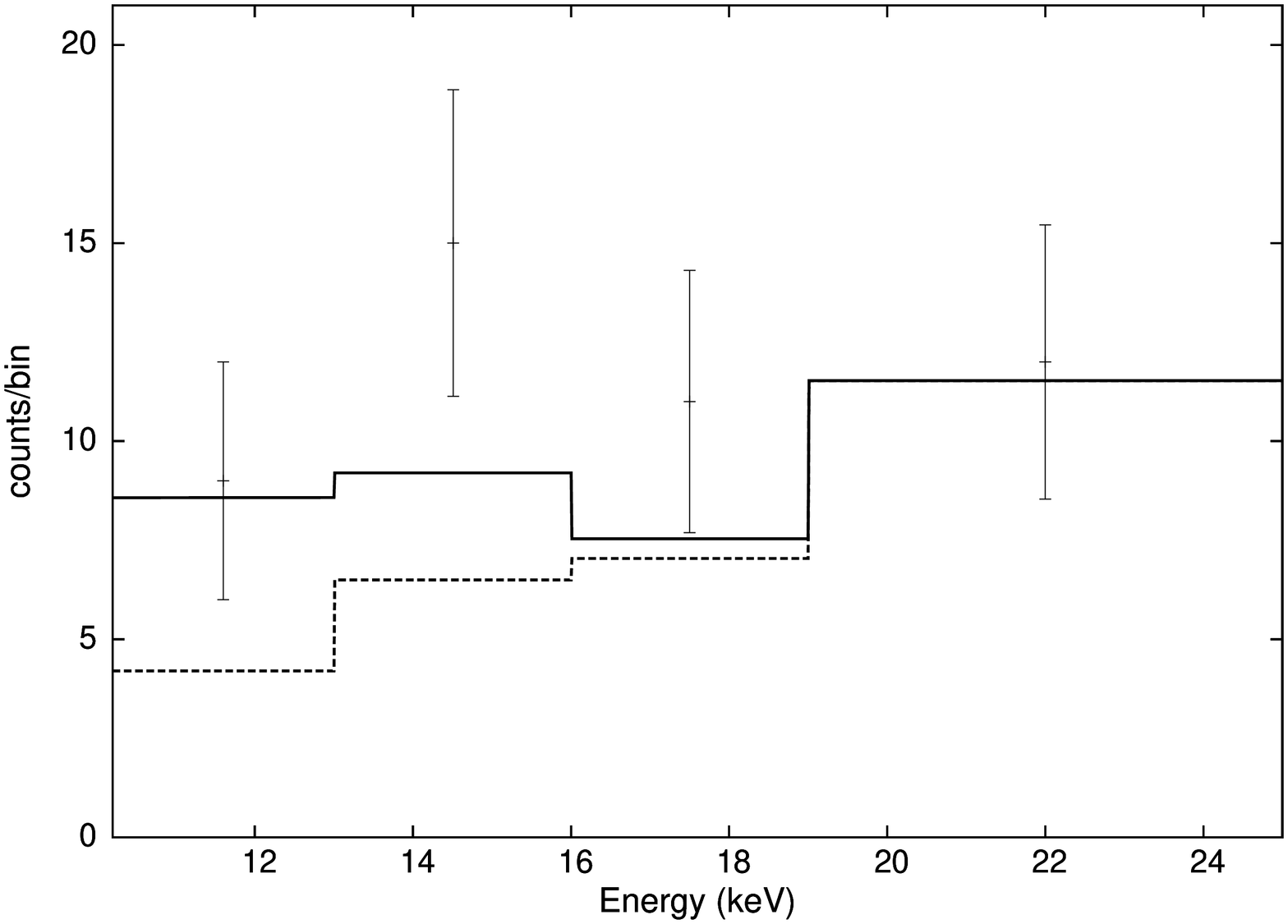,angle=270,width=12.4cm}}
\vskip 0.5cm
\noindent
{\small 
Figure 5.6c:
CRESST-II spectrum for mirror dark matter with
the same parameters as figure 5.6a.
}
\newpage

Interestingly, the two examples show quite different behaviour for the annual modulation
in DAMA at energies below the current threshold.
Even for the point P1, the change in sign
may not occur depending on the halo abundance 
of the lighter components (e.g. $A' \sim O'$ and/or $A' \sim Si'$) 
since the positive contribution to the annual modulation from the lighter components can outweigh the negative contribution
from $Fe'$.
%% more discussion.xxxxx. slow of cogent ok with this model 

\subsection{XENON100 and LUX Constraints}

Null results have been reported by XENON100 \cite{xenon100}, LUX \cite{lux}, CDMS/Ge \cite{cdmsge} 
 and Edelweiss \cite{edel}.
The energy thresholds of CDMS/Ge and Edelweiss are sufficiently high that there is no
serious tension of these null results with mirror dark matter expectations.
The XENON experiments, XENON100 and LUX with nominal energy thresholds of 6.4 keV and 3.0 keV,
do have some tension with mirror dark matter expectations
\footnote{
There are also lower threshold analysis by the XENON10 \cite{xenon10} and
CDMS collaborations \cite{cdmsgelow}.
It has been argued in \cite{collarguts} that neither analysis can
exclude light dark matter (and by extension, mirror dark matter examined here, which
has similar event rates at low recoil energies)
when systematic uncertainties are considered. Also, a reanalysis of the low energy CDMS/Ge 
data in \cite{cf} 
supports [at 5.7 $\sigma$ C.L.] a family of events in the 
nuclear recoil band.
Although the CDMS/Ge data were not included in the analysis of section 5.4, 
the study \cite{tough} indicates that this data
is compatible
with the overlapping region of parameter space given in figure 5.4.
Thus, the low energy CDMS/Ge data \cite{cdmsgelow} may actually be consistent with
CoGeNT's observed low energy excess rate, and adds weight to the dark matter interpretation 
of this excess.}.

\vskip 0.4cm
\centerline{\epsfig{file=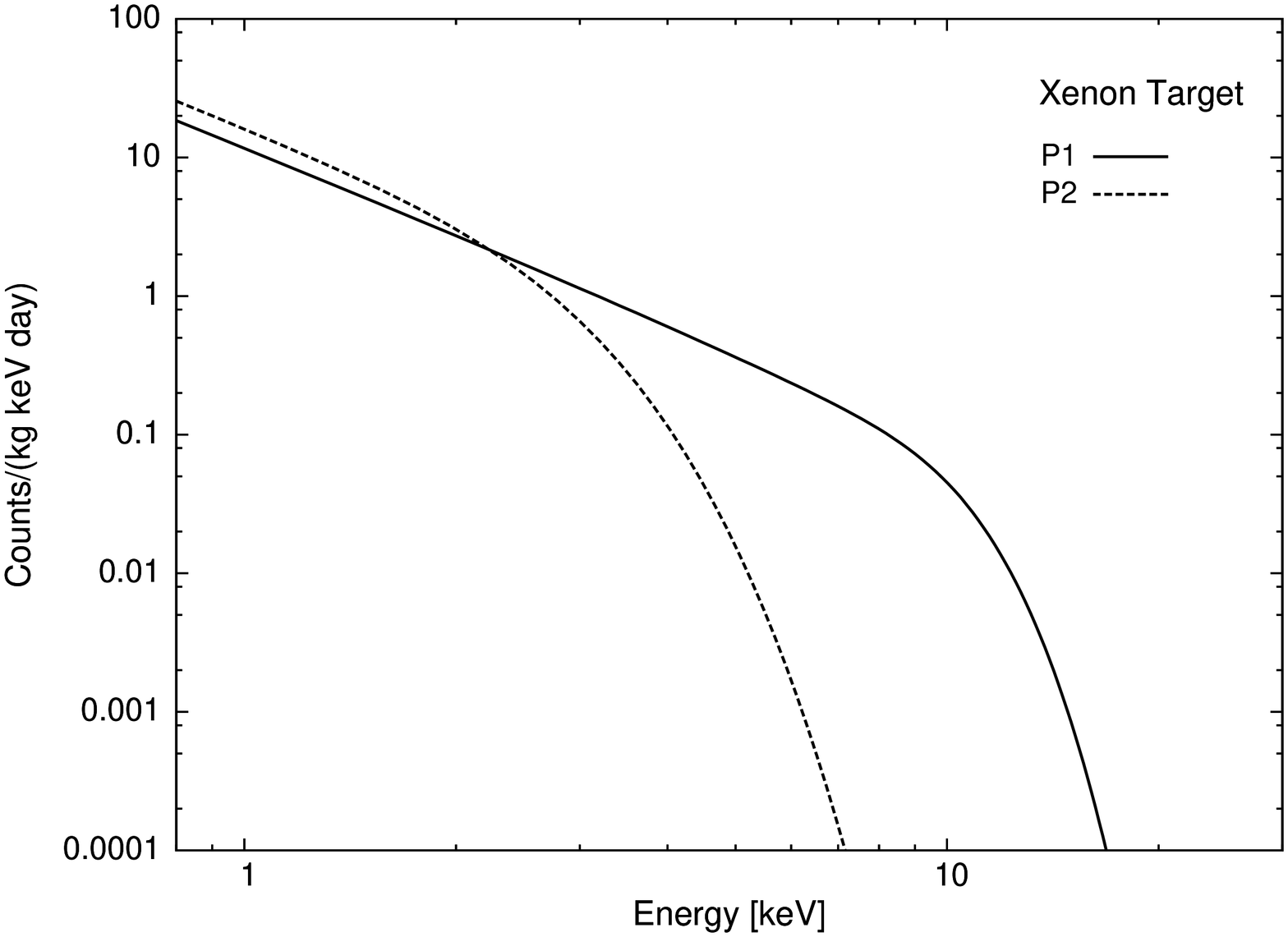,angle=270,width=11.9cm}}
\vskip 0.2cm
\noindent
{\small 
Figure 5.7: Predicted event rate: $dR^0/dE_R$ versus nuclear recoil energy, $E_R$, for a Xenon target for the benchmark points 
P1 [$A'=Fe', \epsilon \sqrt{\xi_{Fe'}} = 2.5\times 10^{-10}, v_{rot} = 200$ km/s] (solid line) 
and
P2 [$A'=O', \epsilon \sqrt{\xi_{O'}} = 6.0 \times 10^{-10}, v_{rot} = 280$ km/s] (dashed line).
}
\vskip 1.4cm

In figure 5.7 we plot the predicted event rate for a Xenon target for the benchmark points, P1 and P2
[Eq.(\ref{p1}) and Eq.(\ref{p2})] discussed in section 5.5.
As the figure shows, the rate falls like $\sim 1/E_R^2$ until the kinematic threshold is reached, after which
the rate falls very rapidly towards zero.
This behaviour is, of course, well understood.
The kinematic energy threshold is 
given approximately by the analytic expression, Eq.(\ref{aa5}).
For the benchmark point, P1,
that is, $Fe'-Xe$ scattering with $v_{rot} = 200$ km/s, 
we estimate this threshold energy to be $E^{threshold}_{Fe'} \approx 10$ keV,
while for the benchmark point, P2,
that is, $O'-Xe$ scattering with $v_{rot} = 280$ km/s, 
we estimate this threshold energy to be $E^{threshold}_{Fe'} \approx 3$ keV.

Unfortunately, it is not easy to
estimate the expected number of nuclear recoil events for the XENON100 and LUX experiments with any confidence.
The key issue is the magnitude of the energy scale uncertainty, which is unknown, 
and the subject of active discussions \cite{collarzzz,damaguts}.
A factor of $\sim 2$ uncertainty is certainly possible.
In view of this situation, one can pose the question: 
What energy threshold would be required for these XENON
experiments
to be consistent with the benchmark points P1 and P2?
An approximate answer can be gleaned from figure 5.7.
The energy threshold needs to be
higher than the kinematic threshold, i.e. $\sim 10$ keV for P1 or $\sim 3$ keV for P2.
A more precise calculation, taking into account the relevant
detection efficiencies, exposure time and detector resolution, shows that
the energy threshold of the XENON100 and LUX experiments needs to be around 12-15 keV for the point
P1 to be allowed at $95\%$ C.L. and 4.5-6 keV for the point P2 to be allowed at 95\% C.L.
This can be compared with the nominal threshold energy
of 6.4 keV and 3.0 keV for XENON100 and LUX respectively.
Obviously the point P2 has only very mild tension with these experiments, while
the tension is more severe for P1
\footnote{
Note that
the example points $P1$ and $P2$ are consistent with
the CDMS/Ge \cite{cdmsge} data (taking a systematic uncertainty in energy scale of 20\%).
Also, the Iodine rate for the parameter points $P1$ and $P2$ is
consistent with constraints from the KIMS experiment \cite{kims}.
}.

Fortunately, there are plans to check more carefully the low energy calibration of these detectors \cite{xc1,xc2}.
Also, forthcoming results from XMAS, PANDA and other experiments should be able to provide an important check
on the XENON constraint.

%%% upto here %%%%%%%

\vskip 1.0cm

\subsection{Diurnal Modulation}

\vskip 0.4cm

Mirror dark matter has one further interesting property due to its self-interactions.
Mirror dark matter captured by the Earth can effectively block
the halo dark matter `wind'. Since the proportion of the halo wind which
is blocked varies during the day due to the Earth's rotation, the interaction rate 
in a direct detection experiment will modulate. This leads to a diurnal modulation signal
with period of a sidereal day \footnote{
Diurnal modulation can also occur due to elastic scattering of dark matter on
the constituent nuclei of the Earth \cite{collardiurnal}.
In that case, the effect is typically small unless the dark matter abundance is 
much less than the reference value: $\rho_{dm} = 0.3 \ {\rm GeV/cm^3}$.}.  
This effect has been studied in \cite{footdiurnal} and we summarize the main conclusions below.

\newpage

How many halo mirror particles can be captured by the Earth? 
Halo mirror particles, $A'$, will occasionally undergo hard scattering with ordinary nuclei and thereby be captured 
by the Earth. They will eventually thermalize with ordinary matter and accumulate in the Earth's core.
Eventually, enough mirror particles will accumulate there so that all halo mirror particles with trajectories
passing near the core will be captured as a result of self interactions. Estimates indicate that
this will occur when the mirror particle column density reaches a `critical' value of around 
$\sim 10^{16}\ {\rm cm^{-2}}$ \cite{footdiurnal}.
At this point, mirror
dark matter will be captured by self-interactions at a rate:
\begin{eqnarray}
{dN \over dt} \sim \pi R_0^2 f_{A'}
\end{eqnarray}
where $f_{A'} \approx v_{rot}\ \xi_{A'}' \ 0.3\ {\rm GeV/cm}^3 /m_{A'}$ 
is the flux of $A'$ mirror particles arriving at the Earth and
$R_0$ is the maximum distance from the Earth's center for which dark matter
can be captured by self-interactions. 
The distance $R_0$ slowly increases with time as more dark matter accumulates within the Earth,
and has been estimated to be currently
around $4,000$ km for a typical mirror metal $A'$ component \cite{footdiurnal}.
It follows that around
\begin{eqnarray}
N &\sim & \int  \pi R_0^2 f_{A'} dt \nonumber \\
&\sim & 10^{39} \left({\xi_{A'} \over 10^{-1}}\right)
\label{666}
\end{eqnarray}
$A'$ particles would have been captured during the last five billion years. This is many orders of magnitude 
within the geophysical limits \cite{slimit}.

The accumulated mirror dark matter within the Earth can potentially shield a detector from halo dark matter.
Whether or not a particular halo dark matter particle gets shielded depends on how close its trajectory takes it
to the center of the Earth. 
Thus we need to know the direction of the dark matter `wind'.

The direction of the Earth's motion through the halo, subtends an (average) angle $\approx 43^o$ with 
respect to the Earth's spin axis.
There is a small annual modulation 
in this angle due to the Earth's motion around the Sun, which can be important, but will be neglected
here.
%%% xxxxxxxxx need to comment a little on this
Another useful angle is the one between the direction of the Earth's motion through
the halo and the normal vector to the Earth's surface at the detector's location. 
This angle, which we label $\psi$, depends on the latitude of the detector's location and also the time of day, $t$:
\begin{eqnarray}
\cos \psi = \cos \theta_{latitude} \sin \omega t \sin 43^o \pm \sin \theta_{latitude} \cos 43^o
\label{42}
\end{eqnarray}
where $\omega = 2\pi/T_d$ with $T_d = 1$ sidereal day (23.934 hours).
Here the $+$ [$-$] sign is relevant for a northern [southern] hemisphere detector, 
and $\theta_{latitude}$ is the north [south] latitude.
%%% figure ??? %%%%%%%%%%%%%%
\vskip 0.3cm
%\centerline{\epsfig{file=fig5last.eps,angle=270,width=12.0cm}}
\centerline{\epsfig{file=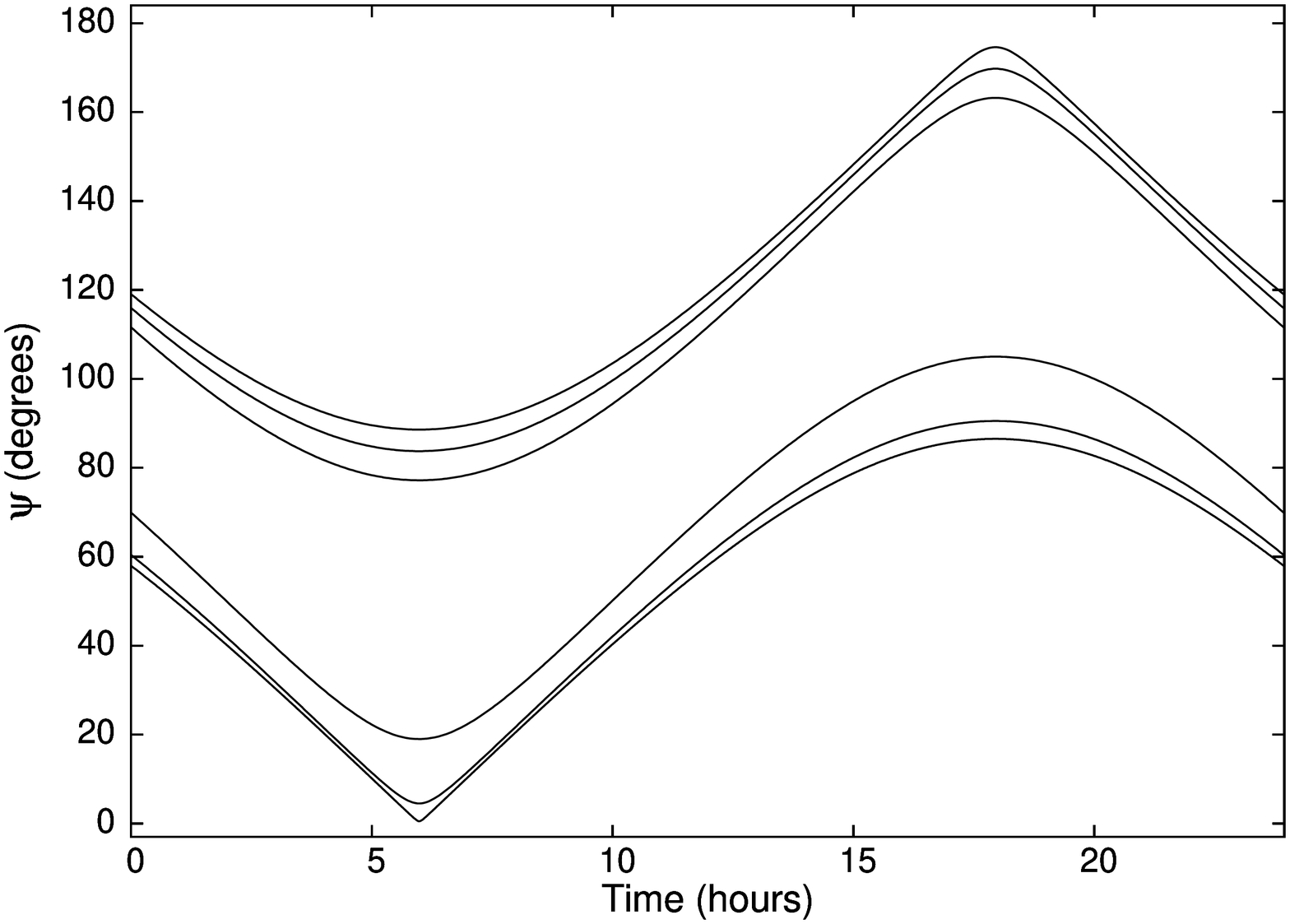,angle=270,width=11.5cm}}
\vskip 0.2cm
\noindent
{\small 
Figure 5.8: The angle $\psi (t)$ [Eq.(\ref{42})]  
between the direction of the Earth's motion through the halo and the normal vector 
to the Earth's surface at the detector location
($0^\circ$ means that the halo wind is coming vertically down, while $180^\circ$ means that the halo wind
is coming straight up through the Earth's core).
The bottom three curves are for (from bottom to top) Sudbury ($\theta_{latitude} = 46.5^\circ \ N$), 
Grand Sasso ($\theta_{latitude} = 42.5^\circ\ N$), Jin Ping
($\theta_{latitude} = 28^\circ\ N$)
while the top three curves are for (from bottom to top) Andes Lab ($\theta_{latitude} = 30.2^\circ \ S$), 
Bendigo ($\theta_{latitude} = 36.7^\circ \ S$), Sierra Grande ($\theta_{latitude} = 41.6^\circ \ S$).
The bimodal distribution means that only detectors located in the southern hemisphere 
are expected to be sensitive to the diurnal modulation effect.
}

\vskip 0.9cm

Since the direction of the halo wind depends on the latitude at the detector's location, it should
come as no surprise that the magnitude
of the diurnal modulation effect also depends critically on this latitude.
In fact the $\psi (t)$ distribution is bimodal depending on whether the detector is located in the northern or southern
hemisphere (see figure 5.8).
For a detector in the northern hemisphere the direction of the dark matter wind is predominately in the downward
direction. Dark matter particles very rarely travel up through the Earth's center. Conversely
for a detector located in the southern hemisphere, dark matter particles mostly travel in the upwards direction,
i.e. they have passed though the bulk of the Earth before reaching the detector. 
It follows therefore that the current experiments, located as they are in the northern hemisphere, 
are relatively insensitive to the diurnal modulation effect. For the DAMA detector located at Gran Sasso,
for example,
the amplitude of the diurnal modulation is estimated to be less than $\sim 1\%$ \cite{footdiurnal}, consistent
with experimental measurements \cite{damadiurnal}.
The situation, of course, changes drastically for a detector located in the southern hemisphere.
There, the diurnal modulation can be maximal. 
This means that if say the CoGeNT or DAMA detector were moved to the southern hemisphere then the diurnal
effect should be observable with only around 30 days of live running (figure 5.9).

\vskip 0.5cm
\centerline{\epsfig{file=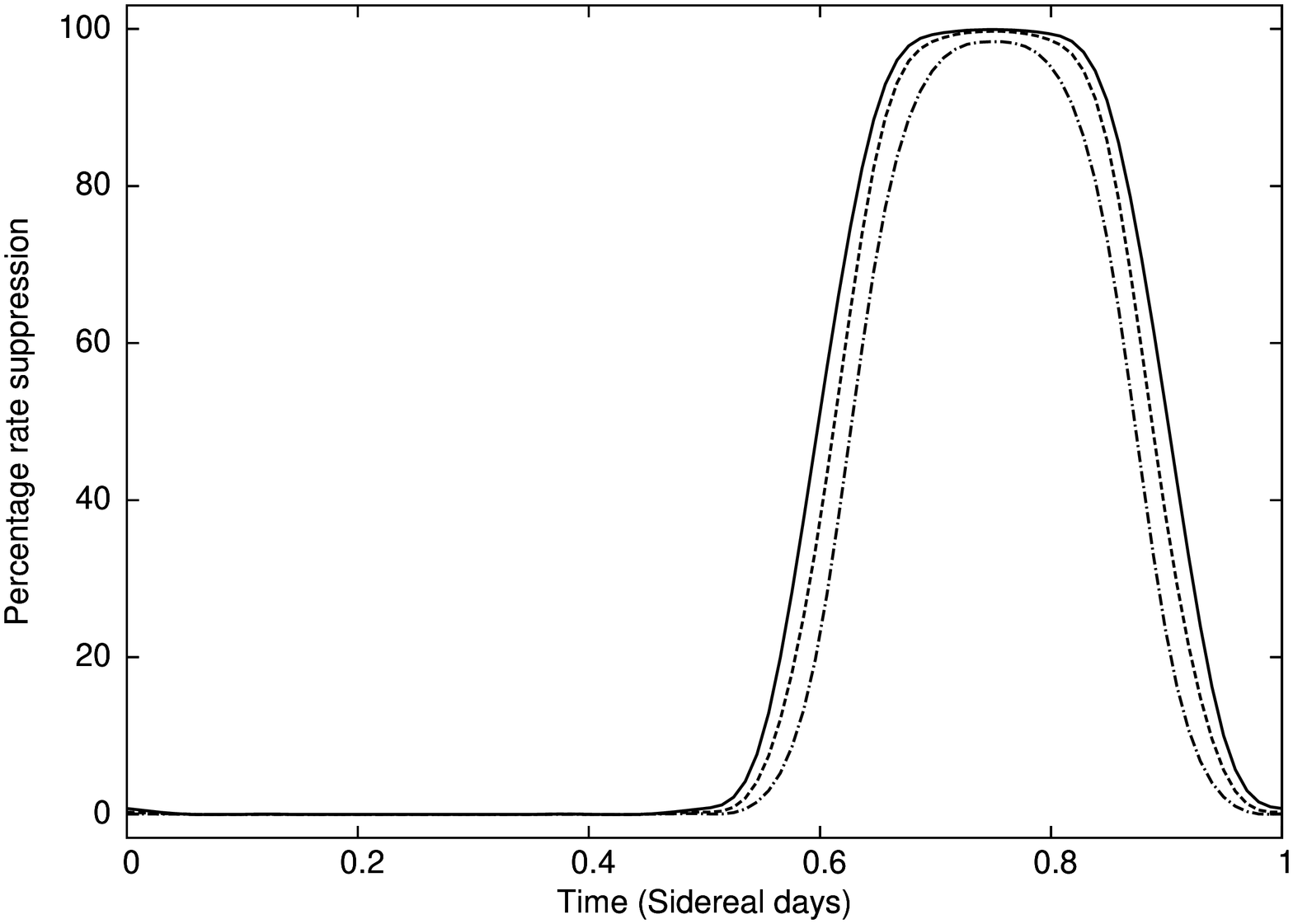,angle=270,width=12.4cm}}
\vskip 0.5cm
\noindent
{\small 
Figure 5.9: Percentage rate suppression due to the shielding of the halo wind by mirror dark matter accumulated in the Earth
for a detector in the southern hemisphere.
The three curves are for (from bottom to top) Andes Lab ($\theta_{latitude} = 30.2^\circ\ S$), 
Bendigo ($\theta_{latitude} = 36.7^\circ\ S$), Sierra Grande ($\theta_{latitude} = 41.6^\circ\ S$).
}
\vskip 1.0cm
%\newpage

\noindent
{\bf DM-Ice}
\vskip 0.4cm
\noindent
The IceCube collaboration plans to check the DAMA annual modulation signal with an experiment 
located in the southern hemisphere at a latitude of $90^\circ \ S$ (the South Pole) \cite{dmice}.
At that location no diurnal modulation is expected.
The shielding of the halo due to captured mirror particles in the Earth
still leads to two important effects. First, the shielding suppresses the dark matter
flux by $\sim 30\%$. Second, the amount of shielding has an annual modulation due to the
change in direction of the dark matter wind caused by the Earth's motion around the Sun.
The phase of the annual modulation caused by this effect is estimated to be $\sim$ April 25 \cite{footdiurnal}.

%%%% paragraph about ice cube xxxxxx

\section{Discussion/Outlook}

Mirror matter, at least as currently understood, appears to be capable of playing the 
role of dark matter in the Universe. Mirror dark matter is exceptionally well motivated
from particle physics. It requires only a single assumption - that nature 
respects space-time parity symmetry. This symmetry is enough to imply the existence
of massive stable cold dark matter particles, without {\it any} additional assumptions.
Furthermore, the properties of the mirror particles: their masses and self interaction
cross-sections are completely fixed. There are no free parameters describing any of 
this physics.
Additionally, the symmetry strongly constrains possible interactions between ordinary matter and
mirror matter. Only one renormalizable interaction, photon - mirror photon kinetic mixing, can be important for astrophysics.

Symmetry principles, even fundamental particle symmetries, can only go so far. 
They cannot tell us the strength of the photon - mirror photon kinetic
mixing, $\epsilon$. They also cannot tell us about the effective initial conditions applicable 
in the early Universe.  Observations and experiments, though, can be relied on to fill this void.
Considering first the initial conditions, cosmological observations such as BBN and CMB
indicate that the radiation energy density is dominated by the contributions from the ordinary particles
and that ordinary baryons comprise only 20\% of the total matter density. That is,
\begin{eqnarray}
T' \ll T \  \  {\rm and} \ \ \Omega_{b'} \approx 5\Omega_b\ .
\end{eqnarray}
Kinetic mixing is also constrained.  Cosmology provides a rough upper limit of $\epsilon \stackrel{<}{\sim} 10^{-9}$.
Importantly there are good reasons to believe $\epsilon$ is nonzero but close
to this upper limit. A value of around $\epsilon \sim 10^{-9}$ is indicated 
if ordinary core-collapse supernovae are responsible for heating mirror dark matter halos of spiral galaxies.
Direct detection experiments also suggest $\epsilon \sim 10^{-9}$. Both applications require spiral galaxy 
halos to have a substantial ($> 1$ \% by mass) mirror metal component. 

This review has focused on three main developments of mirror dark matter during the last decade:
Early universe cosmology, galaxy structure and the application to direct detection experiments.
Early Universe cosmology seems to lead to a consistent picture of the things that can most reliably
be calculated: the CMB anisotropy spectrum and the matter power spectrum in the linear regime.
The formation and early evolution of small-scale structure - the progenitors of galaxies, 
is an unsolved problem
involving nonlinear physics. We have sketched a picture of what might be occurring at early times, but
quantitative details are missing. This is an important topic for future work.

The {\it current} structure of galaxies seems to be a more tractable problem. 
Following an early period of rapid ordinary star formation, the subsequent ordinary supernovae 
have heated and expanded the mirror particles, which were previously
very compact after the initial nonlinear collapse, into a roughly spherical plasma. This plasma,
together with a remnant mirror star subcomponent constitutes the dark matter halo around spiral galaxies today.
These halos, have plausibly evolved to a quasi-stable equilibrium configuration where
the energy dissipated from each volume element is balanced by heating.
The heat source is again supplied by ordinary type II supernovae.
This suggests that kinetic mixing has strength $\epsilon \sim 10^{-9}$, in which case around half of
the core-collapse energy of ordinary supernovae is initially transformed 
into light mirror particles, $e', \bar e', \gamma'$. In the region around each supernova, the energy contained in these mirror particles
is ultimately converted into mirror photons which can then heat the halo
provided that the halo contains a substantial mirror metal component.
This assumed non-trivial dynamics allows the radial profile of the dark matter distribution to be calculated.
The result of such calculations is that the dark matter distribution around spirals  
is approximately quasi-isothermal:
\begin{eqnarray}
\rho (r) \simeq {\rho_0 r_0^2 \over r^2 + r_0^2 }
\ .
\end{eqnarray}
%Here $\rho_0,\ r_0$ is the dark matter central density and core radius.
Calculations also show that the 
core radius, $r_0$, scales with disk scale length, $r_D$, via $r_0 \simeq 1.4r_D$ and  
that the product $\rho_0 r_0$ is roughly $constant$, i.e. independent of galaxy size. 
Such a constrained cored density profile is known to 
provide an excellent description of galactic rotation curves in spirals.

A key test of this framework 
comes from direct detection experiments. If the halos of spiral galaxies are composed (predominately) of a 
mirror-particle plasma, with substantial mirror metal component, then the kinetic mixing interaction will enable 
these components to elastically scatter off ordinary nuclei.
Such dark matter has important  characteristic
features. The halo is multi-component, with possible contributions expected in the range between mirror oxygen 
and mirror iron, i.e. $15 \ {\rm GeV} \le m_{A'} \le 52.5$ GeV.  The velocity dispersion of the metal components is mass 
dependent, and scales like: $v_0 \approx \sqrt{\bar m/m_{A'}} \ v_{rot}$ (where $\bar m \approx 1.1$ GeV is the estimated mean mass
of the particles in the halo and $v_{rot} \approx 220$ km/s is the galactic rotational speed at our location in the Milky Way).
Also noteworthy is that the scattering cross-section is Rutherford-like, leading to $d\sigma/dE_R \propto 1/E_R^2$. 
It turns out that dark matter with these properties can simultaneously explain the positive
direct detection signals reported by DAMA, CoGeNT, CRESST-II and CDMS/Si.
This explanation is not without some tension, mainly with the null results of XENON100 and LUX.

\vskip 0.4cm
\noindent
{\bf Very small scale structure}
\vskip 0.4cm
\noindent
One topic that we have barely touched on is very small scale structure - on stellar mass scales and below.
At the current epoch,
the halo of spiral galaxies is a very hot place. Very little mirror star formation is expected in the halo at the
present time. 
However, some small-scale structure might arise in the disk.
Can mirror dark matter be captured and accumulate 
following interactions with ordinary matter?
Although one can easily check that the rate of mirror dark matter accretion in stars and planets
is currently very low, at an earlier time when the stars were forming,
the capture rate of mirror dark matter was likely much greater.
Let us make a back of the envelope estimate for the amount of mirror dark matter that could
have accumulated in the solar system when it was forming.

The solar system, which naturally is of particular interest,
is believed to have formed from the gravitational collapse of a fragment of
a giant molecular cloud. The collapsing fragment formed a dense core of radius $\sim 0.01$ pc, which ultimately
evolved into the solar system. 
Halo mirror particles can be captured initially due to rare hard scattering processes - possible
due to the kinetic mixing interaction (e.g. $He' \ He \to He' \ He$).
These particles cannot easily evaporate (from further interactions with halo particles) if the core is dense enough.
Captured mirror particles can thermalize with the ordinary matter and migrate towards the central regions of the core.
As their number density increases, so to does the capture rate, as halo mirror particles can be captured due to
their interactions with the local population of captured mirror particles.  There is a critical value for
the mirror-particle column number density, above which all halo mirror particles passing through the collapsing core will
get captured. This critical value has been estimated to be around $\sim 10^{16}\ {\rm cm^{-2}}$ \cite{footdiurnal} 
\footnote{
Reference \cite{footdiurnal} evaluated the critical value for the column density for the 
capture of halo mirror particles striking the Earth, however that physical problem is essentially the same as
for the capture of halo mirror particles hitting the pre-solar nebula.}.
Once mirror dark matter accumulates to the point where it reaches this modest column density, 
%Jof around $10^{16}\ {\rm cm^{-2}}$ 
mirror dark matter can be captured at the rate:
\begin{eqnarray}
{dN_{A'} \over dt} \sim \pi R_0^2 f_{A'}
\label{189}
\end{eqnarray}
where $f_{A'} \approx v_{rot} \xi_{A'} {0.3\ {\rm GeV/cm^3} \over m_{A'}}$ is the flux of $A'$ halo particles,
% hitting the pre-solar nebula, 
and $\pi R_0^2$ is the projected area over which the column density is greater than $10^{16}\ {\rm cm^{-2}}$.
Here $N_{A'}$ is the total number (not number density) of $A'$ particles accumulated.

During the $T \sim 10^{5}$ year formation period of the protosun and nebular disk \cite{mabook}, 
Eq.(\ref{189}) indicates that
the solar system could
accumulate a mass, $M' = \sum_{A'} N_{A'} m_{A'}$, of around:
\begin{eqnarray}
M' \sim 10^{-5}\ m_\odot \ {\pi R_0^2 \over (0.01 \ {\rm pc})^2} \ {T \over 10^5 \ {\rm yr}} \ .
\end{eqnarray} 
Of course this is a very rough estimate, nevertheless it does suggest that 
a substantial amount of mirror matter 
might have accumulated in the vicinity of the solar system during its formation.
If this does indeed happen, then some of this material might have condensed into small mirror matter space-bodies \cite{silplanets} \footnote{
More generally, such small scale mixing of ordinary and mirror matter might possibly have important implications for 
neutron stars \cite{sandin,io}, dark planets \cite{fplanets,fplanets2} etc.}.
It has been argued that there is some evidence for such solar system dark matter objects 
from a variety of observations including: anomalous Earth impact events \cite{foottung,mferos}, 
suppression of small craters on the asteroid, EROS \cite{mferos} etc.  
Perhaps the most fascinating aspect of this remarkable possibility is the potential to extract small
mirror matter fragments from the Earth's surface at an anomalous impact site, as the kinetic mixing
interaction leads to a tiny static force on a small fragment ($<$ 1 cm) which can act against gravity \cite{mfcent}.
See also the earlier reviews, \cite{footreview,footold6}, and the book \cite{bk} for further discussions and 
speculations along these lines.

\vskip 0.4cm
\noindent
{\bf Generic hidden sector models}
\vskip 0.4cm
\noindent
In this review we have focused on a very special hidden sector. Mirror dark matter is 
theoretically singled out because of the enhanced symmetry. The particle content and the parameters describing
the hidden sector are completely fixed in this case.
The only new parameter of importance is the kinetic mixing parameter, $\epsilon$. 
We have shown that such a theory, with suitably chosen initial conditions effective in the very early Universe,
appears to provide an adequate description of dark matter phenomena provided that $\epsilon \sim 10^{-9}$.

Unfortunately, there are no guarantees that nature will share one's aesthetics or one's sense of mathematical beauty.
Moreover, the most successful features of mirror dark matter: early Universe cosmology, galaxy structure and an explanation 
of the DAMA, CoGeNT, CRESST-II and CDMS/Si direct detection experiments, can potentially arise also in 
more generic hidden sector dark matter models. The spectrum of such hidden sector models ranges from models which
closely resemble mirror dark matter to those which have nothing to do with mirror symmetry.
An example 
of a hidden sector theory closely resembling mirror dark matter is where there are 
more than one additional copy of the standard model (as discussed 
in section 2.6). Another example is where mirror symmetry is slightly broken, either spontaneously or softly \cite{chinese}.

More generally, one can consider a fairly generic case where
there is a hidden sector that is comprised of a set of massive fermions: 
$F_1, F_2,...,F_N$ (or indeed bosons, $B_1, B_2 ..., B_N$)
charged under an unbroken $U(1)'$ gauge symmetry. Having the $U(1)'$ unbroken means that there is a massless `dark' photon,
$\gamma_D$, which is of course analogous to the mirror photon, $\gamma'$. Such dark matter can then be dissipative,
with nontrivial halo dynamics, completely analogous to mirror dark matter. 
Naturally, this constrains the mass spectrum of the hidden sector.
Unless some other heating
mechanism can be found, one or more more of the states $F_1,F_2...$
must be light enough so that they can be produced in ordinary supernovae
so that the huge supernovae energy can sustain the halo. Clearly, this again requires the $U(1)_Y - U(1)'$ kinetic mixing interaction to
exist.
%Anyway, the minimal such theory has two fermions, $F_1$, $F_2$, assuming as always (in this review) that the dark matter in the Universe
%results from a particle - antiparticle asymmetry.
As with mirror dark matter, this interaction allows such a theory
to be probed in direct detection experiments.
These experiments then, seem to offer the best hope of uncovering the spectrum of hidden sector particles, 
determining their masses and other properties, and thereby distinguishing mirror dark matter from such a closely
related hidden sector.

%%%%%%%%%%%%%%%%%%%%%%%%%%%%%%%%%%%%%%%%%%%%%%%%%%%%%%%%%%%%
\vskip 1.4cm
\noindent
{\large \bf Acknowledgements}

\vskip 0.2cm
\noindent
This work was supported by the Australian Research Council.

%\newpage

\end{document}